\documentclass[a4paper, twoside, 12pt]{book}

\usepackage{setspace}

\if@twoside
\else 
\fi

\usepackage[english]{babel} 
\usepackage[T1]{fontenc}
\usepackage[ansinew]{inputenc}
\usepackage{multirow}
\usepackage{booktabs}
\usepackage{xfrac}
\usepackage{lmodern} 
\usepackage{url}

\usepackage{graphicx} 
\usepackage[bookmarks=false]{hyperref}
\usepackage{placeins}

\usepackage{amsmath}
\usepackage{amsthm}
\usepackage{amsfonts}

\usepackage{a4wide} 
\usepackage{fancyhdr} 
\usepackage{longtable} 

\usepackage{xcolor}
\definecolor{change}{rgb}{0,0,0}

\usepackage{xifthen}
\usepackage{xkeyval}
\usepackage[final]{changes}

\begin{document}

\newcommand{\kk}{\mathbf{k}}
\newcommand{\rr}{\mathbf{r}}
\newcommand{\qq}{\mathbf{q}}
\newcommand{\abf}{\mathbf{A}}
\newcommand{\BB}{\mathbf{B}}
\newcommand{\bb}{\mathbf{b}}
\newcommand{\CC}{\mathbf{C}}
\newcommand{\HH}{\mathbf{H}}
\newcommand{\GG}{\mathbf{G}}
\newcommand{\MM}{\mathbf{M}}

\newcommand{\low}[1]{\( _{#1} \)}
\newcommand{\etal}{\textit{et al.} }
\newcommand{\mo}{$^{-1}$ }
\newcommand{\mom}{^{-1} }

\newcommand{\tc}{$ T_c $ }
\newcommand{\tcm}{T_c }
\newcommand{\tcmax}{$ T_c^{\textnormal{max}}$ }
\newcommand{\tcmaxm}{ T_c^{\textnormal{max}} }
\newcommand{\tcmf}{$ T_c^{\textnormal{mf}}$ }
\newcommand{\tcmfm}{T_c^{\textnormal{mf}} }
\newcommand{\tconset}{T_c^{\textnormal{onset}} }
\newcommand{\tconsetm}{T_c^{\textnormal{onset}} }
\newcommand{\dd}{\, \mathrm{d}}
\newcommand{\tstar}{$T^*$ }
\newcommand{\cuo}{CuO$_2$ }
\newcommand{\tn}{$T_N$ }
\newcommand{\rln}{$r_{\textnormal{Ln}}$ }
\newcommand{\rlnm}{r_{\textnormal{Ln}} }
\newcommand{\dxy}{$d_{x^2-y^2}$ }
\newcommand{\fmin}{$F_{\textnormal{min}}$ }
\newcommand{\degc}{$^{\circ}$C }

\newcommand{\rttep}{$S(295\textnormal{K})$ }
\newcommand{\rttepm}{S(295\textnormal{K})}
\newcommand{\tep}{$S(T)$ }
\newcommand{\tepm}{S(T)}
\newcommand{\evhsm}{E_{\textnormal{vHs}}}
\newcommand{\evhs}{$E_{\textnormal{vHs}}$ }
\newcommand{\evhsd}{$E_F - E_{\textnormal{vHs}}$ }
\newcommand{\evhsdm}{E_F - E_{\textnormal{vHs}}}

\newcommand{\orderparam}{$\Delta'$ }
\newcommand{\scgap}{$2\Delta_0$ }
\newcommand{\scgapm}{2\Delta_0 }
\newcommand{\hscgap}{$\Delta_0$ }
\newcommand{\hscgapm}{\Delta_0 }
\newcommand{\specgap}{$2\Delta'$ }
\newcommand{\specgapm}{2\Delta'}
\newcommand{\hspecgap}{$\Delta'$ }
\newcommand{\hspecgapm}{\Delta'}
\newcommand{\scgapk}{$2\Delta_0(\mathbf{k})$ }
\newcommand{\scgapkm}{2\Delta_0(\mathbf{k}) }
\newcommand{\hscgapk}{$\Delta_0(\mathbf{k})$ }
\newcommand{\hscgapkm}{\Delta_0(\mathbf{k}) }
\newcommand{\specgapk}{$2\Delta'(\mathbf{k})$ }
\newcommand{\specgapkm}{2\Delta'(\mathbf{k})}
\newcommand{\hspecgapk}{$\Delta'(\mathbf{k})$ }
\newcommand{\hspecgapkm}{\Delta'(\mathbf{k})}
\newcommand{\epg}{$E_{PG}$ }
\newcommand{\epgm}{E_{PG}}
\newcommand{\epgk}{$E_{PG}(\mathbf{k})$ }
\newcommand{\epgkm}{E_{PG}(\mathbf{k}) }

\newcommand{\dos}{$N(E_F)$ }
\newcommand{\dosm}{N(E_F)}
\newcommand{\ek}{$\epsilon(\kk)$ }
\newcommand{\ekm}{\epsilon(\kk) }
\newcommand{\eex}{$E^{ex}[n]$ }
\newcommand{\eexm}{E^{ex}[n] }
\newcommand{\emin}{$E_{\textnormal{min}}$ }
\newcommand{\eminm}{E_{\textnormal{min}} }

\newcommand{\bog}{$B_{1g}$ }
\newcommand{\bogm}{B_{1g} }
\newcommand{\btg}{$B_{2g}$ }
\newcommand{\btgm}{B_{2g} }
\newcommand{\aog}{$A_{1g}$ }
\newcommand{\aogm}{A_{1g}}
\newcommand{\aogbog}{$A_{1g}+B_{1g}$ }
\newcommand{\aogbogm}{A_{1g}+B_{1g} }
\newcommand{\aogbtg}{$A_{1g}+B_{2g}$ }
\newcommand{\aogbtgm}{A_{1g}+B_{2g} }
\newcommand{\ag}{$A_{g}$ }
\newcommand{\agm}{A_{g}}
\newcommand{\cm}{cm$^{-1}$ }
\newcommand{\cmm}{\textnormal{cm}^{-1}}
\newcommand{\degrees}{$^{\circ}$ }
\newcommand{\degreesm}{^{\circ} }
\newcommand{\omax}{\omega_{\textnormal{max}} }
\newcommand{\wmax}{$\omega_{\textnormal{max}}$ }
\newcommand{\wmaxm}{\omega_{\textnormal{max}} }
\newcommand{\musr}{$\mu$SR }
\newcommand{\ns}{$\lambda_0^{-2}$ }
\newcommand{\nsm}{\lambda_0^{-2} }
\newcommand{\bloc}{$\mathbf{B}_{\textnormal{loc}}$ }
\newcommand{\blocm}{\mathbf{B}_{\textnormal{loc}}}
\newcommand{\dbloc}{$D(\mathbf{B}_{\textnormal{loc}})$ }
\newcommand{\bext}{$\mathbf{B}_{\textnormal{ext}}$ }
\newcommand{\ssc}{$\sigma_{\textnormal{SC}}$ }

\newcommand{\dintra}{$d_{[\textnormal{Cu}(2)-\textnormal{Cu}(2)]}$ }
\newcommand{\dintram}{d_{[\textnormal{Cu}(2)-\textnormal{Cu}(2)]}}
\newcommand{\dapical}{$d_{[\textnormal{Cu}(2)-\textnormal{O}(4)]}$ }
\newcommand{\dapicalm}{d_{[\textnormal{Cu}(2)-\textnormal{O}(4)]}}

\newcommand{\fig}{Fig.}
\newcommand{\eq}{Equation}
\newcommand{\refsec}{Sec.}
\newcommand{\tab}{Table}
\newcommand{\refappendix}{App.}

\newcommand{\ybco}{YBa$_2$Cu$_3$O$_{7-\delta}$ }
\newcommand{\ysco}{YSr$_2$Cu$_3$O$_{7-\delta}$ }
\newcommand{\lnbco}{LnBa$_2$Cu$_3$O$_{7-\delta}$ }
\newcommand{\ybasr}{YBaSrCu$_3$O$_{7-\delta}$ }
\newcommand{\cacuo}{CaCuO$_2$ }
\newcommand{\mgcuo}{MgCuO$_2$ }
\newcommand{\srcuo}{SrCuO$_2$ }
\newcommand{\bacuo}{BaCuO$_2$ }
\newcommand{\racuo}{RaCuO$_2$ }

\frontmatter

\begin{titlepage}

\begin{center}



\vspace*{\stretch{1}}
\LARGE{
Ion-size effects on cuprate High Temperature Superconductors
}

\vspace{3cm}

\vspace*{\stretch{1}}
  \large{by\\ Benjamin Patrick Pennington Mallett}\\
\normalsize

  \vfill
          A thesis\\ 
          submitted to the Victoria University of Wellington\\ 
          in fulfilment of the requirements for the degree of\\  
          Doctor of Philosophy \\ 
          in Physics\\
 
 \vfill

  The MacDiarmid Institute \\ 
  Industrial Research Limited \\ 
  and \\ 
  Victoria University of Wellington \\ 
  2013

\end{center}

\end{titlepage}

\begin{center}
\textbf{
	\LARGE{Abstract}
	}
\end{center}

%
%
%

The cuprates are a family of strongly electronically-correlated materials which exhibit high-temperature superconductivity. There has been a vast amount of research into the cuprates since their discovery in 1986, yet despite this research effort, the origins of their electronic phases are not completely understood.  In this thesis we focus on a little known paradox to progress our understanding of the physics of these materials.  

There are two general ways to compress the cuprates, by external pressure or by internal pressure as induced by isovalent-ion substitution.  Paradoxically, they have the opposite effect on the superconducting transition temperature.  This thesis seeks to understand the salient difference between these two pressures.

We study three families of cuprates where the ion size can be systematically altered; Bi$_{2}$(Sr$_{1.6-x}$A$_{x}$)Ln$_{0.4}$CuO$_{6+ \delta}$, ACuO$ _2 $ and LnBa$ _{2-x} $Sr$ _x $Cu$ _3 $O$ _{7-\delta} $ where Ln is a Lanthenide or Y and A=\{Mg,Ca,Sr,Ba\}. 
We utilise a variety of techniques to explore different aspects of our paradox, for example; Raman spectroscopy to measure the antiferromagnetic superexchange energy and energy gaps, Density Functional Theory to calculate the density of states, Muon Spin Relaxation to measure the superfluid density as well as a variety of more conventional techniques to synthesize and characterise our samples.

Our Raman studies show that an energy scale for spin fluctuations cannot resolve the different effect of the two pressures. Similarly the density of states close to the Fermi-energy, while an important property, does not clearly resolve the paradox. From our superfluid density measurements we have shown that the disorder resulting from isovalent-ion substitution is secondary in importance for the superconducting transition temperature.  

Instead, we find that the polarisability is a key property of the cuprates with regard to superconductivity.  This understanding resolves the paradox! It implies that electron pairing in the cuprates results from either (i) a short-range interaction where the polarisability screens repulsive longer-range interactions and/or (ii) the relatively unexplored idea of the exchange of quantized, coherent polarisation waves in an analogous fashion to phonons in the conventional theory of superconductivity. More generally, we have also demonstrated the utility of studying ion-size effects to further our collective understanding of the cuprates.

\chapter*{Acknowledgements}
Throughout this thesis `we' is the pronoun of choice which reflects my deep gratitude to all those who helped with experiments and shared ideas.  
First and foremost I thank my supervisors, Dr. Jeffery Tallon, Prof. Alan Kaiser and Dr. Grant Williams for their vital role in my PhD and career.

In addition Dr. Evgeny Talantsev selflessly guided me through thin-film synthesis. Without Dr. Nicola Gaston I could not have hoped to undertake DFT studies and without Prof. Christian Bernhard the \musr studies could not have happened. Mr. Thiery Schnyder is acknowledged for starting the Bi2201 work. My thanks to Prof. Thomas Wolf for single crystal Ln123 samples and Dr. Edi Gilioli for polycrystalline YSCO samples prepared under high-pressure and high-temperature. 

I could not have completed this work without the support of the MacDiarmid Institute and Industrial Research Limited (`Callaghan Innovation' from Feb. 2013).

Finally, I am extremely grateful for indispensable support and advice from Dr. Shen Chong and Dr. James Storey throughout my PhD.

\tableofcontents 

\clearpage
\addcontentsline{toc}{chapter}{List of Figures}
\listoffigures

\clearpage
\addcontentsline{toc}{chapter}{List of Tables}
\listoftables

\mainmatter
\onehalfspacing       

\chapter{Introduction}
\section{Context}
Superconductivity is a fascinating phenomenon.  Below the superconducting transition temperature, $\tcm$, electrons suddenly overcome their mutual repulsion, pair together with opposite momenta and spin and with phase coherence between pairs, adopt a single quantum mechanical wave function of macroscopic dimensions!  Superconductors transport electric charge with zero resistance when cooled below \tc and as such have a broad range of important applications, from compact and powerful magnets to fault-current limiters.

Until recently, superconductors have needed to be cooled to liquid Helium temperatures ($T=4.2$~K) to operate. Helium is very much a finite resource and \textit{liquid} Helium is very expensive.  Furthermore, designing the liquid Helium cryogenic system is very complicated. This point was highlighted when a Helium leak in 2008 between two of the Large Hadron Collider's 1232 magnets \replaced{contributed to the entire facility having to be shut down}{shut down the entire facility for over a year}. %

However in 1986, Berdnoz and M\"{u}ller discovered the first \cite{bednorz1986} of a whole family of cuprate High Temperature Superconductors (HTS) in Switzerland.  These materials have $\tcm$'s as high as 160~K (at elevated pressure) and can be operated without the need for liquid Helium\footnote{They also have other advantages over the conventional `Low Temperature Superconductors'}.  Making commercial products from HTS has proved very challenging due to their complexity, novel physics and the difficulty in making the material into wire.  Nevertheless, there are now some commercially viable companies selling a broad range of HTS products - for example the New Zealand company \textit{HTS-110 Ltd} (see www.hts-110.com). 

One of the main challenges for commercial HTS currently is to reduce their cost.  This can been done either by more efficient production methods, or by improving the performance of the superconductor itself.  Here, we are interested in the superconductor itself.

The cuprates are a family of strongly electronically-correlated materials which exhibit high-temperature superconductivity.  The cuprates are built up from metal-oxide `layers', \refsec~\ref{sec:cupratestructure}.  Common to all cuprates is at least one corner-shared, square-planar \cuo layer as this is where superconductivity originates below $\tcm$.  The electronic properties of the cuprates are strongly modified by doping charge carriers into the \cuo layer and they have a phase diagram rich in competing and co-existing electronic phases, \fig~\ref{fig:phasediagram}.

A vast amount of research into HTS has been done since their discovery in 1986.  One result is approximately $115000$ papers published\footnote{The number of papers searched by the superconductivity papers database, http://riodb.ibase.aist.go.jp/sprcnd\_etl/DB013\_eng\_top\_n.html.}, another is an acknowledgment of the complexity of these materials and the associated physics, e.g. see \cite{wolf2004}.  Despite the research effort spent on HTS, the origins of their electronic phases are not completely understood, which is perhaps shown most starkly by our inability to theoretically predict a new superconductor \cite{hirschbcs} or to find `the next higher $\tcm$' - the record has remained at $\tcm=164$~K since 1994 \cite{gao1994}.

A wide range of experimental probes has been used to study these materials\footnote{The references cited are recent examples from the literature of each probe being used to study the cuprates or are review articles.}; electronic transport \cite{williams1996}, magnetic structure from neutron studies \cite{letacon2011}, optical studies \cite{sugai2003, bernhard2008}, tricrystal experiments sensitive to the phase of the superconducting wavefunction \cite{kirtley1996}, NMR \cite{williams1998}, high-pressure studies \cite{schillingchapter, lorenzchu}, scanning tunneling microscopy \cite{yeh2010}, polar-Kerr effect \cite{he2011}\ldots Some techniques, such as Angular Resolved Photo-emission Spectroscopy (ARPES) and its derivatives such as Laser ARPES \cite{koralekthesis} or time-resolved ARPES \cite{koralek2011}, have been refined especially to study the cuprates. Key to progressing our understanding of HTS is the correct interpretation of these high quality and wide ranging data. 

There are some important outstanding questions, such as;
\begin{enumerate}
	\item What mechanism(s) cause superconductivity at such high temperatures (\refsec~\ref{sec:pairingmechanisms})?
	\item Perhaps the more interesting question is; what theoretical framework can be used to describe the region between the Mott-Hubbard physics of the un-doped cuprate \added{($p=0$, where $p$ is the number of doped holes per Cu in the CuO$ _{2} $ layer)} and the Fermi-liquid-like overdoped region ($p>0.25$). Or, put another way, how is the superconducting ground state chosen over an anti-ferromagnetic, spin density wave, stripe, Fermi-liquid or pseudogap ground state?  These other ground states are thought of as `competing orders', in that they compete with the superconducting state in regions of the Fermi-surface for the lowest energy state. In strongly correlated materials one would like to be able to predict the electronic ground state\footnote{See \cite{faulkner2010} and \cite{wang2009} and related literature for attempts on this problem.}.
\end{enumerate} 

  Four years ago a new class of HTS, the pnictides, which contain Fe, were discovered \cite{kamihara2008}. The pnictides are only the most recent addition to a long list of strongly-correlated, superconducting materials for which we do not have a satisfactory theoretical understanding \cite{norman2011}. Evidently, the problem of HTS is not limited to several, unusual, materials but is a general unexplained phenomenon in condensed matter physics.

	\section{Synopsis}
		
		\added{The superconducting transistion temperature, $ \tcm $, in the cuprates can be modified by the substitution of isovalent ions of differing ion-size.  This work seeks to understand how the ion-size relates to $ \tcm $. }
		
		\subsection*{\S2: Theory}
		The purpose of this chapter is to introduce and explain the results and concepts necessary to understand the research presented in this thesis. We do not attempt to summarize the vast body of literature relating to high-temperature superconductivity in the cuprates and instead cite some especially good reviews of the topic. 
		
		\subsection*{\S3: Techniques and Samples}
		Chapter 3 explains the techniques and methodologies used throughout this thesis.  Sample synthesis and processing methods are presented in the first sections and then various measurement techniques are presented.  The chapter also presents some basic data on the samples that we have made and studied so as to not encumber the later chapters. 
		
		\subsection*{\S4: Ion-size and Raman spectroscopy studies of Bi2201}
		Using a simple materials variation approach, we study the effect of ion-size and disorder on \tc and on Raman modes in the single-layered cuprate Bi$_2$Sr$_2$CuO$_{6-\delta}$ (Bi2201).  We find that ion substitution can increase \tc despite increasing disorder and conclude that both the ion-size and disorder significantly affect \tc in Bi2201.  We then resolve phonon mode assignments in Bi2201 using simple arguments based on material variation experiments.
		
		We consider the possibility that ion-substitution affects \tc by altering the density of states and explore this possibility in the next chapter.
		
		\subsection*{\S5: Density Functional Theory study of the ion-size effect}
		We perform DFT calculations on undoped ACuO$_2$ for A=\{Mg, Ca, Sr, Ba\} to investigate the effect of ion-size on the electronic properties in this model cuprate system.  Where these materials have been synthesised we find good agreement between our calculated structural parameters and the experimental ones.  There is a peak in the density of states $\sim1$~eV below the Fermi-level and we find that larger ions move the peak closer to the Fermi-level.  This finding is consistent with an interpretation of the ion-size effecting \tc via the density of states.
		
		In addition to the density of states, another purportedly important energy scale is the antiferromagnetic superexchange energy and in Chapter 6 we measure the ion-size effect and external pressure effect on this energy scale.

		\subsection*{\S6: Two-magnon scattering}
		In this chapter we measure the antiferromagnetic superexchange energy, $J$, by two-magnon Raman scattering in undoped cuprates while systematically altering the internal pressure by changing ion-size.  We then compare the internal pressure dependence of $J$ with data in literature for the external pressure dependence of $J$. From these data we can show that $J$ is likely unrelated to $\tcmaxm$: $J$ and \tcmax anti-correlate with internal pressure as the implicit variable and correlate with external pressure as the implicit variable. Thus it is most probable some other physical property is dominant in setting the value of $\tcmaxm$. 
		
		The results from this chapter have various interpretations and consequences which we explore in the following chapters.  For example, many consider that ion-substitution disorder significantly weakens superconductivity.  We explore this suggestion by measuring the superfluid density by Muon Spin Rotation (\musr) in an ion-substituted sample.
		
		\subsection*{\S7: Muon Spin Rotation and the superfluid density}
		Here we present \musr measurements of the superfluid density in the \ybasr cuprate. We find the nice result that \ybasr is a boring, well-behaved material!  The superfluid densities are consistent with those previously reported for pure and Ca-doped \ybco cuprate. Furthermore the suppression of the superfluid density due to Zn doping is fully consistent with that found for \ybco.  The immediate consequence of these experiments is that disorder does not play a significant role in \ybasr as compared with \ybco.  Consequently, the lower \tcmax seen in this compound, and those with lower Sr content, is a genuine `ion-size effect' (or `internal pressure effect').  This is an important conclusion to have demonstrated and it is nice to know the effects studied in one's thesis are genuine!
		
		We also report measurements on the fully Sr substituted \ysco material.  The interpretation of these results is less clear.
		
		\subsection*{\S8: Gap estimates from Raman Spectroscopy}
		We seek to measure energy gaps in the single-electron levels of our ion-substituted cuprate materials. To the best of our knowledge these measurements have not been done as a function of internal pressure before. At the level of accuracy available, we did not find any systematic internal-pressure dependence of the energy gap.  We did find however that the gap values for our ion-substituted samples are consistent with the well-characterised \ybco cuprate.  
		
		
		\subsection*{\S9: Discussion}
		In the final chapter we discuss some broader interpretations of our results and conclude that dielectric properties play a key role in superconductivity in the cuprates. Also, we propose studies to further elucidate ion-size effects on a variety of superconducting and normal-state properties of the cuprates. 

\chapter{Theory}

\subsubsection{Notes}
The purpose of this chapter is not necessarily to summarize the vast body of literature relating to HTS in the cuprates, but rather to introduce and explain the results and concepts we feel necessary to understand the research presented in this thesis.  
%

An excellent introduction to superconductivity and High Temperature Superconductivity is the Chapter by Hott \etal in reference \cite{wolf2004} and the proceeding chapters in that reference are generally excellent reviews of their respective topics.  A useful handbook to HTS, but especially to the materials themselves is  the handbook by Poole \etal \cite{poolehandbook}.  An early, yet detailed overview of the structure of these materials is the book chapter by Hazen \cite{hazen1990}.

\section{Crystallographic structure}
\label{sec:cupratestructure}
\subsection{Generally}

The cuprate family of materials are layered perovskite-oxides with at least one corner-shared, square-planar \cuo layer per unit cell.  For a compilation of the members this large and diverse family see \cite{poolehandbook}.  

\begin{figure}
	\centering
		\includegraphics[width=0.50\textwidth]{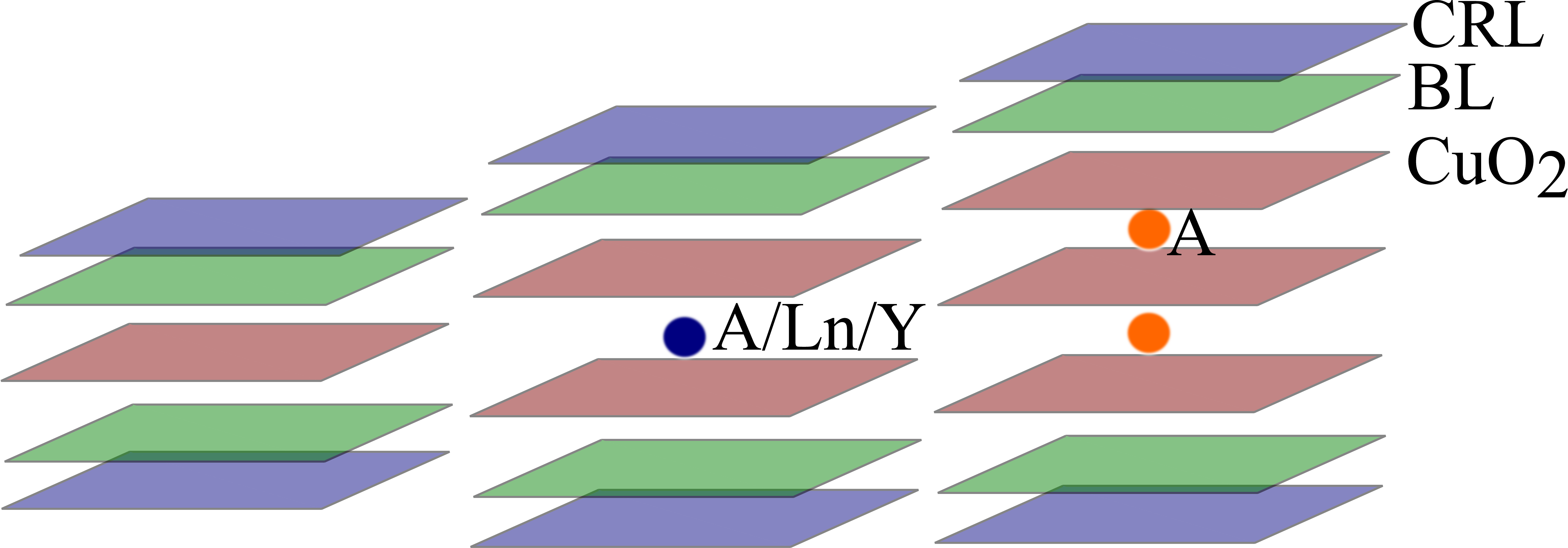}
	\caption[Generic cuprate structure]{A cartoon of the generic cuprate structure.  Single, bi- and tri- CuO$_2$ layer compounds are shown.  `A' represents a group II ion, Ln represents a member of the Lanthanide series.  `BL' is the metal-oxide blocking layer and `CRL' is the metal-oxide charge-reservoir layer.}
	\label{fig:genericcupratestructure}
\end{figure}

\fig~\ref{fig:genericcupratestructure} illustrates the basic components of these materials.  The \cuo layers are structurally and electronically quasi-2dimensional.  The electronic bands associated with this layer reside closest to the Fermi energy, $E_F$, hence superconductivity occurs within these layers\footnote{An obligatory caveat - for any statement about the cuprates there is usually a counter example - is proximity induced superconductivity that occurs on the Cu-O chains in LnBa$_2$Cu$_3$O$_7$, see \refsec~\ref{sec:ybco}.}.  The \cuo layer is sometimes carelessly called the \cuo `plane.'  There is in fact a gradation of planarity, or flatness, amongst cuprate materials with some authors believing this to be crucially correlated with \tcmax \cite{chmaissem1999}.  

A cuprate can be distinguished by the number of \cuo perovskite oxide layers (``n'') per structural repeat unit and if n$>1$ then between these layers sits a group II or group III ion, e.g. Ca$^{2+}$ or Y$^{3+}$, as indicated in \fig~\ref{fig:genericcupratestructure}.  Either side of the outer \cuo layer is a metal-oxide layer, labelled `BL' (it is sometimes called the `blocking layer') in the figure.  The oxygen in this layer, called the `\emph{apical oxygen}' or O(1), plays an important role in the charge transfer between the charge-reservoir layer and \cuo layer it sits between.  The apical oxygen further lifts the degeneracy of the Cu-$d$ orbitals, \refsec~\ref{sec:estructurecuprates}, and may play an important role in the materials dependence of \tc in the cuprates, e.g. \cite{pavarini2001, sakakibara2012}.

`CRL' represents the metal-oxide charge-reservoir layer which may have variable oxygen content and can act as a doping site for the material.  The ability to electronically dope the \cuo layer is crucial to the plethora of interesting electronic properties the cuprates exhibit. Finally, not all of these layer-types are necessarily present in a given cuprate.

A significant research effort was needed for the synthesis of these complicated, defect prone, materials to be reliable and reproducible \cite{erb1997, wolf2004} and it is not uncommon for phase separation or cation disorder to be a major complication in the interpretation of data, see e.g. \cite{boyer2007}. 



\subsection{Ion-size}
`Ion-size' is a concept for the physical space occupied by a nucleus and its ``core'' electrons - those electronic states which are orthogonal to electronic states associated with neighbouring ions.  This is a `region of exclusion' as no orbital overlap is permitted in this region (electrons associated with neighbouring ions are excluded from occupying this region).  An alternate view of ion-size is what I call `cavity size' which views ion-size as the crystallographic volume occupied by the ion.  This fine distinction in conception is probably best left to the theorists and does not affect the results of this thesis.

The concept of ion-size is important for targeted materials engineering. Normally isovalent ions of similar size can be readily substituted in a given material without affecting the material's stability.  A variation in ion size however will induce an effect somewhat analogous to pressure, as discussed below \refsec~\ref{sec:intpressuretheory}. 

Ion-size is dependent on the valence of the ion, its atomic number and its co-ordination.  Empirical values for most ion sizes have been tabulated by Shannon \cite{shannon}.  In \tab~\ref{tab:ionsizes} we tabulate some ion sizes relevant to this work.


\begin{table}
	\centering
		\begin{tabular}{ll|ll} \toprule
			Ln$^{3+}$ or Y$^{3+}$ (VIII) & $r$ [$\times 10 ^{-10}$ m]  &  A$^{2+}$ (IX) & $r$ [$\times 10 ^{-10}$ m]  \\ \midrule
			Lu & 0.977 &  Mg & 0.90 \\
			Yb & 0.985 &  Ca & 1.18 \\
			Ho & 1.015 &  Sr & 1.31 \\
			Y & 1.019 &  Ba & 1.47 \\
			Dy & 1.027 &  Ra & 1.59 \\
			Gd & 1.053 &   & \\
			Eu & 1.066 &   & \\
			Sm & 1.079 &   & \\
			Nd & 1.109 &   & \\
			La & 1.160 &   & \\
			\bottomrule
			
		\end{tabular}
	\caption[Ion sizes]{The `ion-size' of various ions of valence and co-ordination (VIII or IX) relevant for the materials we study.  Data are from Shannon \etal \cite{shannon}.}
	\label{tab:ionsizes}
\end{table}

\subsection{Y123 and siblings}
\label{sec:ybco}

\begin{figure}
	\centering
		\includegraphics[width=0.45\textwidth]{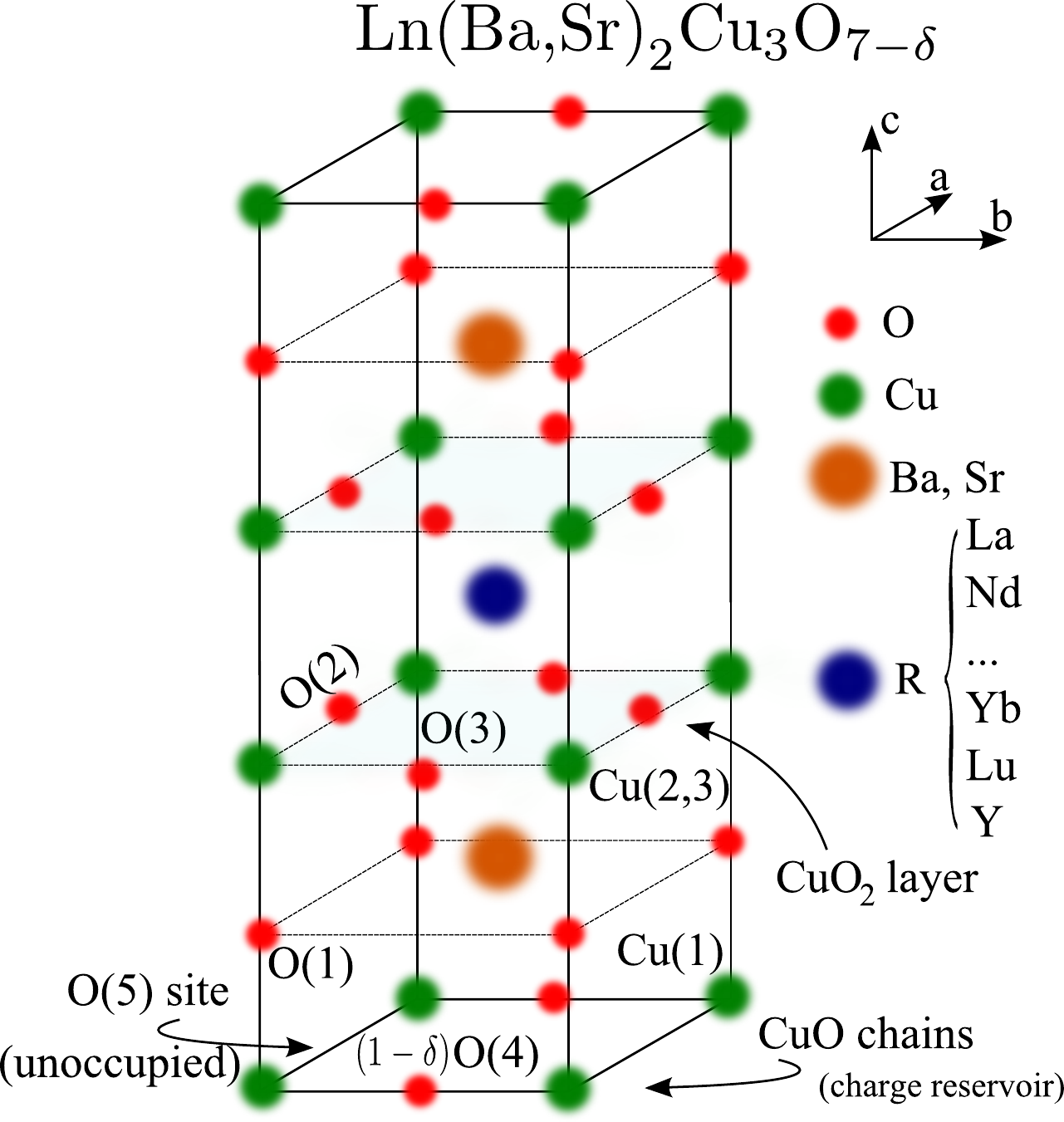}
	\caption[Ln123 crystal structure]{The crystal structure of Ln123. Space group 123.1 (or \replaced{P4/mmm}{P/4mmm}) for small $\delta$ (the exact value of $\delta$ depends on the Ln ion and Sr concentration) and the lower symmetry orthorhombic space group 47.1 (or P/mmm) at larger $\delta$. Generally, oxygen does not occupy the O(5) site.}
	\label{fig:ln123xtal}
\end{figure}

Most of the work in this thesis relates to LnBa$_{2-x}$Sr$_x$Cu$_3$O$_{7-\delta}$ where $0\leq\delta \leq 1$, $0\leq x \leq 2$ and Ln = \{La, Nd, \ldots, Yb, Lu\} is an element of the Lanthanide series of the periodic table of elements\footnote{Barring Promethium and Holmium!}. The chemical formula initially looks a mess but it is nothing more than the well-known high-temperature superconductor YBa$_2$Cu$_3$O$_{7}$ with the possibility of several ion substitutions and variable oxygen content. The crystal structure of this `system' or `family' is shown in \fig~\ref{fig:ln123xtal}.  

\subsubsection{Some terminology} 
\ybco is often written in short hand as YBCO or Y123 (reflecting the 1:2:3 ratio of Y:Ba:Cu).  Throughout this work we use the latter because it can be transparently generalised to Ln123 = LnBa$_2$Cu$_3$O$_{7-\delta}$, or for example Nd123 = NdBa$_2$Cu$_3$O$_{7-\delta}$.  This notation also makes it easy to distinguish Y123 from YBa$_2$Cu$_4$O$_{8}$ (Y124) and Y$_2$Ba$_4$Cu$_7$O$_{15-\delta}$ (Y247), although we have not studied these materials in this work. Both $\delta=0$ and $\delta=1$ are special cases and so are given their own short hand; Y123O6 and Y123O7.  

The case where $x > 0$ has been notated as YBSCO.  I find this dense so generally write the chemical formula in full in these situations.  For $x=2$, YSCO is somewhat common. 

\subsubsection{Structural features}
\fig~\ref{fig:ln123xtal} shows the crystal structure of Ln123.  With reference to the generic cuprate crystal structure shown in \fig~\ref{fig:genericcupratestructure} we identify the role that each layer of ions takes;

	\emph{\cuo layers}:
Composed of Cu(2), and O(2) and O(3) sites with Wycoff positions $2q$, $2r$ and $2s$ respectively for the space group P/mmm.  The valence of these ions depends on the hole carrier concentration per Cu\replaced{ (the doping level $p$)}{, $p$,} in the layer. 

Ln123 has two, equivalent \cuo layers. The energies associated with quasi-particles in this layer reside closest to the $E_F$.  Doping holes, $p$, into this layer leads to rich electronic structure and the magnetic-moments associated with this layer at low $p$ give rise to static and dynamic magnetic order, see \refsec~\ref{sec:phasediagram}.  

	\emph{Ln ions}:
The Y site has Wycoff position $1h$ for the space group P/mmm.  Y$^{3+}$, Ln$^{3+}$=\{La,\ldots,Lu\} or Ca$^{2+}$ occupies this site between the two \cuo layers.  Initially it was a surprise how little effect the Ln species had on the superconducting properties of the \cuo layer, \replaced{for example see reference}{e.g.} \cite{cardona1988}, especially given the strong magnetism of some Ln ions which in a BCS picture effectively dephases Cooper-pairs.  As such, the coupling between the Ln$^{3+}$ magnetism and \cuo layer has received quite some interest \cite{golnik1987, golnik1988, clinton1995, allenspach2000}.  These effects are observed at low temperatures and do not appear to influence the superconducting properties. In fact, the only observable effect of the Ln $f$-electrons in this thesis (other than via the ion-size), is \added{that} of the Nd$^{3+}$ crystal-field in low temperature Raman spectroscopy studies, \refsec~\ref{sec:ramangapsnd123}.

Cationic substitution of Ca$^{ 2+}$ for Y$^{ 3+}$ or Ln$^{ 3+}$ is a method of \added{hole }doping the \cuo layers.  There is a slight decrease in the highest obtainable \tc with this method \cite{schlachter1999} attributable to disorder or ion-size effects or both.

Other than the valence, the Ln property of key importance is the ion-size. This systematically alters the crystal structure (precisely how is discussed in detail below) via what can be thought of as ``internal pressure'' or the \emph{ion-size effect}. We discuss these concepts in more detail in \refsec~\ref{sec:intpressuretheory}.


	\emph{BaO layers and apical oxygen}:
	\label{sec:apical}
Composed of Ba and O(1) sites with Wycoff positions $2t$ and $2q$ respectively for the space group P/mmm.  It is possible to partially or fully substitute Sr for Ba \cite{licci1998, gilioli2000} to effect changes in ``internal pressure''.  O(1) is called the apical oxygen capping as it does the square pyramid co-ordination of the Cu. The apical oxygen has an important \cite{ohta1991, pavarini2001, sakakibara2012}, but not fully understood, role in the electronic properties of the cuprates.

	\emph{CuO$_{1-\delta}$ layers}:  
Composed of Cu(1) and (1-$\delta$)O(4) and O(5) (a defect) sites with Wycoff positions $1a$, $1e$ and $1b$ respectively for the space group P/mmm.  This layer acts as the `charge reservoir'.  Doped O onto the O(4) site removes negative charge from the \cuo layer, via the apical oxygen, increasing the hole density \emph{if} successive, linear lengths of O-Cu-O form \cite{tallonchapter}.  This additional structural feature of one-dimensional CuO `chains' gives the name ``chain layer''.  Conventionally the chains are defined to lie along the $b$-axis.  Chains are formed if the doped O occupies the O(4) site.  Occupation of the O(5), `chain disorder', reduces the hole concentration in the \cuo layer. 
Single crystals of this family are `twinned' if some lengths of CuO chains are perpendicular to others, or put another way, the $a$ and $b$ axes are swapped within regions of the crystal. `De-twinned' crystals indicate the chains run parallel throughout the whole crystal and this can be achieved by applying uniaxial stress.
	
With sufficiently long chain-lengths (small $\delta$ and low O(5) occupation), \musr studies have shown proximity-induced superconductivity occurs on these Cu-O chains \cite{tallon1995}.  Amongst other effects, this significantly increases the superfluid density (or condensate density), \ns, measurable by \musr.






LnBa$_2$Cu$_3$O$_6$, or Ln123O6, has no oxygen in the chain layer.  In this state the chains are sometimes referred to as `empty' and `inert'. To remove oxygen from the chain layer we anneal our samples at 600\degc in a pure Ar gas and then quench, see \refsec~\ref{sec:aranneal}. 

\subsection{Ln123 crystallographic systematics}
\label{sec:ln123systematics}
We owe much to the excellent study by Guillaume \etal \cite{guillaume1994}.  Many of the results presented in this section are from their systematic investigation of Ln123O7 and Ln123O6 by neutron diffractometry. 
	
\begin{figure}
	\centering
		\includegraphics[width=1.00\textwidth]{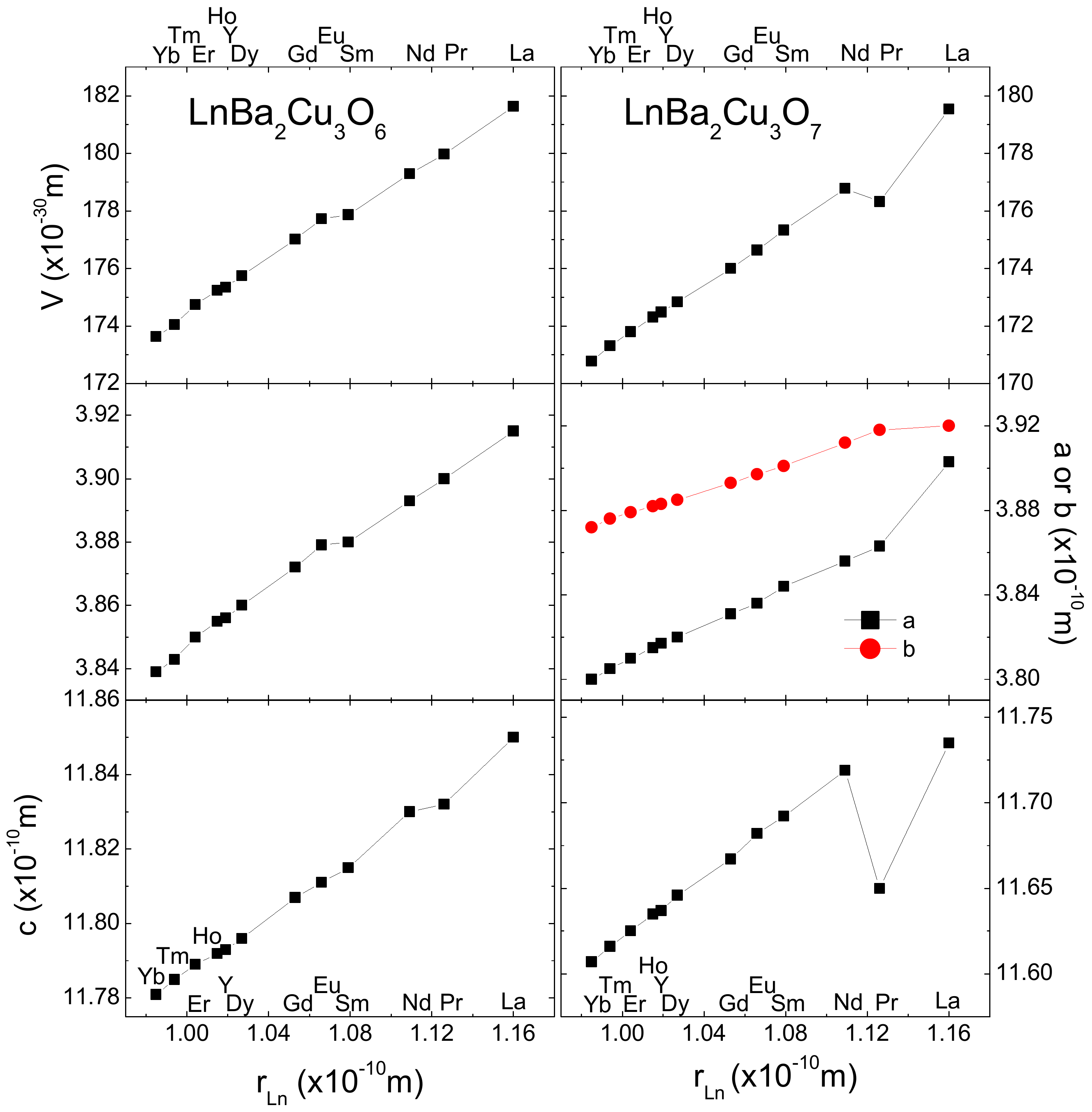}
	\caption[Unit-cell parameters of Ln123]{\label{fig:ln123lattice}Unit-cell parameters, $a$, $b$ and $c$ and the unit-cell volume $V$ for Ln123O6 (tetragonal) and Ln123O7 (which is orthorhombic because oxygen occupation the O(4) site breaks the $a$-$b$ symmetry) as a function of Ln (or Y) ion-size, $\rlnm$.  }
\end{figure}

In \fig~\ref{fig:ln123lattice} we plot the dependence of lattice parameters on the Ln/Y ion-size, $\rlnm$, in Ln123.  Larger $\rlnm$ result in larger lattice parameters.  This is what underlies what we call the ``internal-pressure'' effect, or simply the ``ion-size effect''. 

\begin{figure}
	\centering
		\includegraphics[width=1.00\textwidth]{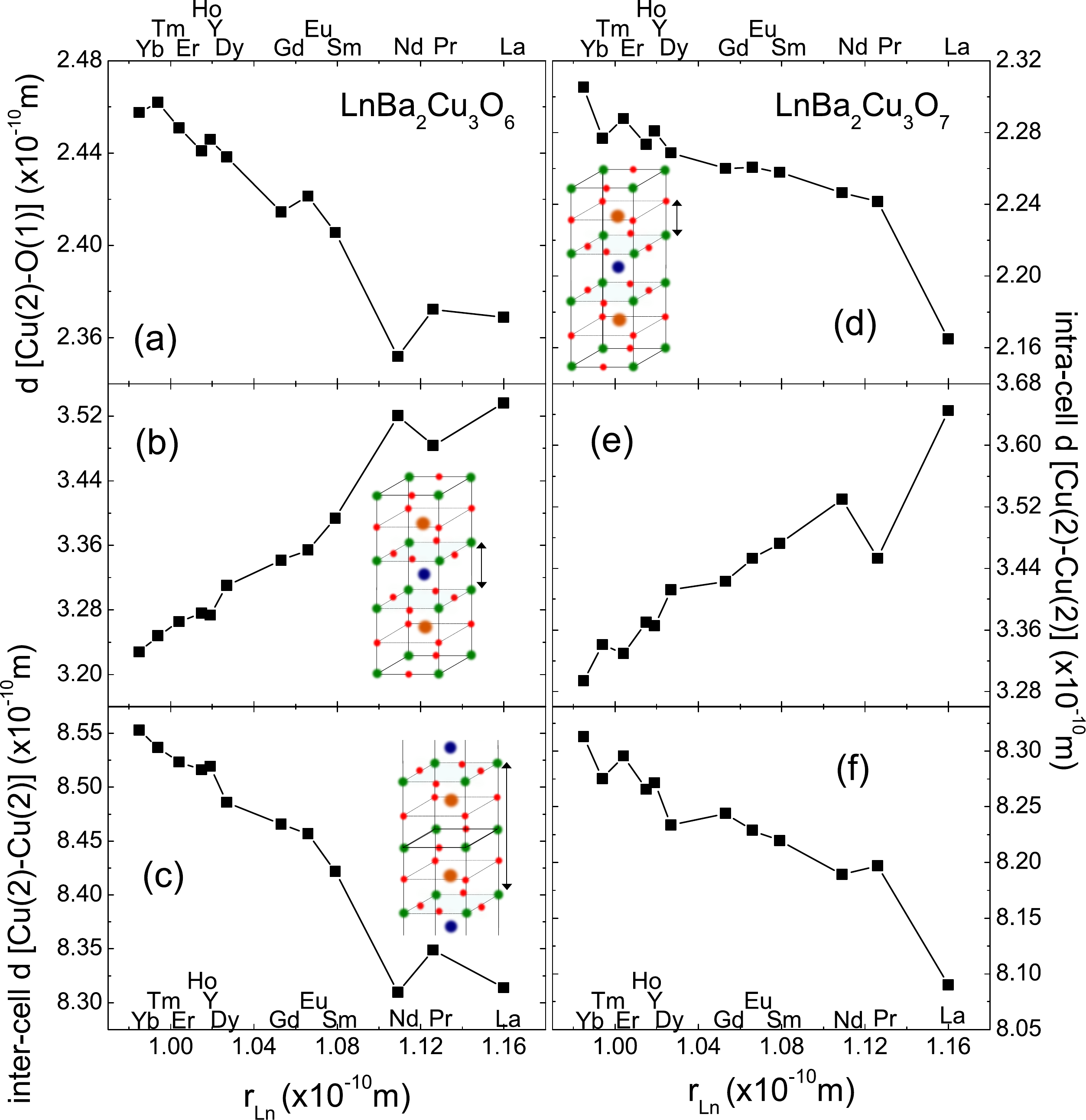}
	\caption[Various Ln123 bond-lengths]{\label{fig:ln123structure}The dependence of the apical oxygen bond length (a,d), intra-cell \cuo separation (b,e) and inter-unit cell \cuo separation (c,f) on the Ln ion-size for undoped (a,b,c) and fully oxygen-loaded crystals (d,e,f).}
\end{figure}

\begin{figure}
	\centering
		\includegraphics[width=0.83500\textwidth]{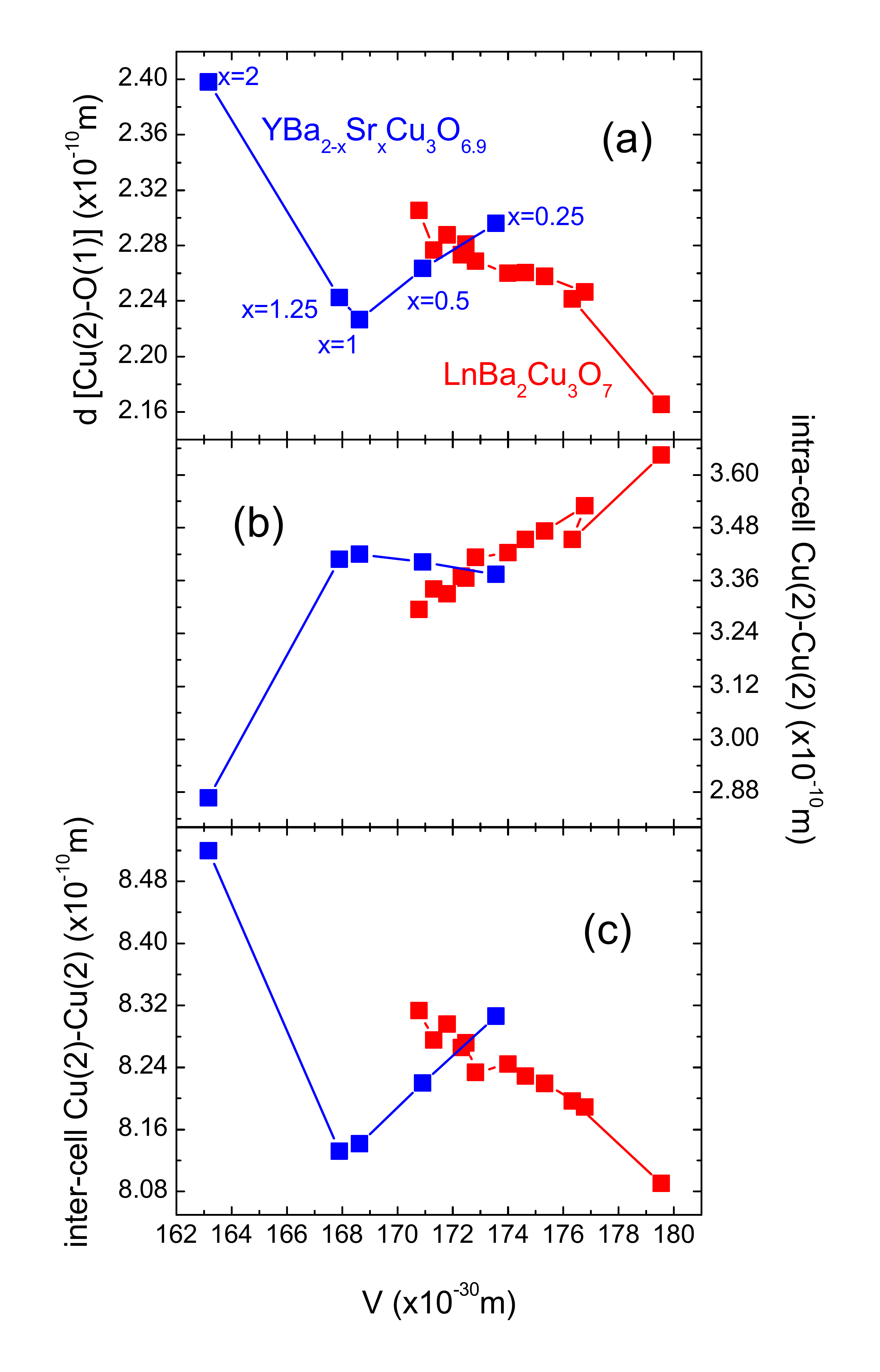}
	\caption[Various bond-lengths for Ln123O7 and Sr-doped Y123O7]{\label{fig:lnybasrstructure}The dependence of the apical oxygen bond length (a), intra-cell \cuo separation (b) and inter-unit cell \cuo separation (c) on the unit cell volume, $V$, for Ln123O7 and, using data from references \cite{licci1998, gilioli2000}, YBa$_{2-x}$Sr$_x$Cu$_3$O$_{7-\delta}$.}
\end{figure}

\fig~\ref{fig:ln123structure} shows the dependence of the apical oxygen bond length (a,d), inter-cell \cuo separation (b,e) and intra-unit cell \cuo separation (c,f) on the Ln ion-size for undoped (a,b,c) and fully oxygen-loaded crystals (d,e,f).  In \fig~\ref{fig:lnybasrstructure} we also plot these bond lengths for the YBa$_{2-x}$Sr$_x$Cu$_3$O$_{7-\delta}$ materials using additional data from \cite{licci1998, gilioli2000}. \added{These data serve to show that the crystallographic strain introduced by the ion substitution is not `hydrostatic' in the sense some intra-unit cell dimensions do not simply scale with the change unit-cell parameters shown in \fig~\ref{fig:ln123lattice}.  We now discuss in more detail the effect of the ion substitution on various structural parameters.}
	
The intra-unit-cell \cuo layer separation, $\dintram$, \added{is shown in panels (b,e) of \fig~\ref{fig:ln123structure}. This distance between the \cuo layers shortens} with decreasing $\rlnm$.  The Josepheson coupling between the two \cuo layers in the Ln123 unit cell is strongly influenced by $\dintram$.  Larger $\dintram$ results in weaker coupling meaning the \cuo becomes ``more 2-dimensional''.  This energy scale can be estimated from the Josephson-like plasma mode measured by infrared c-axis spectroscopy \cite{yu2008}.

On the other hand, the bond length between the in-layer Cu \added{(Cu(2) and Cu(3))} and apical O \added{(O(1))}, $\dapicalm$, is lengthened with decreasing $\rlnm$ \added{as shown in panels (a,d) of \fig~\ref{fig:ln123structure}}. Similarly, the inter-unit-cell \cuo layer separation also increases with decreasing $\rlnm$ \added{as shown in panels (c,f)}.  This can be pictured by separating Ln123 into two blocks; a CuO$ _{2} $-Ln-\cuo block and a BaO-CuO$_{1-\delta}$-BaO block.  Decreasing $\rlnm$ is can then be thought to increasingly separate these two blocks from each other. 
	
The Cu(2)-O(2,3) bond angle does not change significantly with Ln substitution \cite{guillaume1994}.  This angle is sometimes called the `buckling angle'. 	There are suggestions the buckling angle in the \cuo layer is correlated with \tc in the cuprates, see \cite{chmaissem1999}\footnote{This reference concludes that the buckling of the \cuo layer may be a response to lower the free energy as the Fermi-energy passes through a peak in the DOS.} and references therein.  Out of interest, there is also evidence \tc depends on the As-Fe-As bond angle in the pnictide superconductors \cite{ishida2009}.  Apparently the optimum angle there is 109\degrees. 

As the $\rlnm$ decreases, charge transfer between the \cuo layer and CuO chains is enhanced.  This has been ascribed to the decrease in $c$-axis length and concurrent decrease in separation between these layers \cite{wijngaarden1999} but we feel that the primary cause is the that the small Lanthenide ions promote better long-range ordering in the chains.  Occupying of the otherwise vacant O(5) sites is promoted by larger Lanthenide ions.   This results in a \replaced{lower}{higher} doping, \(p\), for given \(\delta\) for \replaced{larger}{smaller} Ln ions. 
Pressure-induced charge transfer is an important effect to consider for which a separate section is devoted below, see \refsec~\ref{sec:extpressuretheory}.

%
%
%
%
%
%
%
%

\section{Electronic structure, phase diagram and gaps}
\subsection{Electronic structure of undoped cuprates}
\label{sec:estructurecuprates}
 Because of their layered structure the cuprates are often electronically highly anisotropic and are some of the best experimental approximations for 2D magnetic systems that we know \cite{dejongh}.

\begin{figure}
	\centering
		\includegraphics[width=0.75\textwidth]{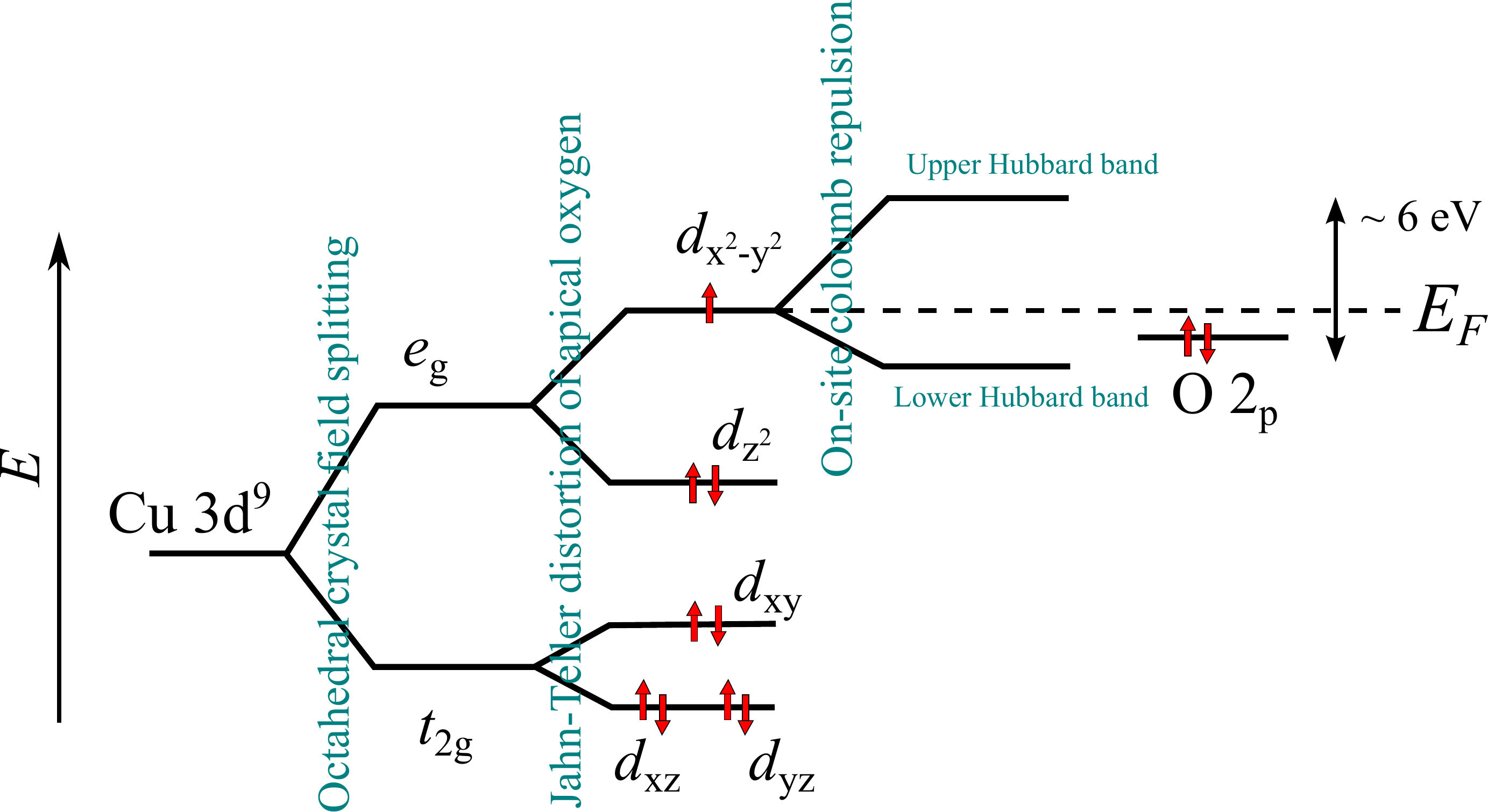}
	\caption[Splitting of the electronic levels of the undoped \cuo layer]{Various degeneracy lifting distortions that give rise to the splitting of electronic levels associated with the undoped \cuo layer.}
	\label{fig:electronicconfig}
\end{figure}

\fig~\ref{fig:electronicconfig} shows how we arrive at the electronic configuration of the Cu ion in the \cuo layer of undoped cuprates \cite{lee2006}.  The valence of Cu in the \cuo layers of undoped cuprates is 2+ with an electronic $d^9$ configuration giving nine electrons to place in the $d$-orbital shell. The degeneracy of the Cu $d$ orbital is firstly split by the oxygen octahedral crystal field into upper $e_g$ (d$_{x^2-y^2}$ and d$_{z^2}$) and lower $t_{2g}$ (d$_{xy}$, d$_{xz}$, d$_{yz}$) bands. A strong Jahn-Teller effect, caused by apical oxygen distortion of the oxygen octahedra, then splits these two bands further and lowers the energy of the d$_{z^2}$ orbital. As each orbital can accommodate two electrons of anti-aligned spins, the d$_{x^2-y^2}$ band is left half filled with one electron. 

Conventional band theory predicts that a half-filled band at the Fermi-level results in a metal. However with these strongly correlated materials it is energetically a poor choice to have two electrons occupy the same \dxy orbital because of strong on-site Coloumb repulsion between the electrons.  Mott-Hubbard physics is more appropriate for this situation and the band is split by quite a large energy, a Mott-Hubbard 
gap of energy $U\approx6$ eV. 
The Cu \dxy electrons are no longer free to move resulting in an insulating material. 

The $p_\sigma$ band of the oxygen in the \cuo layer is now nearest to the Fermi-energy. The lobes of the Cu \dxy electron wavefunction align with the O $p_\sigma$ lobes. The Cu electrons can virtually hop, with significant probability, to the O $p_\sigma$ band.  Hence, Cu electrons lower their energy by anti-aligning their spins to allow both adjacent Cu electrons to virtually hop to the oxygen band.

We can discuss this last idea in terms of exchange interactions. The exchange interaction is a purely quantum mechanical effect which comes from the Pauli exclusion principle for identical particles. For interacting Fermions the total wavefunction must be anti-symmetric (the Pauli exclusion principle), thus, when the spatial (or orbital) wavefunction symmetry is symmetric for a two Fermion system, the spin symmetry must be \emph{anti}symmetric (i.e. the spin-singlet state; $S=0$ where $S$ is the total spin of a two particle system) and vice versa (i.e. the symmetric spin-triplet state; $S=1$). The energy difference between these two spin-symmetry states is usually denoted $J$, the exchange energy.  The interaction is extremely short ranged in space as it occurs only where the wavefunctions of the two Fermions overlap `appreciably'. 

\emph{Super}exchange \cite{anderson1950} is where the exchange interaction is mediated through an intermediary ion, usually  a `non-magnetic' anion (e.g. O) between two cations (e.g. Cu), see e.g. \cite{goodenough1955}. In the cuprates the magnetic moment associated with the spin of the unpaired \dxy electron interacts with the neighbouring Cu \dxy electron via the O $p_\sigma$ orbital.  As mentioned, this interaction is possible due to the spatial overlap of the \dxy and $p_\sigma$ orbitals.  

Thus, rather than a normal metal the ground state of a cuprate with no doped charge carriers in the \cuo plane is an antiferromagnetically ordered, charge-transfer insulator. $J$ is the energy difference between a parallel and anti-parallel alignment of spin moments on neighbouring Cu(2) ions in the \cuo layers and is of order $0.12$ eV.  This can be compared with the charge-transfer gap, $\sim2$ eV, and the superconducting gap, $\scgapm \sim0.025$ eV.

\subsection{Phase diagrams}
\label{sec:phasediagram}

The physics of the cuprates really comes alive once the materials are electronically doped. This can be either electron-doping or hole-doping.  The hole-doped cuprates are more extensively studied and are what we consider exclusively in this work.

\begin{figure}
	\centering
			\includegraphics[width=0.835\textwidth]{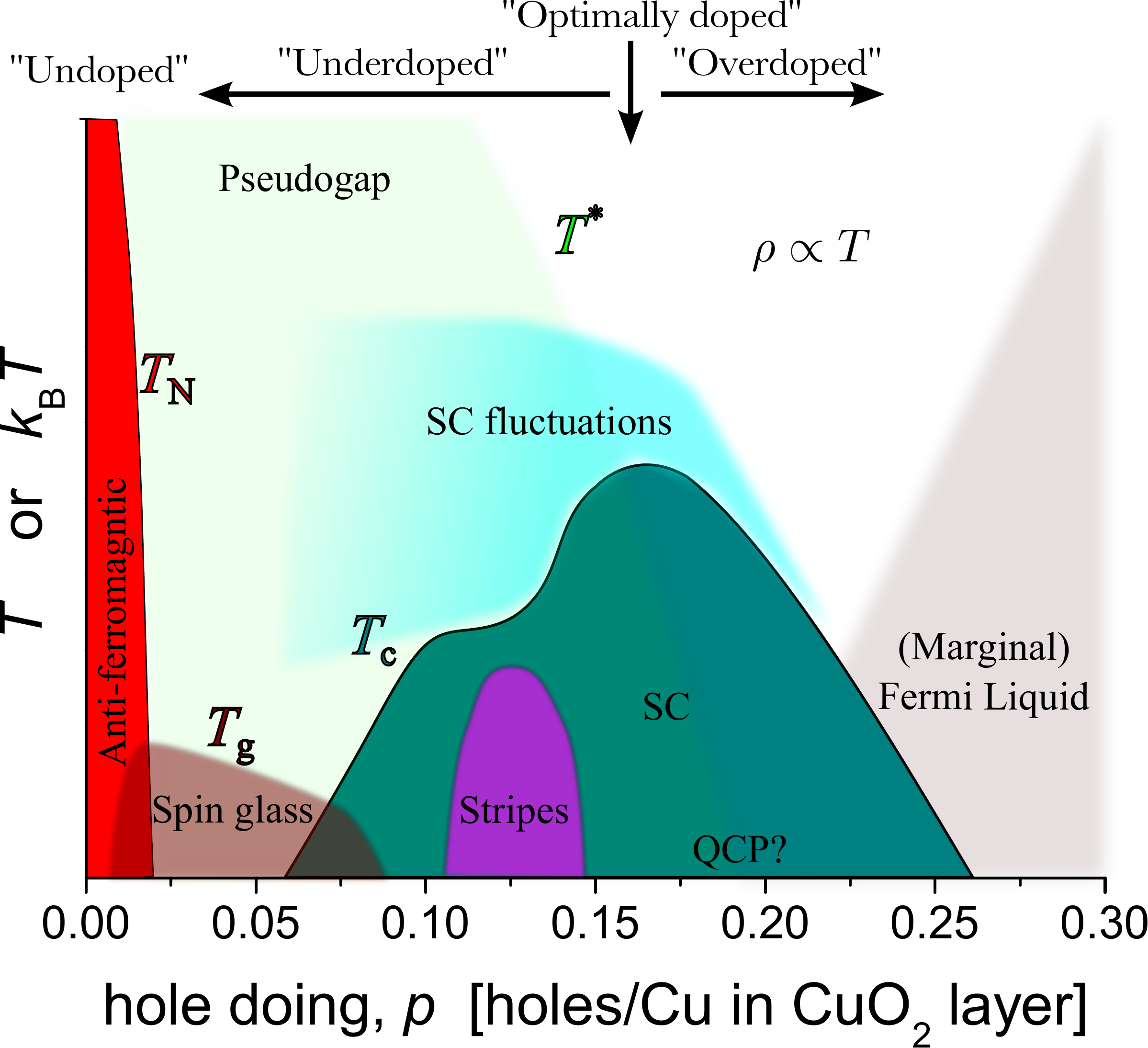}
	\caption[Phase diagram of the cuprates]{An indicative phase diagram for hole doped cuprates compiled from various sources \cite{tallonchapter, kivelson2008, varma2010}. At low doping the cuprates are charge-transfer insulators with the Cu(2,3) magnetic moments antiferromagnetically ordered.  The long-range, inter-unit cell AF order is rapidly destroyed as $p$ is increased.  At low temperatures a spin glass phase is observed.  Superconductivity (dark blue shaded region) first appears around $p\approx 0.06$ with the onset temperature of this macroscopic phase coherence, \tc, increasing to its highest temperature, \tcmax by $p=0.16$.  \tc then decreases again as indicated making a `dome'. In the region of \tcmax almost all cuprates follow $\tcm/\tcmaxm \approx 1-82.6(p-0.16)^2$ \cite{tallon1993}. Above \tc superconducting fluctuations are observed \cite{tallonfluctuations, tallon2011}, sometimes up to remarkably high temperatures (180 K) \cite{dubroka2011}.  At still higher charge carrier concentration, $p$, Fermi-liquid behaviour is observed.  Around $p=\sfrac{1}{8}$ ``stripe'' charge ordering results as a compromise between (dynamic) AF order and Coloumb repulsion \cite{tranquada1997}.  The `pseudogap' phase dominates the underdoped region.  It is a ground state property that competes with superconductivity.  The temperature below which it is observed, $T^*= \epgm/k_B$ decreases with increasing $p$.  By $p\approx0.16$ $T^*\approx \tcm$ and as $p\rightarrow 0.19$ $\epgm = k_BT^* \rightarrow 0$.  The pseudogap is discussed in more detail below in section \ref{sec:pseudogap}.  QCP? marks the location of putative quantum critical point \cite{tallon2003, varma2010, sachdev2011} where the pseudogap closes at $p=0.19$.  In a `cone' above QCP? the resistivity, $\rho$, above \tc is linear in temperature up to at least $T=1000$ K \cite{wolf2004} - a signature of self-similarity associated with critical phenomenon.  Note the terminology annotated at the top of the diagram to describe the doping level of a cuprate.}
	\label{fig:phasediagram}
\end{figure}

\fig~\ref{fig:phasediagram} illustrates an indicative phase diagram for hole doped cuprates, sometimes called the ``universal phase diagram'' to emphasize the scaled quantitative similarity of it between the members of the hole-doped cuprates\footnote{There is however good evidence now that some members of the family have distinct doping-phase diagrams, such as Bi2201 \cite{schneider2005, ando2000}. }.  The figure has been compiled from various sources \cite{tallonchapter, kivelson2008, varma2010}. Clearly the phase behaviour is rich (and this is a partly simplified phase diagram as well)!  The complicated phase diagram is partly a result of these materials having been studied in such detail, but mostly a result of strong electronic correlations allowing an array or electronic orders that jostle with each other for dominance at a particular $T$ and $p$.

At low doping the cuprates are charge-transfer insulators with the Cu(2,3) magnetic moments antiferromagnetically ordered, as discussed above.

As $p$ is increased the long-range, inter-unit cell antiferromagnetic order is rapidly destroyed.  At low temperatures, below $T_G$, spin fluctuations may freeze out forming a spin-glass. A spin-glass state corresponds to thermally meta-stable microscopic regions of magnetic order which are macroscopically disordered.  Such a magnetic state can be observed in AC magnetisation measurements or by with zero-field muon spin-relaxation ($\mu$SR) \cite{ofer2006}.

Superconductivity (dark blue shaded region) first appears around $p\approx 0.06$ with the onset temperature of this macroscopic phase coherence, \tc, increasing to its highest temperature, \tcmax by $p=0.16$.  \tc then decreases again as indicated making a `dome'. In the region of \tcmax almost all cuprates follow ${\tcm}/{\tcmaxm} \approx 1-82.6(p-0.16)^2$ \cite{tallon1993}. Above \tc superconducting fluctuations are observed \cite{tallonfluctuations, tallon2011}, sometimes up to remarkably high temperatures (180 K) \cite{dubroka2011}.

Around $p=\sfrac{1}{8}$ a particular type of charge ordering results as a compromise between (dynamic) 2D antiferromagnetic order and Coloumb repulsion.  This fascinating physics is given the name `stripes' \cite{tranquada1997}.  In some choicely doped cuprates these charge and spin density waves become static and destroy superconductivity \cite{tranquada1997}. 


The `pseudogap' phase is a ground state property that competes with superconductivity.  In the underdoped region of the phase diagram ($p<0.16$ as annotated in \fig~\ref{fig:phasediagram}) the pseudogap `opens' at a temperature higher than $\tcm$, $T^*>T_c$ and has a significant effect on the superconducting properties. It is strongly doping dependent with the approximate behaviour $\epgm \approx J (1-p/p_{\textnormal{crit}})$ \cite{loram2001}. $p_{\textnormal{crit}}=0.19$ is the so-called critical doping where the pseudogap closes, $\epgm \sim k_BT^* \rightarrow 0$.  The pseudogap is discussed in more detail below \ref{sec:pseudogap}.


\subsection{The normal-state Fermi surface}

\begin{figure}
	\centering
		\includegraphics[width=0.80\textwidth]{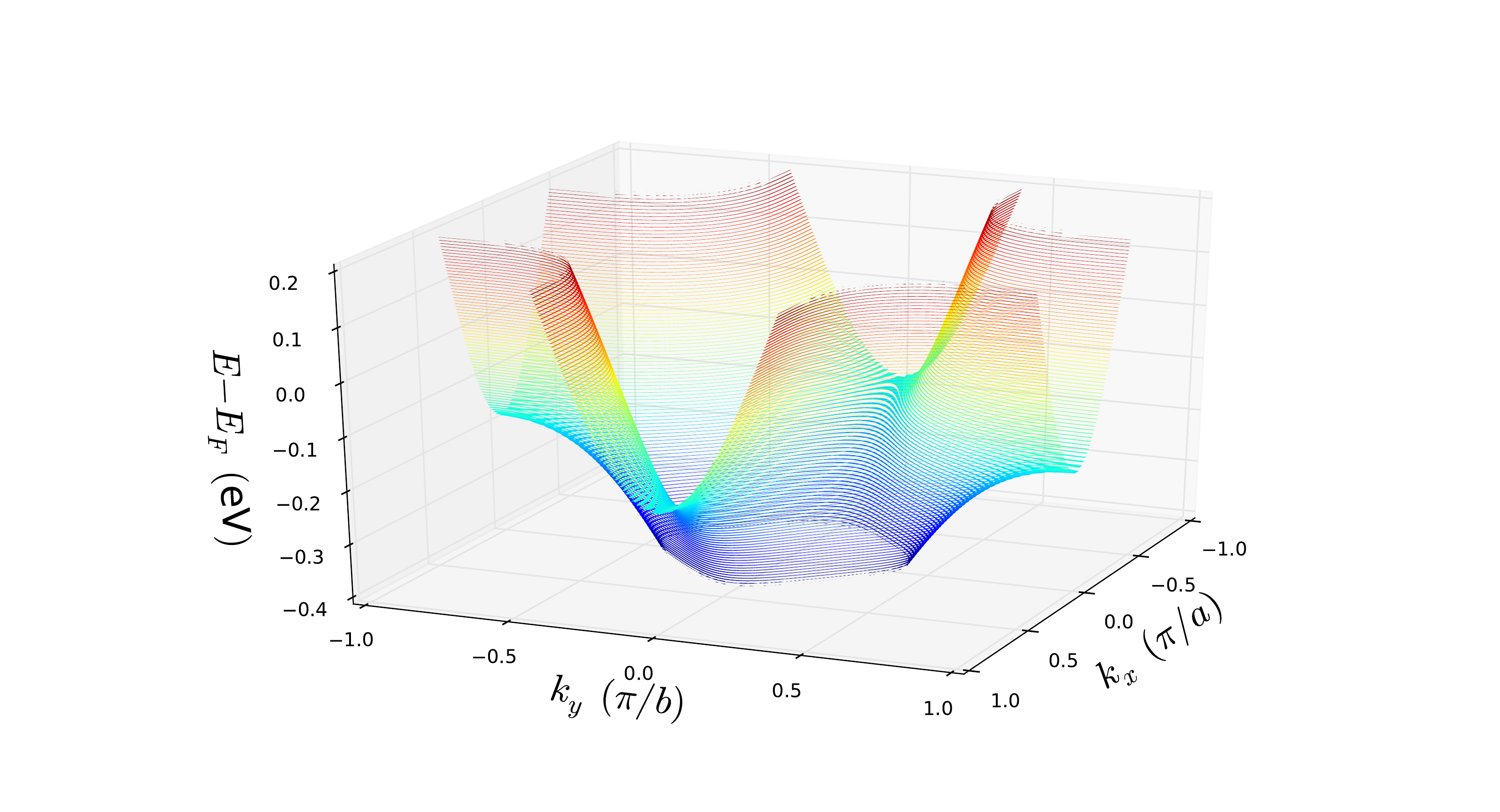}
	\caption[Generic electronic dispersion for the cuprates]{Electronic dispersion of the \cuo layer calculated using tight binding parameters derived from ARPES data \cite{kaminski2006}.}
	\label{fig:Fermisurface}
\end{figure}

Using \added{Angular Resolved Photo-emission Spectroscopy} (ARPES) it is possible to measure the electronic dispersion of the \cuo layers, see e.g. \cite{kondo2004}.  As the \cuo layers are quasi-2D so their dispersion can be easily pictured, as shown in \fig~\ref{fig:Fermisurface}.  The dispersion has 90\degrees symmetry reflecting the square-planar, 4-fold symmetry of the \cuo layer.  The ``nodes'' are at $(\sfrac{\pi}{2}, \sfrac{\pi}{2})$ and symmetry related points, the ``antinodes'' at $(0,\pi)$, see \fig~\ref{fig:bz} for a simplified schematic diagram of the Fermi-contour.  With increasing hole concentration the Fermi-level moves lower in energy relative to this dispersion - a so call `rigid band' shift.  At $p\approx 0.27$ the Fermi-level for the single-layer cuprates reaches the saddle point in the electronic dispersion at the anti-nodes resulting in a van Hove singularity (vHs) in the density of states, see e.g. \cite{ashcroftmermin}.  Increasing hole doping further, the Fermi-surface, counter-intuitively, becomes electron-like \cite{kaminski2006}. 

	\subsection{Gaps}
	\label{sec:gaps}
		\subsubsection{Superconducting energy gap}
		\label{sec:scgap}
In the superconducting state an energy gap of width $\scgapm$ opens in the single-electron states symmetrically about $E_F$. \scgap represents the energy required to break a Cooper-pair.  In weak-coupling BCS theory  for a $d$-wave gap (see below) the magnitude of \scgap at $T=0$~K is proportional to the (mean-field) transition temperature\added{, $ \tcmfm $, }as;
		\begin{equation}
		\scgapm = 4.28k_B\tcmfm 
		\end{equation}
\noindent \added{where $ k_{B} $ is the Boltzmann constant.} The prefactor $4.28$ relates only to weak-coupling, for strongly-coupled superconductors the prefactor is larger.  The appropriate value for the cuprates is still a topic of debate which is partly obscured by the fact that \tcmf can be significantly larger than \tc in the cuprates \cite{dubroka2011}. 

\begin{figure}
	\centering
		\includegraphics[width=0.660\textwidth]{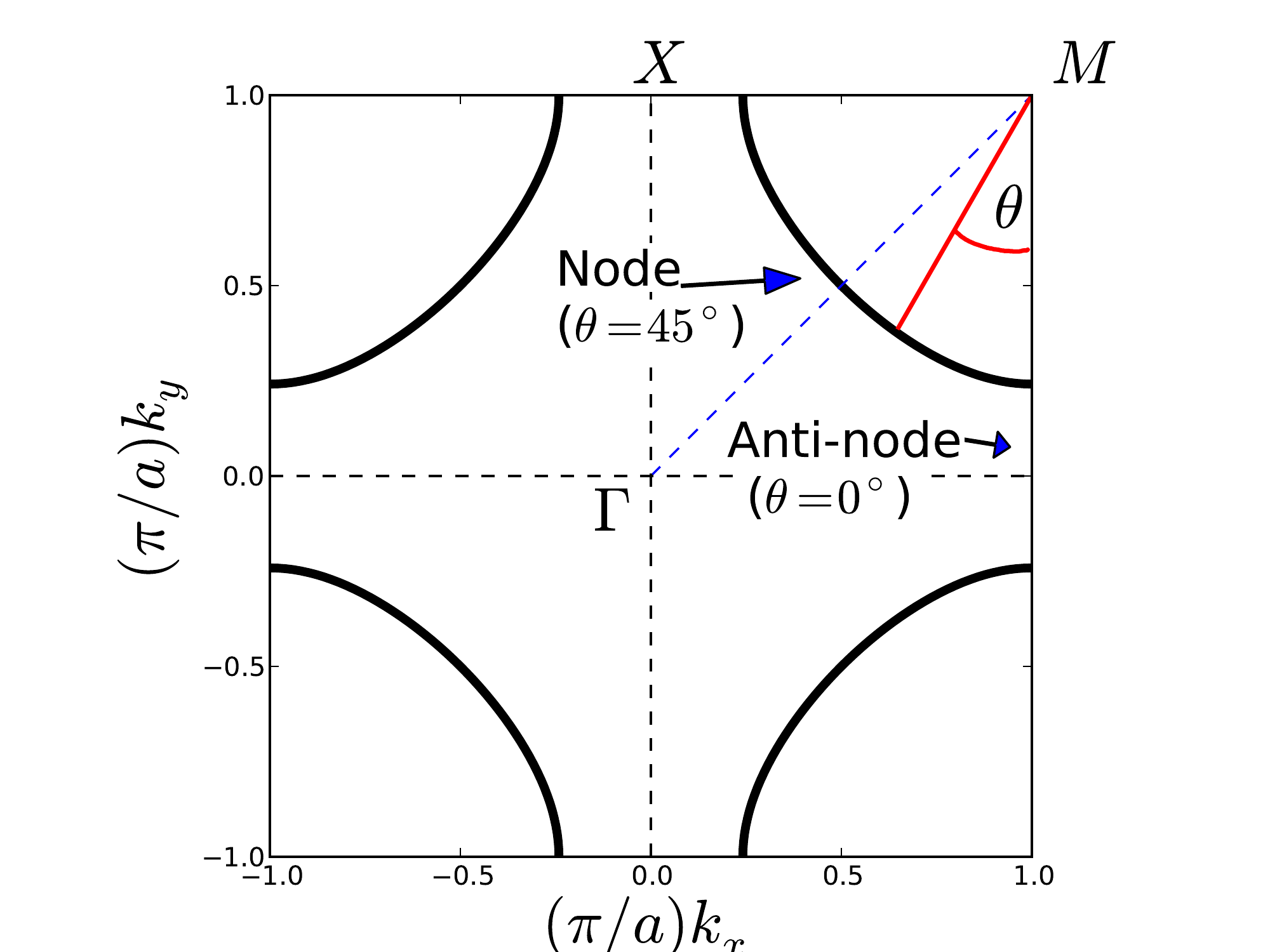}
	\caption[Generic Fermi-contour for the cuprates]{Schematic plot of the Fermi-contour in the 2D Brillouin zone.  \added{The red line indicates how the angle $ \theta $ is defined.}}
	\label{fig:bz}
\end{figure}

In the cuprates the superconducting gap is observed to be anisotropic around the Fermi-surface. \scgap is the amplitude of the gap which has \dxy symmetry;

\begin{equation}
\hscgapkm = \frac{1}{2}\hscgapm [\cos(\kk_x a) - \cos(\kk_y b)] 
\label{eq:scgap}
\end{equation}

\noindent where $x=\pi/a$ and $y=\pi/b$ are unit vectors in the 2D Brillouin zone and $a$, $b$ unit cell lengths, parallel to the Cu-O bonds in the \cuo layer. Normally the $a=b$ simplification is made.  Alternatively we can write the angular dependence of the \dxy gap as $\hscgapm(\theta) = \hscgapm \cos(2\theta)$ where $\tan (\theta) = \kk_x/\kk_y$ is the angle along the Fermi surface as shown in \fig~\ref{fig:bz}.  In this case the nodes are at 45\degrees and anti-nodes at 0\degrees. 

\begin{figure}
	\centering
		\includegraphics[width=1.00\textwidth]{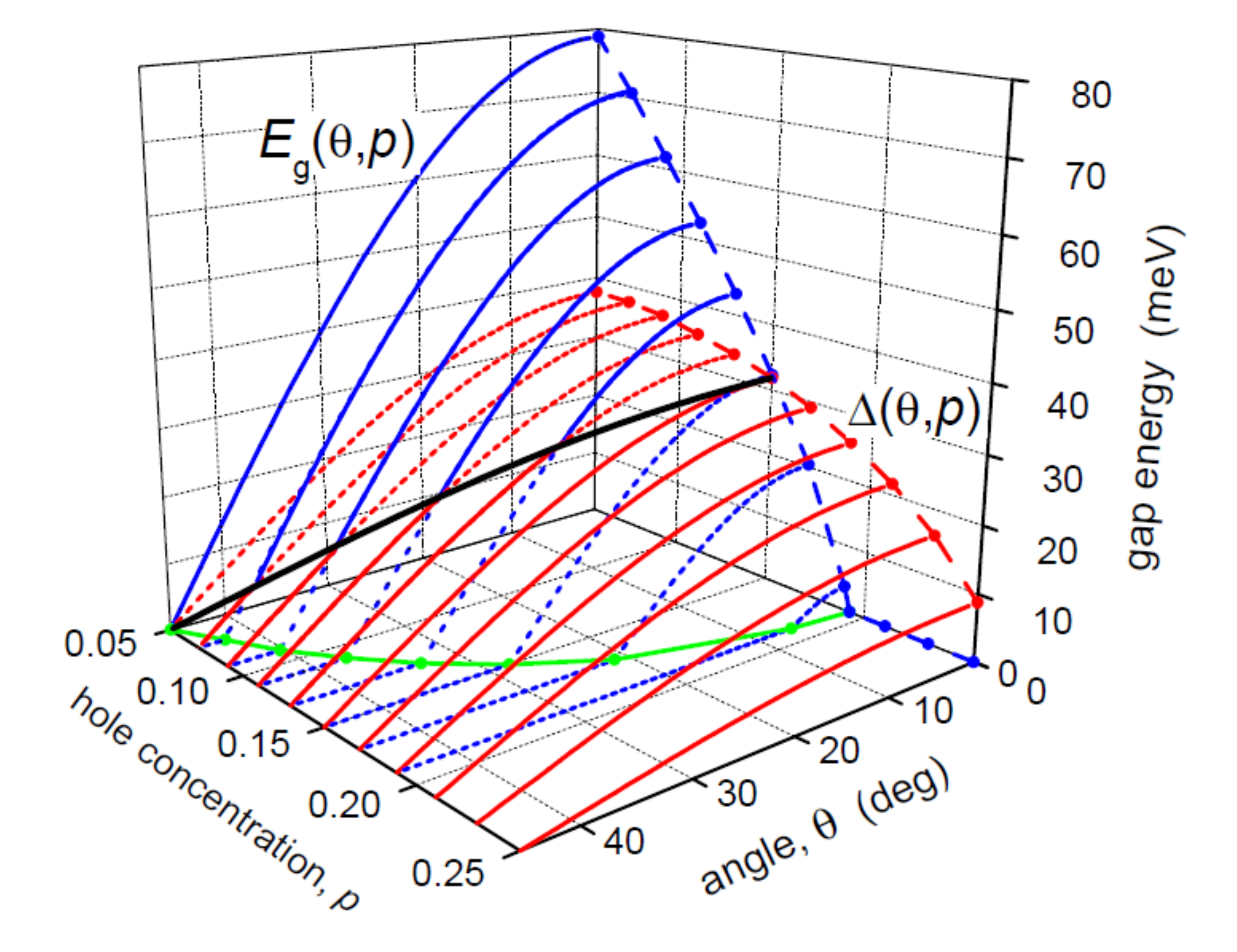}
	\caption[Doping and angular dependence of the superconducting gap and pseudogap]{Figure reproduced from the work of Tallon \etal \cite{tallonarXiv}. Phenomenology of the doping and angular dependence of the pseudogap energy, \epg (blue curve) and superconducting gap energy, \hscgap (red curve). $\theta = \arctan [\kk_x/\kk_y]$ measures an angle around the normal-state Fermi-surface with 45\degrees equivalent to the nodes and 0\degrees the antinodes. As is customary, $p$ is holes per Cu in the \cuo layer. Dotted curves show the expected progression of one gap in the absence of the other. \added{Thus the green line indicates the boundary where the pseudogap vanishes.} Note that in the cuprates the \hscgapk has $d$-wave symmetry (\eq~\ref{eq:scgap}) so that $\hscgapkm \rightarrow 0$ at the nodes.}
	\label{fig:gaps}
\end{figure}

An illustration of \eq~\ref{eq:scgap} is shown in \fig~\ref{fig:gaps} (this Figure is reproduced from \cite{tallonfluctuations}) by the red curve as a function of doping and $\theta$.  In this figure the blue curve is our phenomenological  understanding of the pseudogap, as is discussed in the next section.  The maximum energy of both \hscgapk and \epgk on the (normal state) Fermi-surface is at the antinodes, whilst the \dxy symmetry of \hscgapk means $\hscgapkm \rightarrow 0$meV at the nodes. 

Finally, for completeness, the temperature dependence of \hscgap near \tc in the simple BCS theory and Ginzberg-Landau theory is;
\begin{equation}
\hscgapm(T) = \hscgapm [1-t]^{1/2}
\end{equation} 
\noindent where the reduced temperature $t$ is defined as $t\equiv T/T_c$.

		\subsubsection{Pseudogap}
		\label{sec:pseudogap}
This pseudogap has been the topic of much debate for many years.  Originally it was seen as a downturn in the temperature dependence of the NMR spin susceptibility at temperature well above \tc \cite{johnston1989} and soon after in specific heat \cite{loram1994}.  We can now locate signatures of the pseudogap in most physical properties; e.g. resistivity \cite{naqib2005}, magnetic susceptibility \cite{naqib2009}, ARPES \cite{yoshida2009}, STM \cite{lawler2010}, Raman spectroscopy \cite{opel2000}, \musr \cite{khasanov2008}, thermopower \cite{storey2013}, $c$-axis transport \cite{bernhard2008}, infrared ellipsometry \cite{dubroka2010}, and the nature of the pseudogap state is still an topic of active research and debate \cite{yrzpaper, borne2010, he2011, nakayama2011, lawler2010}.  

Some view the pseudogap as phase-incoherent Cooper-pairs \cite{daou2010, wen2009}, a ``preformed pairs'' or SC-precursor-state idea. However, the experimental evidence, that I consider overwhelming, rules in favour of the pseudogap \emph{co-existing and competing} with the superconducting state. As a selection see \cite{tallonchapter, tallon2001, bernhard2008, yoshida2009, dubroka2010, alloul2010, okada2010} and references therein. Note that superconducting fluctuations also exist above $T_c$ \cite{lee2006, dubroka2010, dubroka2011} which probably leads to confusion between precursor pairs and the pseudogap. The effect of a pseudogap however can be observed in ground-state SC properties such as the superfluid density, $ \nsm $, in the limit $T\rightarrow 0$\deleted{K} \cite{tallon2003superfluid}.

In essence the pseudogap is a particle-hole asymmetric gap in the single-electron levels (the `density of states' or DOS) around $E_F$.  This understanding results from the success of the YRZ model \cite{yrzpaper}.

%

The phenomenological $\kk$-dependence of the pseudogap to the best of our knowledge is illustrated in \fig~\ref{fig:gaps}. It is similar to \hscgapk being largest at the anti-nodes, however it drops to zero more quickly leaving an ungapped region around the nodes above $\tcm$ - the so-called `Fermi arcs'.  Below \tc the SC gap, $\hscgapm$, opens solely on these Fermi arcs.  The Fermi arcs as observed in ARPES (e.g. \cite{kanigel2006}) are demarked by the green contour in \fig~\ref{fig:gaps}. This figure also shows the doping dependence of the pseudogap as does \fig~\ref{fig:phasediagram}.

The pseudogap is only weakly temperature dependent, if at all.  It is also relatively insensitive to moderate disorder, for example Ca doping for Y in Y123 \cite{naqib2009}, Zn doping on the Cu site \cite{naqib2005, kim2010} or electron irradiated samples studied by Alloul \etal \cite{alloul2010}.  \epg is however enhanced by Ni doping, a magnetic ion, on the Cu site \cite{pimenov2005}. 

The emerging physical picture of the pseudogap is that of a Fermi-surface reconstruction, possibly caused by short-range, fluctuating AF correlations which strongly scatter quasi-particles \cite{naqib2005, storey2008pggroundstate, harrison2007, yrzpaper}. In this case, the anti-ferromagnetic exchange energy, $J$, at $p=0$ is the key energy scale for the pseudogap \cite{loram2001}. The effect of Ni doping then would be to pin these spin fluctuations to promote Fermi-surface reconstruction.

		\subsubsection{Spectral gap}
		\label{sec:specgap}

In spectroscopy, \replaced{there is depletion of spectral-weight below a certain energy in the superconducting state relating to the opening of the superconducting energy gap,}{a pair-breaking energy gap is observed} which we call the spectral gap, $\specgapm$. In a $d$-wave superconducting state and in the absence of a pseudogap, $\specgapm=\scgapm$ at the anti-nodes but still vanishes, $\specgapm\rightarrow 0$, at the nodes \cite{tallonarXiv}.  These relations are modified by a finite pseudogap energy, $\epgm$, \cite{storey2007fermi, tallonarXiv} so that;
\begin{equation}
\specgapm [\bogm] = 2 \sqrt{\hscgapm^2 + \epgm ^2}
\label{eq:specgaps}
\end{equation}
\begin{equation}
\specgapm [\btgm] \approx \frac{3}{2}\sqrt{\hscgapm^2-\left( \sfrac{2}{3} \epgm \right) ^2}
\label{eq:specgaps2}
\end{equation}

Here \bog and \btg represent Raman scattering geometries that measure scattering from regions of the Brillouin zone around the anti-nodes and nodes, respectively, as shown in \fig~\ref{fig:b1gb2gweight}.  There is a contribution to \specgap from both \hscgap and \epg but because of their distinct angular dependence around the Fermi-surface their respective contributions vary around the Fermi-surface.  \eq~\ref{eq:specgaps2} is approximate only due to the different $\theta$-dependence of \hscgap and $\epgm$.  If they were to have the same $\theta$-dependence, then the term in the square root is replaced by $\hscgapm^2-\epgm^2$.

\section{Pressure effects and a dichotomy!}
\label{sec:pressureeffects}

(La,Ba)$_2$CuO$_4$ was the first superconducting cuprate to be discovered by Bednorz and M\"{u}ller and it had $T_c = 35$ K.  A very important early observation was that under external pressure, this \tc increased further \cite{kurisu1987}.  This is in fact a very general property of the cuprates: \tcmax increases under external pressure.

Paul Chu and his group took this idea and experimented with simulating this external pressure with ion substitution - the so called ion-size effect or `internal pressure'.  The La in (La,Ba)$_2$CuO$_4$ is a prime target for such an experiment\replaced{.}{!} Because of its $f$-electrons the ion-size of La is much larger than Y (see e.g. \fig~\ref{fig:ln123lattice}), which sits above it on the periodic table of elements and so they began exploring Y$_2$O$_3$, BaO, CuO combinations.  What they accidentally discovered doing this is the famous \ybco which has $\tcmaxm = 93.5$ K.  The crystal structure of Y123 is actually quite different from the (La,Ba)$_2$CuO$_4$ they were guided by; it has extra CuO chains, two CuO layers and the Y ion in a different site compared with La. Although they had not increased \tc simply with internal-pressure, it was that idea that led them to the discovery of the extremely significant Y123 superconductor.

A vast amount of research into HTS has been carried out since their discovery in 1986 \cite{bednorz1986}.  Correctly interpreting the high quality data from this research is key to progressing the understanding of HTS.  In particular we focus on a little known but central puzzle: \tcmax \emph{decreases} with `internal pressure', where ionic substitution is used to decrease the lattice parameters.  However, \tc \emph{increases} under external pressure \cite{lorenzchu, schillingchapter, wijngaarden1999, schlachter1999}.  What is (are) the important difference(s) affecting \tcmax between these two types of `pressure' in the cuprates?  Answering this question will lead us to a better understanding of the important material properties affecting \tc and to new insights into the HTS puzzle. 

	\subsection{External pressure}
	\label{sec:extpressuretheory}
	A comprehensive discussion of pressure effects in HTS can be found in the review by Schilling \cite{schillingchapter}. 

		Pressure-induced charge transfer (PICT) is the first of many complications caused by the application of pressure that one must consider \cite{schillingchapter}.  Negative charge is transferred from the \cuo layers to the charge reservoir layers because of better electronic coupling.  This causes an increase in $p$ on the \cuo layers with increasing pressure.  This is seen experimentally and in DFT calculations \cite{khosroabadi2004}.
		
		In addition to PICT, the \tcmax of cuprates is also universally increased under external pressure \cite{lorenzchu, schillingchapter, wijngaarden1999, schlachter1999}. At $P=0$ optimal doping the initial increase in the \tcmax with pressure for the cuprates \cite{schillingchapter, klehe1993} is ${d\tcmaxm}/{dP}\sim1$ K.GPa$^{-1}$. For example, for Y123 it is estimated \cite{neumeier1993} as $0.96$ K.GPa$^{-1}$ whereas for single- bi- and tri-layer Hg-based cuprates it is $1.75 \pm 0.05$ K.GPa$^{-1}$ \cite{schillingchapter}.  At higher pressures, the \tc of samples that are optimally doped at $P=0$ will decrease as PICT starts to overdope the \cuo layer.  To counteract the effect of PICT and measure the pressure dependence of $\tcmaxm$, it is necessary to use samples that are progressively more underdoped at $P=0$ as one goes to higher pressures.

\begin{figure}
	\centering
		\includegraphics[width=0.660\textwidth]{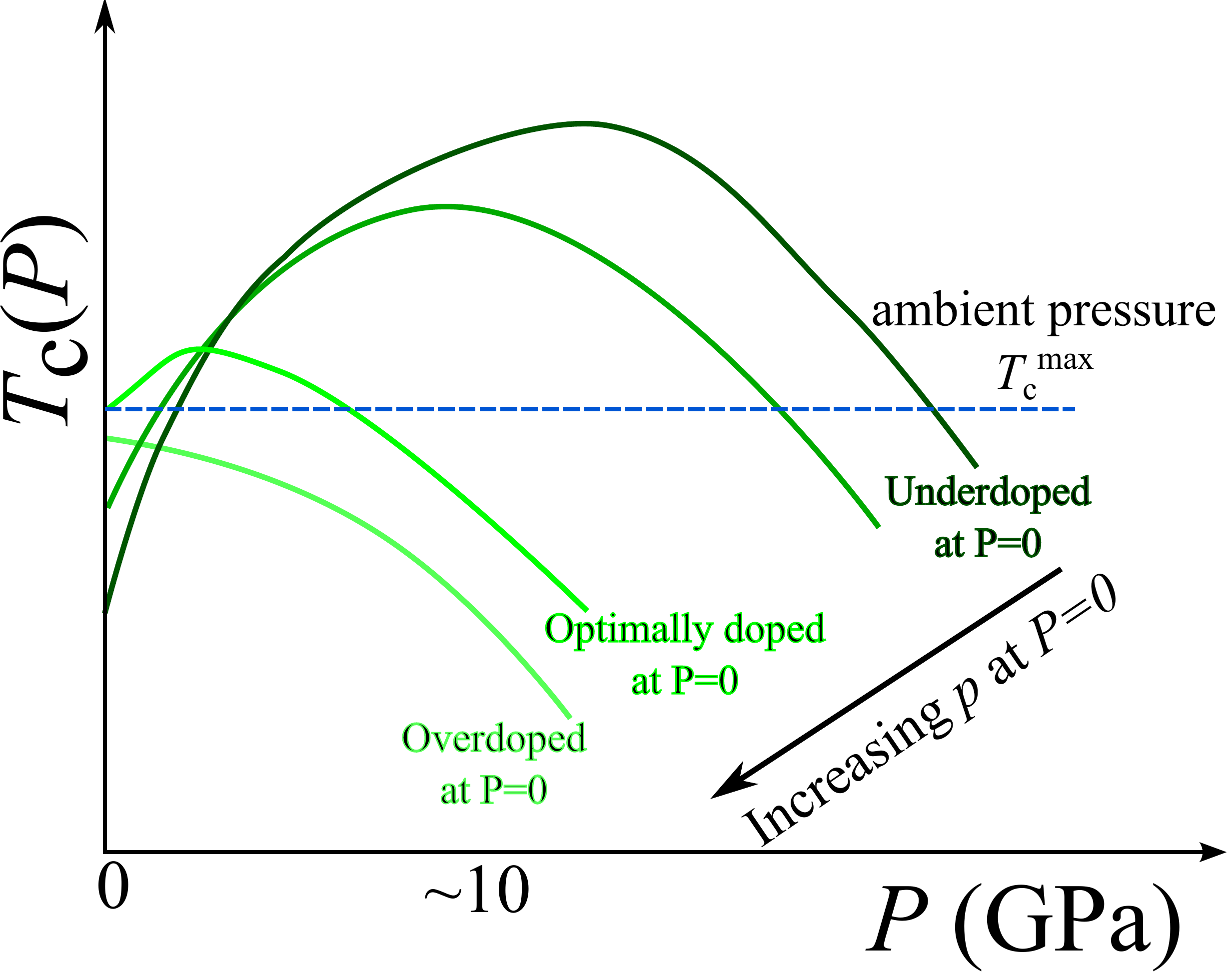}
	\caption[Illustrated pressure dependence of \tc ]{ An illustration of the pressure dependence of \tc for several $P=0$ doping states from underdoped to overdoped. Doping of the \cuo layer increases with pressure because of PICT but there is also an \emph{intrinsic} increase in $\tcmaxm$. }
	\label{fig:pict}
\end{figure}
		
		We illustrate these ideas in \fig~\ref{fig:pict}. \fig~\ref{fig:pict} shows the inferred change in \tc with pressure for several $P=0$ doping states from underdoped to overdoped.  Because of PICT the horizontal axis also measures $\Delta p$.  Although this behaviour has yet to be systematically mapped out experimentally, it can be inferred, especially for Ln123
, from published data \cite{mcelfresh1988, mori1991, sadewasser2000,  schillingchapter}.  
		
		

		An important question is what causes this \emph{intrinsic} increase in $\tcm$, as distinct from a variation in \tc because of a PICT doping variation?  One possibility is a reduction in fluctuations due to stronger $c$-axis coupling. An alternate possibility is that the enhanced polarisability of the material is important.  This idea, which results partly from the research presented in this thesis, is discussed further in \refsec~\ref{sec:polarisabilitydiscussion}.

	\subsection{Internal pressure}
	\label{sec:intpressuretheory}
	
There are two cuprate systems we have studied which show this decrease in \tc due to `internal pressure', Bi$_{2}$Sr$_{1.6-x}$(Ba,Ca)$_{x}$Ln$_{0.4}$CuO$_{6+\delta}$ (Bi2201) and LnBa$_2$Cu$_3$O$_{7-\delta}$ (Ln123) where Ln represents a member of the Lanthanide series (La, Ce, \ldots, Lu).  In Ln123 the increase in \tcmax as Lu123 goes to La123 has long been known \cite{veal1989, lindemer1994}.  The Ln ion-size increases from Lu to La \cite{shannon} and results in shorter in-plane bond lengths \cite{guillaume1994} and larger effective internal pressures on the CuO$_2$ layers \cite{marezio2000}.  
	
A recent paper by Gao \etal reports that \tc is increased by Ba substitution for Sr in Sr$_{2-x}$Ba$_x$CuO$_{3+\delta}$ \cite{gao2009}.  They find $\tcmaxm=98$K for $x=0.6$ - an impressive value for a single-layered cuprate. The authors correlate the \tc with the Cu(2)-O(2,3) bond length across a wide variety of single-layered cuprates and find their new compound fits the trend; materials with higher \tc values have longer Cu-O bond lengths.  Soon afterwards an article by Geballe and Marezio questioned these conclusions, citing issues with determining the correct lattice from these multi-phase materials \cite{geballe2010}.  Nevertheless, the work by Gao \etal appears to be a demonstration of the same ion-size effect that we are discussing\footnote{It would appear a PICT explanation of the increase in \tc cannot be excluded.}.
	
\begin{figure}
	\centering
		\includegraphics[width=0.7500\textwidth]{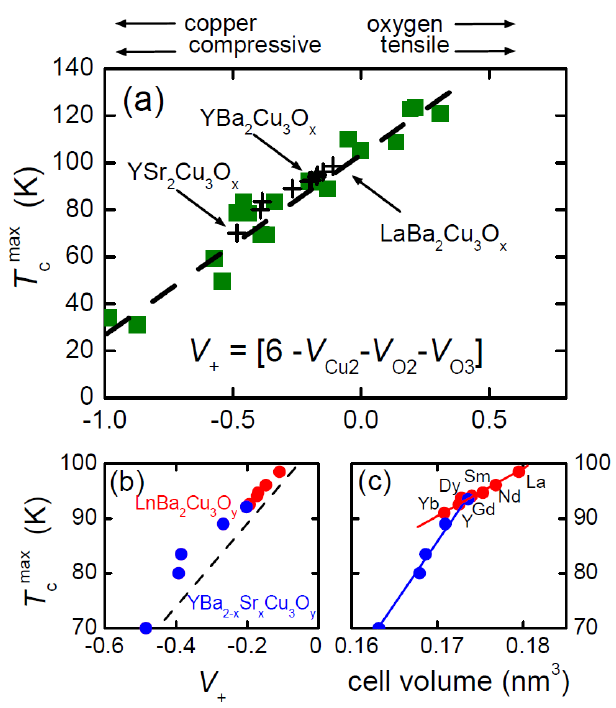}
	\caption[Correlation of \tc with the composite bond valence sum $V_+$]{(a) $T_c^{\textnormal{max}}$, plotted as a function of the bond valence sum parameter $V_+ = 6 - V_{\textnormal{Cu}(2)} - V_{\textnormal{O}(2)} - V_{\textnormal{O}(3)}$ as discussed in the text. Green squares are as previously reported in \cite{tallon1990}.  Black crosses are for LnBa$_2$Cu$_3$O$_y$ (Ln = La, Nd, Sm, Gd, Dy and Yb) and  YBa$_{2-x}$Sr$_x$Cu$_3$O$_y$ ($x = 0$, 0.5, 1.0, 1.25 and 2) and these same data are shown in more detail in panel (b). (c) These same data are shown as a function of unit-cell volume rather than $V_+$.}
	\label{fig:bvs}
\end{figure}

This `internal pressure' effect can been seen more broadly in the cuprates by considering bond valence sums \cite{tallon1990}; Fig.~\ref{fig:bvs} shows $T_c^{\textnormal{max}}$ plotted against the composite bond valence sum (BVS) parameter, $V_+ = 6 - V_{\textnormal{Cu}(2)} -V_{\textnormal{O}(2)} -V_{\textnormal{O}(3)}$, from reference~\cite{tallon1990} (green squares). 
	
BVSs are a strategy that crystallographers use to describe valence states of ions based on the degree of nearest-neighbour co-ordination around that ion.  A BVS is calculated using the formula;
	\begin{equation}
	V_i = \sum_j{\exp\left(\frac{r_0-r_{ij}}{0.37}\right)}
	\end{equation}
	\noindent where $r_0$ is a constant for a given anion-cation pair and tabulated by Brown and Altermatt \cite{brown1985}, while $r_{ij}$ is the inter-atomic distance.  Summing over all of the co-ordinated anions, for example, should return the formal oxidation state of a given cation.  Departures from this value indicate mixed valence but can also indicate stress on a bond.

$V_+$ is actually a composite BVS - combining BVS of O and Cu from the \cuo layer. In this case, $V_{\textnormal{Cu}(2)}$, $V_{\textnormal{O}(2)}$ and $V_{\textnormal{O}(3)}$ are the planar copper and oxygen BVS parameters and the plot reveals a remarkable correlation of $T_c^{\textnormal{max}}$ across single-, two- and three-layer cuprates. 
We may write $V_+ =(2-V_{\textnormal{O}(2)})+(2-V_{\textnormal{O}(3)}) - (V_{\textnormal{Cu}(2)}-2)$.  Hence $V_+$ is a measure of doped charge distribution between the Cu and O orbitals \cite{brown1985}. But $V_+$ is also a proxy for the stress on the Cu-O bond in the \cuo plane \cite{brown1989}, as noted at the top of the figure.  Broadly, a stretched \cuo plane has more positive $V_+$ and higher $ \tcmaxm $. However, as shown by \fig~\ref{fig:bvs}(c) it is more subtle than a simple dependence on volume or \cuo bond length as $V_+$ also includes contributions from the apical oxygen.

Crucially, this plot reveals that all cuprates follow a systematic behaviour. \emph{There are no anomalous outliers}. It is common to regard La$_{2-x}$Sr$_x$CuO$_4$ as anomalous due to its propensity for disorder. But the left-most data point in Fig.~\ref{fig:bvs} shows that it is entirely consistent with the other cuprates. It remains then to determine just what this $V_+$ parameter encapsulates so systematically.

To this plot we now add new data for the compounds, LnA$_2$Cu$_3$O$_{y}$, as Ln is varied and, in the case of Ln = Y, A = Ba$_{2-x}$Sr$_x$ where $x = 0$, 0.5, 1, 1.25 and 2 as \added{black crosses}. We use the structural refinements of Guillaume \etal \cite{guillaume1994}, Licci \etal \cite{licci1998}  and Gilioli \etal \cite{gilioli2000} and calculate $V_+$ in the same way as previously \cite{tallon1990}. Notably, the global correlation is also preserved across this model system, reflecting the progressive expansion of the lattice as ion size is increased.  \added{These same data are shown in more detail in panel (b) of \fig~\ref{fig:bvs}}. Using the bulk compressibility these volume changes may be converted to an effective internal pressure and \fig~\ref{fig:bvs} thus summarises a general feature of the cuprates, namely that internal pressure decreases $T_c^{\textnormal{max}}$. 

In contrast, as noted, it is well known that external pressure increases $T_c^{\textnormal{max}}$ \cite{schillingchapter}. It should be explicitly stated the analogy between external pressure and internal pressure is imprecise.  For example, large ion-size changes do not isotropically shorten lattice parameters as discussed previously, \ref{sec:ln123systematics}.  Nevertheless, such differences allow us to pose our question again; what is (are) the salient difference(s) affecting \tcmax between these two pressures in the cuprates?

For illustrative purposes, this question may be discussed within a weak-coupling BCS framework \cite{bcspaper}, which recent work has shown can describe the cuprates once the competing pseudogap phase, see e.g. \cite{tallon2001, wolf2004}, and superconducting fluctuations are considered \cite{tallon2011, tallonarXiv}. Specifically, the weak coupling relation for d-wave symmetry is;

\begin{equation}
\label{eq:bcsdwaveweak}
	k_B T_c^{\textnormal{mf}}=0.935\hbar \omega\!_B \exp \left[ \frac{-1}{\dosm V} \right]
\end{equation}

\noindent where \tcmf is the mean-field superconducting transition temperature, \(\omega\!_B\) is the pairing boson energy scale, \dos is the density of states (DOS) integrated around the Fermi surface at the Fermi-level and \(V\) the pairing potential.  

On the underdoped side the DOS is progressively depleted by the opening of the pseudogap \cite{storey2008, storey2008pggroundstate} (with energy scale $E_g$), and, on the overdoped side, is enhanced by the proximity of the van Hove singularity (vHs) \cite{storey2007}. A full treatment of the problem would therefore explore the comparative effects of  internal and external pressure on the key variables, \(\omega\!_B\), $V$, $\dosm$, $E_g$.  Internal pressure effects on fluctuations have already been investigated in the LnBa$_2$Cu$_3$O$_{7-\delta}$ system and found not to be significant compared with chain disorder \cite{williams1996}.

\section{Raman Spectroscopy and two-magnon scattering}

Raman scattering is the inelastic scattering of light.  Raman spectroscopy measures the intensity of this inelastic scattered light as a function of energy difference relative to the excitation source energy.  This energy difference is usually expressed as a frequency, $\omega$, and is conventionally quoted in units of inverse wavelength $\cmm=0.124$~meV.  Lasers are used to provide an intense source of well-defined energy.  Scattered light is collected by some optics and the elastically scattered light, which is several orders of magnitude larger than the Raman signal, is rejected using either a series of diffraction gratings and slits, or by holographic notch filters.  

What the cabin-feverish experimentalist in the Raman lab sees on their CCD (via some electronics) is a combination of noise, Rayleigh scattered light and, with luck, some \emph{inelastically scattered} light proportional to the scattering efficiency of the material, $\sfrac{\dd^2 \sigma}{\dd \Omega \dd \omega}$. \added{This expression for the scattering efficiency involves the scattering cross-section, $ \sigma $, per unit solid angle of detected light, $ \dd \Omega $, per unit frequency, $\dd \omega $, of a particular excited state.}

There are three basic components to $\sfrac{\dd^2 \sigma}{\dd \Omega \dd \omega}$ that we will discuss;
\begin{enumerate}
	\item Scattering from phonons.  Phonons generally result in sharply defined `peaks' in the spectra.  
	\item Electronic Raman scattering (ERS). ERS from intra-band excitations is generally a broad continuum reflecting the density of states.  However, features in the DOS, such as the pile up of states either side of a superconducting gap, or a van Hove singularity can be observed.
	\item Two-magnon Raman scattering.  In the cuprates this has a distinctive broad, asymmetric line shape.
\end{enumerate}

Generally, the scattering efficiency is expressed in terms of a susceptibility tensor\added{, $ \chi $,} that can be derived from, or at least reflects, the phononic, electronic or magnetic structure of the material.  \added{In general it is frequency dependent.}  The Raman tensor, $\mathbf{R}$, is one example of such a susceptibility tensor.  A feature of Raman spectroscopy that is important to this work is the ability to measure different elements of the susceptibility tensor, which correspond to different symmetries of the material, by adjusting the polarisation of both the incident and detected scattered light with respect to the crystal;
\begin{equation}
\frac{\dd^2 \sigma}{\dd \Omega \dd \omega} \propto |\mathbf{e}_i \cdot \mathbf{\chi} \cdot \mathbf{e}_s|^2
\end{equation}

\noindent where $\mathbf{e}_i$ and $\mathbf{e}_s$ are the polarisation unit vectors of the incident and scattered light respectively.

\subsection{Scattering from phonons}
The scattering efficiency for phonons can be expressed as \cite{weber2000};

\begin{equation}
\frac{\dd^2 \sigma}{\dd \Omega \dd \omega} \propto |\mathbf{e}_i \cdot \mathbf{\chi} \cdot \mathbf{e}_s|^2\left[ n(\omega,T)+1 \right] F(\omega)
\end{equation}

\noindent  where $n(\omega,T)=[\exp(\frac{\hbar \omega}{k_B T})-1]^{-1}$ is the \replaced{Bose-Einstein thermal occupation factor}{familiar Bose-Einstein distribution} \cite{cardona1999}.  $F(\omega)$ is the line shape of the phonon excitation.  Generally it is Lorentzian. However, \label{sec:fanolineshape} the Fano effect occurs when discrete excitations, such as a phonon with well-defined energy, coherently couple to continuous excitations \cite{cardona1999}.  It is due to interference between the two excitations.  In our particular case the phonons, particularly the phonon with \bog character at $\approx330$~\cm in Ln123, are coupled to the electronic continuum.

The effect results in a modified line shape of the discrete excitation.  This Fano line shape is described by;
\begin{equation}
 \label{eq:fano}
\frac{(q+\eta)^2}{1+\eta^2}
\end{equation}

\noindent where $q$ is the `Fano parameter' and $ \eta \equiv (\omega_p - \omega)/\Gamma $. $\omega_p$ is the peak centre, which is shifted from the energy shift of maximum intensity, \wmax, due to the asymmetry of the Fano line shape as $\omega _{\textnormal{max}}=\omega _p + \Gamma / \hbar q$, and $\Gamma$ is the half-width-at-half-maximum (HWHM) in units of energy. $\eta$ is dimensionless.

This can be compared with the Lorentzian line shape;
\begin{equation}
 \frac{1}{1+\eta^2}
\end{equation}

If $|q| \gg |\eta|$ the Fano line shape reduces to a Lorentzian shape.  $q$ is inversely proportional to the coupling between the discrete and continuous excitations.

\subsection{Electronic Raman Scattering}
\label{sec:ramaners}
Devereaux and Hackl \cite{devereaux2007} have written a comprehensive review covering the theory of electronic Raman scattering and exemplary results from this technique.

\begin{figure}
	\centering
		\includegraphics[width=0.45\textwidth]{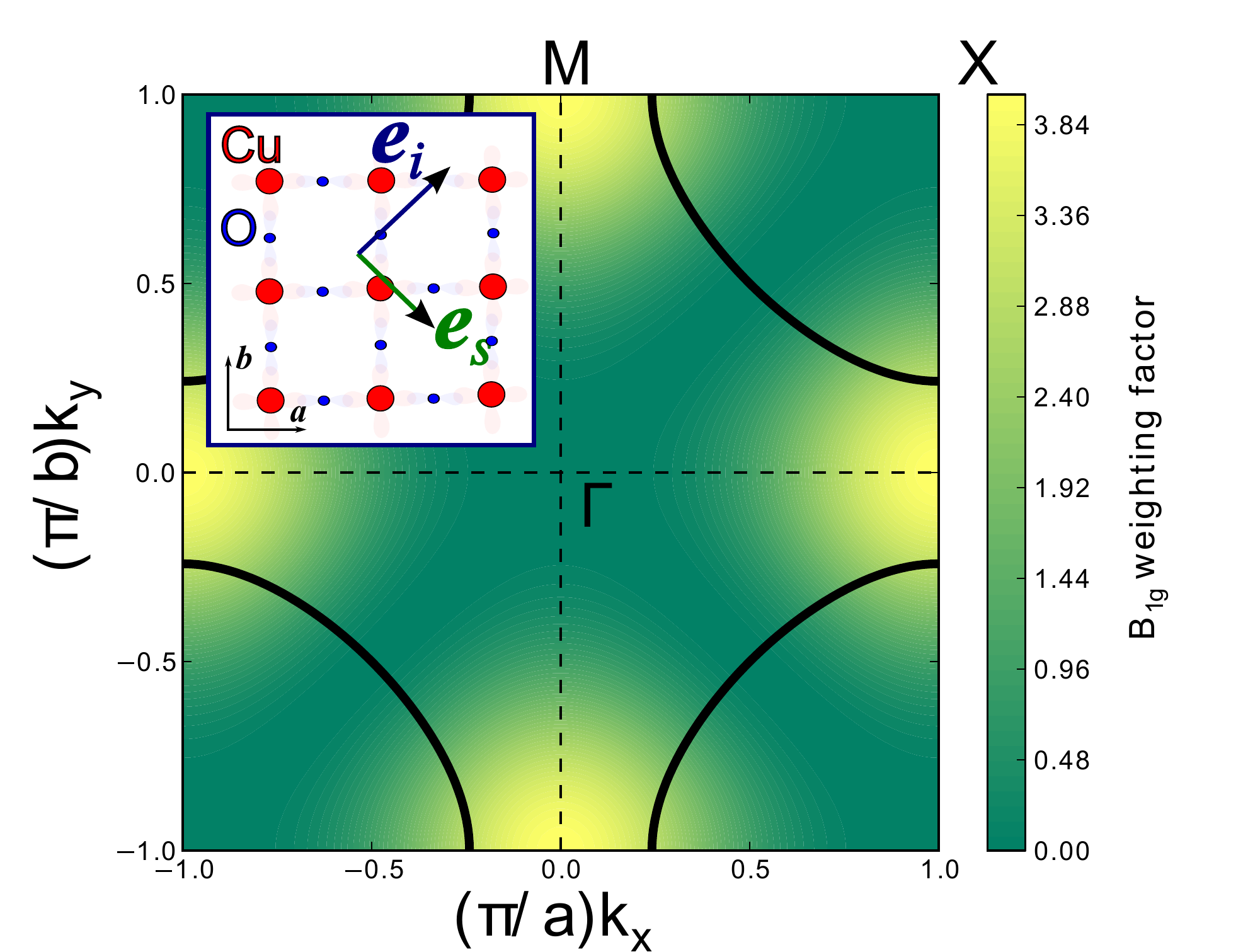}
		\includegraphics[width=0.45\textwidth]{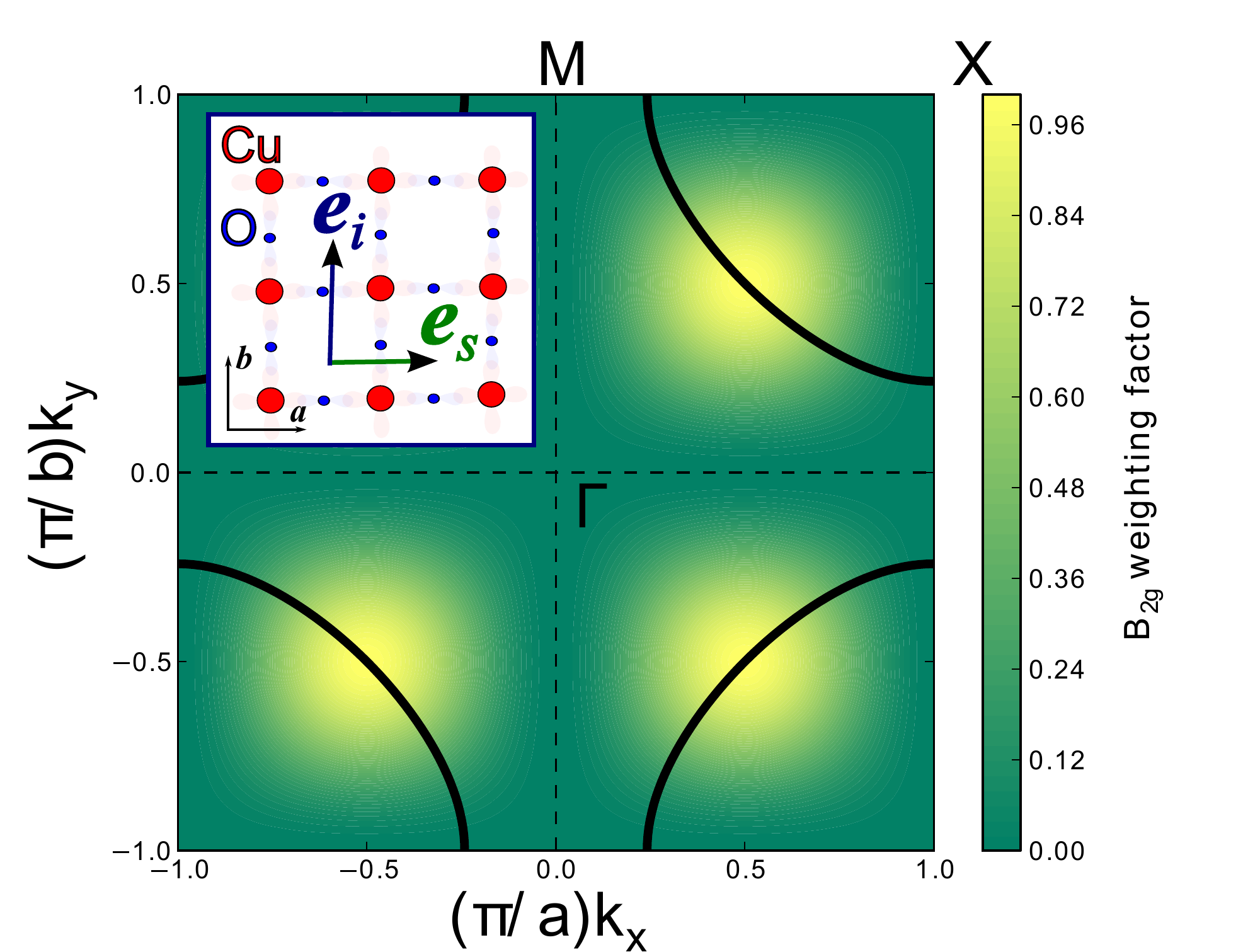}
	\caption[\bog and \btg scattering geometry factors]{The \bog and \btg scattering geometry factors superimposed on a representative Fermi-contour of the cuprates. The inset of each plot sketches the incident and scattered polarisation direction with respect to the Cu and O ions in the \cuo layer.  \bog measures scattering preferentially from the anti-nodes, whilst \btg measures nodal scattering.}
	\label{fig:b1gb2gweight}
\end{figure}

The scattering efficiency is related to the imaginary part of the Raman response function $\chi_{\gamma \gamma}(\qq, \omega)$ by \cite{weber2000};
\begin{equation}
\frac{\dd^2 \sigma}{\dd \Omega \dd \omega} = -\left(\frac{\omega_s}{\omega_0}r_0^2 \frac{\hbar}{\pi}  \right) \left[ n(\omega,T)+1 \right] \chi''_{\gamma \gamma}(\qq, \omega)
\end{equation}

\noindent where $\chi''_{\gamma \gamma}(\qq, \omega)$ is the imaginary part of the susceptibility tensor $\chi = \chi' + i\chi''$ and is the quantity of interest for probing the redistribution of spectral weight due to the opening of a gap, \specgap. Thus, when comparing spectra we must multiply them by $\left[ n(\omega,T)+1 \right]^{-1}$ using the appropriate temperature for the spectra. $\gamma$ is the so called Raman vertex that weights what region of the electronic dispersion is measured and depends on the scattering geometry.  
Two important examples are \bog and \btg scattering geometries and are illustrated in \fig~\ref{fig:b1gb2gweight};

The \bog vertex has the form
\begin{equation}
\gamma_{\bogm}(\kk) = \gamma_{\bogm} (\cos k_x - \cos k_y)
\end{equation}

\noindent and the \btg vertex has the form 
\begin{equation}
\gamma_{\btgm}(\kk) = \gamma_{\btgm}\sin k_x \sin k_y
\end{equation}

These two functions are shown in \fig~\ref{fig:b1gb2gweight} as false-colour plots superimposed on a representative cuprate Fermi-surface. 

When including a superconducting gap function, \scgapk, into the expression, the imaginary part of the Raman response at $T=0$ is the density of states weighted by a scattering geometry factor, $\gamma$, and scaled by a factor containing \scgapk \cite{storey2007fermi};
\begin{equation}
\chi''(\omega)= \int{ \frac{\dd ^2 k}{4\pi^2} \partial(\omega - 2E(\kk))\frac{|\Delta(\kk)|^2}{E(\kk)^2}|\gamma(\kk)|^2  }
\label{eq:ramanresponse}
\end{equation}

\noindent where $E(\kk)=\sqrt{\ekm^2 + \scgapkm^2}$ and $\ekm$ is the bare normal-state electronic dispersion.

\begin{figure}
	\centering
		\includegraphics[width=1.00\textwidth]{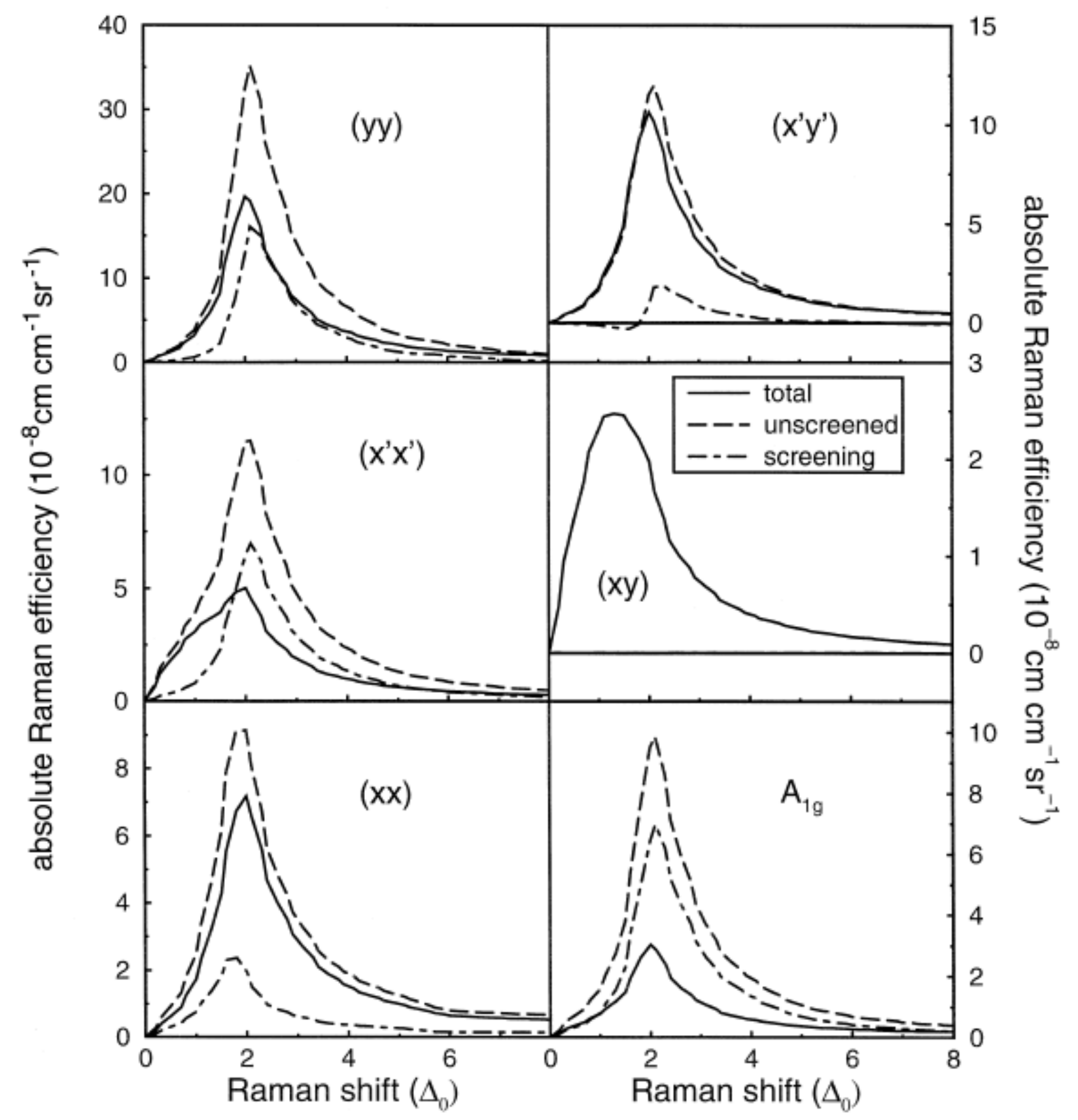}
	\caption[Theoretical electronic Raman scattering efficiency of superconducting Y123]{ A figure from \cite{strohm1997} showing theoretical electronic scattering efficiencies obtained by BZ integration for optimally-doped Y123 in the superconducting state.  Each of the six panels contains the calculated total, unscreened and screened absolute ERS efficiency (quantities that are difficult to deduce experimentally).}
	\label{fig:cardona1999y123scatteringefficiencies}
\end{figure}

Scattering efficiencies in various geometries in the superconducting state were calculated by Strohm and Cardona \cite{strohm1997} and their \fig~2 is reproduced here in \fig~\ref{fig:cardona1999y123scatteringefficiencies}.

\subsection{Two-magnon scattering}
\label{sec:twomagtheory}

A magnon is a propagating spin-excitation.  Alternatively in a classical picture they can be thought of as spin-waves.  A magnon can be thought of as the spin analogue to phonon-related ion-displacements.  A damped magnon, as would be the case resulting from spin-flip scattering from charge carriers, is called a paramagnon.  As discussed, in the cuprates these spins are coupled and in the case of nearest-neighbour coupling with energy $\sim J$. Such coupling is necessary for a spin-excitation to propagate through the spin-lattice. $J$ is of interest to us here and it can be measured using Raman spectroscopy\replaced{.}{!}   

\begin{figure}[htbp]
	\centering
		\includegraphics[width=1.00\textwidth]{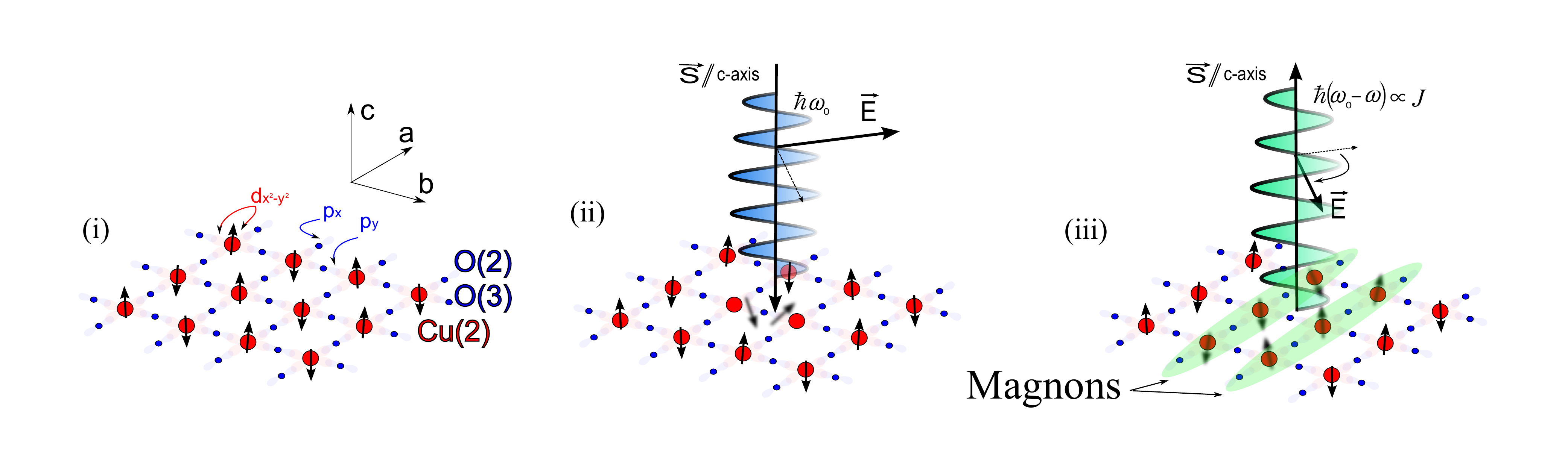}
	\caption[Schematic diagram of the two-magnon scattering process]{Schematic diagram of the simplified \bog two-magnon scattering process; (i) the undoped \cuo layer has long-range anti-ferromagnetic order. (ii) An absorbed photon excites a spin to doubly occupy an adjacent site.  (iii) Relaxation occurs by the opposite spin filling the empty site accompanied by the emission of a red-shifted photon.  Two magnons are created from this two-spin flip process with a rotation in polarisation of the Raman scattered photon and energy loss proportional to $J$.}
	\label{fig:twomagsketch}
\end{figure}

The technique of Raman two-magnon scattering is illustrated in \fig~\ref{fig:twomagsketch}.  An absorbed incident photon from the laser excites a spin to doubly occupy an adjacent site.  Relaxation occurs by the opposite spin filling the empty site accompanied by the emission of a red-shifted photon.  Two magnons are created from this two-spin flip process with a rotation in polarisation of the Raman scattered photon and an energy loss proportional to $J$.  Because two magnons are created only the sum of their crystallographic momenta $\kk$ must satisfy the conservation of momentum requirement that $\kk_i \approx \kk_s$ for Raman spectroscopy, where $\kk_i$ and $\kk_s$ are respectively the incident and scattered light wave vectors.  This is the reason why we are able to measure spin excitations that predominantly occur around $(\pi,0)$ in the Brillouin zone. 

The most common and tractable theoretical treatment of Raman scattering from magnons is the Loudon-Fleury theory\footnote{This is also referred to as the Fleury-Loudon-Elliott theory\cite{freitas2000}} \cite{loudonfleury}. An extremely useful\deleted{n} result for us relates the Raman shift, commonly denoted $\omega$, of the peak maximum with the antiferromagnetic exchange constant, $\omega_{max}\approx2.75J_{\textnormal{eff}}\approx3.2J$. A similar relation is found from several, more modern, theories of two-magnon scattering \cite{chubukov1995, li2012feedback} and so we use the position of the two-magnon scattering peak to estimate $J$ in our work.

\label{sec:cupratetwomag}

\begin{figure}
	\centering
		\includegraphics[width=0.66\textwidth]{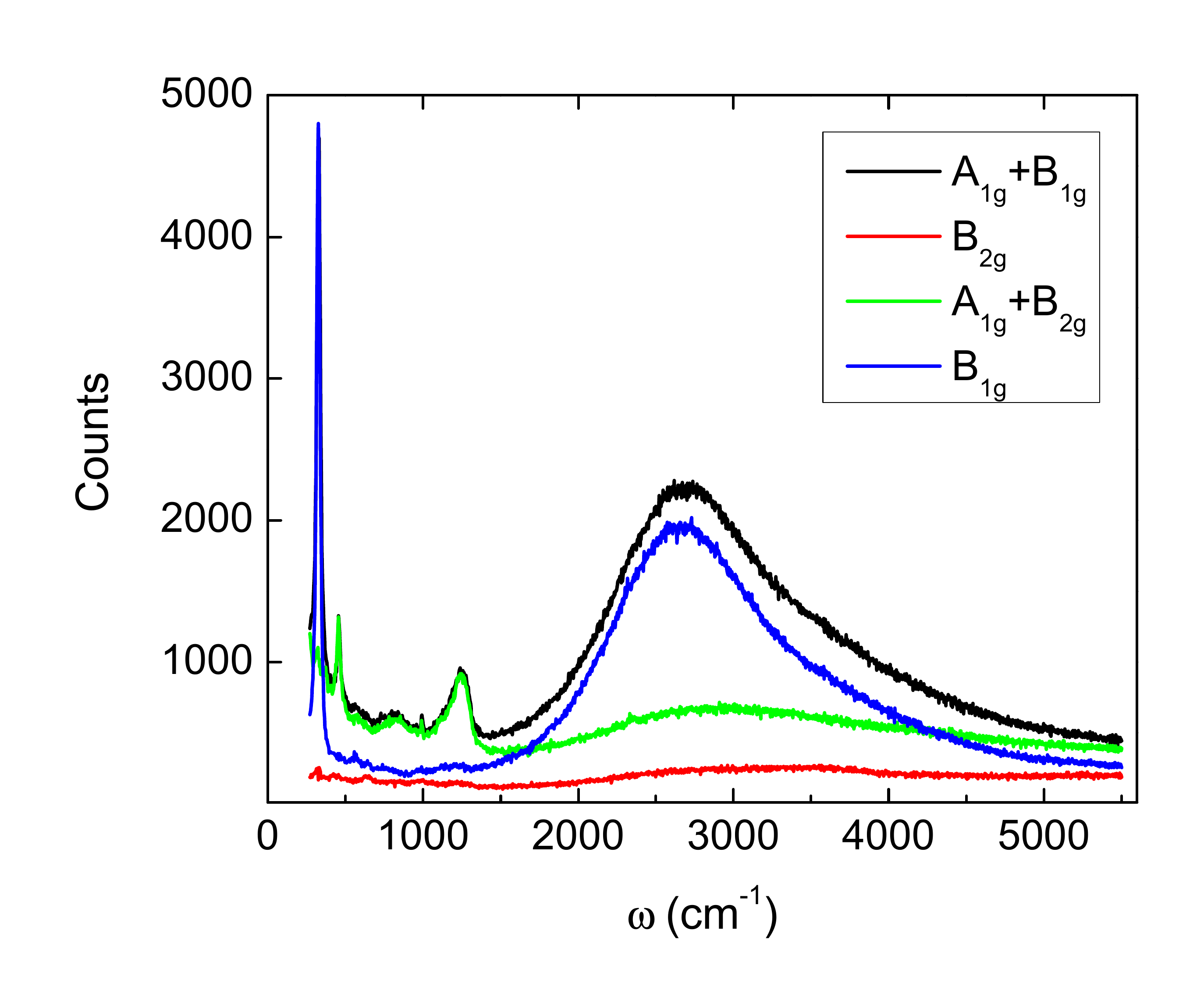}
	\caption[Typical Raman spectra showing two-magnon scattering]{Typical Raman spectra showing two-magnon scattering. The material is the undoped SmBa$_2$Cu$_3$O$_6$ and the scattering geometries $ \aogbogm $, $ \btgm $, $ \aogbtgm $ and $ \bogm $ are indicated in the legend.}.
	\label{fig:twomagallsymm}
\end{figure}

Typical experimental Raman spectra that show a two-magnon scattering response around $ 3000 $~\cm are shown in \fig~\ref{fig:twomagallsymm} for the \emph{un}doped Sm123O6. Also visible are several phonon modes below $1500$~$ \cmm $ which are not of immediate interest to our two-magnon studies.  A cuprate two-magnon peak\footnote{It is possible to talk about \emph{a} cuprate two-magnon peak as the position, width and shape of the peak is similar for all cuprate materials, see \cite{freitas2000} and references therein.} has the following basic features;

(i) \textit{Position} The maximum intensity is $\wmaxm \sim 3000$~cm$^{-1}$ with the \wmax depending mainly on the doping (presumably by spin-flip scattering) and super-exchange path length: the Cu(2) to O(2,3) to Cu(2) bond length.  Using the result $\wmaxm\approx 3.2J$, this corresponds to a nearest-neighbour superexchange energy of the order $J\sim 1090$~cm$^{-1}$ $=1570$~K $=0.135$~eV. 

(ii) \textit{Laser wavelength dependence} Raman experiments commonly use lasers in the visible light frequencies whose energy, $\sfrac{hc}{\lambda_i}\approx 2.5$~eV, is close to the charge-transfer band gap low-energy edge, $\approx 1.8$~eV \cite{lee2006,tohyama2002}.  While the shape of the peak changes little with these photon frequencies \cite{singh1989,sulewski1990}, the intensity of the peak is seen to go through a single maximum \cite{chubukov1995}.  The position of this maximum (in incident photon-frequency space) is close to the upper rather than lower edge of the charge transfer gap - an unexpected experimental result which may be explained by a triple-resonance mechanism \cite{chubukov1995} or an excitonic mechanism \cite{hanamura2000} or by an exact (computational) treatment of the Hubbard Hamiltonian \cite{tohyama2002} which argues that the resonant enhancement of the two-magnon scattering intensity results from the spatial distribution of charge carriers created from photo-excitations.  

(iii) \textit{Asymmetry} Higher frequencies are better represented leading to a generally asymmetrical peak-shape.  The peak-shape changes little with incident photon wavelength between 666~nm and 454~nm \cite{chubukov1995, freitas2000} .

(iv) \textit{Width} The two-magnon peak is very broad, extending $\sim 1000$ \cm either side of the maximum.

(v) \textit{Polarisation selection} In both $A_{1g}$ and $B_{1g}$ scattering geometries the two-magnon peak is observable. \added{The intensity in $A_{1g}$ was measured by Lyons \etal to be $\sim 50$\% of the $B_{1g}$ intensity in La$_{2}$CuO$ _{4}$ \cite{lyons1989}. Our own results on Ln123O6 do not show such intense two magnon scattering in $A_{1g}$ however it is still observable, see \refsec~\ref{sec:a1g}.   Here the intensity is the measured counts per-second and the $A_{1g}$ spectra are constructed from subtracting either \bog or \btg spectrum from combined \aogbog or \aogbtg spectrum (see \refsec~\ref{sec:twomagmeasurements}) respectively that have been measured under identical experimental conditions.}
Two-magnon scattering is not seen for $B_{2g}$ geometry.   

The Loudon-Fleury theory has several deficiencies in the description of two-magnon scattering in the cuprates.  Firstly it cannot describe the fact that comparable two-magnon intensity is seen in $A_{1g}$ geometry as well as $B_{1g}$. It cannot describe the large width of the peak and its asymmetry is also outside the scope of this theory. Also curious is that in the incident photon frequency dependence of the two-magnon peak intensity, a single maximum is located $\sim 1$~eV above the lower charge-transfer gap edge \cite{chubukov1995} - placing it at the top of the charge-transfer band.  

Work on extensions to Loudon-Fleury theory and novel approaches have been partially successful in explaining the two-magnon peak of the cuprates. See the excellent summary of \cite{freitas2000} and recent work by Li \etal \cite{li2012feedback} and Piazza \etal \cite{piazza2012}.  
\section{Density Functional Theory}


Density Functional Theory (DFT) is an exact method for calculating the ground state of a system \emph{ab initio}.  Practical implementation of DFT however requires several approximations which we will now discuss.

We want to calculate how the ground state (lowest energy) electronic and crystal structure is affected by ion-size substitution.  To do this we need to solve Schr\"{o}dinger's equation for the specific materials to find the wavefunction, $\Psi$, that is a solution for the lowest energy, $ \eminm $;

\begin{equation}
H\Psi = E \Psi
\label{eq:schrodinger}
\end{equation}

We use the result from the Hohenberg-Kohn-Sham theorem which proves that the ground-state energy of a physical system is a unique functional of the particle density, $n_0(\rr)$.  The particle density is given by $n(\rr)=|\left< \Psi|\rr \right>|^2$, hence $n_0(\rr)\Leftrightarrow \Psi_0$ and the origin of the name Density Functional Theory. However, solving the equations for the many-body interactions is intractable for all but the simplest systems. The Kohn-Sham approach to overcome this difficulty is to reduce the interacting many-body system to a single-electron problem with an effective exchange-correlation functional of $n(\rr)$, \eex \cite{kohn1965, hafner2007, hafner2008}. The energy of the system, $E$, can then be expressed as a functional of the particle density, $n(\rr)$, which can be iteratively refined to find the lowest energy (within a specified range). $E[n]$ is written as a sum of various terms;
\begin{equation}
E[n] = T[n]+ E^H[n]+ \eexm + \int{V(\rr)n(\rr)\dd \rr}
\label{eq:energyfunctional}
\end{equation}
\noindent where $T[n]$ is the kinetic energy functional, $E^H[n]$ an electron-electron repulsion energy term, $V(\rr)$ is an external potential and \eex is the exchange-correlation energy which is included to account for reformulation of the many-bodied Schr\"{o}dinger's equation to a single-electron one.  Of these four components of the Hamiltonian the exact form of the kinetic energy and exchange energy is not known.  

Next a method called the Local Density Approximation (LDA) is used.  The LDA method is to divide the system into small enough regions such that each region can be thought of as non-interacting, homogeneous electron gas. For example, for a non-interacting, homogeneous electron gas the kinetic energy functional is known from the Thomas-Fermi theory;

\begin{equation}
T[n] = \int{\alpha n(\rr)^{\sfrac{5}{3}}\dd\rr}
\label{eq:thomasfermi}
\end{equation}
\noindent \added{where $ \alpha = \sfrac{3\hbar^2}{10m} (3\pi^2)^{2/3} $ and $ m $ is the mass of a free electron.}

There are many possible forms of $\eexm$, each with the purpose of exactly reformulating the many-electron Schr\"{o}dinger's equation. One of the simplest is an LDA approach such that \eex is an integral over space of the exchange-correlation energy of a homogeneous electron gas of density $n(\rr)$, $\eta_{\textnormal{ex}}[n(\rr)]$ \cite{hafner2007, hafner2008};

\begin{equation}
\eexm = \int{n(\rr)\eta_{\textnormal{ex}}[n(\rr)] \dd \rr}
\label{eq:exchange}
\end{equation}
\noindent where the exchange correlation energy can be expressed as $\eta_{\textnormal{ex}}[n(\rr)]= -\frac{3e^2}{4\pi}(3\pi ^2 n(\rr))^{\sfrac{1}{3}}$. 

A further correction to this is the L(S)DA+U approach, which includes a Hubbard-like on-site repulsion between electrons.

\eq~\ref{eq:thomasfermi} and \ref{eq:exchange} show that even if the local detail of $n(\rr)$ is not faithfully reproduced in our calculations, the LDA can still predict the correct ground state provided that the integrals over space are accurate. Put another way, `first-order' errors in the exact form of $\eta_{\textnormal{ex}}$ become `second-order' errors in the energy \eex because of the integration over space. 

In the Kohn-Sham approach $\Psi$ is written as a weighted sum of (orthogonal) basis functions, $\psi_i$; $\Psi = \sum_i{c_i\psi_i}$. There are a variety of options for the appropriate choice of basis functions, for example one could choose basis functions that are believed to represent physical states in the crystal, such as atomic orbital-like basis functions, e.g. \dxy.  Such basis functions may not be easiest to work with computationally - leading to slower code - but may have a clearer physical interpretation.  Computationally the easiest basis functions to work with for a periodic crystal are plane waves, $\psi_i=\exp(-i\kk_i.\rr)$.  Plane waves are implemented in the Vienna ab initio Simulation Package (VASP), which is a computer program that implements DFT calculations\footnote{VASP is able to do other simulations as well, such as Molecular Dynamics (MD), see the VASP homepage for details http://cms.mpi.univie.ac.at/vasp.} \cite{kresse1993, kresse1994, kresse1996,  kresse1996a}.  

An issue with using plane-waves as a basis set is that a large number are needed to accurately describe $\Psi$ close to ion sites because $\Psi$ changes rapidly there. A large basis set leads to long computation times. One solution - and the one we use here - is to replace the `true' potential around an ion site with a pseudopotential that nevertheless reproduces the `true' potential outside of a specified cut-off radius from the ion site. 

In particular we use the GGA-PW91 pseudopotentials in the VASP library \cite{perdew1992, perdew1993}.  These are Projector Augmented Wave (PAW) pseudopotentials (see e.g. \cite{bloch1994, kresse1999, hafner2008}), which are a development of the basic pseudopotential idea that more accurately accounts for the potential inside the cut-off radius, with the Generalised Gradient approximation. PAW+GGA is a good compromise between accuracy and calculation speed.


%
%
%


\section{Disorder, $ \tcm $, and the superfluid density}
\label{sec:sfintro}		
		

The superfluid density, $ n_s $, \added{is the density of Cooper-pairs in a superconductor. $ n_s $ is directly related to the London penetration depth, $\lambda$, which is the length scale over which magnetic flux can penetrate into the superconductor.  These quantities are related as} \cite{ashcroftmermin}:

\begin{equation}
n_s = \frac{m^*}{\mu_0 e^2} \lambda^{-2} \Rightarrow \lambda^{-2} \propto \frac{n_s}{m^*}
\label{eq:superfluid}
\end{equation}

\noindent \added{$ e $ is the charge of an electron, $ \mu_0 $ is the permeability of free space} and $m^*$ is the effective mass of a carrier and cannot be determined from $ \mu $SR.  Also, the cuprates are extremely isotropic so that the $c$-axis penetration depth is much greater than the $ab$-axis penetration depth,  $\lambda_c \gg \lambda_{ab}$. In polycrystalline samples where one gets and average over the $c$ and $ ab $ directions, we measure $\lambda^{-2} \approx \lambda_{ab}^{-2}$ and this is the quantity we refer to throughout as the `superfluid density', as is common in the literature, though strictly it is only proportional to $n_s$.	 Finally, often we discuss the superfluid density in the low temperature limit, $\lambda^{-2}(T=0)\equiv \nsm$.

		 The superfluid recruits its electrons from a fraction of the normal-state electrons. Indeed, from \cite{tallon2003superfluid} we quote the relation, which is valid below $p=0.19$:
\begin{equation}
\nsm = \mu_0e^2\left\langle v_x^2 N(E) \right\rangle
\label{eq:superfluidfermi}
\end{equation}

\noindent where $v_x$ is the projection of the Fermi-velocity supercurrent direction and we average the density of states, $N(E)$, over an energy $\pm \Delta_0$ about $E_F$. Modifications to the density of states are thus reflected in \ns and therefore both stripe-order \cite{bernhard2001}, the pseudogap \cite{bernhard2001, tallon2003superfluid, khasanov2008} have observable effects on \ns and are discussed below.
		 
		 In addition, disorder and pair-breaking effects suppress $ \nsm $.  In fact, it is well established that \ns is especially sensitive to disorder \cite{sunmaki, bernhard1996, tallon2005} in the cuprates due to the phase symmetry of the superconducting order parameter. 

    The pioneering work of Uemura \etal \cite{uemura1989} found a correlation between \tc and \ns in the underdoped cuprates; $\rho_s \propto T_c$ with the same gradient for six different cuprates.  This linear relation has come to be known as the Uemura line.  Since this 1989 paper, Uemura has shown a correlation between \tc and \ns for a wide variety of unconventional superconductors, e.g. \cite{uemura1991,uemura2003,uemura2009} and he argues these observations imply real-space-paired bosons rather than the $\kk$-space-paired Cooper pairs. 
    
However, $\rho_s \propto T_c$ is only accurate for well underdoped cuprates and the proportionality constant differs between members of the cuprates \cite{tallon2003superfluid}.  Indeed, the relationship between \tc and \ns as a function of increasing doping proves to not be linear \cite{tallon2003superfluid, sonier2007}. Close to optimal doping \ns increases faster than \tc and in the overdoped region both \ns and \tc decrease again (the exception is \ybco due to a \ns contribution from oxygen rich, superconducting CuO chains).  Plotting \tc against \ns results in a horseshoe, or boomerang, shape \cite{bernhard1994}.
    
The authors of \cite{tallon2003superfluid} show that instead, across three disparate cuprates from under- to over-doping, 
	\[
	\lambda_0^{-2} T_c \propto S(T_c) \propto z_{crit}
	\] 
\noindent where $S(T_c)$ is the electronic entropy at \tc and $z_{crit}$ is the Zn concentration required to just suppress \tc to zero.  This succinct result shows the importance of the pseudogap energy on the superconducting properties by tying \ns to the electronic entropy, a ground state property that is suppressed by the pseudogap \cite{loram1994}.  The `Uemura line' of proportionality between \tc and \ns is then understood rather as a consequence of the decreasing pseudogap energy with increasing doping, which results in sublinear behaviour.

Transverse-field muon spin relaxation (TF-\musr) is an ideal technique to accurately determine the London penetration depth, $\lambda$ \cite{uemura2003}, and this technique will be discussed in \refsec~\ref{sec:musr}.

\section{Pairing mechanisms}
\label{sec:pairingmechanisms}

A fascinating physical aspect to superconductivity is that below $T_c$, electrons suddenly overcome their mutual repulsion, pair together and with phase coherence between pairs, adopt a single quantum mechanical wave function of macroscopic dimensions! The physics that causes them to pair is referred to as the pairing mechanism.

In the low-temperature superconductors it is (almost \cite{hirschbcs}) universally believed that phonons mediate a retarded attractive interaction between metallic electrons as laid out by Bardeen, Cooper and Schrieffer (BCS theory) \cite{bcspaper}.  For the cuprates there are complicated and sophisticated theories covering a wide variety of alternative possible pairing mechanisms as well as extensions to the Eliashberg-BCS theory, see for example the review article by Wolf \etal \cite{wolf2004}. 

In the same year as Bednorz and M\"{u}ller discovered the first high-\tc cuprate \cite{bednorz1986}, three papers were published showing models in which spin-mediated electron pairing occurred with a resulting  $d$-wave superconducting state \cite{miyake1986, scalapino1986, beal1986}.  The authors did not mention this could be the mechanism in the just discovered cuprates\footnote{They were instead seeking to understand superconductivity in the heavy-Fermion superconductors and organic superconductors, the ``Bechgaard salts''.} but it was soon realised that this spin-wave mediated pairing was a good candidate for superconductivity in the cuprates, e.g. \cite{bickers1987}.  These papers find a positive pairing potential (by convention, that is one that leads to electron pairing) close to an anti-ferromagnetic, but not ferromagnetic\footnote{Apart from some obscure possibilities, see footnote in \cite{miyake1986}.}, spin density wave (SDW) state.  Superconductivity occurs in the even-parity, aniostropic singlet channel and the gap function has \dxy symmetry.

In most unconventional superconductors, superconductivity occurs close to a magnetic instability - that is a point in phase space close to where some form of magnetic ordering occurs.  These magnetic-to-superconducting transitions are tuned in different materials by altering either the doping state, pressure or the magnetic field \cite{sachdev2011}.   The similarity of the magnetic excitation spectrum for both the cuprates and pnictides has been interpreted as evidence for a pairing mechanism of magnetic origin \cite{uemura2009, yu2009magneticresonance, letacon2011, scalapino2012}.  

There appears to be a broad consensus now that the pairing mechanism in the cuprates is probably mediated by a spin-fluctuations \cite{moriya2003, abanov2008, uemura2009, yu2009magneticresonance, letacon2011, li2012feedback, scalapino2012}.  However, a continuing challenge has been the apparent low spectral weight of associated spin fluctuations, as measured by inelastic neutron scattering \cite{maksimov2010}. Recent studies using inelastic x-ray scattering seem to locate the missing weight by identifying intense \emph{para}magnon (damped magnon) excitations across the entire superconducting phase diagram  \cite{letacon2011}.  Furthermore, with Eliashberg techniques (an extention to BCS theory) le Tacon \etal calculate the \tc of \ybco to be $\approx 170$~K  using their experimentally determined  paramagnon dispersion - easily accounting for the experimental value $\tcm = 90$~K, especially if one recognises that \tcmf may be significantly above $\tcm$.   

\emph{In this situation, a key energy scale for pairing is, to leading order, the antiferromagnetic exchange interaction, $J$ }\cite{abanov2008, letacon2011}.

Pairing mechanisms are discussed further at the end of the thesis, \refsec~\ref{sec:pairingmechanismdiscussion}.

\chapter{Techniques and Samples}
\section{Introduction}
\label{sec:techniquesintro}
\label{sec:cupratesythnandchar}

The purposes of this chapter are to (i) explain the techniques and methodologies used throughout this thesis and (ii) present some basic data on the samples we have studied so that it does not encumber the later chapters. 

\begin{figure}
	\centering
		\includegraphics[width=1.0\textwidth]{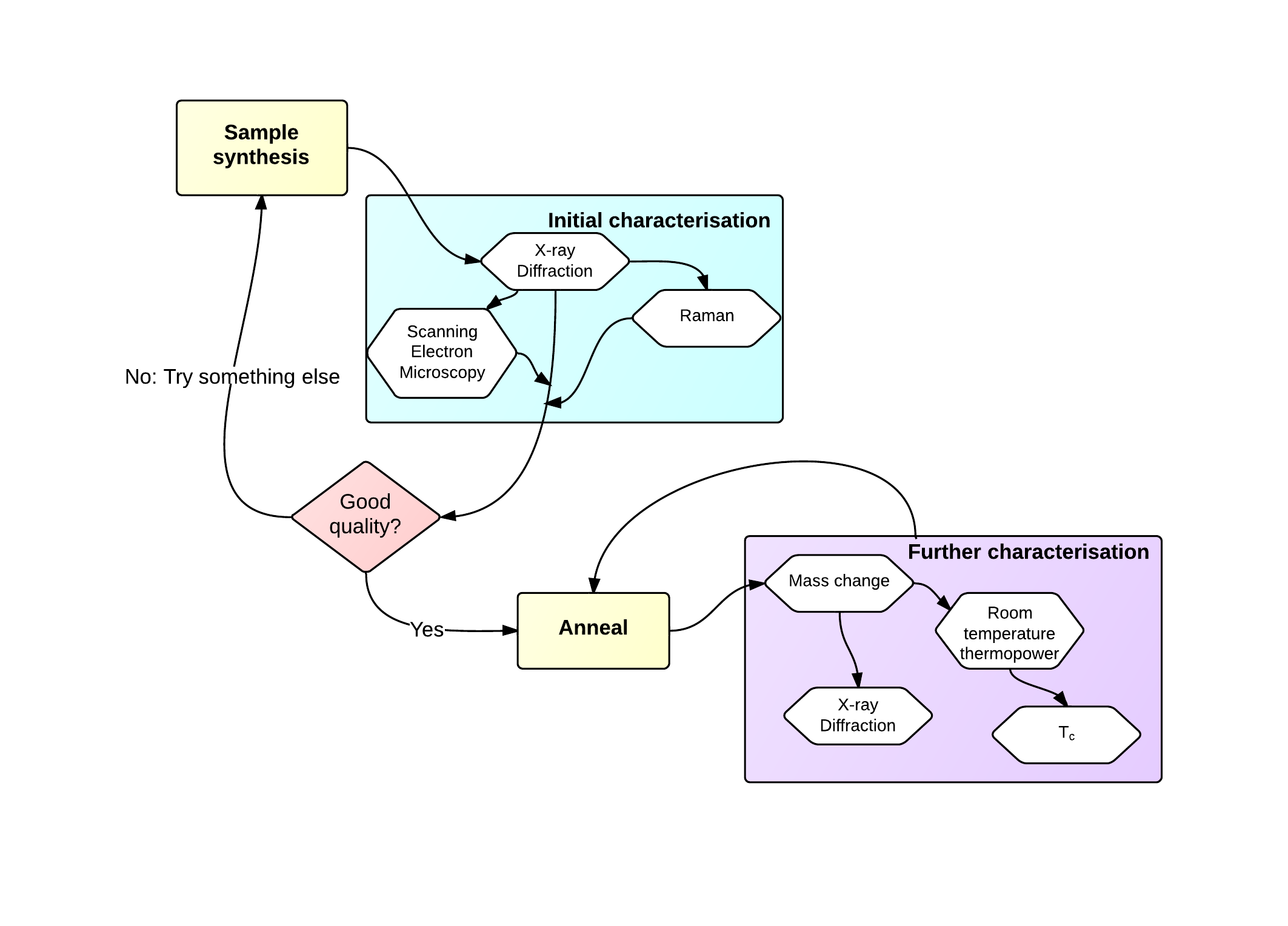}
	\caption[Standard sample synthesis procedure.]{An illustration of the standard procedure followed when making new samples - especially polycrystalline samples.}
	\label{fig:standardops}
\end{figure}

\fig~\ref{fig:standardops} illustrates the standard synthesis and characterisation procedure used for most of our samples.  Each of the steps is discussed in the following sections.

\section{Sample synthesis}

\subsection{Solid-state synthesis}
\label{sec:solidstatesynth}

Solid-state synthesis is a common technique for making polycrystalline material.  It involves mixing and then reacting solid precursor materials to form your material.  For the cuprates, the precursor materials are almost always metal- oxides, nitrates, or carbonates as these are stable at room temperature.  To make these materials react, one must (i) increase the temperature (i.e. increase their thermal energy) (ii) decrease the separation between the precursor agents (which reduces the energy barrier for reaction) and (iii) perhaps alter the gas under which they are reacted.  As implied by the name of the technique, the temperature should be kept (just) below the melting temperature of the constituent materials so that the precursor does not melt into the sample holder and upset the stoichiometry of the constituent metals.  Thus to decrease the separation between the precursors we grind them together, in a mill or with mortar and pestle, to a fine powder and then press the powder to a pellet under high pressure in a stainless-steel die. This step is often repeated 3-4 times during synthesis.  The chemistry may be specific to the materials involved, but generally it seems high O$_2$ content in the reacting-gas increases the reaction rate  - which is not always desirable and sometime leads to unwanted solid-phases. \added{From experience, the melting temperature of a cuprate is approximately 1200\degc in O$_{2}$ and lower in lower oxygen partial-pressure.}

By way of example, polycrystalline YBaSrCu$_3$O$_{7-\delta}$ pellets are synthesized by starting with high-purity Y$_2$O$_3$, BaCO$_3$, SrCO$_3$, CuO powder\footnote{Which are commercially available from e.g. Sigma Aldrich.}. 
\begin{enumerate}
	\item The powders are weighed out so that there is the desired 1:1:1:3 molar ratio of Y:Ba:Sr:Cu. The oxygen content is controlled later by the annealing process, \refsec~\ref{sec:annealingprocess}.  
	\item The powders are then thoroughly mixed with mortar and pestle. Isopropanol can be added at this stage to aid mixing of the powders.  We have also used acetone instead of isopropanol, particularly for the synthesis of NdBa$_2$Cu$_3$O$_{7-\delta}$, with no observable change in the results.  The powder is then dried and pressed at 20-30~MPa in a stainless-steel die to produce pellets of 10~mm diameter. 
	\item Next the pellets are `decomposed' at 750\degc under dry air, a process which removes much of the C and excess O. Normally the pellets are placed on a Yittria-stabilised Zirconia container when put into the furnace and preferably seated on an inert substrate such as MgO.
	\item Next the pellets are broken and ground to a fine powder with mortar and pestle. As before, isopropanol can be added at this stage.  The powder is pressed into pellets again and this time reacted at 950\degc in dry air. For \ybasr there is no particular temperature ramping sequence. Other materials can require careful temperature sequences, which are generally found in the literature. Even so, there is almost always an period of trial-and-error before the appropriate synthesis conditions are found.
	\item The previous step is repeated 2-4 more times.  Each repetition increases the homogeneity of the final product. Sometimes these reactions are at incrementally higher temperatures and for incrementally longer times.
\end{enumerate}

Generally the synthesis process - reaction temperatures, times, O$_2$ partial pressure - for our materials, or similar materials, can be found in the literature.  As mentioned above however, there is almost always a period of trial-and-error before the appropriate synthesis conditions for us to make the material are found.  These depend on such things as the particle sizes in the precursor powders.

For Bi2201 we were in such a position.  

We prepared samples of Bi$_{2-l}$Pb$_{l}$Sr$_{1.6-x}$A$_x$Ln$_{0.4}$CuO$_{6+\delta}$ for A = Ba or Ca and Ln = La, Nd, Sm, Eu, Gd.  For $l=0$ we make Ba with $x=0.1$ and $0.2$ and Ca for $x=0.2$.  With $x=0$ and Ln = La we also make samples with $l=0.0$, $0.2$, $0.35$ and $0.45$. Typically we react at 840\degc in a dry air atmosphere.  We vary the doping state, $p$, via the oxygen content, $\delta$, by annealing at various temperatures and oxygen partial pressures.  By doing so we can achieve a maximum \tc of 30~K (the onset of diamagnetism is 2~K higher than this).  The highest reported \tc for Bi$_2$Sr$_{1.6}$La$_{0.4}$CuO$_{6+\delta}$ is $T_c=35.1$~K as reported by Kim \etal \cite{kim2008}.  Despite replicating their synthesis conditions, several times, we still only achieve a $\tcmaxm \approx 30$~K.  Doubtless, replicating their higher \tc result is a matter of fine tuning the many variables that are involved in the synthesis process of such a complicated material as Bi2201.

Finally, a note on impurity substitution; Zn is considered to substitute onto the Cu(2) site preferentially to the Cu(1) site \cite{bhalla2005} and so a 0.1:1 Zn-Cu ratio implies a concentration of $\sfrac{3}{2} \times 0.1$ or 15\% on the \cuo layer \emph{for bi-layer} Ln(Ba,Sr)$_2$Cu$_3$O$_y$. Moderate Zn substitution does not affect the average doping state either \cite{tallon1995, naqib2005} 

A typical example of a sample synthesis and characterisation process is given in \fig~\ref{fig:standardops}

\subsubsection{Nd123 synthesis}
\label{sec:nd123synth}

Here, we wish to synthesize polycrystalline NdBa$_2$Cu$_3$O$_{7-\delta}$, which is reported to have ideally $\tcmaxm=96$~K, and then to carry out a set of thermopower and \tc measurements in order to determine the annealing conditions for optimal doping.  See \refsec~\ref{sec:annealingconditions} for an example relating to a similar material.  

We believe that the large ion-size of Nd$^{3+}$ (\added{with co-ordination number }VIII), $1.109\times 10^{-10}$~m, and concurrent smaller Cu(2)-O(1) bond length mean that the energy difference for the Nd ion occupying the `Y' site or the `Ba' site becomes small (see \fig~\ref{fig:ln123xtal} and \fig~\ref{fig:ln123structure} for an illustration).  Nd$^{3+}$ occupation on the Ba-site is a known issue when synthesising this material \cite{williams1996} and this impurity will donate electrons to the \cuo layer causing the material to be more underdoped than desired.  The issue is even more severe in La123 and is the primary reason why we choose to make Nd123 rather than La123 polycrystalline samples.  


Sintering Nd123 in low partial O$_2$ pressure discourages Nd occupation of the Ba site.  Synthesis conditions are reported by Williams \etal \cite{williams1996} and MacManus-D\added{r}iscoll \etal \cite{macmanus1997}.  We did not have much success following these, see \fig~\ref{fig:tcvsrttepln123}. Instead after many attempts we describe below what we found to be the optimal synthesis conditions.

We use high purity\footnote{99.95\%+ purity, supplied by Sigma Aldrich.} Nd$_2$O$_3$, BaCO$_3$ and CuO precursor powders in a 1:2:3 ratio of Nd:Ba:Cu.  Before being weighed out the Nd$_2$O$_3$ powder was fired at 280\degc for 3 hours in order to remove any adsorbed H$_2$O. The precursor powders were weighed out, thoroughly mixed in acetone and pressed into pellets. These pellets are placed in an Au container basket and placed into the furnace.  At 350\degc the furnace is evacuated and backfilled with 0.1\%O$_2$ in N$_2$ gas.  The temperature is then increased to 850\degc, without any overshoot, and sintered for five hours with a moderate flow rate of 0.1\%O$_2$ in N$_2$ gas.  

The pellets are then reground in acetone, pressed into pellets again and sintered at 865\degc in 0.1\%O$_2$ in N$_2$ overnight. The pellets are again reground, but not in acetone this time, pressed into pellets again and sintered at 872\degc in 0.1\%O$_2$ in N$_2$ for 72 hours.

After suitable annealing, see \ref{sec:annealingconditions}, this produced a Nd123 sample with $\tcm=94$~K, $\tconsetm=96.5$~K and $\rttepm = 15.5 \pm 0.3$~$\mu$V.K$^{-1}$, indicating significant underdoping and hence the reduced $\tcm$.

Some experiments which did not produce good quality Nd123 for us include:
\begin{enumerate}
	\item Providing excess Ba by starting with stoichiometry Nd:Ba:Cu = 1:2.25:3.
	\item Providing deficient Nd by starting with the stoichiometry Nd:Ba:Cu = 0.9:2:3.
	\item Using metal-nitrate precursors rather than metal-carbonate precursors.
	\item Adding a small amount of Zn in the ratio Nd:Ba:Cu:Zn = 1:2:3:0.006.  Zn acts as a fluxing agent and is reported to reduce the melting point of \ybco by approximately 2 K/(\% of Cu) \cite{youwenxu1990}.
	\item `Pure' N$_2$ sintering atmosphere, with a concurrent lower temperature of 830\degc (the material melts at higher temperatures in the lower O$_2$ partial pressures).
\end{enumerate}

\subsection{Single crystal synthesis}
 
Although I did not attempt to grow any single crystals in this work, single crystals grown by Thomas Wolf at the Karlsruhe Institute for Technology (KIT) of Ln123 were used extensively.  These high-quality LnBa$_2$Cu$_3$O$_{7-\delta}$ (Ln=Lu, Yb, Dy, Gd, Eu, Sm, Nd, La) single crystals were flux grown in Y-stabilized zirconia crucibles under reduced oxygen atmosphere, where necessary, to avoid substitution of Ln ions on the Ba site.

\subsection{High-pressure, high-temperature synthesis}

Bulk, polycrystalline, fully Sr substituted \ysco cannot be grown by the usual solid-state synthesis techniques described above, \refsec~\ref{sec:solidstatesynth}, presumably because the \ysco phase is only meta-stable. Instead, polycrystalline samples of \ysco must be prepared in O$_2$ at 3 GPa of pressure and at 1050\degc with KClO$_3$ as an oxidant.  In order to achieve such high pressures (we are capable of 2~MPa only in our labs) a multi-anvil apparatus is needed - similar in concept to the Diamond Anvil Cell.  As such, these samples were prepared by Dr. Edi Gilioli at IMEM, CNR in Parma, Italy.  The ratio of precursor materials (Y$_2$O$_3$, SrO$_2$, CuO) to KClO$_3$ is 1:0.45.  A description of the synthesis method is found in references \cite{licci1998, gilioli2000}.

\ysco is clearly difficult to synthesize.  Unfortunately a range of attractive measurements (e.g. Resonant Inelastic X-ray Scattering, Raman spectroscopy, Ellipsometry) require single crystal samples, or at least aligned thin films.  Making such samples will require a big materials-chemistry effort which was beyond the scope of this work.  That being said we were able to prepare some \ysco in a film on a SrAlLaO$_4$ substrate at atmospheric O$_2$ pressure. This particular substrate was chosen for its close lattice match to \ysco. The method used for making these films is described below, \refsec~\ref{sec:yscothinfilmsynth}.

\subsection{Thin-film synthesis using metal-oxide deposition}
\label{sec:modprocess}

We make thin films using a metal-oxide deposition (MOD) process.  Here salts, e.g. Y-TFA\footnote{Yittrium tri-fluoro acetate.}, Cu-OHP\footnote{The chemical formula is an industrial secret! \added{Although we note that the Cu-acetate salt may be used instead of Cu-OHP.}}, Sr-acetate, are weighed out then dissolved in dry methanol.  Each new salt is added to the solution and completely dissolved before the next salt is added.  We add $\sim 3$\% of volume propionic acid once to assist dissolution.  A sonicator is used to fully mix the solution and dissolve the salts after each new salt is added.  For our samples, the maximum length of time needed in the sonicator fully to dissolve the salts is 20 minutes.

The most common substrate used is the trademarked RABiTS substrate which is an industry standard for `2G' (`second generation' - YBa$_2$Cu$_{3}$O$_7$) wire.  It is a Ni-W alloy base with three 75 nm buffer layers, Y$_2$O$_3$, YSZ and CeO$_2$ to ensure good lattice matching.  The substrate is spin coated with the solution in a water- and oxygen-free atmosphere inside a glove-box.  The typical Y123 film thickness is $\sim 800$ nm, although this depends on the spin speed and solution concentration. The samples are transported from the glove box to the decomposition furnace inside a sealed container and the decomposition process started immediately.  Although the un-decomposed films are briefly exposed to air whilst being loaded into the decomposition furnace, we were able to good quality films with this process and so take this method to be satisfactory.

The film is decomposed in flowing O$_2$ and H$_2$O vapour (a source of H) up to $450^{\circ}$C to remove all organic material.  The optimal specific temperature-time profile will change depending on the solution.  After the decomposition process the only non-oxide remaining should be Ba-OF (strictly speaking (Y,Ba)-OF). Ba-OF converts to BaF$_2$ at $T=575\pm 25^{\circ}$C, a process which takes place during the initial phase of the next process.

The sintering sequence (or reaction sequence) - where the desired Y123 phase is formed - is done under a partial H$_2$O pressure.  During the initial stage of the sintering Ba-OF converts to BaF$_2$ which in turn reacts above 720~$^{\circ}$C with the water present; BaF$_2$+H$_2$O$\rightarrow$BaO + 2HF (which is vented to the atmosphere).  This allows the remaining BaO to participate in the Y123 phase formation.  The temperature ramp rate, O$_2$ and H$_2$O pressure and sintering temperature (standard is $788^{\circ}$C) are critical for growing epitaxial films (i.e. where the Y123 c-axis aligned vertically).  The temperature ramp rate is controlled by the speed at which the sample is moved into the hot zone of the furnace.  It is important to ensure the optimal BaF$_2$:BaO ratio at each temperature in the sequence, which is in turn important to ensure the YBCO crystal growth is seeded from the correct (001) face off the (aligned) substrate to result in $c$-axis aligned\footnote{And for critical current densities this process can be fine-tuned so that there is good $a$ and $b$ axis alignment also.} epitaxial films.

There is an additional complication in our material. To make YBa$_{2-x}$Sr$_x$Cu$_3$O$_{7-\delta}$ we use Sr-acetate and Ba-TFA salts in the solution.  After the decomposition process Ba-OF and SrO are present.  This is a problem as during the reaction sequence the SrO is immediately available to react with Y$_2$O$_3$ and CuO to form a Sr-rich impurity phase whereas the BaF$_2$ must first decompose into BaO before it can react with Y$_2$O$_3$, SrO and CuO.  Nevertheless we are able to synthesize good quality, well aligned YBa$_{1.75}$Sr$_{0.25}$Cu$_3$O$_{7-\delta}$ thin films for Raman spectroscopy measurements.  The quality of films with higher values of $x$ were not as good.  Presumably a concerted effort to optimise the sintering conditions would sufficiently improve the thin film quality and epitaxy.  

A summary of this synthesis process can be found in a review by Obradors \etal \cite{obradors2012}.

\subsubsection{\ysco thin film synthesis on SrLaAlO$_4$ substrates}
\label{sec:yscothinfilmsynth}

This section describes how we grow thin films of \ysco on single crystal substrates, in both (100) and (001) orientations, of SrLaAlO$_4$. 

SrLaAlO$_4$ has a tetragonal unit cell\footnote{with I4/mmm space group, similar to the P4/mmm space group of Y123O$_6$.} with $c=12.636\times10^{-10}$~m $a=b=3.756\times10^{-10}$~m, very similar to the \ysco $a$ and $b$ unit cell parameters reported as $a=b=3.7855\times10^{-10}$~m \cite{gilioli2000}.  This close lattice matching promotes the growth of the desired \ysco phase.  It is hoped this will also promote epitaxial growth of the \ysco.

Y-TFA, Cu-OHP and Sr-acetate salts are mixed, spin coated on the substrate and decomposed as per our standard MOD process (see above, \ref{sec:modprocess}).  The standard reaction conditions used to grow the Y123 phase did not produce YSr$ _{2} $Cu$ _{3} $O$ _{7-\delta} $, or indeed any crystalline phases.  This is perhaps not too surprising as it was found that high temperatures and O$_2$ pressures are required to grow polycrystalline \ysco \cite{gilioli2000}.  We therefore carried out a series of reactions at 950$^{\circ}$C, 1000$^{\circ}$C and 1050$^{\circ}$C in pure O$_2$ with each followed by XRD analysis, to determine suitable reaction conditions.

\begin{figure}
	\centering
		\includegraphics[width=0.45\textwidth]{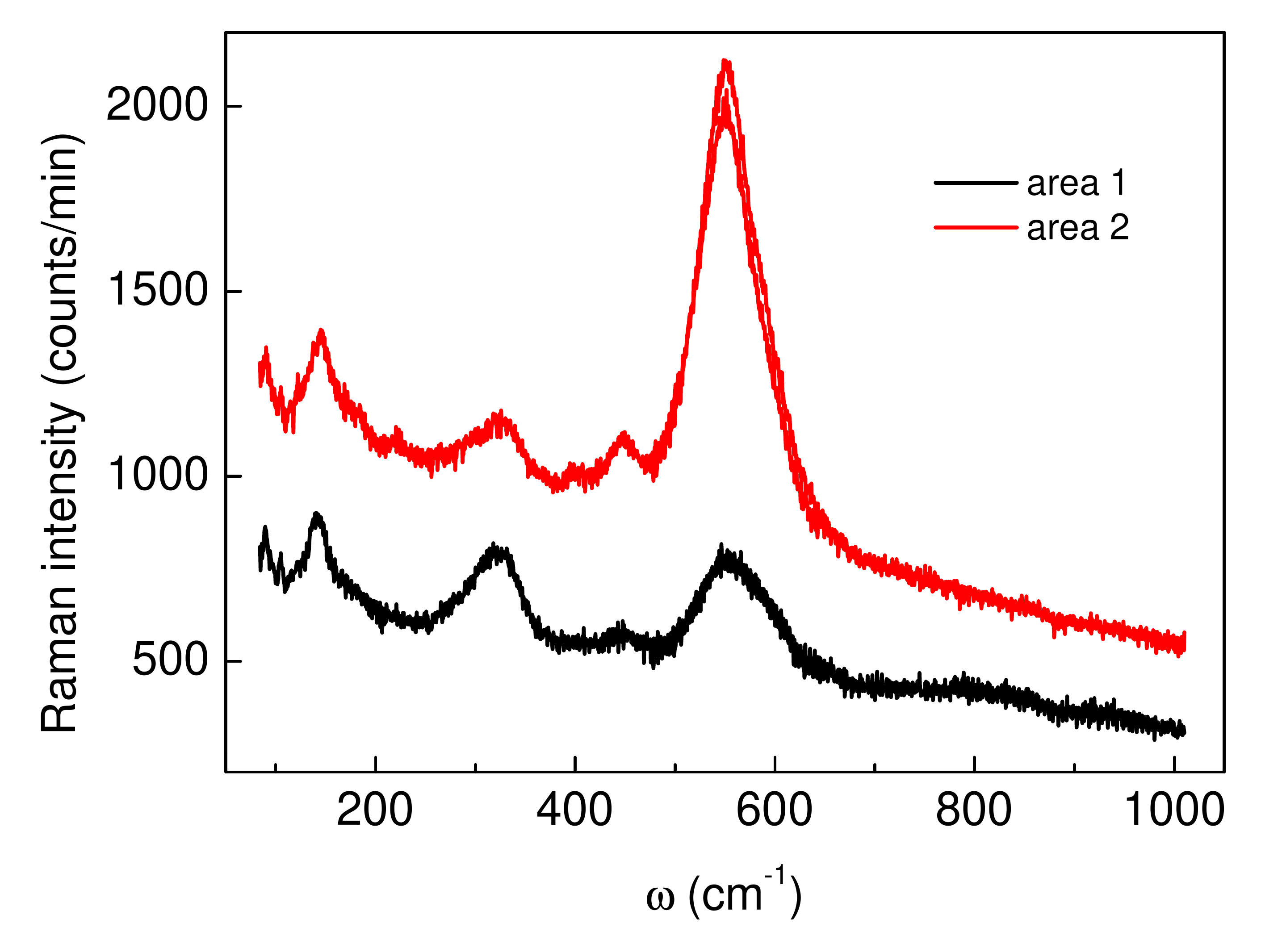}\
		\includegraphics[width=0.45\textwidth]{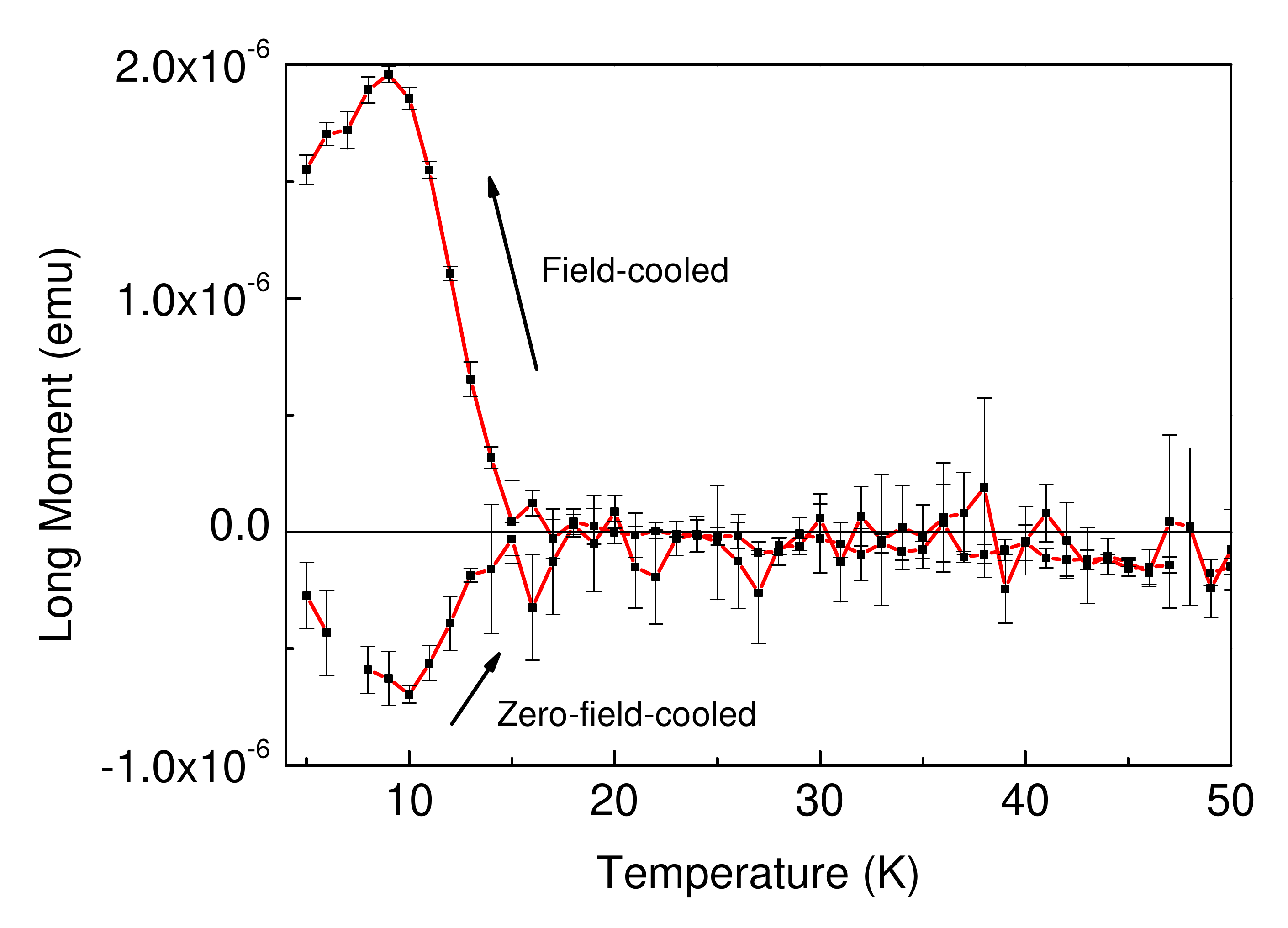}
	\caption[Raman spectrum and magnetisation data for a thin-film of YSCO.]{(Left) Raman spectra from two different, representative, regions of a \ysco film on a SrLaAlO$_4$ (001) substrate after being conditioned at 1000\degc in pure O$_2$ for 20 minutes. Each is made from two spectra joined at $\approx 550$~\cm - as can be seen by the slight intensity discrepancy there. Taken on the LabRam spectrometer with the 514.5~nm laser and a x100 objective.  Note that the phonon spectrum here is entirely consistent with the (unorientated) \ybco phase.  (Right) Zero-field-cooled and field-cooled DC magnetisation of the same film as measured in the SQUID magnetometer.  $\HH = 20$~Oe (2~mT).  These data show a likely transition to a superconducting state at $\tcm=14$~K when zero-field cooled. }
	\label{fig:yscofilm}
\end{figure}

It was found that processing a decomposed \ysco film on SrLaAlO$_4$ (001) at 1000\degc in pure O$_2$ for 20 minutes, followed by a rapid cool did indeed produce the \ysco phase!  The presence of the \ysco phase was confirmed by Raman spectroscopy as shown in \fig~\ref{fig:yscofilm}(Left).  Only phonon modes associated with the (unorientated) \ybco structure are present - the fact that the 550~\cm mode is relatively intense indicates the \ysco phase is multi-domain and unaligned. Interestingly they are at similar energies to the equivalent \ybco modes.  Note however that it is not a \ybco sub-phase we observe as might be the case with a partially Sr-Ba substituted film.  This film has no Ba, only Sr. 

Furthermore, a weak diamagnetic transition was observed (as might be expected from a thin film) for this film at the low temperature of $\tcm=14$~K upon zero-field-cooling (ZFC) as shown in \fig~\ref{fig:yscofilm}(Right).  Upon field-cooling (FC) the diamagnetism is destroyed.  These observations are consistent with a weak superconducting transition at $T_c=14$~K.  We believe this to not originate from the SrLaAlO$_4$ substrate for two reasons (i) at $T=4$~K we measure no magnetism from the bare substrate down to the resolution of the SQUID magnetometer; $10^{-7}$~emu (ii) there is no reported superconductivity in this material.

\section{Sample preparation and annealing}

After our samples have been synthesized they are taken through a process of annealing.  

The materials we work with have variable oxygen content.  In the case of \ybco and related materials O can be incorporated at the O(1) or, undesirably, at the O(5) site, see \fig~\ref{fig:ln123xtal}.  In Bi2201, O can be incorporated in the BiO$_{1-\delta}$ layer or interstitially, see \fig~\ref{fig:bi2201xtal}.  To control the level of oxygen we anneal our samples at carefully controlled temperatures (or a sequence of temperatures) in a controlled O$_2$ partial pressure then quenched.  The standard procedure and two special cases are described below.

\subsubsection{Standard annealing process}
\label{sec:annealingprocess}

For this task, we always use vertically orientated furnaces, the reason for which will be clear soon.  The sample is held in a YSZ container which is suspended from a hooked wire inside the furnace tube, held in place by strong rare-earth magnets on the outside of the furnace tube. When annealing in an Ar atmosphere an Au basket is used instead of a YSZ container as Au does not contain any O that may be taken up by the sample.


Initially the sample is not raised to the centre of the heating coils (the `hot zone') - this must wait until the correct gas has been backfilled into the furnace tube.

The furnace tube is sealed and pumped on to a moderate vacuum.  The vacuum valve is then closed and the desired gas (e.g. 2\% O$_2$ in N$_2$, or pure O$_2$) introduced into the tube.  The tube is evacuated again and then again, the desired gas introduced.  This is the process of `backfilling', or as it is called when the PPMS and SQUID fill the sample space with He, `purging'.  This step is repeated three times.  After the last backfill, the desired gas is pushed through the tube with a slow flow rate (the purpose being to mitigate air leaking into the sealed furnace tube).  

Only now is the sample raised to the centre of the heating coils, using the strong rare-earth magnets outside the tube to pull the basket up. Typically the sample is held at the annealing temperature overnight (for polycrystalline samples) or over three days (for single crystals) to give sufficient time for oxygen to diffuse from/into the sample. 

When removing the sample one tries to avoid exposing it to an O$_2$ partial pressure different to that in which it was annealed, at elevated temperatures.  The solution for polycrystalline samples is dramatic!  Once the furnace tube is unsealed, the polycrystalline samples are dropped rapidly from the centre of the heating coils and into liquid N$_2$, a process called `quenching'.  This rapidly cools the samples, preventing any further significant oxygen diffusion.  Finally the sample is dried once removed from the liquid N$_2$ to prevent H$_2$O condensation on the surface.  

For single crystals there is a significant chance a quench into liquid N$_2$ will destroy the crystal, and so in this case the sample is instead rapidly lowered as far as possible from the hot-zone and the gas flow-rate increased.  The furnace is not unsealed until the crystal cools to room-temperature.

\subsubsection{``O$_7$'' anneal}
With this process we aim to incorporate as much O as possible.  The process is outlined in \tab~\ref{tab:o7anneal} described below;

The samples are placed on Au foil and placed inside the pressure furnace - called the `Bomb'.  The Bomb is firmly sealed, flushed with O$_2$ several times and heated to 570\degc over two hours.  The temperature is then decreased to 400\degc over two hours.  The O$_2$ pressure is then increased to 10 or 12 bar.  A conventional pressurized gas cylinder is the source of the above atmospheric pressure O$_2$.  We then cool the sample to 300\degc over 72 hours before turning off heating coil.  The sample not removed or O$_2$ pressure released until the furnace is close to room-temperature.

\begin{table}%
\centering
\begin{tabular}{l||c|c|c|c}
	T (\degc) & $0 \rightarrow 570$   &  $570 \rightarrow 400$  &  $400 \rightarrow 300$  & $300 \rightarrow 28$ \\ 
	Time taken & 2 hrs                 &  2 hrs                   &  72 hrs                 &  furnace-cool ($\sim 18$ hrs) \\ 
	O$_2$ pressure* & 1 bar ($100$kPa)   &  1 bar          &       10-12 bar   & 10-12 bar  \\
\end{tabular}
\caption[The ``O$ _{7} $'' annealing process]{An annealing process designed to incorporate as much O as possible into the sample.  For the purposes of this work it is referred to as the ``O$_7$'' anneal. *Above atmospheric pressure.}
\label{tab:o7anneal}
\end{table}

\subsubsection{Ar anneal}
\label{sec:aranneal}
With this process we aim to remove as much O as possible from the material. An alternative name for the process is the `O$_6$' anneal.  

The sample, held in a Au basket, is introduced into the furnace at $\approx 200$\degc.  
Next the furnace is sealed and pumped to a moderate vacuum. The vacuum valve is closed and Ar admitted into the chamber (Ar is less likely to have trace O$_2$ impurities than N$_2$ gas). The furnace is again evacuated and then refilled with Ar.  This is known as `backfilling' and the process is repeated 3-4 times to ensure the furnace is filled with Ar only.  After backfilling we maintain a slow flow rate of Ar through the furnace to mitigate air leaking into the sealed furnace (the inside of the furnace will be at a slightly higher pressure than outside).   The temperature is then increased to 600\degc and held there for 24 hours (for polycrystalline samples where diffusion is faster) to 72 hours (for single crystals).  During the longer anneals the furnace is backfilled every day.  

At the end of the anneal, the Ar flow rate is significantly increased, the sample rapidly lowered from the centre of the furnace (the `hot-zone') and then left to cool to room temperature before the furnace is opened.  

Samples annealed this way must be stored in a desiccated environment under weak vacuum.

\subsection{Annealing temperatures for studied samples}
\label{sec:annealingconditions}
\label{sec:annealing}

\begin{figure}
\centering
 \includegraphics[width=0.715\textwidth]{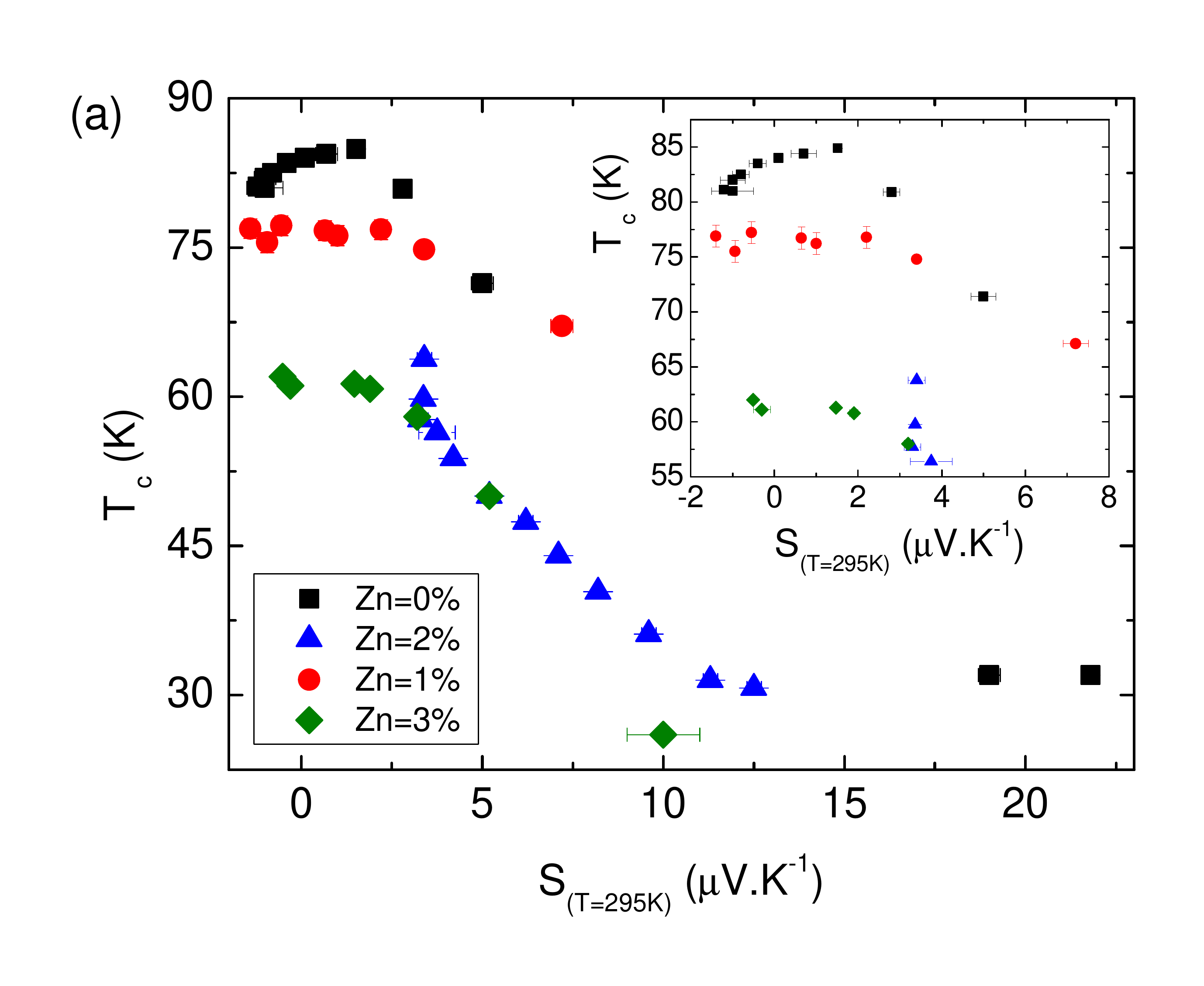}
 \includegraphics[width=0.715\textwidth]{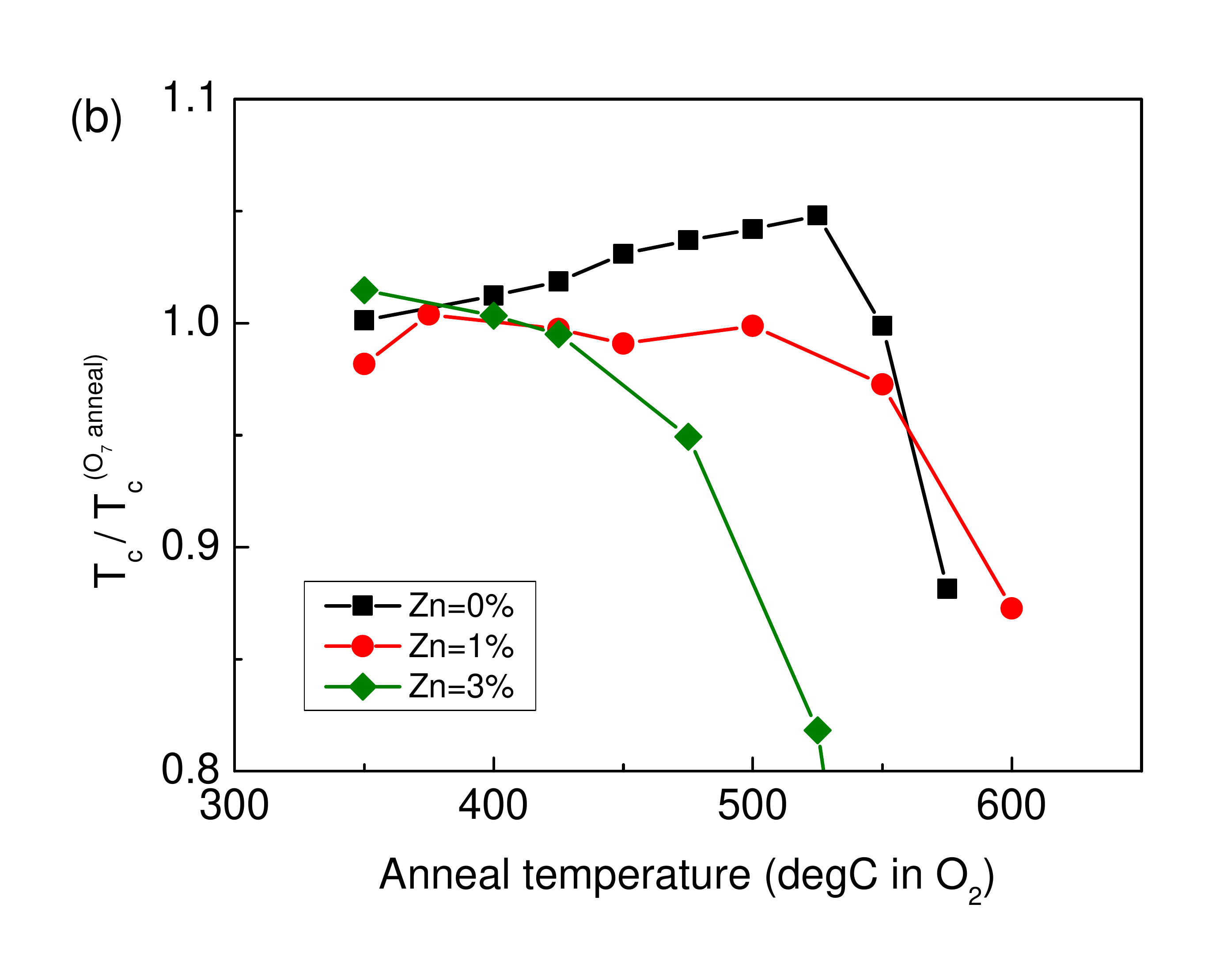}
	\caption[\tc vs \rttep or annealing temperature for YBaSrCu$_{3-z}$Zn$_z$O$y$.]{\label{fig:tcvsrttepybasr} (a)\tc plotted against the room temperature thermoelectric power, $\rttepm$, for YBaSrCu$_{3-z}$Zn$_z$O$y$ with $z=0.00$, $0.02$, $0.04$, $0.06$.  Systematically increasing the annealing temperature, see \tab~\ref{tab:ybasrannealingconditions}, is used to reduce the oxygen content with a corresponding increase in $\rttepm$. \tc is estimated from magnetisation measurements, see \refsec~\ref{sec:measuringtc}. The inset shows the region around optimal doping in closer detail. (b) A plot of \tc normalised to the \tc value measured after an `O$_7$' anneal, see \tab~\ref{tab:o7anneal}, plotted against the subsequent annealing temperature in O$_2$ for three Zn concentrations indicated in the legend.  Note that for higher Zn concentration the highest \tc value is obtained after lower annealing temperatures and therefore presumably higher oxygen content and larger $p$.  }
\end{figure}

Samples of YBaSrCu$_{3-z}$Zn$_z$O$_y$ for $z=0.00$, 0.02 0.04 and 0.06 were synthesized by the method described above in \refsec~\ref{sec:solidstatesynth} and followed the sintering temperatures reported by Licci \etal \cite{licci1998}. Zn is known to substitute preferentially for Cu(2,3) \cite{bridges1993} - the Cu site in the \cuo layers.  Hence we have a set of 1\%, 2\% and 3\% Zn doped samples.  In \tab~\ref{tab:ybasrannealingconditions} and \fig~\ref{fig:tcvsrttepybasr} we show the results of a systematic annealing study of these samples.  From these data we know the annealing conditions to obtain specific $p$ and corresponding \tc values. For example, these data show the annealing conditions required for `optimal' doping, \tc=\tcmax, for each Zn concentration. \tc, $y$ and $p$ values for $z=0.0$ can also be found here \cite{licci1998}. We were not able to fully dope the $z=0.04$ sample to determine \tcmax.

\begin{table}
	\centering
		\begin{tabular}{clllll}  \toprule
		 Zn & Annealing temperature (\degc) & \tc(K) & $\rttepm$ ($\mu$V.K$^{-1}$) & $y$ & $p$ \\ \midrule \midrule
		 \multirow{11}{*}{z=0.0} & ``O$_7$'' & 81 & -1.0 & 7.00 & 0.184 \\ 
		  & 350 & 81.1 & -1.2 & 6.99 & 0.183 \\ 
		  & 400 & 82.0 & -1.0 & 6.98 & 0.180 \\ 
		  & 425 & 82.5 & -0.8 & 6.97 & 0.179 \\ 
		  & 450 & 83.5 & -0.4 & 6.96 & 0.174 \\ 
		  & 475 & 84.0 & 0.1 & 6.95 & 0.171 \\ 
		  & 500 & 84.4 & 0.7 & 6.93 & 0.168 \\ 
		  & 525 & 84.9 & 1.5 & 6.91 & 0.160 \\ 
		  & 550 & 80.9 & 2.8 & 6.88 & 0.135 \\ 
		  & 575 & 71.4 & 5.0 & 6.95 & 0.115 \\ 
		  & 600(Under air.) & 32 & 19.0 & 6.68 & 0.075 \\
		  & 600(Under 2\% O$_2$ in N$_2$) & 32 & 21.8 & 6.67 & 0.073 \\ \midrule
		 \multirow{8}{*}{z=0.02} & ``O$_7$'' & 76.9 & -1.4 & 6.99 & 0.167 \\ 
 			& 350 & 75.5 & -0.94 & 7.00 & 0.176 \\ 
		  & 375 & 77.2 & -0.55 & 6.98 & 0.160 \\ 
		  & 425 & 76.7 & 0.65 & 6.95 & 0.151 \\ 
		  & 450 & 76.2 & 1.0 & 6.94 & 0.147 \\ 
		  & 500 & 76.8 & 2.2 & 6.90 & 0.152 \\ 
		  & 550 & 74.8 & 3.4 & 6.87 & 0.141 \\ 
		  & 600 & 67.1 & 7.2 & 6.83 & 0.120 \\ \midrule
		 \multirow{10}{*}{z=0.04} & ``O$_7$'' & 56.4 & 3.8 & 6.98* & - \\ 
 			& 350 & 57.7 & 3.3 & 7.00* & - \\ 
		  & 400 & 53.8 & 4.2 & 6.97 & - \\ 
		  & 425 & 50.0 & 5.2 & 6.96 & - \\ 
		  & 450 & 47.4 & 6.2 & 6.95 & - \\ 
		  & 475 & 44.0 & 7.1 & 6.93 & - \\ 
		  & 500 & 40.4 & 8.2 & 6.92 & - \\ 
		  & 525 & 36.1 & 9.6 & 6.90 & - \\ 
		  & 550 & 31.5 & 11.3 & 6.88 & - \\ 
		  & 575 & 30.7 & 12.5 & 6.88 & - \\ \midrule		  
		 \multirow{7}{*}{z=0.06} & ``O$_7$'' & 61.1 & -0.3 & 7.00 & 0.173 \\ 
 			& 350 & 62.0 & -0.5 & 6.99 & 0.160 \\ 
		  & 400 & 61.3 & 1.5 & 6.94 & 0.148 \\ 
		  & 425 & 60.8 & 1.9 & 6.92 & 0.144 \\ 
		  & 475 & 58.0 & 3.2 & 6.90 & 0.132 \\ 
		  & 525 & 50.0 & 5.2 & 6.86 & 0.112 \\ 
		  & 575 & 26.0 & 10 & 6.80 & 0.076 \\ 
		  \bottomrule

		\end{tabular}
		\caption[Annealing conditions, $y$, \rttep and \tc values for YBaSrCu$_{3-z}$Zn$_z$O$y$.]{\label{tab:ybasrannealingconditions} Annealing temperatures under pure O$_2$ for polycrystalline YBaSrCu$_{3-z}$Zn$_z$O$y$ (unless otherwise noted). \tc is estimated from magnetisation, see \refsec~\ref{sec:measuringtc}, $\rttepm$ is an average of measurements on different pellets. $y$ is estimated from the mass change after each anneal and assumes the ``O7'' anneal results in $y=7$. $p$ is estimated from the parabolic relation $T_c/T_c^{\textnormal{max}}=1-82.6(p-0.16)^2$ (which assumes, amongst other things, optimal doping at $p=0.16$holes/Cu) \cite{tallon1993}. Reasonable uncertainties are $\pm 0.5$~K in \tc, $\pm 0.1$ $\mu$V.K$^{-1}$ in $\rttepm$, $\pm0.03$ in $y$ and $\pm 0.005$ holes/Cu in $p$.  *Because of the high thermopower values and lower than expected \tc values, we believe this is not fully oxygenated, or if $y=7$, then significant O(5) site occupation causing underdoping.}
\end{table}

\tc plotted against the number of holes per Cu in the \cuo layer, i.e. the `doping' $p$, has a maximum shaped like a parabola at $p\approx 0.16$  We call this doping, where \tc=\tcmax, `optimal doping' and it gives us a common point at which to compare Ln123. Sometimes $p$ is estimated from this parabolic relation, $T_c/T_c^{\textnormal{max}}=1-82.6(p-0.16)^2$ (which assumes, amongst other things, optimal doping at $p=0.16$ holes/Cu) \cite{tallon1993}. However, determining optimal doping solely by measuring \tc is problematic as the small decrease in \tc from \tcmax could imply either under- or over-doping.  The thermo-electric power (TEP) at room temperature, $\rttepm$, is however a monotonic function of doping \cite{oct}, being positive for under-doping and going negative for over-doping - see \refsec~\ref{sec:tep}. For optimal doping in the cuprates \rttep is typically $+2$~$\mu$V.K$^{-1}$.

\begin{figure}
\centering
 \includegraphics[width=0.75\textwidth]{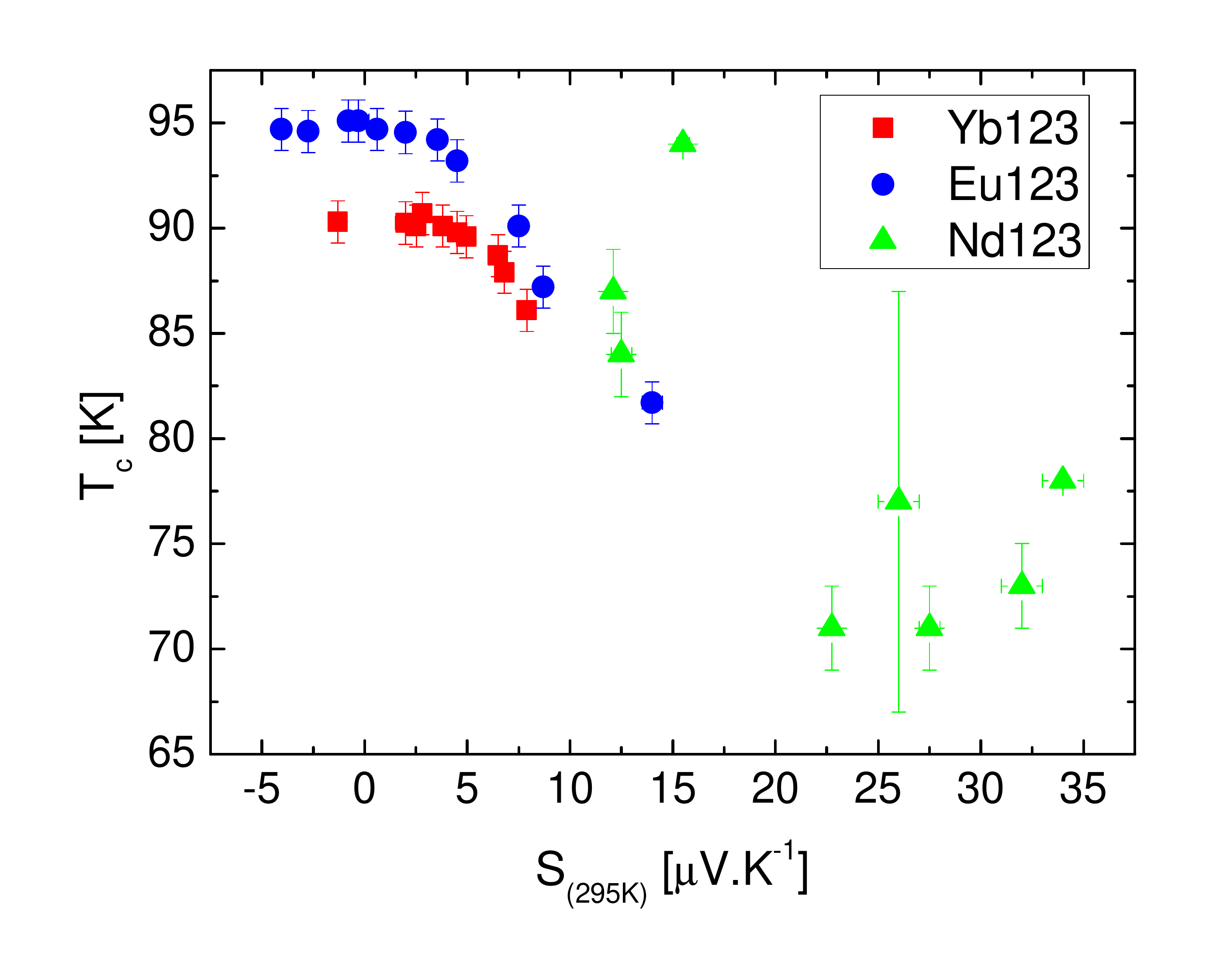}
	\caption[\tc vs \rttep for polycrystalline Ln123 with Ln=Yb, Eu, Nd.]{\label{fig:tcvsrttepln123} \tc plotted against the room temperature thermoelectric power, $\rttepm$, for Yb123, Eu123 and Nd123.  We reduce the oxygen content by systematically increasing the annealing temperature, see \tab~\ref{tab:ln123annealingconditions}, which causes an increase in $\rttepm$. \tc is estimated from magnetisation measurements, see \fig~\ref{fig:tcestimates}.}
\end{figure}

To determine the annealing conditions to optimally dope each of Nd123, Eu123, Yb123 we measure \tc vs. $\rttepm$ on polycrystalline samples made by standard solid-state reaction methods \ref{sec:solidstatesynth} using known sintering conditions \cite{williams1996}.  Our results are shown in \fig~\ref{fig:tcvsrttepln123} and tabulated in \tab~\ref{tab:ln123annealing}.  For cuprates without Cu(1)-O(4) chains, close to optimal doping $\rttepm$ is linearly proportional to $p$ \cite{oct} but for Y123 we expected an additional, and positive, contribution to \rttep from the metallic chains as they are filled.  The chain contribution cannot be easily distinguished in polycrystalline samples. 

\begin{table}
	\centering
		\begin{tabular}{llllll}  \toprule
		 Ln & Annealing $T$ (\degc) & \tc(K) & $\rttepm$ ($\mu$V.K$^{-1}$) & $y$ & $p$ \\ \midrule \midrule
		 \multirow{9}{*}{Yb} & ``O$_7$'' & 90.3 & 2.0 & 7.00 & 0.168 \\ 
		  & 350 & 90.1 & 2.5 & 6.99 & 0.169 \\ 
		  & 400 & 90.7 & 2.8 & 6.99 & 0.160 \\ 
		  & 450 & 90.1 & 3.8 & 6.97 & 0.151 \\ 
		  & 475 & 89.8 & 4.5 & 6.96 & 0.149 \\ 
		  & 500 & 89.6 & 5.0 & 6.95 & 0.148 \\ 
		  & 525 & 88.7 & 6.5 & 6.93 & 0.144 \\ 
		  & 550 & 87.9 & 6.8 & 6.91 & 0.141 \\ 
		  & 575 & 86.1 & 7.9 & 6.88 & 0.135 \\ \midrule
		 \multirow{10}{*}{Eu} & 120 kbar O$_2$ & 94.6 & -2.8 & 7.00 & 0.168 \\ 
		  & ``O$_7$'' & 95.1 & -0.3 & 7.00 & 0.160 \\ 
		  & 350 & 95.1 & -0.8 & 6.99 & 0.160 \\ 
		  & 400 & 94.7 & 0.6 & 6.98 & 0.153 \\ 
		  & 450 & 94.6 & 2.0 & 6.97 & 0.152 \\ 
		  & 475 & 94.2 & 3.6 & 6.96 & 0.149 \\ 
		  & 500 & 93.2 & 4.5 & 6.94 & 0.144 \\ 
		  & 525 & 90.1 & 7.5 & 6.91 & 0.135 \\ 
		  & 550 & 87.2 & 8.7 & 6.89 & 0.128 \\ 
		  & 575 & 81.7 & 14.0 & 6.86 & 0.119 \\ 
		  \bottomrule

		\end{tabular}
	\caption[Annealing conditions, $ y $, \rttep and \tc values for Yb123 and Eu123.]{\label{tab:ln123annealingconditions} \label{tab:ln123annealing} The annealing temperatures under pure O$_2$ for LnBa$_2$Cu$_3$O$_y$ as described in section \ref{sec:annealing}. \tc is estimated from magnetisation (see section \ref{sec:measuringtc}), $\rttepm$ measured on a home-built rig (each value represents an average of typically 3 measurements, each with different heating conditions or on different pellets). $y$ is estimated from the mass change after each anneal and assumes the ``O7'' anneal (\ref{sec:annealing}) results in full oxygen loading ($y=7$). $p$ is estimated from the parabolic relation $T_c/T_c^{\textnormal{max}}=1-82.6(p-0.16)^2$ (which assumes, amongst other things, optimal doping at $p=0.16$holes/Cu) \cite{tallon1993}. Reasonable uncertainties are $\pm 0.5$K in \tc, $\pm 0.1$ $\mu$V.K$^{-1}$ in $\rttepm$, $\pm 0.03$ in $y$ and $\pm 0.005$ holes/Cu in $p$.  Together, these data show the annealing conditions required for `optimal' doping, \tc=\tcmax. \tc values and annealing conditions for various Ln123 can also be found here \cite{chikumoto1997,wong2006}.}
	
\end{table}

As can be seen from \fig~\ref{fig:tcvsrttepln123} Nd123 has high \rttep values indicating our Nd123 is underdoped, which in turn implies Nd$^{3+}$ occupation of the Ba$^{2+}$ site - a known issue.  See \refsec~\ref{sec:nd123synth} for discussion of our attempts to synthesize Nd123 with low Nd$^{3+}$ occupation of the Ba$^{2+}$ site.  The method to identify optimal doping used for Yb123 and Eu123 will obviously not work in this case.  We know from the literature that Nd123 is difficult to overdope \cite{veal1989} in general. Our approach to `optimally dope' Nd123, and La123, then is to use the ``O$_7$'' annealing process \ref{tab:o7anneal} to incorporate as much oxygen as possible.  We note however that for our best Nd123 polycrystalline sample $\tcm=94$~K and $\tconsetm=96.5$~K which are comparable to the highest reported values in the literature \cite{veal1989}.

%
%
%

\section{X-ray Diffraction }

X-ray diffraction (XRD) is a common technique for examining crystal structure. A collimated beam of (parallel) X-rays are directed onto the sample at a specific angle of incidence.  This incident radiation is scattered from electrons in the material which are most densely located in the outer `electron clouds' of ions in the crystal.  The intensity of scattered radiation is proportional to that density.  

The basic tenet of this technique is that for certain, specific angles of incidence and reflection depending on the crystal symmetry, there is constructive scattering off planes of ions in the crystal.  Quantitatively, let $\kk' -  \kk = \GG$, where $\GG= h\abf+k\BB+l\CC$ is a reciprocal lattice vector, $(hkl)$ are the Miller indices and $\kk'$ and $\kk$ are the scattered and incident X-ray wave vectors.  This condition combined with the momentum conservation (since the photon has such a tiny momentum) $|\kk'|=|\kk|$, gives the Laue formulation for constructive scattering (reflection) from a crystal lattice, $2\kk\cdot\GG +G^2 = 0$.  This is equivalent to the more commonly quoted Bragg condition; $2d\sin(\theta)=n\lambda$, where $d$ is the distance between crystal planes, $\lambda$ the wavelength of the radiation, $\theta$ the angle between the crystal plane and incident radiation and $n$ is some integer.

The resulting diffraction pattern reflects the periodicity, and internal structure, in the crystal.  These are some excellent resources describing this common technique, from the basic \cite{argonnexrdtutorial}, to the more advanced \cite{mccusker1999}.  


In this work, XRD is used to characterise the samples we make, as illustrated in \fig~\ref{fig:standardops}. 
Polycrystalline samples are ground to a fine powder which is then placed level in an Aluminium holder.  There are several reasons why it is better to measure a powder than a sintered pellet; (i) signal-to-noise is better (ii) from experience the composition and phases at the surface of a pellet may not be representative of the bulk whereas the powder from ground pellet ensures an average composition is being measured and (iii) if prepared properly the crystallites are known to be randomly-orientated.  On the other extreme, XRD on single-crystals requires much more geometrical care when setting up the measurement. Because they are very well-orientated, constructive scattering peaks can be easily missed with the point detectors we use.  All XRD measurements are carried out at room temperature and pressure.  A typical measurement takes 1 hour. 

We use Cu-K$_{\alpha}$ X-rays and a Bruker ``D8'' diffractometer in parallel beam (Debye-Scherrer) geometry.  

\subsection{ Examples }

\begin{figure}
	\centering
		\includegraphics[width=0.85\textwidth]{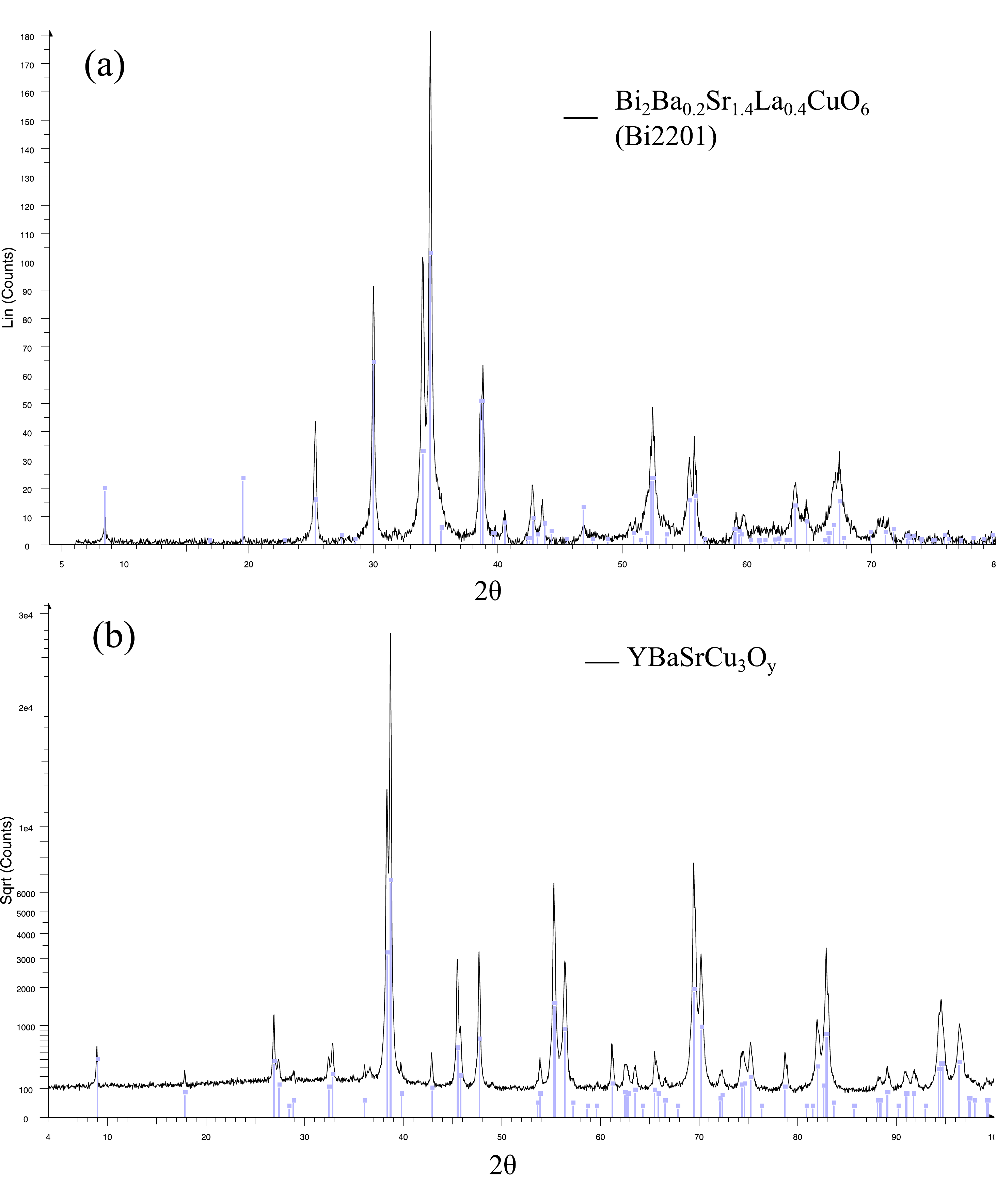}
	\caption[Typical XRD patterns for Bi2201 and YBaSrCu$ _{3} $O$ _{y} $.]{Typical XRD patterns obtained from our powder samples of (a)  Bi$_2$Ba$_{0.2}$Sr$_{1.4}$La$_{0.4}$CuO$_{6+\delta}$ and (b) YBaSrCu$_3$O$_y$ (note the $y$-axis square-root scale now despite both having 10~s count-times per step).  Light blue lines mark the expected crystal reflection $2\theta$ and approximate relative intensity for similar materials. The signal to noise ratio of these XRD scans is not good enough for a Rietveld analysis. Appropriate data for this analysis requires more than a day of counting time (and careful sample preparation) if using a lab based source (whilst on a synchrotron only a few minutes would be needed). Instead, these data are used to measure the lattice parameters of the material and identify any crystals of different composition and symmetry (impurity phases).}
	\label{fig:xrd}
\end{figure}

\fig~\ref{fig:xrd} shows typical XRD data used to check the phase purity of samples we synthesized.  Bi2201 has especially weak X-ray diffraction reflecting poor long-range crystallinity. 

We primarily draw from a database of previous refinements on similar materials\footnote{For example, there is no database entry for Bi$_2$Ba$_{0.2}$Sr$_{1.4}$La$_{0.4}$CuO$_{6+\delta}$ as we are the first to synthesize it, thus we compare our diffraction patterns with Bi$_2$Sr$_{1.6}$La$_{0.4}$CuO$_{6+\delta}$.} (e.g. Y123 or pure Bi2201).  From these patterns we can identify diffraction peaks with particular reflections from the unit cell, e.g. (001) reflection or (115) reflection, refine the lattice parameters and identify additional diffraction peaks. With this method however we do not get any quantitative information from the intensity of the diffraction peaks - for that we must perform a Rietveld refinement. 

\begin{figure}
	\centering
		\includegraphics[width=0.85\textwidth]{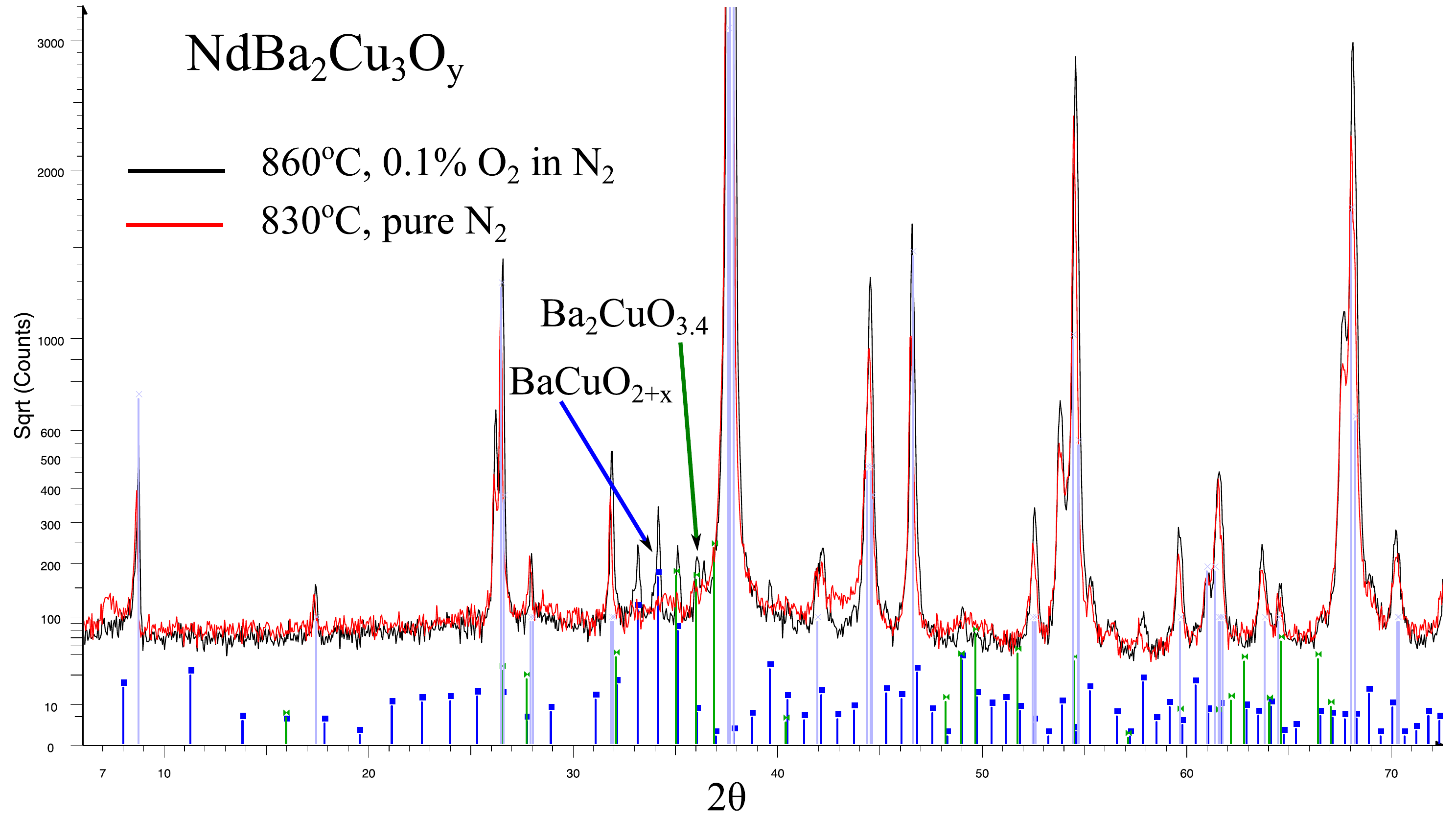}
	\caption[Powder XRD patterns of Nd123 attempts.]{Powder XRD patterns from two attempted NdBa$_2$Cu$_3$O$_y$ samples.  Light blue lines mark the expected crystal reflection $2\theta$ and approximate relative intensity for YBa$_2$Cu$_3$O$_y$, while the other colour lines represent possible impurity phases. From these we identify Cu and Ba-rich phases such as Ba$_2$CuO$_{3+\delta}$ and BaCuO$_{2+\delta}$ as impurities - probably forming as a result of excess of Ba and Cu after Nd occupation of the Ba site in the `Y123' phase. Interestingly, the Nd123 sample sintered under the fully reducing atmosphere of pure N$_2$ (red curve) had lower \tc and larger \rttep values than the samples with more Ba- and Cu-rich impurity phases.  The reflection in this sample at $2\theta \approx 7.1^{\circ}$ is from an unidentified phase which must have a large repeat unit of $\sim 1.4$~nm.}
	\label{fig:nd123xrd}
\end{figure}

If there are impurity phases visible in the XRD pattern, they can often be identified by searching the database for appropriate possibilities.  An example of a diffraction pattern with impurity phases is shown in \fig~\ref{fig:nd123xrd} for (mostly) NdBa$_2$Cu$_3$O$_y$.  Crystallographically, the material is very similar to \ybasr shown in \fig~\ref{fig:xrd}, displaying the same reflections, but with larger lattice parameters.  However it is difficult to prevent the Nd ion occupying the Ba site.  When this occurs there is an excess of Ba, Cu and O to form the `123' phase and so impurity phases such as CuO and Ba$_2$CuO$_3$ form which we then observe in the diffraction pattern.

		\section{Thermopower}
		\label{sec:tep}

Thermopower, or thermo-electric power (TEP), is the voltage developed across a material due to a temperature gradient. More formally, the thermopower is the voltage difference between point A and point B, $\Delta V = V_A-V_B$, divided by the temperature difference between point A and B, $\Delta T = T_A-T_B$, in the limit where $\Delta T/T$ is small;
\[
S_{AB} = -\frac{\Delta V}{\Delta T}
\]

\noindent Conventionally $S$ is quoted in the convenient units $\mu$V.K$^{-1}$ as $\Delta V$ is usually $\sim \mu$V. 


\subsection{Room-temperature thermopower}
\label{sec:rttep}

Room-temperature thermopower, $\rttepm$,  measurements are made on a custom-built rig. 
The sample is placed between two polished Cu plates. By applying pressure we make a good thermal and electrical contact between the sample and the Cu plates.  Cu leads connect the Cu plates to a nano-Volt amplifier, which is powered by batteries to minimise electrical noise.  A heater coil is wound around the bottom Cu plate and a thermocouple measures the temperature difference between the plates.


To begin the process, 10 measurements of $\Delta V$ are made without a current to the heater coil giving $\Delta T \approx 0$~K (although the actual $\Delta T$ is measured).  The heater is then left on for 2 minutes followed by another set of 10 measurements of $\Delta V$ and $\Delta T$.  The current to the heating coil is adjusted so that $\Delta T \approx 2.5$~K after the two minutes heating time.  

To estimate \rttep a linear fit is made to all the data points and a $1.6$~$\mu$V.K$^{-1}$ correction for the room-temperature Cu voltage leads subtracted from the best fitting gradient.  
%
%
%
%
%

\section{Magnetisation}

DC magnetisation is the linear response of a material to a static applied magnetic field, $\BB = \mu_0(\HH + \MM_{\textnormal{DC}}) = \mu_0\HH(1+\chi)$\added{, where $ \BB $ is the total magnetic flux, $ \HH $ is the applied magnetic field and $ \MM_{\textnormal{DC}} $ is the DC magnetisation, $ \chi $ the magnetic susceptibility and $ \mu_0 $ the permeability of free-space}.  The superconducting state is characterised by the Meissner effect whereby magnetic flux is expelled from the superconductor. Thus, for the case of weak fields\footnote{Such that the applied field is less than the lower critical field of the material, $\HH\leq \BB_{c1} $, assuming that there are no demagnetisation effects (which are related to the geometry of the sample).}, $\chi = -1$.  This provides a nice, clear signal that we use to characterise \tc.

The magnetic moment of a sample is often measured in units called `emu' - electromagnetic units. The following relation is used to convert from these units to a volume susceptibility;
\[ 
\chi_V = \frac{\mu_0m^{[emu]}\rho\times 10^3}{mass\times H} 
\] 
\noindent where \(\rho\) is the sample density in g.cm$^{-3}$, mass is measured g, $H$ in Tesla, and the magnetic moment, $m$, in emu.

We measure the magnetic susceptibility of our samples using either Vibrating Sample Magnetometry (VSM) or for more sensitive measurements we use a Superconducting QUantum Interference Device (SQUID).  These are now both standard techniques for magnetisation measurements and we use commercial Quantum Design SQUID and VSM systems\footnote{The Quantum Design VSM is part of a Physical Properties Measurement System (PPMS) that we use also for resistivity measurements.} and LakeShore Design VSM systems. The Quantum Design SQUID Magnetometer is capable of measuring magnetic moments down to $10^{-7}$~emu at temperatures between 2~K and 400~K in magnetic fields up to 7~T. 

We use magnetisation as the primary technique to measure $ \tcm $.  We note here the potential difficulties with extracting reliable \tc values from resistivity data.  In polycrystalline and multi-domain single crystal HTS the resistivity will fall to zero only when there is an unbroken path of superconducting material\footnote{Of sufficient quality and size so that the probing current does not exceed the critical current density - though this is not an issue for HTS with the low currents used to measure resistivity.} between the voltage probes.  In inhomogeneous samples, for example, oxygen may not be evenly distributed throughout the sample leading to a path through material of varying doping and higher \tc than most of the rest of the sample; an overestimation of the bulk \tc would result.  More often however, the resistivity has a tail below \tc due to weak links between grains.  This tail grows with increasing current density or applied field.  In this case, \tc defined as where the electrical resistance falls to zero would underestimate the bulk $\tcm$.

Measuring \tc by magnetisation can avoid these issues as the magnitude of the diamagnetic signal from the Meissner effect shows when the bulk material has become superconducting.  In this respect, weak link behaviour between grain boundaries in polycrystalline samples can be clearly seen if present in the magnetisation data.  In polycrystalline materials, smaller single crystallites are weakly linked to their neighbouring crystallites within the ceramic sample \cite{levy1994}.  Pressing the powders into pellets before sintering at high temperatures is an attempt to strengthen the links between grains and to aiding the growth in size of the crystallites through diffusion-assisted grain-growth during sintering \cite{schandbook}.  Weak link behaviour can be seen in DC susceptibility vs. temperature data by two distinct gradients, a higher temperature intra-grain superconductivity onset and lower temperature inter-grain currents shielding the external magnetic field\footnote{In AC magnetisation measurements two distinct peaks in $\chi ''$ show these same two crossovers \cite{gomory1997}.}.

There are some magnetic effects arising from the superconducting state which may confuse the determination of $ \tcm $.  The thermal energy at temperatures around \tc for HTS is often large enough to significantly suppress SC below its mean-field value, see e.g. \cite{tallonfluctuations}.  Consequently SC fluctuations are significant \emph{above} \tc and manifest, in this case\footnote{Superconducting fluctuations also lead to a decrease in resistivity above \tc which must be carefully distinguished from a pseudogap related downturn in resistivity by suppressing the SC fluctuations in a strong magnetic field.  Alternatively superconducting fluctuations at a given temperature may be suppressed by unitary scatterer substitution for Cu (e.g. Zn) which suppresses \tc as well, but not the pseudogap \cite{naqib2005,kim2010}}, in remnant diamagnetism above \tc \cite{mosqueira2007}.  These effects however are generally subtle and were not an issue in this work.

Cuprates, being type II superconductors, display irreversible magnetisation at temperatures below $ \tcm $.  Different magnetic behaviour is seen if one's sample is cooled below \tc in no magnetic field, referred to as zero-field cooled (ZFC), relative to if it is cooled below \tc in a non-zero magnetic field, referred to as field-cooled (FC).  This difference is due to the pinning of vortices as they attempt to enter the superconductor below $\tcm$.
In ZFC, there is no flux within the material when the external magnetic field is initially turned on.  Supercurrents form on the \textit{surface} of the sample to repel magnetic flux from entering the bulk superconductor.  We determine our \tc estimates from ZFC magnetisation data.   



\subsection{Estimating \tc}
\label{sec:measuringtc}
\begin{figure}
	\centering
		\includegraphics[width=0.70\textwidth]{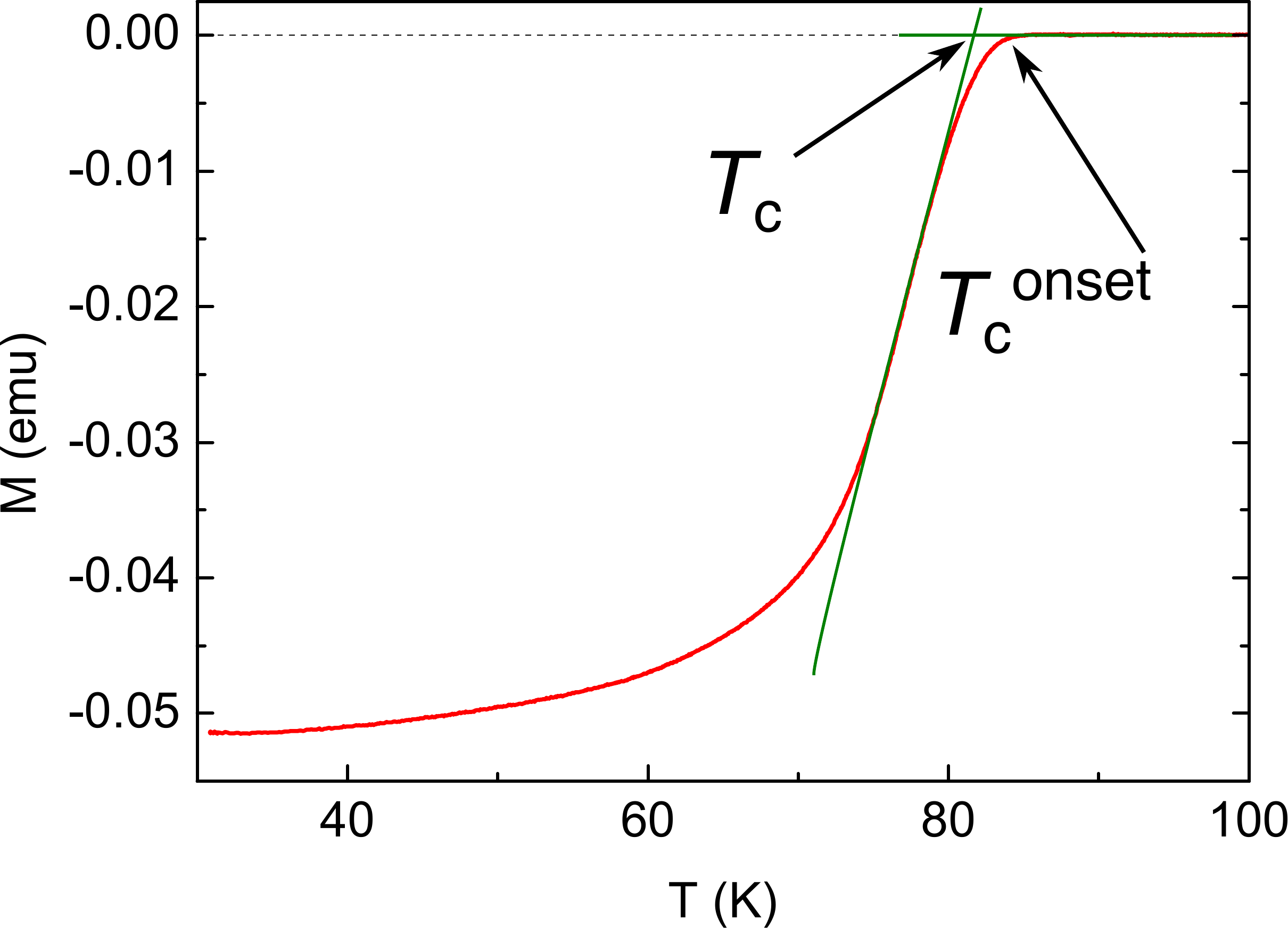}
	\caption[An illustration of two ways to estimate $\tcm$.]{An illustration of two ways to estimate $\tcm$.  The solid red curve is ZFC magnetisation data and the green lines are illustrative extrapolations.}
	\label{fig:tcestimates}
\end{figure}

There are several ways of estimating \tc from magnetisation measurements;
\begin{enumerate}
\item From the onset of the diamagnetic downturn in magnetisation vs temperature data under a small DC field, typically 20~Oe ($ 0.2 $~mT).  The downturn is of course due to the Meissner effect.  In this work we refer to this temperature as $\tconsetm$. 
\item From taking the intercept of the extrapolated normal state background magnetisation with the extrapolated steepest slope of the diamagnetic signal in $M(T)$ data.  This is the method used to measure \tc throughout this work.  The first method listed leads to higher \tc values but can be misleading; detecting a downturn in magnetisation depends on the sensitivity/noise in the measurements. 
\item From the peak in the imaginary part of the susceptibility from an AC magnetisation measurement.
\end{enumerate}
\added{An illustration of the first two methods listed above is shown in \fig~\ref{fig:tcestimates}. }

Resistivity measurements are another common way of estimating $ \tcm $.  \tc can be defined as where $\rho (T) \rightarrow 0$, or by extrapolating to zero the steepest slope in $\rho(T)$, or where $\rho(T)$ is 10\% of its normal-state value before the onset of the SC transition.
On samples with sharp superconducting transitions all these methods of measuring \tc give very similar results, but for poorer quality transitions they differ. 

%
%

%
%
%
%
%
%
%
%


\section{Muon Spin Relaxation}

\label{sec:musr}
\label{sec:musrtheory}

In Muon Spin Relaxation\footnote{This is sometimes called Muon Spin Resonance instead.} ($ \mu $SR) one directs a polarised beam of positively charged muons onto the sample. The implanted positive muons are repelled by the positively charged ions in the crystal which causes them to preferentially occupy an interstitial site in the lattice (although we cannot control which interstitial site is preferentially occupied, it is believed to be adjacent to the apical oxygen). The muons are strongly coupled to the local magnetic field, $\mathbf{B}_{\textnormal{loc}}$, causing the muon's magnetic moment to precess at a frequency proportional to the local field.  This is Larmor precession and the angular frequency is given by $\omega_{\mu}=\gamma_{\mu}\mathbf{B}_{\textnormal{loc}}$ where $\gamma_{\mu} = 0.8516\times 10^{9}$ rad.s$^{-1}$T$^{-1}$ is the gyromagnetic ratio of a muon \cite{reotier1997}.  Note the similarities between the fundamental concepts of the \musr and the NMR technique.  

\begin{figure}
	\centering
		\includegraphics[width=0.65\textwidth]{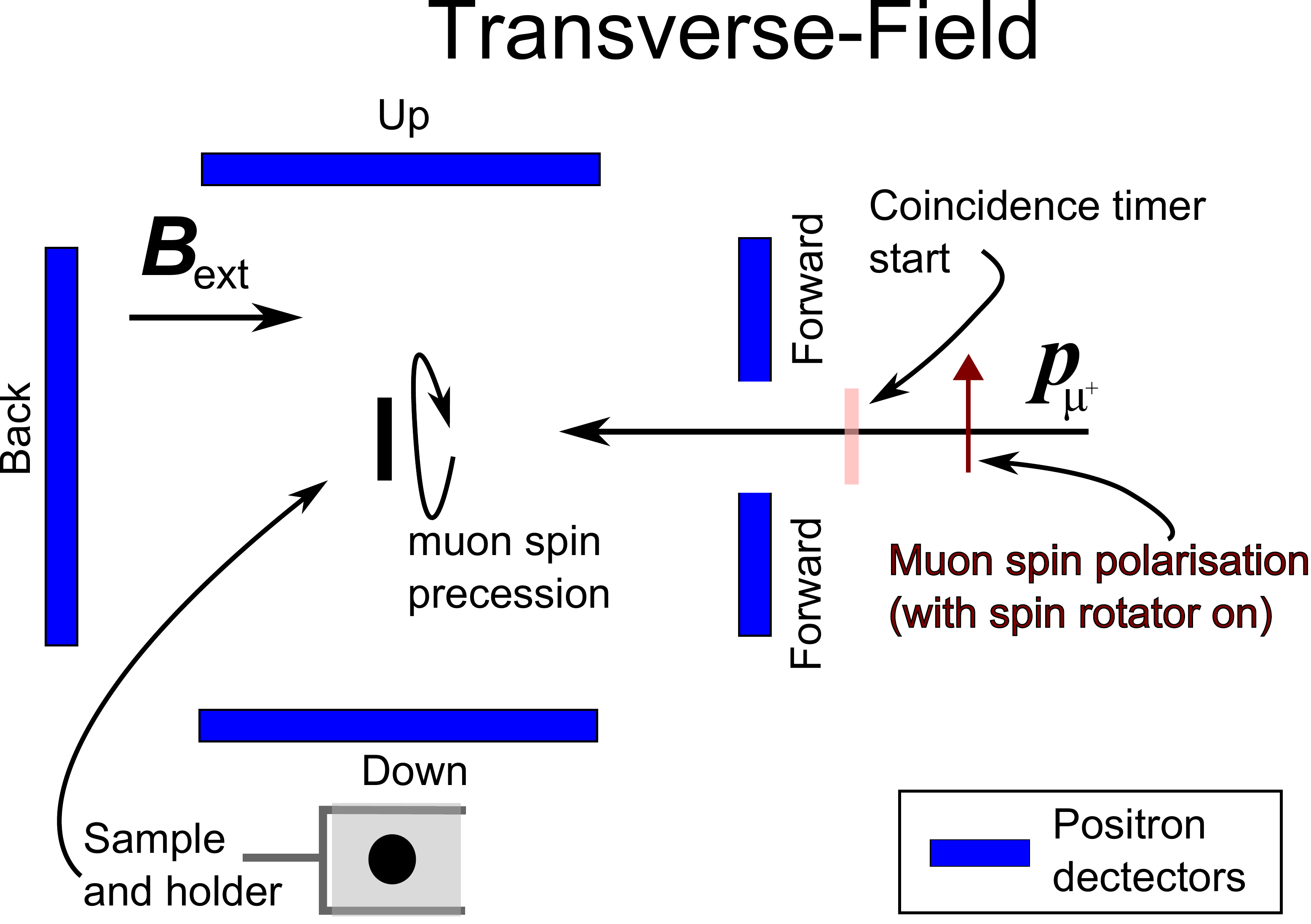}
	\caption[A schematic diagram a \musr experiment.]{\label{fig:musrsetup} A schematic diagram of the experimental set-up at the GPS beamline (used to make our \musr measurements).  Normally, the spin of the muon is anti-parallel to its momentum, however, GPS is equipped with spin-rotating magnets which flip the muon spin near perpendicular to its momentum - as annotated. A magnetic field up to 6000 G ($0.6$ T) at the sample can then be applied, around which the muons will precess. In this geometry, to make a TF measurement we would measure the difference in the number positron detections resulting from muon decay after a given time $t$ between the Up and Down detectors.  $t=0$ is when a muon is detected passing through the coincidence timer in the Front positron detector.  At the bottom a sample (black circle) is sketched taped (light grey square) to a silver holder.}
\end{figure}

When the muon decays, with a half-life of $2.2$ $\mu$s, it emits a positron with momentum preferentially along the muon's spin direction.  It is these positrons we detect. Conveniently, they have a kinetic energy large enough to not interact with the sample or holder so by observing the direction in which the positron was emitted we can infer, with sufficient statistics, what was the $\mathbf{B}_{\textnormal{loc}}$ that the muon experienced.  The direction of emission is determined by comparing the positron count rate from detectors on opposing sides of the sample, e.g. above and below, or in front and behind, see \fig~\ref{fig:musrsetup}.  Then, a depolarisation function $P_{\alpha}$ is used to represent the probability of emission in a certain direction $\alpha$.  This is the parameter we discuss throughout.  The asymmetry function is defined as 
\begin{equation}
P_{\alpha}(t) = \frac{F(t)-B(t)}{F(t)+B(t)}
\label{eq:asymm}
\end{equation}
\noindent where $F(t)$ and $B(t)$ are the raw counts (positron detection events) measured by the forward and backward detectors (or more generally, the opposing detectors) after a time $t$ from the muon injection. At $t=0$, $P_{\alpha}(t)$ takes the value $A_0$ which is typically $\approx 0.21$.  Normally one defines $\alpha=Z$ as the direction of the external magnetic field $\BB_{\textnormal{ext}}$. In zero field however, $P_Z$ is usually the direction of the muon spin polarisation. See \fig~\ref{fig:musrsetup} for a picture of these geometries.

By way of example, if all muons experience exactly the same, static field, $\mathbf{B}_{\textnormal{loc}}$ at an angle $\theta$ to the muon spin $\mathbf{S}_{\mu}$, then the Larmor equation gives;
\[
P_{\alpha}(t)=\cos^2\theta+\sin^2\theta \cos(\omega_{\mu}t)
\]

This is of course not generally the case.  Generally there is a spatial distribution of static fields $\mathbf{B}_{\textnormal{loc}}$ that the muons experience and/or fluctuations which cause a time variation in $\mathbf{B}_{\textnormal{loc}}$.  In either case, our muons are depolarised.  Depolarisation because of a field distribution \dbloc is described as dephasing, whilst from fluctuations it is described as relaxation.  Our depolarisation function in these cases looks like;
\[
P_{\alpha}(t)= \int{S_{\mu,\alpha}(t,\mathbf{B}_{\textnormal{loc}}) D(\mathbf{B}_{\textnormal{loc}}) \dd \mathbf{B}_{\textnormal{loc}} }
\]
\noindent where $S_{\mu,\alpha}(t,\mathbf{B}_{\textnormal{loc}})$ is a projection of $\mathbf{S}_{\mu}$ along $\alpha$ and is a function of time (Larmor precession) and $\mathbf{B}_{\textnormal{loc}}$ (via the Larmor frequency).  One of the simplest field distributions, \dbloc, and the one we use to analyse our data, is the Gaussian distribution.

Now, let us consider the case where $\BB_{\textnormal{ext}}=0$T and $\left\langle \blocm \right\rangle =0$ and we measure the difference between the `up' and `down' dectectors relative to the muon spin polarisation, $P_Z(t)$.  If \dbloc is isotropic and Gaussian then there is an analytical formula for $P_Z(t)$ called the Kubo-Toyabe function \cite{kubo1966, kubo1981}:
\begin{equation}
P_Z(t)=\frac{1}{3}+\frac{2}{3}(1-\sigma^2t^2)\exp\left(-\frac{1}{2}\sigma^2t^2 \right)
\label{eq:kubo}
\end{equation}
\noindent where $\sigma^2/\gamma_{\mu}^2$ is the variance of the magnetic field distribution which depolarises the muons.  Generally we expect this form of $P_Z(t)$ from magnetic fields caused by the weak nuclear moments in a sample.

I have found a useful way to think of the asymmetry $P_{\alpha}(t)$ is as a Fourier transform of the local field distribution.  A broad distribution leads to a rapid decay of asymmetry in the time domain (in which we measure), on the other hand a single local field value, as you might get from a large external field (in the absence of vortices), will lead to an oscillating signal (depending on the relative orientation of $P$ and $\BB$).  Within this picture the amplitude of a Fourier component (field value or frequency $\omega_{\mu}$) is understood as the fraction of muons experiencing that component - which can then be taken as a volume fraction of the sample with that local field value.  The local field distribution is often what we are interested in and is the physically intuitive quantity to grapple with.  

We are primarily concerned with measuring the superfluid condensate density, \ns, and so the key step in all of this is that the depolarisation rate is due to $D(\mathbf{B}_{\textnormal{loc}})$, which in a superconductor is in turn related to the vortex lattice.  Outside the vortex core the field decays as $B(r)=\frac{\Phi_0}{2\pi\lambda^2}K_0(r/\lambda)$ where $\Phi_0$ is the flux quantum and $K_0(r/\lambda)$ the zeroth-order Bessel function, see pg. 553 of reference \cite{poolehandbook}. While the vortex core radius is $r_c=\xi \ll \lambda$ (for the cuprates), the London penetration depth scales the distance over which the field drops to zero outside of the vortex core.  For the cuprates the separation between adjacent vortices is smaller than $\lambda$ for \bext$>0.1$ T, in which case the second moment of the field distribution, $\Delta \BB^2=\left< (\BB - \left< \BB \right>^2 )^2\right>$ is proportional to the penetration depth \cite{uemura1989}.  

For a static, Gaussian distribution of $\mathbf{B}_{\textnormal{loc}}$ in sufficiently large \bext, then $\sigma \propto \Delta\mathbf{B}_{\textnormal{loc}}$ where $\sigma$ is the depolarisation rate \cite{reotier1997}:
\begin{equation}
P_X(t)=A_0 \exp \left( - \frac{\sigma^2 t^2}{2} \right) \cos(\gamma_{\mu}\BB_{\textnormal{loc}}t+\delta)
\label{eq:gaussian}
\end{equation}

\noindent Here $A_0$ is the volume fraction with this magnetic phase.  Note the convention we have adopted where a factor of $1/2$ appears in the exponential (thus some data for $ \sigma $ reported in the literature needs to be corrected by a factor of $\sqrt{2}$). Recall $\gamma_{\mu}\mathbf{B}_{\textnormal{loc}}=\omega_{\mu}$ is a frequency. This is valid for \textit{polycrystalline} samples of \ybco (where the vortex lattice is 3D) but for single crystals, or Bi-based HTSs\footnote{Depending on the anisotropy of the compound, which is Pb doping and hole doping dependent, Bi-based SCs have a pancake vortex lattice above a certain magnetic field which can take quite low values (100-300 Oe) anisotropic samples.}, more care must be taken to account for the vortex lattice structure and perhaps a more sophisticated model would be required \cite{reotier1997}.  Indeed, from the Fourier transform of our $P(t)$ time spectra we see a nice Gaussian distribution of local fields with its mean value below \tc slightly shifted below $\BB_{\textnormal{ext}}$ (see \fig~\ref{fig:tfraw}).

In detail, we have the following relation between $\lambda_{ab}^{-2}$ and the depolarisation rate $\sigma$ from Bernhard \etal \cite{bernhard1995musr}; 
\begin{equation}
\lambda_{ab}^{-2}=12.72\sigma
\label{eq:sfdensity}
\end{equation}
\noindent Here $\lambda_{ab}$ has units of $\mu$m and $\sigma$ has units of $\mu$s$^{-1}$.  Thus for a typical value of $ \lambda = 160 $ nm at optimal doping, $ \sigma $ is about 3.1 $ \mu $s$ ^{-1} $.


So, let us now follow the string all the way out of the labyrinth;
\[
\sigma \propto \Delta\mathbf{B}_{\textnormal{loc}} \propto \lambda_{ab}^{-2} \propto n_s/m^*
\]


\section{Raman Spectroscopy}


We use commercial T64000 and LabRam Raman spectrometers. For lasers we use the 514.5~nm (green) and 458~nm (blue) lines of an Ar ion laser and the 633~nm (red) laser line from a \replaced{HeNe}{He-Cd} laser.  Measurements are performed in backscattering geometry.  Except for variable-temperature measurements, we use confocal microscopes (with x50 or x100 objective lenses) to focus the laser and collect the Raman-scattered light.  A $\sfrac{\lambda}{2}$ half-plate is used to orientate the polarisation of the incident laser beam.  Liquid N$_2$ cooled CCDs\footnote{Charge-Coupled-Detectors} count the diffracted photons where the diffraction grating is chosen depending on the measurement.  For example, to measure the broad two-magnon scattering signal it was desirable to use a low, 300 lines/mm grating in order to measure a large range of $\omega$ within a single spectrum.  These spectrometers detect light of a certain polarisation more efficiently, the T64000 has more than a factor of 10 difference between the efficiency of horizontal- and vertically-polarised light detection. Therefore the experiments are set up to have the exit polariser in the appropriate alignment. 

\subsection{Two-magnon measurements}
\label{sec:twomagmeasurements}

\begin{figure}
	\centering
		\includegraphics[width=0.75\textwidth]{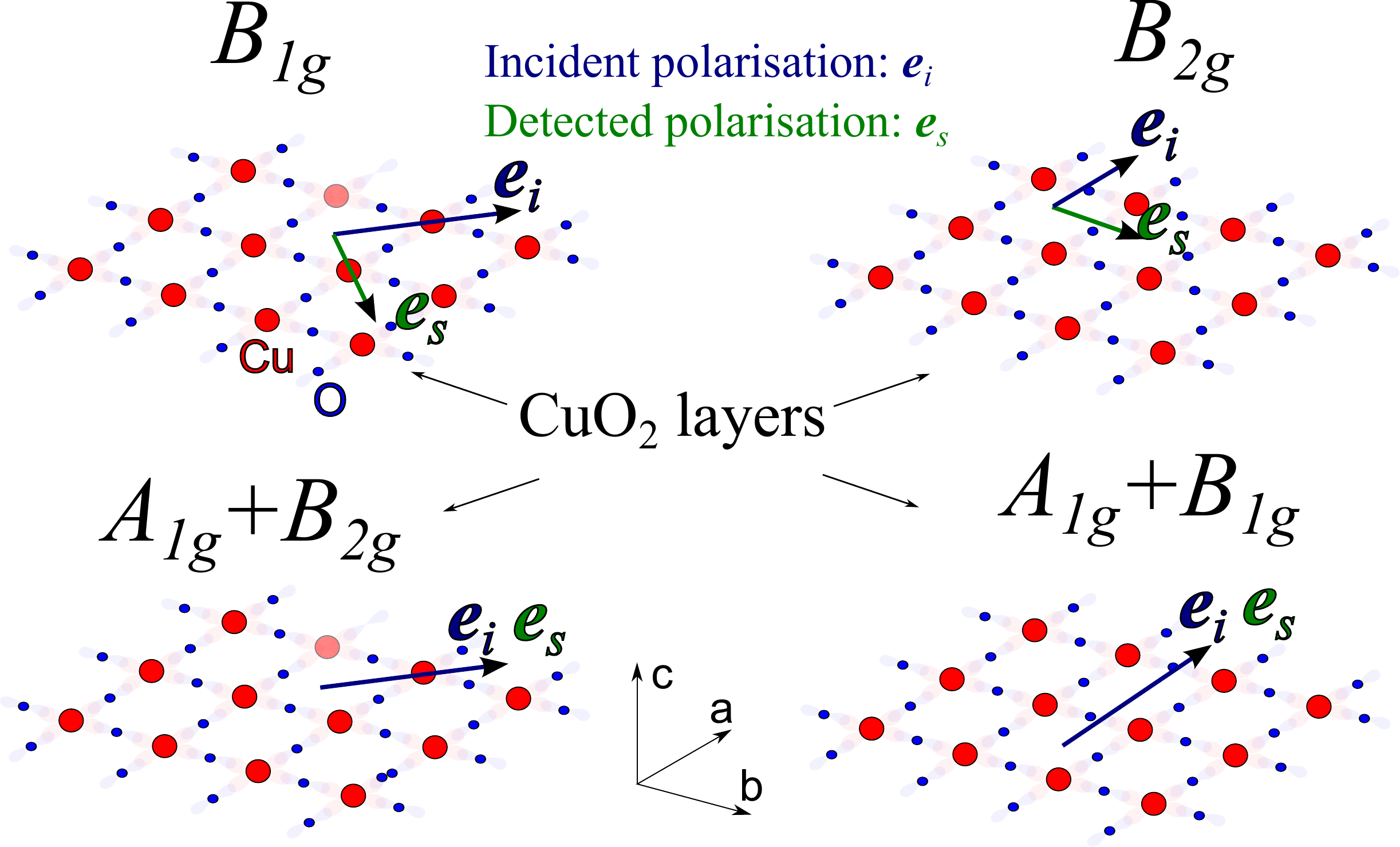}
	\caption[Schematic diagram of Raman active symmetries and the required polarisation configurations.]{Schematic diagram of various Raman active symmetries and the polarisation configurations to obtain them. The two-magnon peak is manifest in \bog and to a lesser extent in $ A_{1g} $.  These relations are strictly correct only for tetragonal material. }
	\label{fig:polarisationconfig}
\end{figure}

The relative polarisation of the incoming laser and detected scattered light must be considered in a two-magnon experiment.  As mentioned earlier, in the cuprates appreciable two-magnon response is seen in both the $A_{1g}$ and $B_{1g}$ channels but not in the $B_{2g}$ channel. Selecting the polarisation configuration allows a check on whether a genuine two-magnon peak is present in the data and can be manipulated to select out the background for subtraction.  
Magnons are excited from incident light traveling parallel to the $c$-axis\footnote{\(\mathbf{S}\equiv\mathbf{E}\times\mathbf{B}\) is parallel to the $c$-axis unit-cell vector.} so that the electric field gradients are in the $a,b$ plane. Following the notation of Sugai \etal \cite{sugai2003}, let $a$ and $b$ denote polarisation parallel to the $a$ and $b$ crystallographic axes, which lie along the Cu-O-Cu directions, respectively, while $x$ and $y$ denote polarisation rotated $ 45^{\circ} $ from $a$ and $b$. Next, we represent incident polarisation parallel to $a$ and scattered polarisation parallel to $b$ as $(a,b)$.  The Raman-active symmetries and polarisation configurations are thus $(x,y)$ for $ \bogm $, $(a,b)$ for $ \btgm $, $(x,x)$ for \aogbtg and $(a,a)$ for $ \aogbogm $. These are sketched in \fig~\ref{fig:polarisationconfig}.
  
Single-crystal samples have the feature of well-defined crystallographic axes on length scales that the x50 objective microscope probes.  Although there are clearly different single crystal domains visible under the microscope, it is possible to find a homogeneous region to measure from.  
To change between $B_{1g}$ and $B_{2g}$ scattering geometry, for example, the polarisation of the incident laser beam must be rotated from 45\degrees to 0\degrees relative to the Cu-O bond length direction.  In practice the orientation of the single crystal, both the c-axis direction and orientation of the polarisation with respect to the Cu-O bond length, can be ascertained by inspecting the phonon modes with reference to polarisation-specific spectra reported in the literature, for example \cite{osada1997,sugai2003}.  We found it was sufficient to manually rotate the crystal approximately 45\degrees to swap between \bog and \btg (or \aogbtg and \aogbog).

The most reliable way to unambiguously detect a two-magnon peak was by measuring \btg and then \bog geometries.  Both have detection polarisation orthogonal to the incident polarisation resulting in weak `background' electronic scattering and most phonon peaks are greatly reduced in intensity.  The two-magnon peak, if present, is therefore a major feature of the spectrum.  Most other experimental parameters can easily be kept constant by this modification, such as: spot position on the sample, temperature, incident laser power.    In practice though all four configurations are measured to confirm the (non)existence of the two-magnon peak, see \fig~\ref{fig:sm123allsymm} for an example.  In the underdoped Ln123 samples the two-magnon peak is easily observed at room temperature.  
 
%

\subsection{High-pressure measurements}

\label{sec:dac}

A Diamond Anvil Cell (DAC) is capable of producing very high hydrostatic pressures, well beyond $10$~GPa\footnote{In an ideal situation up to $200$~GPa}, whilst allowing spectroscopic measurements to be carried out \added{because of the transperency of diamond}.  A pressure medium surrounds the sample and is used to ensure pressure \added{on the sample} is isotropically (hydrostatically) distributed.

Background subtraction is vital for our two-magnon measurements when using a DAC. Diamond has a strong Raman active phonon mode at $1332$~\cm \cite{solin1970} which is pressure dependent and can even be used as a pressure calibrant, though less reliably than an \textit{in situ} ruby chip.  Higher-order phonon modes are significant and can be seen between $2100$~\cm and $2670$~\cm, precisely the area of interest \added{for two-magnon scattering}.  In addition, the greater the impurities or C vacancies, the stronger the fluorescence seen between $\sim 1700$~\cm and $4000$~$ \cmm $. Again, precisely the area of interest for two-magnon scattering.  We use a 4:1 methanol:ethanol pressure medium but due to the vibration energy of H modes coinciding with the two-magnon scattering energy range, Ar or He would be preferable \cite{kawada1998} if the facilities are available.  Some possible methods for subtracting these background contributions, and why they did not work, are;

\begin{itemize}
\item Location variation;  Take a spectrum with the laser spot on sample and then subtract off a spectrum taken with the laser focused on a blank part of the chamber.  \textit{It is difficult to reliably find blank part within the chamber which is commonly $\sim\!200$ $\mu$\textnormal{m} in diameter and shrinks (initially) under pressure. Also the chamber becomes opaque to optical wavelengths above about $4$ \textnormal{GPa}. In addition there is a huge signal from stray ruby crystals which are used as pressure calibration within the chamber.  This is an issue when coming to measure the two-magnon peak in cuprates ($2000$ to $4500$ \textnormal{$ \cmm $}) as the signal from the tail of the ruby peaks (centred $\approx\!5000$ \textnormal{\cm} with the $514.5$ \textnormal{nm} laser) swamps the two-magnon peak and/or changes the character of the background signal compared with other parts of the chamber.}
\item Polarisation variation; Take a spectrum first with a parallel polarisation configuration, and then with crossed polarisation.  Subtract one from the other (depending on the relative orientation of the sample's \cuo layer). \textit{Unfortunately, the response from diamond seems to also be polarisation dependent}.
\item Rotation variation; Take a spectrum and then rotate the DAC 45\degrees and re-take the spectrum. Subtract one from the other. \textit{Again, diamond unfortunately seems to have a different response depending relative orientation of the diamond to the laser polarisation, just as for two-magnon scattering}.
\end{itemize}

For the reasons listed above we were not able to subtract the large fluorescence signal from our DAC data.  

We moved to the 633 nm laser line and found the luminescence background reduced by a factor of 1000.  However, two-magnon scattering is also weak at this laser wavelength (it is resonant around 458 nm) and again we were not able to observe it.

\subsection{Variable-temperature measurements}
\label{sec:ramanvtexptechniques}
For these measurements the sample is mounted on a Cu-plate which is in good thermal contact with a Cu-cold finger cooled by a closed-cycle cryogenic system (a fridge).  This allows us to access temperatures between $8.5$~K and $295$~K.  These elements are encompassed in a chamber, with optical windows, that is pumped down to $1.4$~$\mu$Torr by a turbo-pump\footnote{With diffusion pump systems oil-droplets can back-diffuse into the sample chamber.}.  With the cryogenic apparatus in the way we can no longer do micro-Raman because the focal lengths of microscopes are too short.  Instead, the polarised incident beam is focused by a conventional 15~\replaced{cm}{mm} focal length lens and an intermediate mirror reflects the beam onto the sample in as close to backscattering geometry as possible. 

\begin{figure}
	\centering
		\includegraphics[width=1.00\textwidth]{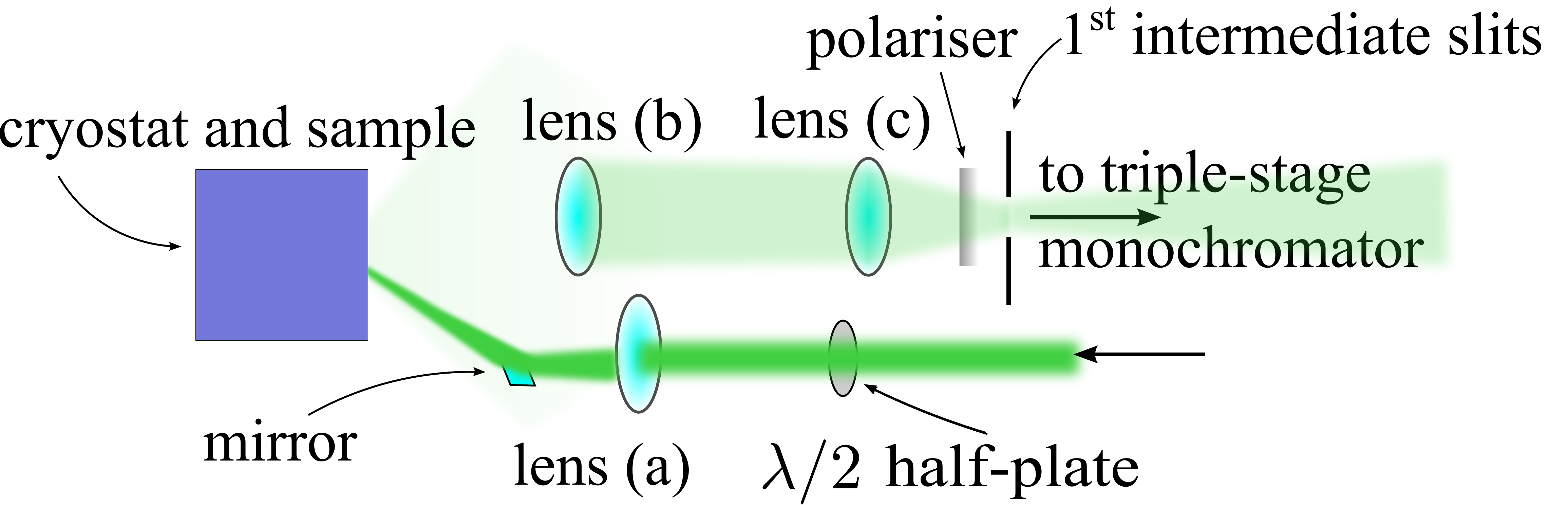}
	\caption[Schematic diagram of the experimental set up for variable-temperature Raman spectroscopy measurements.]{Schematic diagram of the experimental set up for variable-temperature Raman spectroscopy measurements.}
	\label{fig:ramanlowtempsetup}
\end{figure}

A schematic diagram of the experimental set-up for variable-temperature Raman measurements is shown in \fig~\ref{fig:ramanlowtempsetup}.

Next, we need to collect scattered light from our sample and create an image of it to send into the spectrometer.  We use a lens of focal length $f_1=10$~\replaced{cm}{mm} to collect some of the divergent scattered light and focus it into a parallel path.  A second lens captures this parallel beam and focuses it through an entrance slit into the spectrometer.  The numerical aperture of this second lens, $f_2=17.5$~\replaced{cm}{mm}, \footnote{which is related to the angle of the cone of scattered light collected, $\theta$, and the refractive index, $n$, by, $N\!A = n\sin\theta$ . In our case $n=1$.} is matched to the size of the collection mirror in the spectrometer. The magnification of our image is easily calculated, $m={f_1}/{f_2}$. If $m$ is too large, reducing the entrance slit size, which is done to  reject stray light and excess Rayleigh scattered light, will cut off signal from the sample.  It really is a fine balancing act so that although we collect very little of the scattered light with this experimental set up, it is close to the optimal set up (barring radical changes to the entire system).

By using a $\sfrac{\lambda}{2}$-half plate to rotate the polarisation of the incident laser beam we can measure Raman spectra in the usual $\bogm$, $\btgm$, $\aogbogm$, $\aogbtgm$ symmetries. \added{Note that switching between \btg and \bog geometries, for example, requires manually rotating the single crystal \textit{ex-situ}, whereas switching between \btg and \aogbog geometries, for example, requires only a rotation of the $\sfrac{\lambda}{2}$-half plate.} \deleted{Note}The detection polarisation direction is kept fixed because the T64000 triple monochromator is approximately ten times more sensitive to vertically-polarised light than horizontally-polarised light.

\chapter{Bi2201: Ion-size and Raman spectroscopy studies}
%
%
\label{ch:bi2201}
\subsubsection{Summary}

Using a simple materials variation approach, we study (i) the effect of ion-size and disorder on \tc and (ii) Raman modes in the single-layered cuprate Bi$_2$Sr$_2$CuO$_{6-\delta}$ (Bi2201).

We apply \emph{negative} internal pressure by Ba substitution for Sr in Bi2201. We find that this substitution can increase $T_c$ despite increasing disorder. That is, \tcmax is in fact better correlated to the average ion-size, rather than disorder, on the Sr site. We conclude both the ion-size and disorder significantly affect $T_c$ in Bi2201. This is a new and important interpretation of the data and will be built on in subsequent chapters. 

We also present Raman spectroscopy measurements on Bi$_{2-x}$Pb$_x$Sr$_{1.6}$Ln$_{0.4}$CuO$_{6-\delta}$ where we alter Ln and $x$ and compare the results with measurements on the bi- and tri-layer Bi-based cuprates, Bi2212 and Bi2223.  We argue from the simple interpretation of these material variation experiments that the $120$~\cm mode in Bi-based cuprates has a significant Bi vibration contribution and insignificant Sr vibration contribution. Neither vibration significantly contributes to the lower $70$~\cm mode in Bi2201 however we do see a consistent shift to lower frequencies with increasing \cuo layers of this mode in the Bi-based cuprates.

\subsubsection{Motivation}

In \refsec~\ref{sec:pressureeffects} we introduced the differing effects of internal pressure (as caused by isovalent ion substitution) and external pressure on superconductivity in the cuprates. As mentioned, it is unclear what the salient difference between these two sources of unit-cell compression is.  In this section however we merely use this observation to \emph{increase} \tcmax through \emph{negative} internal pressure effected by substituting Ba for Sr in Bi2201. By doing so we elucidate an unexplored contribution to the magnitude of \tc in the Bi2201 system. 

The hypothesis that we wish to explore here is that it is the positive internal pressure which contributes mainly to the reduction in \tcmax upon Ln substitution in the Bi2201 system. A simple test on this hypothesis is to substitute the larger Ba ion for Sr to induce ``negative internal pressure'' (i.e. increase the unit cell volume) and measure the resulting doping dependence of $\tcm$. In this case disorder, as commonly characterised in this system by the variance quantity $\sigma^2 = \left\langle  r^2 \right\rangle - \left\langle r \right\rangle^2$ \cite{fujita2005} where $r$ is ionic radius of the Sr-site ion from \cite{shannon}, is actually increased \added{because of the difference in ion-size, $ r $, between the Ba and other (Sr and Ln) ions on the Sr-site (see \fig~\ref{fig:bi2201xtal} for diagram of the Bi2201 crystal structure)}.  If disorder plays a dominant role \tcmax should decrease, while \tcmax will be increased if our internal pressure hypothesis is correct and significant. 


%
%

\section{Ion substitution effects on \tc}
\subsection{Introduction}

\begin{figure}
	\centering
		\includegraphics[width=0.40\textwidth]{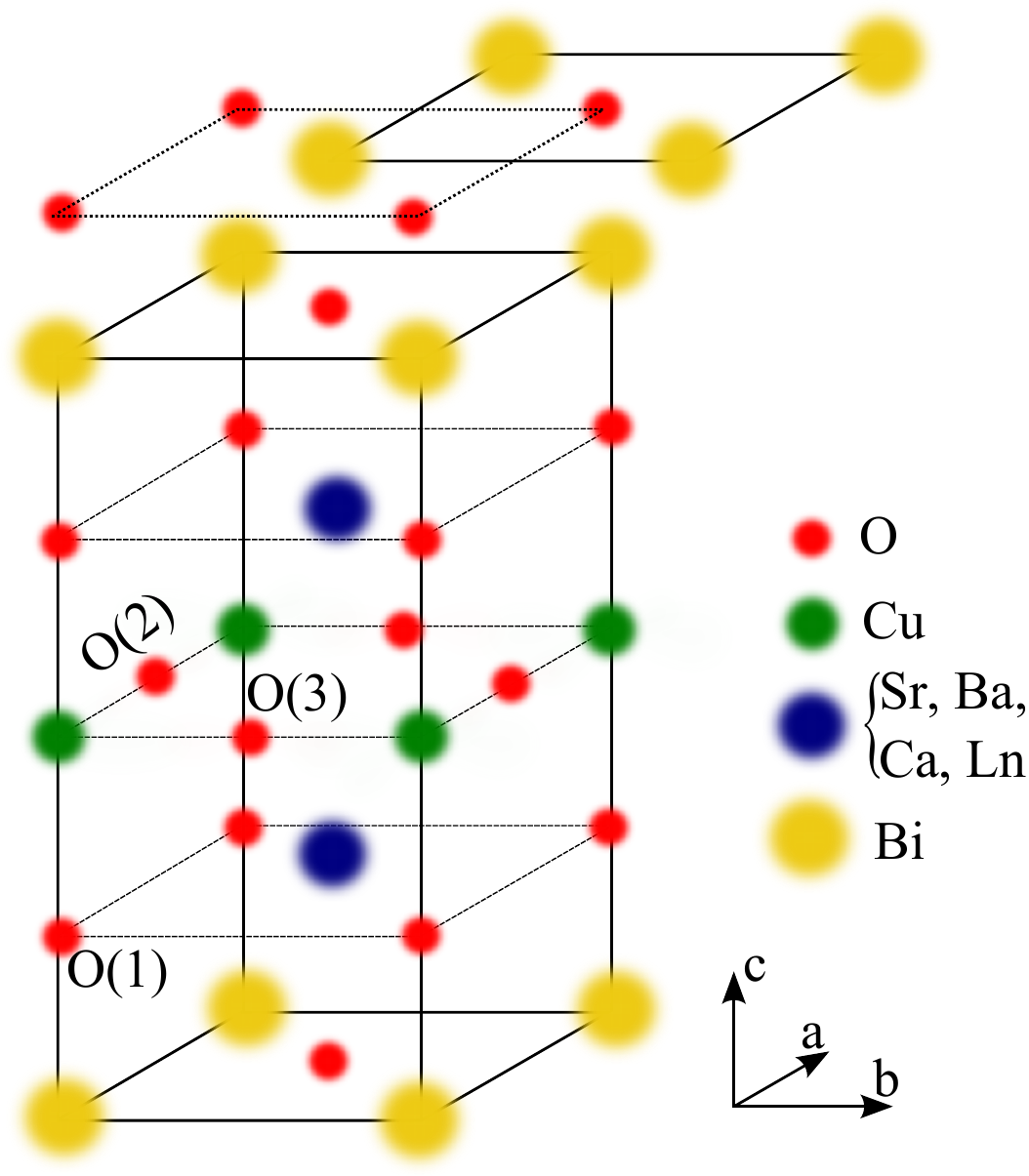}
	\caption[Schematic diagram of the Bi2201 crystal structure.]{Schematic diagram of the Bi2201 crystal structure.}
	\label{fig:bi2201xtal}
\end{figure}

Bi$_{2}$Sr$_{2-y}$Ln$_{y}$CuO$_{6+\delta}$ (Bi2201), where Ln may be any lanthanide rare-earth element, has received much attention recently \cite{fujita2005, kim2010, schneider2005, sato2009, boyer2007, okada2010, wen2009, yang2006, hashimoto2009}. Bi2201 has a single CuO$_{2}$ layer residing between SrO layers which are, in turn, situated between BiO$_{1+ \delta}$ layers with variable oxygen content, see \fig~\ref{fig:bi2201xtal}. Doping, $p$, is often controlled by varying the Ln ratio, $y$, on the Sr site and it was observed that smaller Ln ions suppress the \tc for any particular doping \cite{nameki1994}. Many authors have attributed this decrease in $T_c$ to disorder on the Sr site (the ``$A$-site'') \cite{kim2010, sato2009, okada2008, fujita2005} resulting from the difference in ion-sizes between Sr and Ln and quantify this disorder using $\sigma^2$. 

Superconductivity in the cuprates occurs on the \cuo layer and as such intra-layer impurities or disorder \added{(e.g. Zn or Ni ions on the Cu site in the \cuo layer)} rapidly suppresses superconducting properties such as $T_c$ \cite{tallon1997,kim2010} and the superfluid density \cite{tallon2005,kim2008}. Inter-layer disorder is generally less effective at suppressing these properties and its effectiveness is dependent on the type of disorder \cite{eisaki2004}.  For this reason we seek to the test hypothesis that the ion-size, rather than disorder, is the variable with the most important effect on \tc. As we have discovered, the Bi2201 material is amenable to such an experiment as the large Ba ion can be partially substituted for Sr, as indicated in \fig~\ref{fig:bi2201xtal}.

It has also been found that smaller Ln ions enhance the pseudogap in Bi2201 as measured by ARPES \cite{okada2006, okada2008, okada2010}, possibly because of a larger $J$ arising from a shorter superexchange path-length\footnote{As extensively discussed in Chapter~\ref{ch:twomag}.}. The pseudogap competes with superconductivity \cite{tallon2001} and so a larger pseudogap energy will further suppress $T_c$.  We believe this must be considered as an additional (rather than primary) effect because \tcmax itself is reduced by decreasing Ln ion size.


\subsection{Results and Discussion}

Bi2201 has two doping channels, the Sr$^{2+}$/Ln$^{3+}$ ratio and excess oxygen in \added{the} BiO$_{1+ \delta}$ blocking layer. Unlike others, we anneal at different temperatures and oxygen partial pressures to reversibly alter the doping state through the oxygen content. Most other studies in this system have simply utilised the as-prepared oxygen content and altered the Sr/Ln ratio to control doping.
The experimental procedure follows standard techniques as described in the previous chapter.

Room-temperature thermopower measurements, $\rttepm$, are used as a proxy for the doping state $p$ - which is very tricky to directly measure in these systems. The well-known \added{Obertelli-Cooper-Tallon }(OCT) relation between $p$ and \rttep \cite{oct} is invalid for Bi2201 \cite{ando2000, konstantinovic2003} and in general the precise relation between \rttep and $p$ is not known for Bi2201 - although it is known that \rttep monotonically decreases with increasing $p$ as is familiar from the OCT relation. In the special case of Ln=La (Bi$_{2}$Sr$_{2-y}$La$_{y}$CuO$_{6+ \delta}$) however, we have derived a working relation between these two quantities from the work of Ando \etal \cite{ando2000};
\begin{equation}
p=-0.026+0.18\exp\left(\frac{-\rttepm}{110}  \right)
\label{eq:prttepbi2201}
\end{equation}
\noindent Furthermore, this relation is verified by the independent measurement technique of $p$ of Schneider \etal \cite{schneider2005}. 

\begin{figure}
	\centering
		\includegraphics[width=0.75\textwidth]{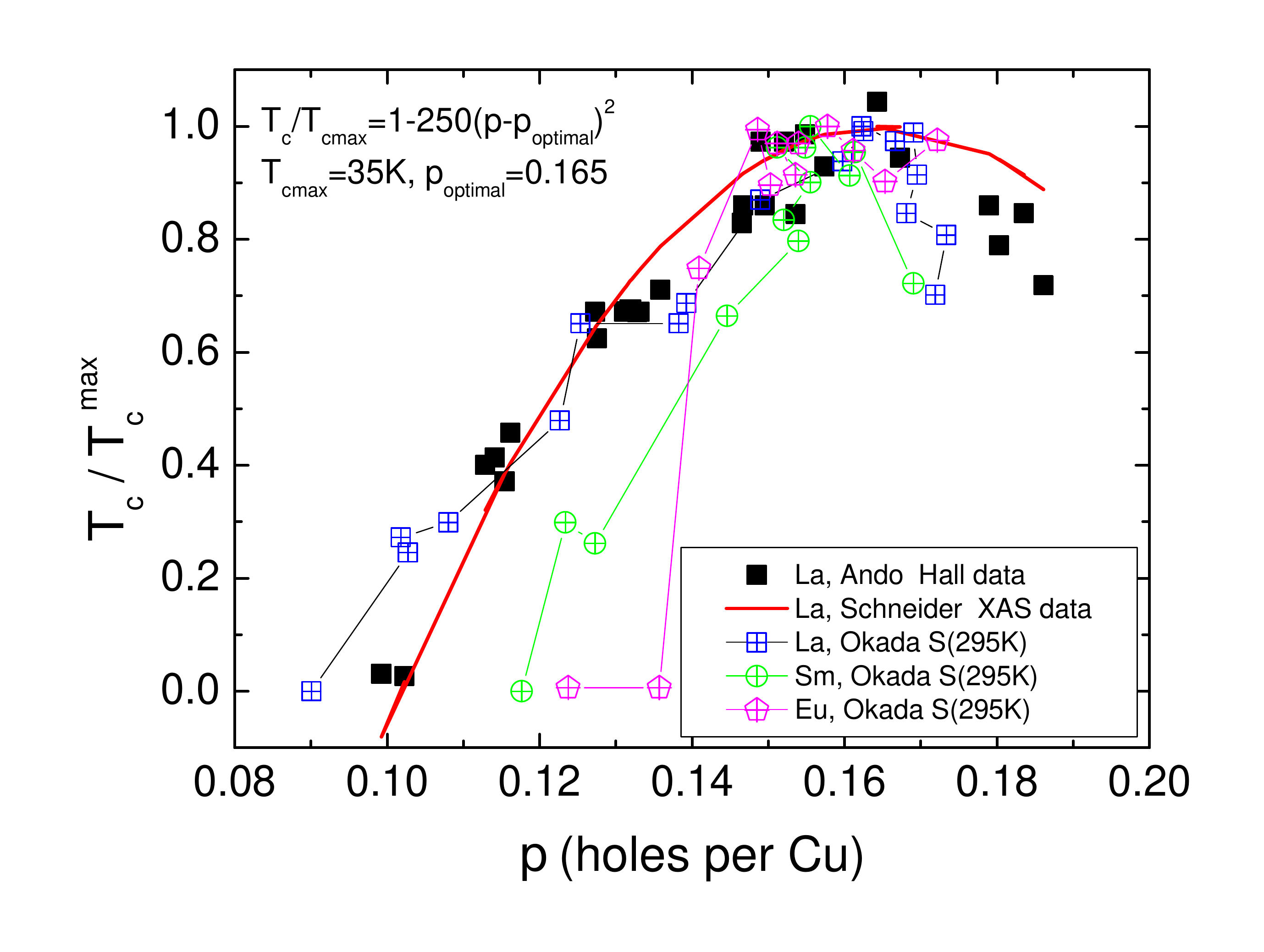}
	\caption[\tc vs. $ p $ for Bi2201 from the literature for La, Sm and Eu substitutions.]{A collection of normalised \tc values of Bi$_{2}$Sr$_{2-y}$Ln$_{y}$CuO$_{6+ \delta}$ (Ln=La, Sm or Eu as annotated in the legend) vs. $p$ taken from literature. Ando \etal (black square data points) determine $p$ from Hall-effect measurements, testing the validity of their results by comparing with La214 Hall-effect data \cite{ando2000}.  From this work we also derive the working relation between $p$ and \rttep shown in \eq~\ref{eq:prttepbi2201}.  Independent confirmation of these $p$ values come from the X-ray Absorption Spectroscopy measurements of Schneider \etal \cite{schneider2005} (Red line).  From these two data sets we derive the modified parabolic dependence of $T_c/T_c^{\textnormal{max}}$ with parameters shown in the top left of the figure.  Applying \eq~\ref{eq:prttepbi2201} to data from Okada \etal \cite{okada2006} results in the three other data sets plotted.  The Ln=La data are all consistent, but not the smaller ion-size data.  The discrepancy for smaller ions is probably because of a genuinely smaller doping range over which superconductivity is observed but may also be because \eq~\ref{eq:prttepbi2201} is not valid for Ln$\neq$La.}
	\label{fig:rtvsp}
\end{figure}

Once this relation is applied to the smaller Ln ions, the materials describe a more narrow parabolic doping dependence of \tc as shown in \fig~\ref{fig:rtvsp}.  This is likely because of a genuinely smaller doping range over which superconductivity is observed - despite possible pressure-induced-charge-transfer (PICT) $y$ seems to be a good measure of $p$ \cite{fujita2005}, though it may also be because \eq~\ref{eq:prttepbi2201} is not valid for Ln$\neq$La.  As mentioned, we use thermopower measurements, rather than $p$ estimates, throughout. 

\begin{figure}[tb]
	\centering
		\includegraphics[width=0.750\textwidth]{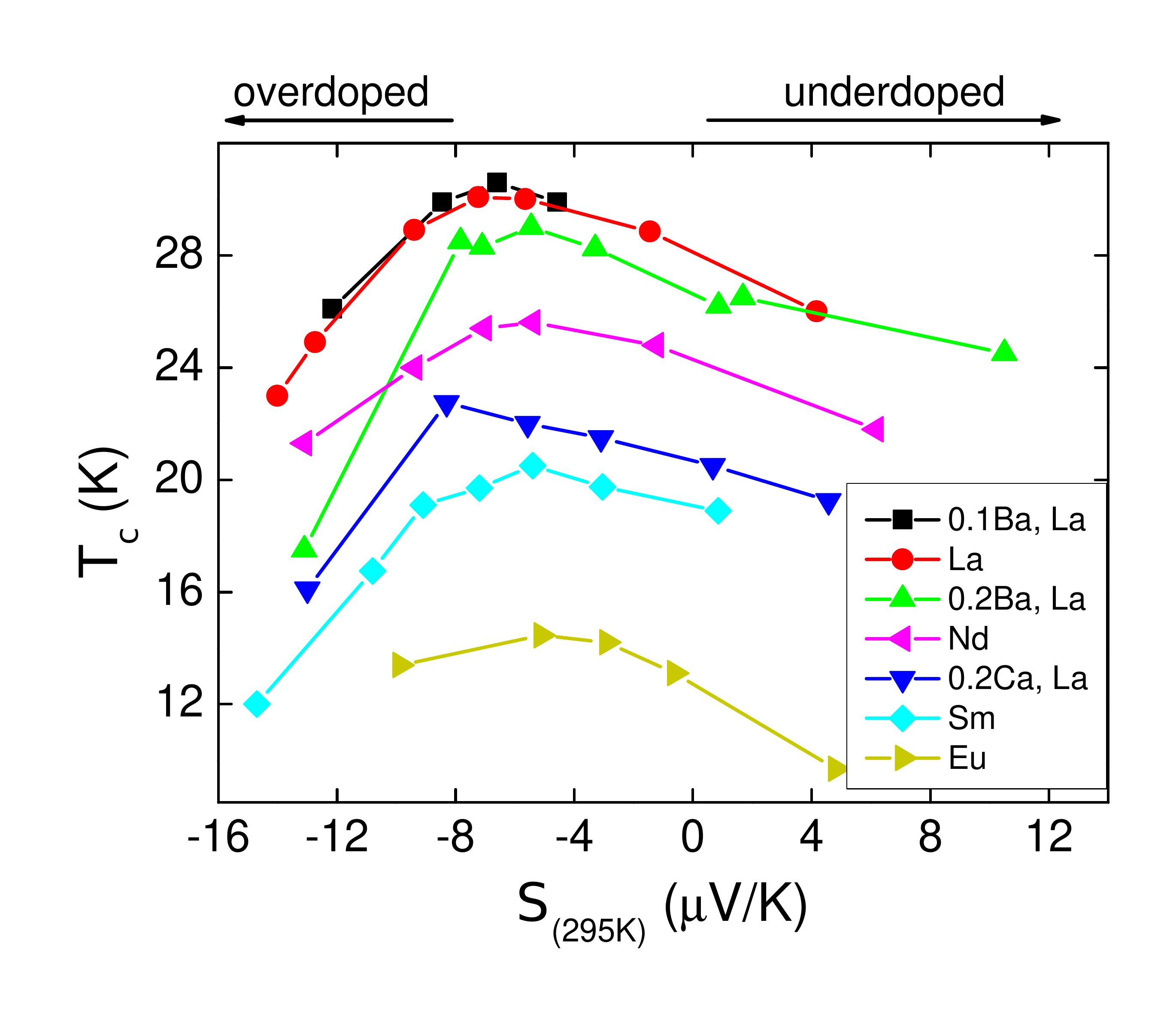}
	\caption[\tc vs. \rttep for Bi$_{2}$(Sr$_{1.6-x}$A$_x$)Ln$_{0.4}$CuO$_{6+ \delta}$ with A=Ba or Ca.]{\label{fig:tcvsrttep} \tc plotted against the room temperature thermopower, $\rttepm$, for Bi$_{2}$(Sr$_{1.6-x}$A$_x$)Ln$_{0.4}$CuO$_{6+ \delta}$ where A=Ba or Ca and in the legend the Ln ion is specified.  Note the comparable \tc values of the Ba substituted samples with the standard Bi$_{2}$Sr$_{1.6}$Ln$_{0.4}$CuO$_{6+ \delta}$ material.  Higher $\rttepm$ values represent underdoping as annotated. The onset of superconductivity is typically observed between 1 and 3~K higher than the values plotted here.}
\end{figure}

\fig~\ref{fig:tcvsrttep} shows \tc values for our model system with various Ln substitutions with different oxygen content. The data is plotted as \tc versus $\rttepm$ for a range of doping states from underdoped (right-hand side, $\rttepm > -2$~$\mu$V/K) to overdoped (left-hand side, $\rttepm < -8 $~$ \mu$V/K). As other groups have noted \tcmax falls rapidly with decreasing ion size. To this we add our new data for Ba and Ca substituted samples of composition Bi$_{2}$(Sr$_{1.6-x}$A$_{x}$)La$_{0.4}$CuO$_{6+ \delta}$ , where A = Ba or Ca.  Samples of the Bi$_{2}$(Sr$_{1.5}$Ba$_{0.1}$)La$_{0.4}$CuO$_{6+ \delta}$ and Bi$_{2}$Sr$_{1.6}$Eu$_{0.4}$CuO$_{6+ \delta}$ materials were made by Thierry Schnyder, \added{a joint Victoria University of Wellington - \'{E}cole Polytechnique F\'{e}d\'{e}ral de Lausanne (EPFL) Master's student who studied in our group}, and the data presented here for these samples \replaced{were}{was} taken by him. We find that \tcmax is in fact slightly increased by moderate Ba substitution on the Sr site despite the increasing disorder, while higher Ba substitution then decreases \tcmax. 

We note however that our samples did not display \replaced{superconducting transition temperatures}{$ T_c $s} as high as others have reported \cite{kim2010} despite replicating their sintering conditions. Unfortunately this leaves the possibility that the best possible quality Ba-free sample might still have a higher \tcmax than best possible quality Ba-doped sample.  We would need to of course synthesize a Ba-doped Bi2201 with a \tcmax higher than the highest reported value for Ba-free Bi2201 to make the conclusions of this section more reliable.  One possibility in this regard is 10-25\% substitution of Bi by Pb. The Bi-O$_x$ rock salt layer size mismatch with the rest of the unit cell causes an incommensurate modulation over several unit cell lengths, see e.g. \cite{tallonchapter, nameki1993, nameki1994}.  The unit-cell repeatability can be returned when overdoped with oxygen or when Pb is partially substituted for Bi in Bi$_{2-x}$Pb$_x$Sr$_{1.6}$La$_{0.4}$CuO$_{6+ \delta}$ as the (Bi,Pb)-O$_x$ unit increases in size and fits into the unit cell \cite{williams2000}. This tends to improve the sample quality, for example, sharper \tc transition widths and less super-lattice modulations, see below \ref{sec:bi2201ramanintro}.

\begin{figure}[tb]
	\centering
		\includegraphics[width=0.45\textwidth]{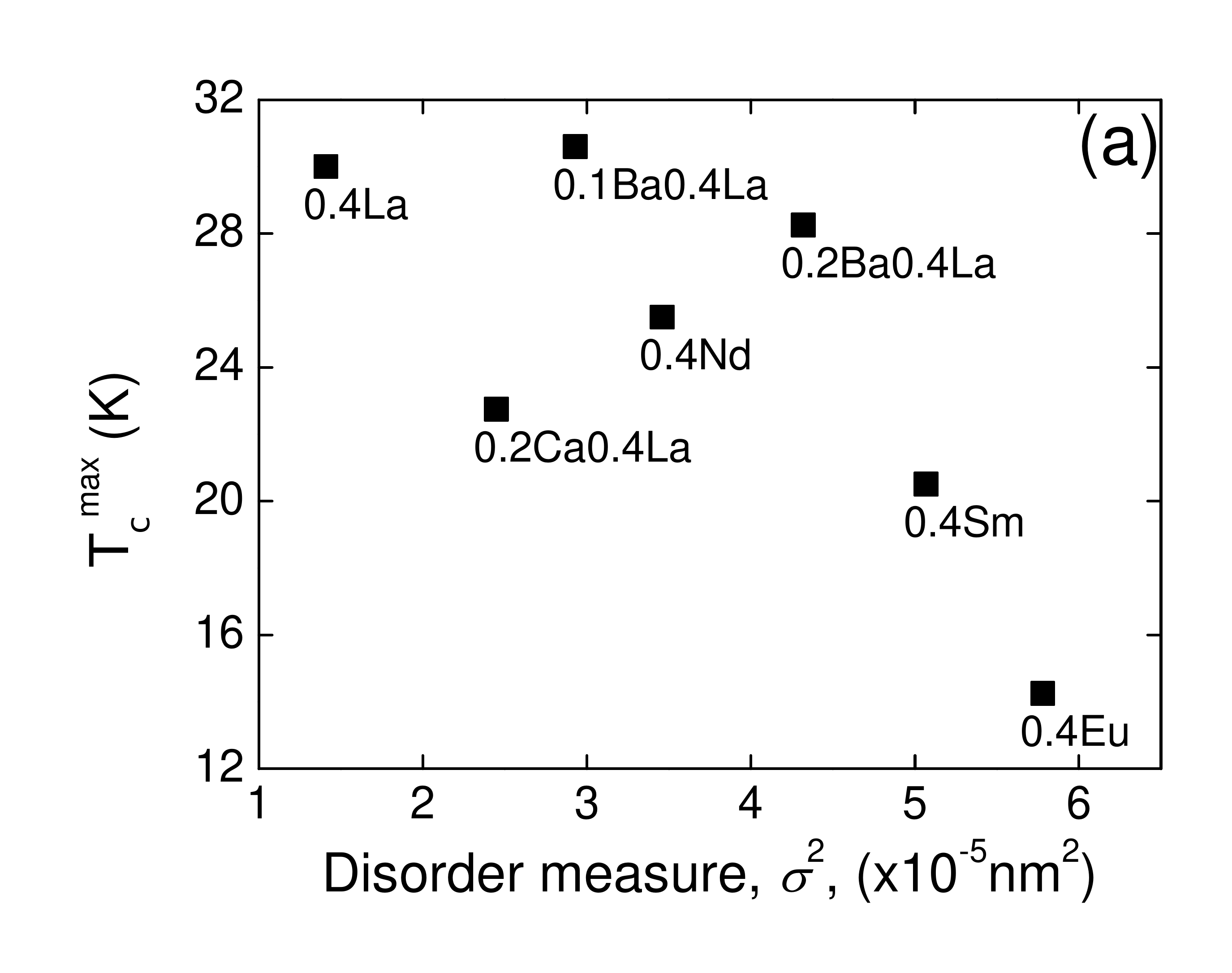} \includegraphics[width=0.45\textwidth]{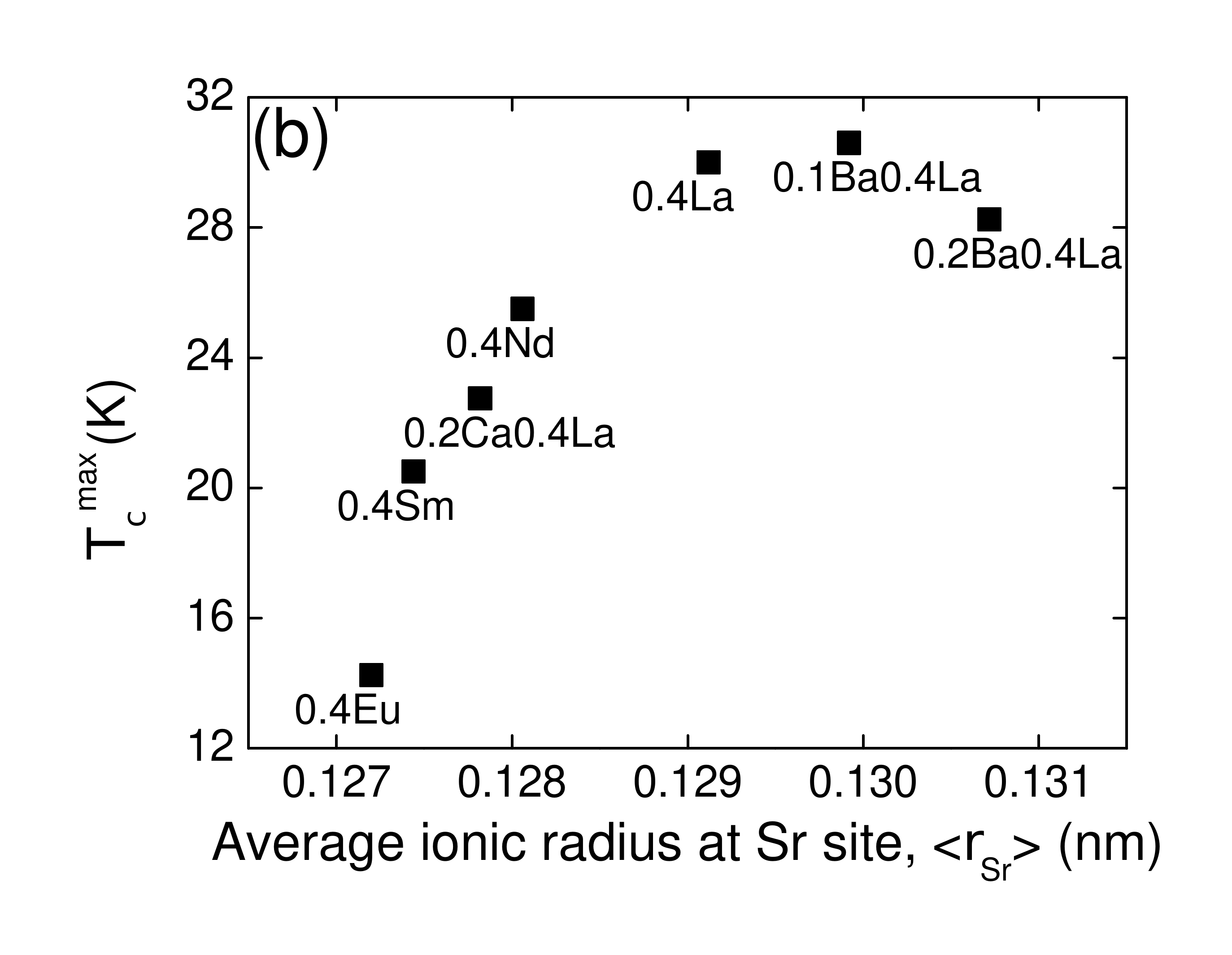} 
	\caption[$ \tcmaxm $, disorder and ion-size in Bi2201.]{(a) The maximum \tc, \tcmax, of our Bi2201 samples plotted against a standard measure of disorder in this system, $\sigma^{2}$. (b) The same \tcmax data plotted against the average ion-size at the Sr site, $\left\langle r\right\rangle$.  Collectively these plots show the importance of ion-size on the value of \tcmax in this system.  Apparently, with \textit{negative} internal pressure one can increase the \tcmax of a cuprate.}
	\label{fig:tcvssigma}
\end{figure}

That ion-size affects $T_c$ through internal pressure can be more clearly seen by the data plotted in \fig~\ref{fig:tcvssigma}.  Recall that a common measure for disorder used in the literature is the quantity $\sigma^2 = \left\langle r^2 \right\rangle - \left\langle r \right\rangle^2$ where $r$ is the ion-size at the Sr site. As an example, for the case of Bi$_{2}$Sr$_{1.6}$La$_{0.4}$CuO$_{6+ \delta}$, $\left\langle r \right\rangle=1.6r_{\textnormal{Sr}}+0.4r_{\textnormal{La}}$ and $\left\langle r^{2} \right\rangle=1.6r_{\textnormal{Sr}}^{2} +0.4r_{\textnormal{La}}^{2}$ where $r_{\textnormal{La}}$ and $r_{\textnormal{La}}$ are the ionic radii of Sr$^{2+}$ and La$ ^{3+} $ respectively from Shannon \cite{shannon}.  The superfluid density is more sensitive than \tc to disorder and so a more direct measure of the effect of disorder on the superconducting properties could be obtained by measurements of the superfluid density using, for example, the muon spin-relaxation technique.  We use this approach to study disorder effects in the YBaSrCu$ _{3} $O$ _{7-\delta} $ system in Chapter \ref{ch:musr}.
If disorder were the only mechanism suppressing $T_c$, we would expect the plot of \tcmax vs $\sigma^{2}$ shown in \fig~\ref{fig:tcvssigma}a to show a clear correlation between the two quantities. It does not. As \fig~\ref{fig:tcvssigma}b shows there is in fact a clearer relation between \tcmax and the average ion size at the Sr site. \fig~\ref{fig:tcvssigma} thus demonstrates the influence and importance of internal pressure on $T_c$.

It is likely that both disorder and internal pressure have an effect on $T_c$ upon ion-size substitution in our samples; disorder weakly suppressing $T_c$ and negative internal pressure increasing $T_c$ in the Ba doped samples. Indeed decreasing the internal pressure further by going to 0.2Ba substitution does not further increase \tcmax in this system, as shown in \fig~\ref{fig:tcvssigma}b.

\section{Investigation of Raman phonon modes}

\subsubsection{Introduction}
\label{sec:bi2201ramanintro}

Raman spectroscopy is an important tool for studying the vibrational properties and characterizing the superconducting order parameter \cite{sugai2003} and pseudogap \cite{storey2007} of high temperature superconducting cuprates.  For example, it can be used to characterize their doping level and in Bi-based cuprates the Bi:Pb ratio (where Pb is substituted for Bi) \cite{williams2000, kakihana1996, williams2007}. However there are several conflicting assignments of the low frequency phonon modes in the Bi$_2$Sr$_2$CaCu$_2$O$_{8-\delta}$ (Bi2212) system \cite{kakihana1996} and these modes have not been extensively studied in the Bi$_2$Sr$_2$CuO$_{6-\delta}$ system.  

Our approach to studying the nature of these modes is simple; we make ion-substitutions and observe the effect on the low-frequency modes.  Using standard synthesis and characterisation techniques described earlier, we prepare Bi$_{2-x}$Pb$_x$Sr$_{1.6}$La$_{0.4}$CuO$_{6-\delta}$ for $x=0.0$,0.2,0.35 and 0.45. 

Raman measurements were made at ambient temperature using a T64000 confocal microscope Raman spectrometer in back-scattering geometry.  We use the $633$~nm line from a HeNe laser with power $\leq 1$~mW focused to a spot by a x50 objective (N.A.=0.75). A 1800 lines/mm diffraction grating was used to obtain high resolution measurements of the low-frequency modes. 

\begin{figure}[tb]
	\begin{center}
			\includegraphics[width=0.66\textwidth]{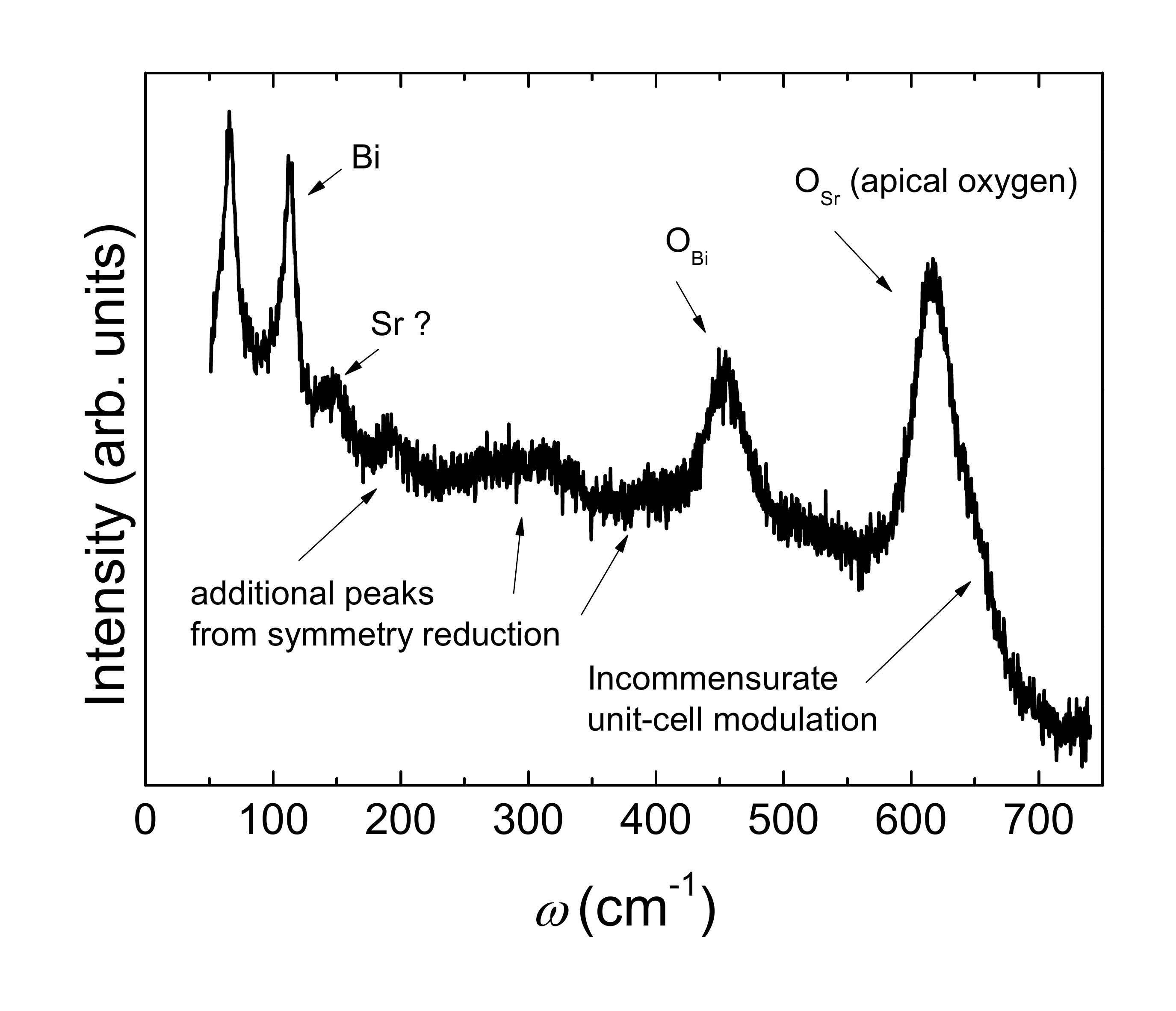}
			\caption[Typical Raman spectrum of Bi2201.]{An annotated spectrum for optimally doped polycrystalline Bi$_2$Sr$_{1.6}$La$_{0.4}$CuO$_{6-\delta}$ indicating the predominant nature of each phonon mode.}
			\label{fig:fullscan}
	\end{center}
\end{figure}

\fig~\ref{fig:fullscan} shows a typical spectrum for optimally-doped Bi$_{2}$Sr$_{1.6}$La$_{0.4}$CuO$_{6-\delta}$ constructed from two high-resolution frames joined at $\omega\!\approx\!410$~$ \cmm $.  The Raman signal was weak for all samples studied which necessitated long integration times.  

The peaks at $ \approx\!450$~\cm and $ \approx\!620$~\cm are predominantly due to vibrations of oxygen ions in the BiO$_{1-\delta}$ layer, O$_{\textnormal{Bi}}$, and the apical-oxygen ions in the (Sr,Ln)O layer, O$_{\textnormal{Sr}}$, respectively \cite{williams2000, kakihana1996, falter2003} - refer to \fig~\ref{fig:bi2201xtal} for a picture of the crystal structure.  The shoulder peak at $\approx\!650$~\cm is likely due to incommensurate modulation of the Bi2201 `unit cell' \cite{williams2000, osada1997} and it disappears when we substitute Pb on the Bi site ($x=0.2$,0.35 and 0.45).

Only four Raman active modes are predicted for stoichiometric, tetragonal Bi2201 \cite{falter2003}. The additional peaks observed between $420$ \cm and $150$ \cm are due to local symmetry reduction from Ln substitution, oxygen non-stoichiometry and the orthorhombicity observed with XRD \cite{kakihana1996, osada1997}.  The un-labelled peak at $ 65 $~\cm is discussed in the following section along with the assignment of the $ 120 $~\cm mode.

This work has been published here \cite{mallett2012}.

\subsection{Low frequency modes} 

\begin{figure}
	\centering
		\includegraphics[width=0.63\textwidth]{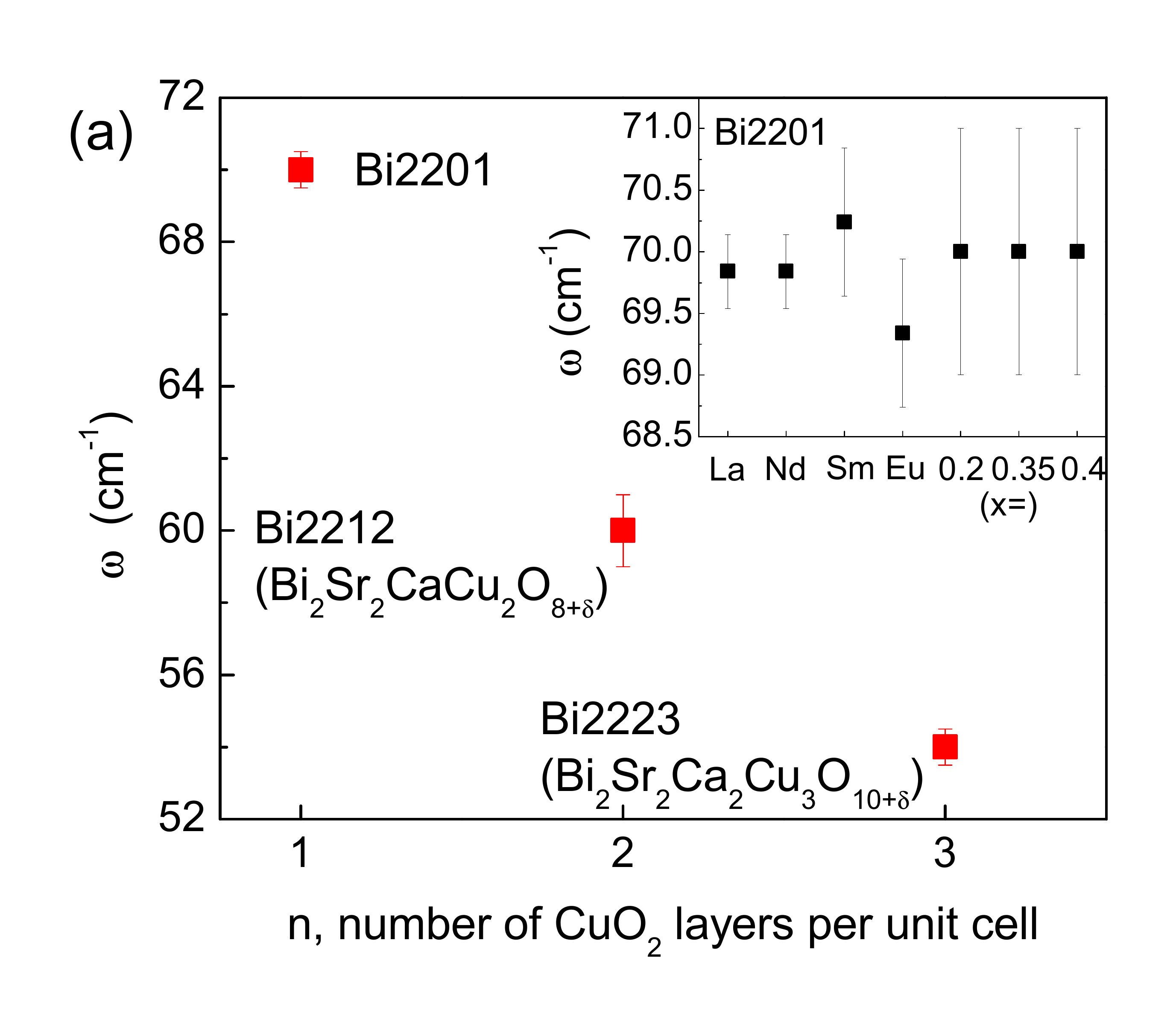} 
		\includegraphics[width=0.63\textwidth]{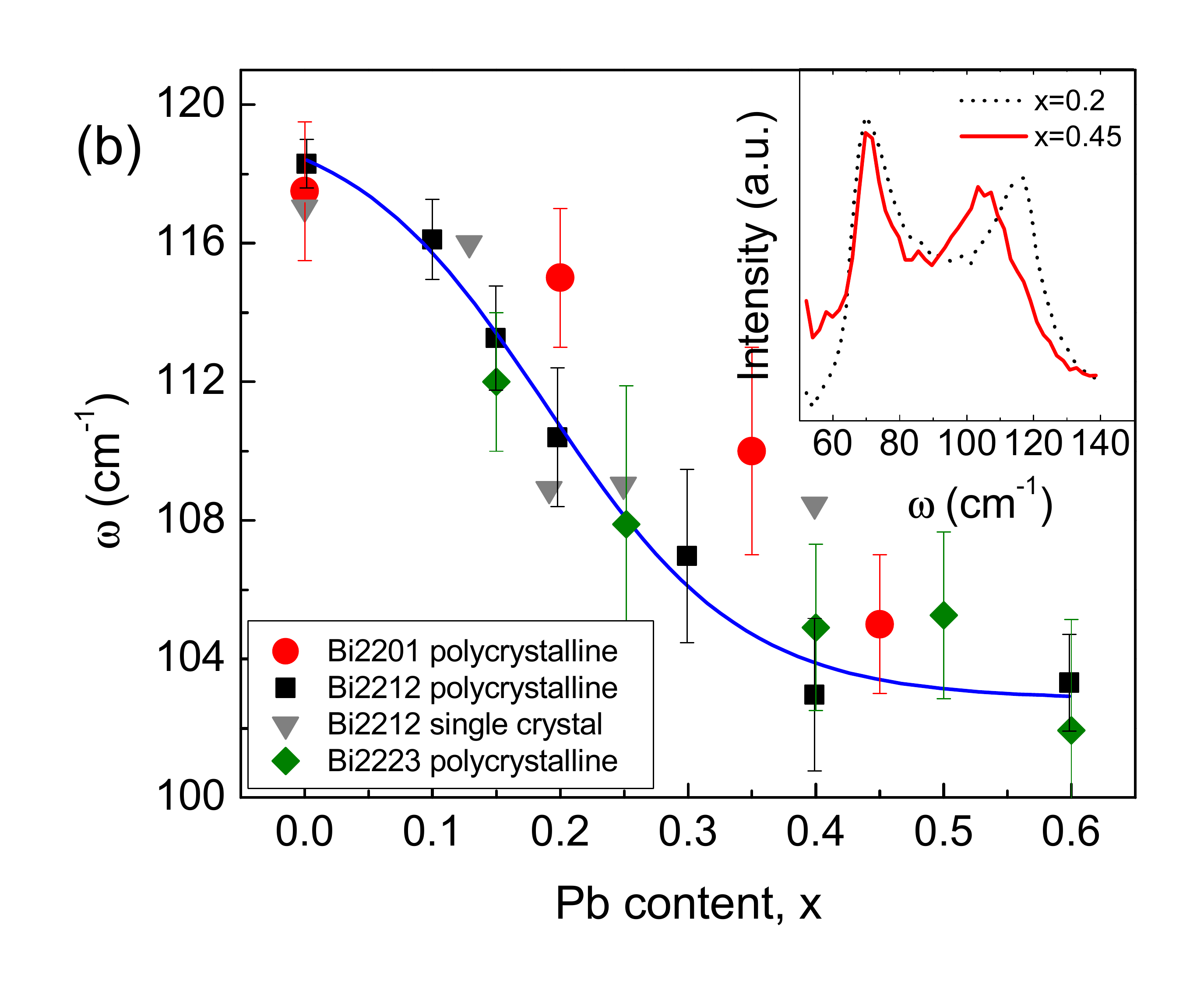}
	\caption[Dependence of the low frequency Raman modes on various ion substitutions.]{(a) The position of the $\omega _1=70\pm 0.5$ \cm peak for Bi2201 with data for Bi2212 and Bi2223 from \cite{williams2007}. Inset; Peak positions for the $70$~\cm mode for \added{Bi2201 with} $x=0$, Ln=\{La, Nd, Sm, Eu\} and various Pb concentrations\deleted{of our Bi2201 samples}. (b) Positions of the $\omega_2\sim 118$ \cm mode in Bi2201 are plotted against Pb content, $ x $, \replaced{as red circles}{(black circle)}.  Also plotted are reproduced data for Pb substituted Bi2212 (black squares and downward-triangles) and Bi2223 (green diamonds) from reference \cite{williams2007}.  The line is a fit to $\omega=a_0+a_1\left[1-\tanh\left(\frac{x-x_0}{a_2}\right)\right]$ and a guide to the eye \cite{williams2007}. Inset; Raw spectra for Bi2201 with two different $x$ values\deleted{, this data was taken using a 600 lines/mm grating}.}
	\label{fig:lowwmodes}
\end{figure}

As mentioned, the two most prominent low-frequency modes observed in Bi2212, labelled $\omega_1$ and $\omega_2$, have received conflicting assignments \cite{kakihana1996}.  In Bi$_{2-x}$Pb$_x$Sr$_{1.6}$La$_{0.4}$CuO$_{6-\delta}$ they occur at $\omega_2=117.5\pm2 $ \cm for $x=0$, $\omega_2=115\pm2 $ \cm for $x=0.2$, $\omega_2=110\pm2 $ \cm for $x=0.35$, $\omega_2=105\pm2 $ \cm for $x=0.45$  and  $\omega_1=70\pm0.5 $ \cm independent of $x$.  These data are plotted in \fig~\ref{fig:lowwmodes}. We note Sato \etal report Raman data on single crystals of optimally-doped Bi$_{2-x}$Sr$_{1.6}$Ln$_{0.4}$CuO$_{6-\delta}$ which show this peak slightly higher at $\omega_2=121\pm 1$ \cm. In fact, we find their peak positions are consistently $\sim 3$~\cm higher than ours.

The peak at $\omega_1=70\pm 0.5$ \cm does not shift with doping, Pb concentration $x$, or Ln ion substitution, as shown in \fig~\ref{fig:lowwmodes}a (inset).  We compare \added{these data with} the position of this peak in \added{the bi-layer} Bi2212 and \added{tri-layer} Bi2223 \added{members of the Bi-based cuprate family (the general chemical formula for this family is Bi$ _{2} $Sr$ _{2} $Ca$ _{n-1} $Cu$ _{n} $O$ _{2n+4+\delta} $ where $ n $ is the number of \cuo layers)} as reported in \cite{williams2007}. \replaced{These data are}{and}  shown in \fig~\ref{fig:lowwmodes}a \added{and reveal} that \replaced{the frequency of this mode}{it} decreases with \deleted{the }increasing number of \cuo layers. Thus, changing the local geometry about the Bi site (through oxygen content in the BiO$_{1-\delta}$ layer, Ln ion size and Pb substitution) does not affect the frequency of this mode in Bi2201. However the larger Bi2212 and Bi2223 unit cells do. These two observations discourage an assignment of this peak to a mode with a strong Bi contribution. 

The mode at $\omega_2\sim 120$ \cm does not shift with doping level or Ln.  Increasing Pb concentration however does shift the mode to lower frequencies.  Again we can compare this result with that found in Bi2212 and Bi2223 where the position of this mode was found to be dependent only on Pb concentration \cite{williams2007}.   We reproduce this data in \fig~\ref{fig:lowwmodes}b along with our data on Bi2201.  While the mode frequencies for $x=0.2$ and $x=0.35$ are slightly higher than for Bi2212 and Bi2223 of equivalent Pb concentrations, the overall trend is replicated.  Hence this mode would appear to have an insignificant Sr contribution in Bi2201 but significant Bi contribution in the three Bi-based cuprates Bi2201, Bi2212 and Bi2223.

These results also further demonstrate the utility of Raman measurements in rapid characterization of Bi-based cuprates.

\subsection{O(2)$_{\textnormal{Sr}}$ mode}
\begin{figure}
	\begin{center}
	\( \begin{array}{cc}
		\includegraphics[width=0.45\textwidth]{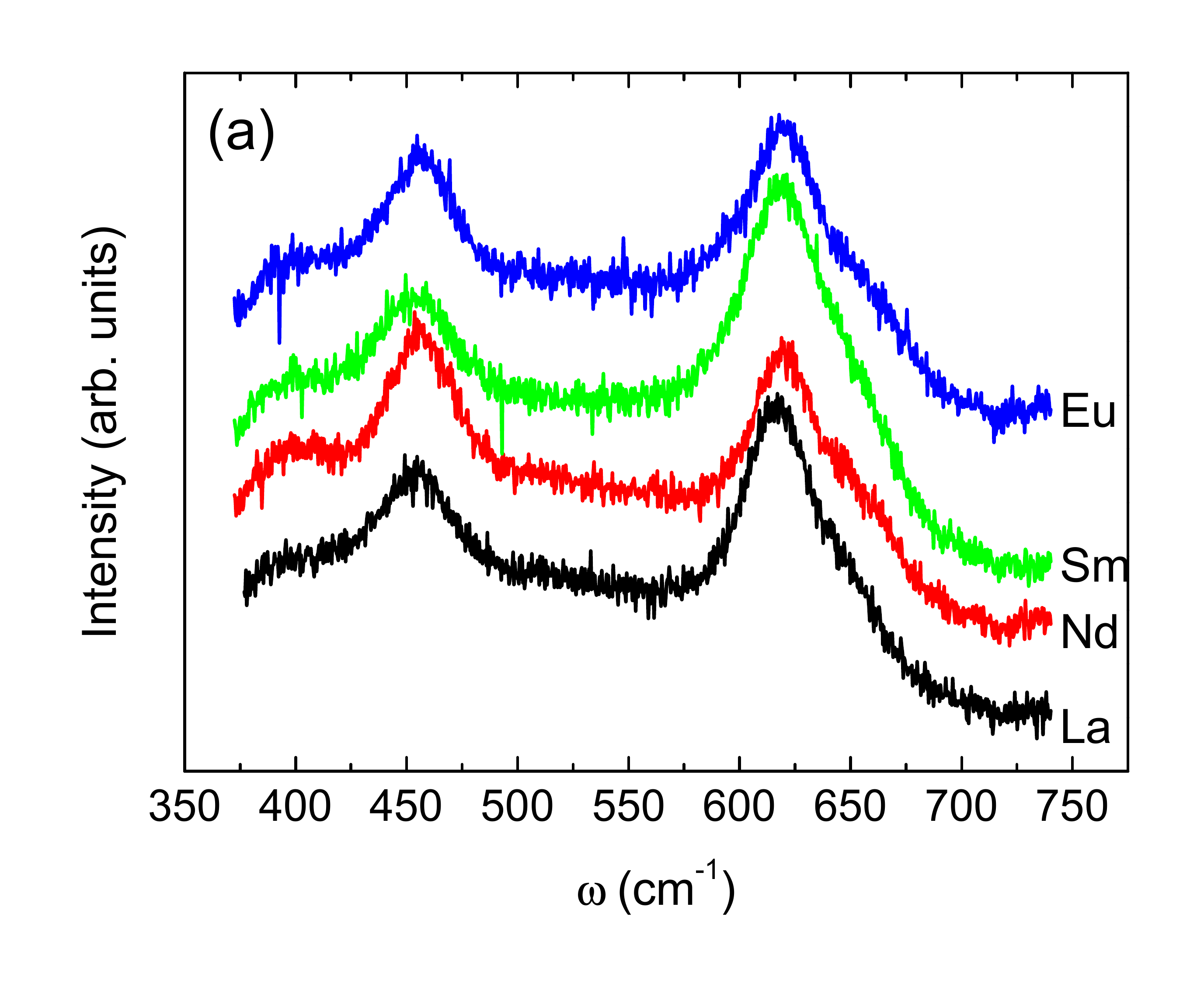} & \includegraphics[width=0.45\textwidth]{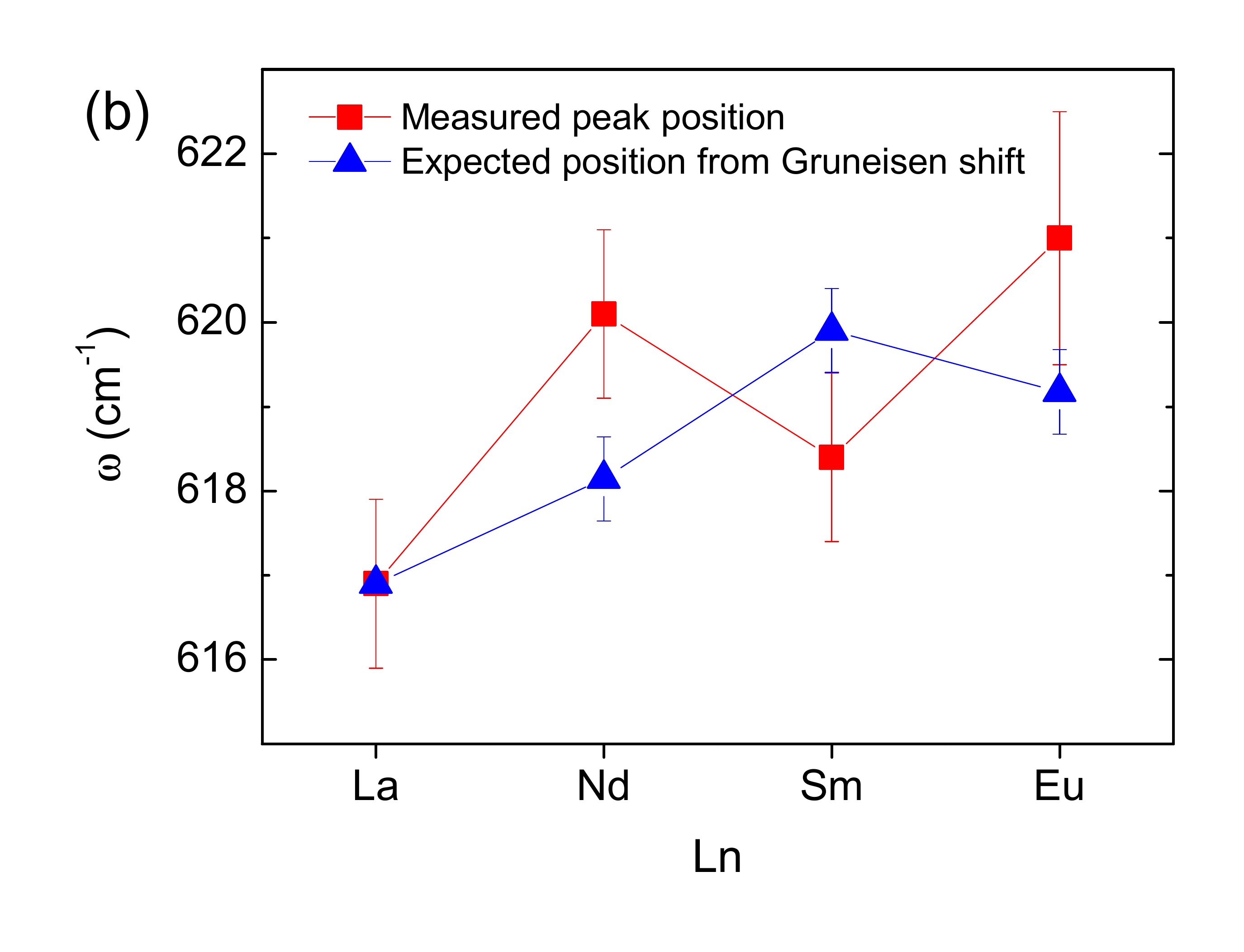}
		\end{array}
		\)
	\caption[Ion-size dependence of apical oxygen phonon mode in Bi2201]{(a) Representative high-resolution spectra for optimally-doped Bi$_2$Sr$_{1.6}$Ln$_{0.4}$CuO$_{6-\delta}$ with, from bottom to top, Ln=La, Nd, Sm and Eu. (b) The measured position of the O(2)$_{\textnormal{Sr}}$, apical oxygen mode plotted as squares and the predicted shift from mode Gr\"{u}neisen scaling is plotted as triangles.}
	\label{fig:apical}
	\end{center}
\end{figure}

The O(2)$_{\textnormal{Sr}}$ site, also known as the apical oxygen, is the nearest neighbour to the Sr site and its relative position to the \cuo plane can be related to the hole concentration \cite{tallon1991}. We therefore measured the O(2)$_{\textnormal{Sr}}$ mode frequency for the series of optimally-doped Bi$_{2}$Sr$_{1.6}$Ln$_{0.4}$CuO$_{6-\delta}$ for Ln=La, Nd, Sm and Eu.  The results are plotted in \fig~\ref{fig:apical}. As \fig~\ref{fig:apical}b shows there was no systematic shift in the peak frequency due to the Ln ion-size.  We compare the small shifts of this peak, $\sim 3$~\cm, where the FWHM is $\sim 40$~\cm, with those expected from mode Gr\"{u}neisen scaling: $\gamma_i=-\frac{\delta\ln(\omega_i)}{\delta\ln(V)}$ where $V$ is the unit cell volume and $\gamma_i=0.4$ is a mode Gr\"{u}neisen parameter appropriate for the similar mode in Bi2212 \cite{osada2001}.  This basic analysis may explain the shifts we observe within experimental error (Fig.~\ref{fig:apical}b). We note in previous studies on Bi2212 and Bi2223, where this mode decreased with increasing hole concentration, the predicted Gr\"{u}neisen shift could not explain their data \cite{williams2000, osada1997}.  Structural refinement data across this series is lacking which prevents a more detailed analysis based on the bond lengths of the apical oxygen to its neighbours.

\section{Discussion}
\label{sec:vhshypothesis}
The next step in these investigations would be to collect appropriate XRD scans for a full Rietveld refinement analysis across the series.  From these data \added{one can determine intra-cell bond lengths and from these bond lengths the} $V_+$ \added{bond-valence-sum }parameter \added{introduced in \refsec~\ref{sec:intpressuretheory}.  Recall from \fig~\ref{fig:bvs} that $ V_+ $ has been shown to correlate with \tcmax across all known cuprates \cite{tallon1990} and so it would be interesting to test whether the correlation still holds for our ion-substituted Bi2201 materials.} \replaced{In addition, the full structural refinement of our materials would provide information on the ion-substitution induced changes of important bond-lengths, such as the apical oxygen bond length, and structural parameters such as the \cuo layer buckling angle}{can be determined as well as any ion-substitution induced cyrstallographic distortion - $\sigma ^2$ may not be the relevant measure of disorder as we make substitutions of Ba or Ca}. \added{These structural parameters potentially have a significant effect on the superconducting properties, as discussed in \refsec~\ref{sec:ybco}. Furthermore, the apical oxygen phonon mode shown in \fig~\ref{fig:apical} would be expected to closely follow the apical oxygen bond-length and this would be useful to test.} 

\added{Another useful data set for our novel Bi2201 materials that could be gathered using readily accessible equipment would be resistivity measurements for a range of dopings. From the resistivity, it may be possible to reliably estimate the pseudogap energy as has been done recently by Kim \etal on their ion-substituted Bi2201 samples \cite{kim2010}.  It would be interesting to measure what effect \textit{negative} internal pressure has on the pseudogap energy, $\epgm = k_B T* $.}

\begin{figure}
	\centering
		\includegraphics[width = 0.45\textwidth]{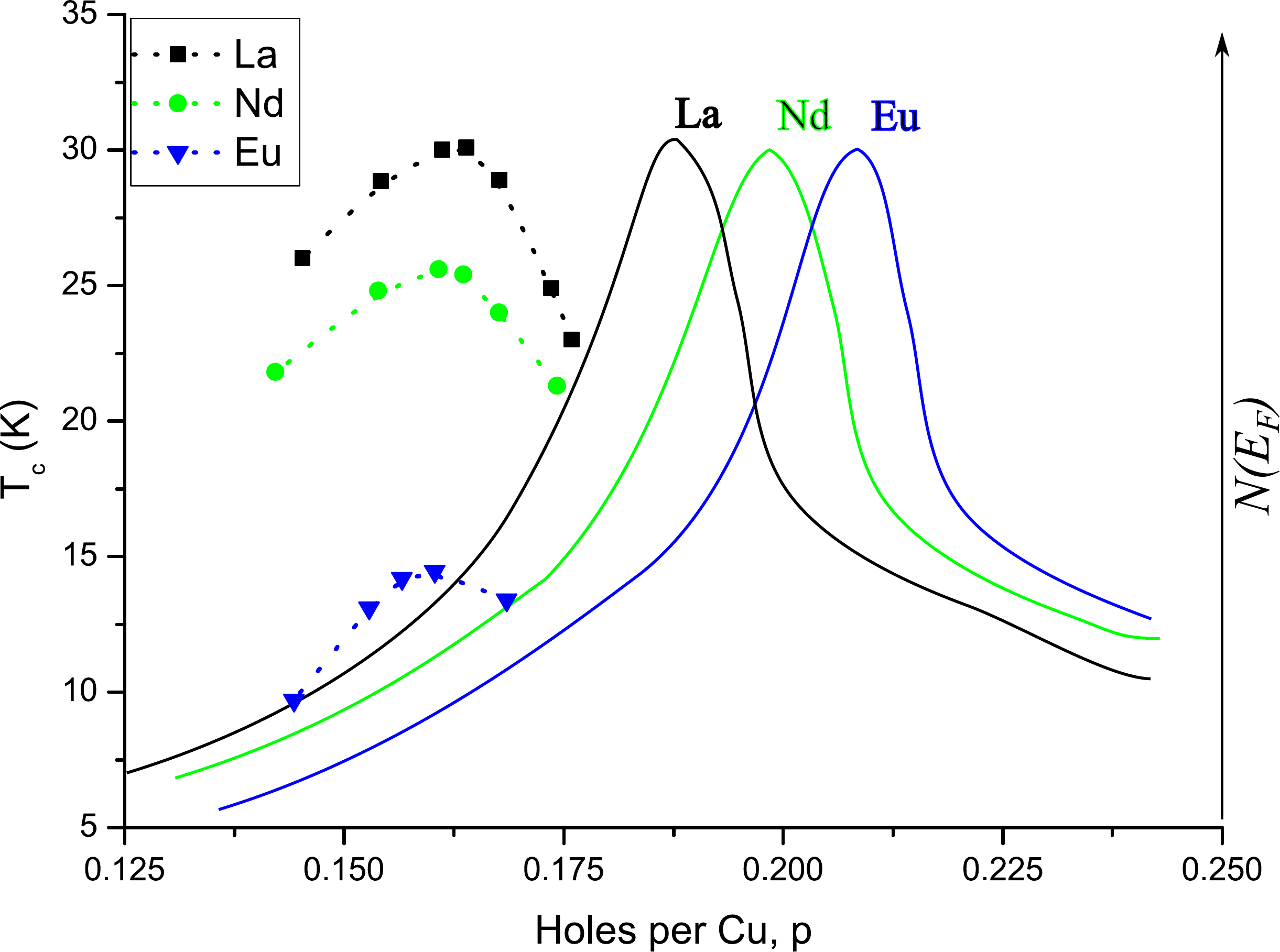}
	\caption[Sketch of van Hove singularity hypothesis for the ion-size effect in Bi2201.]{A sketch illustrating the hypothesis that distortion of the Fermi-surface from ion substitution shifts the location of the van Hove singularity (vHs) in the DOS. \added{Symbols connected by dotted lines are observed \tc values for Bi$_{2}$Sr$_{1.6}$Ln$_{0.4}$CuO$ _{6+\delta} $ with Ln=\{La, Nd and Eu\}. The solid lines correspond to the right-hand axis and are an hypothesised DOS at the Fermi-energy, $ N(E_F) $, for each Ln. These lines show that at optimal doping $ N(E_F) $ increases for larger ion-size due to the closer proximity of the vHs.} In the simple BCS theory \tc is very sensitive to $ N(E_F) $.}
	\label{fig:vhshypothesis}
\end{figure}

\added{We now briefly consider the measurements of \tc as a function of ion-size in our novel Bi2201 materials.} If we take the view that the ion-size effect is significant, how can we explain its effect \added{on} $T_c$?  One hypothesis is indirectly via the density of states (DOS). The hypothesis is that ion-size substitution results in a distortion of the Fermi-surface such that the van Hove singularity (vHs) in the (electronic) DOS moves closer in energy to the Fermi-level, that is $(E_F-\evhsm ) \rightarrow 0$\added{, at optimal doping for larger ion-size}. Visually, this hypothesis is sketched in \fig~\ref{fig:vhshypothesis}.  This figure shows measured \tc values \added{as symbols connected by dotted lines } \replaced{corresponding}{relating} to the left axis for Bi$_{2}$Sr$_{1.6}$Ln$_{0.4}$CuO$ _{6+\delta} $ with Ln=\{La, Nd and Eu\}\deleted{ as solid symbols.  Inferred `superconducting domes' from these \tc data points for each Ln are shown by a thin lines.}  Also on shown in this figure is an hypothesised DOS at the Fermi-energy\added{, $ N(E_F) $, shown as solid lines.  The solid lines corresponding to the right-hand $y$-axis and show that at optimal doping $ N(E_F) $ increases for larger ion-size due to the closer proximity of the vHs.}

The associated, perhaps dramatic, changes in the DOS can have a significant influence on $T_c$. \replaced{For example, in a simple, weak-coupling BCS picture we have $T_c \propto \exp[-1/(N(E_F)V)]$ where $V$ is the pairing potential as shown previously \eq~\ref{eq:bcsdwaveweak}. Thus, within this framework at least a small increase in $ N(E_F) $ could significantly increase $ \tcm $.}{, e.g. in a weak-coupling BCS framework $T_c$ is exponentially dependent on the DOS () as;  }

We explore this idea further in the next chapter using density functional theory calculations and also note that a fuller discussion of ion-size effects must wait until we have considered our investigations on the Ln(Ba,Sr)$_2$Cu$_3$O$_y$ system.

\section{Conclusions}

In this chapter we used a simple materials variation approach to study (i) the effect of ion-size and disorder on \tc and (ii) Raman modes in the single-layered cuprate Bi$_2$Sr$_2$CuO$_{6-\delta}$ (Bi2201).

We simulated \emph{negative} internal pressure by Ba substitution for Sr in Bi2201 and found that this can increase $T_c$ despite increasing disorder. That is, \tcmax is in fact better correlated to the average ion-size, rather than disorder, on the Sr site, although it is possible that ion-substitution induced disorder is a significant additional effect that suppresses $ \tcm $. 

We also presented Raman spectroscopy measurements on Bi$_{2-x}$Pb$_x$Sr$_{1.6}$Ln$_{0.4}$CuO$_{6-\delta}$ where we alter Ln and $x$ and compared the results with measurements on the bi- and tri-layer Bi-based cuprates, Bi2212 and Bi2223.  We argued from the simple interpretation of these material variation experiments that the $120$~\cm mode in Bi-based cuprates has a significant Bi vibration contribution and insignificant Sr vibration contribution. Neither vibration significantly contributes to the lower $70$~\cm mode in Bi2201 however we do see a consistent shift to lower frequencies with increasing \cuo layers of this mode in the Bi-based cuprates.



\chapter{Density Functional Theory study of the ion-size effect}
\label{ch:dft}
\subsubsection{Summary}
We perform DFT calculations on undoped ACuO$_2$ for A=\{Mg, Ca, Sr, Ba\} to investigate the effect of ion-size on the electronic properties in this model cuprate system.  Where these materials have been synthesised we find good agreement between our calculated structural parameters and the experimental ones.  There is a peak in the density of states $\sim 1$~eV below the Fermi-level and we find that larger ions increase the size of the peak and move it closer to the Fermi-level.  This is consistent with an interpretation of the ion-size affecting \tcmax via the density of states.

\section{Introduction}

In this chapter we seek to correlate structural distortions from ion-size substitution with changes in the normal-state electronic structure of the materials in an effort to further understand the observed effects of ion substitution. Electronic structure and crystallographic structure are intimately related and one could imagine that structural distortions induced by isovalent ion substitution can have a significant effect on the former.   

\begin{figure}
	\centering
		\includegraphics[width=0.33\textwidth]{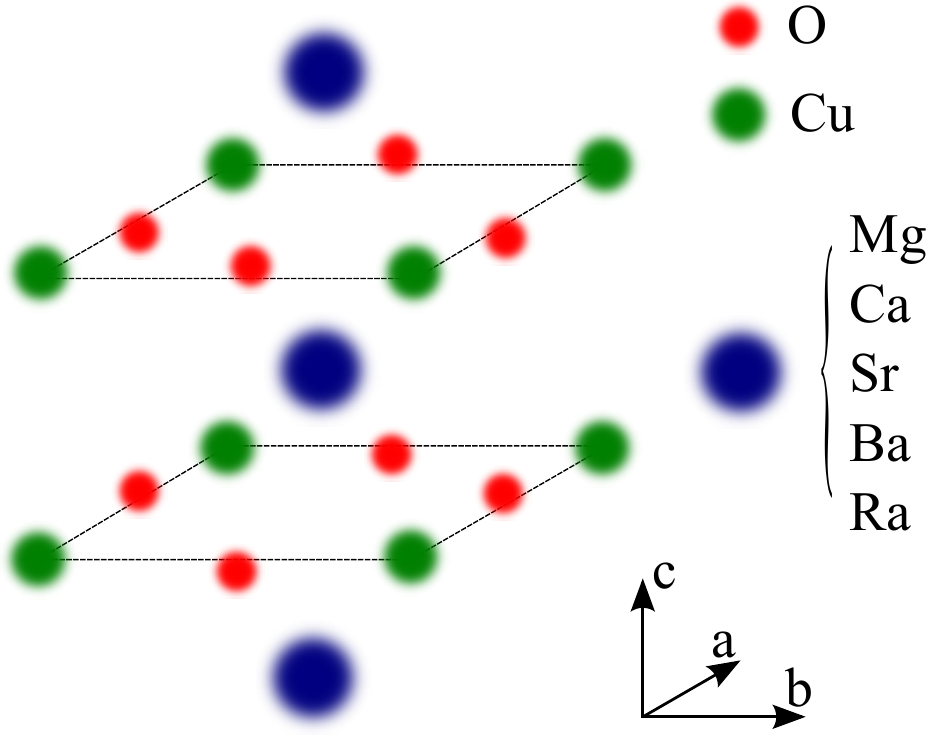}
	\caption[The infinite-layer cuprate, ACuO$ _{2} $, crystal structure.]{Idealised crystal structure of ACuO$_2$ where A=\{Mg,Ca,Sr,Ba,Ra\}.  The space group is P4/mmm and atomic positions are A$=(\sfrac{a}{2},\sfrac{a}{2},\sfrac{c}{2})$, Cu$ = (0,0,0)$, O$=(\sfrac{a}{2},0,0)$ and O$=(0,\sfrac{a}{2},0)$ where $a=b$ and $c$ are the lattice parameters.}
	\label{fig:ifxtal}
\end{figure}

The family of materials we choose to study are the so-called infinite layer cuprates.  These are the simplest HTS cuprates known.  They have the chemical formula ACuO$_2$ with A=\{Mg,Ca,Sr,Ba\} and an illustration of their crystal structure is shown in \fig~\ref{fig:ifxtal}.  This system is ideal for a computational study: it has a simple tetragonal crystal structure, there are comparatively few electrons per unit-cell and a wide range ion-size variation is possible. 

As such there have been many previous computational studies of ACuO$_2$, the majority on \cacuo \cite{anisimov1991, hatta1992, agrawal1993, andersen1996, massidda1997, wu1999, singh2010}. 
In distinction, we are primarily interested in systematic \emph{trends} in the electronic structure as A varies from Mg to Ba. 


We characterise the electronic structure by the electronic dispersion, $\ekm=E(\kk)-E_F$ under the reasonable assumption that $\kk$ is a good quantum number for our materials.   
We are also interested in the density of states.  The density of states of the $n^{\textnormal{th}}$ band is related to $E_n(\kk)$ directly; 
\begin{equation}
N_n(E)=\int{\frac{\kk}{4\pi^3}\delta (E-E_n(\kk))}
\label{eq:dos}
\end{equation}
\noindent where $\delta(E-E(\kk))$ is the delta function.  Summing over all bands will give the total density of states, $N(E)$.  Within a BCS theory of superconductivity the density of states at the Fermi-level, $ \dosm $, is a key electronic parameter \cite{bcspaper, surma1983, aleksandrov1989}, irrespective of the specific pairing mechanism \cite{abbpaper}.

$N_n(E)$ can also be represented as a function of $E_n(\kk)$ as \cite{ashcroftmermin};
\begin{equation}
N_n(E)=\int_{S(E)}{\frac{\dd S}{4\pi^3}\frac{1}{|\nabla E_n(\kk)|}}
\label{eq:dos2}
\end{equation}
\noindent where $S(E)$ is the surface of $E(\kk)$ of energy $E$ and $\nabla E(\kk)$ is the gradient of $E(\kk)$ (or equivalently \ek). The purpose of expressing $N(E)$ in this way is to see that at extremes (and inflection points) of \ek the integrand of \eq~\ref{eq:dos2} diverges. These points are called ``van Hove singularities'' (vHs) and at such points, depending on dimensionality of the system, $N(E)$ will diverge. In practise such a divergence may be cut-off by disorder or strong coupling of the \cuo layers along the $c$-axis.

To calculate the electronic structure of our materials we use density functional theory (DFT) implemented with the Vienna Ab-Initio Simulation Package \cite{vasp, kresse1993, kresse1994, kresse1996, kresse1996a} (VASP).  Recent reviews of the use of DFT implemented with VASP in condensed matter physics are given by Hafner \cite{hafner2007, hafner2008}. Density Functional Theory (DFT) is a method to find the charge density, $n(\rr)$, corresponding to the minimal energy of a physical system.  To perform the calculation considering the interaction of every electron in the system is computationally infeasible and so some approximations need to be made.  Firstly the Kohn-Sham approach replaces the many-electron problem by a single-electron problem with an exchange-correlation potential between electrons. We next use a Generalised Gradient Approximation (GGA) to derive the form of the exchange-correlation potential and kinetic energy in the single-electron Hamiltonian.  In particular, we use the GGA-PW91 scheme developed by Perdew \etal \cite{perdew1992, perdew1993}. GGA is an extension to the Local Density Approximation (LDA). 
Finally we use a Projector Augmented Wave (PAW) scheme for efficient, accurate calculation of the electronic wavefunctions \cite{bloch1994, kresse1999}.  An excellent introduction to the DFT, LDA and various pseudo-potentials are the VASP workshop lectures which can be found at www.vasp.at/index.php/documentation. 

Armed with the wavefunction of the system, a panoply of physical properties can be determined.  Needless to say, however, experimental evidence trumps computer simulations.  One of the first tasks is to assess if these calculations can reproduce experimental results where they are available. The paucity of experimental studies of the ACuO$_2$ system (especially for A$\neq$Ca) is another reason to opt for a DFT study; it is very difficult to synthesize these materials with good quality for experimental studies. 


\section{Experiment}
\subsection{Calculated vs experimental crystal structures}

\begin{figure}
	\centering
		\includegraphics[width=1.00\textwidth]{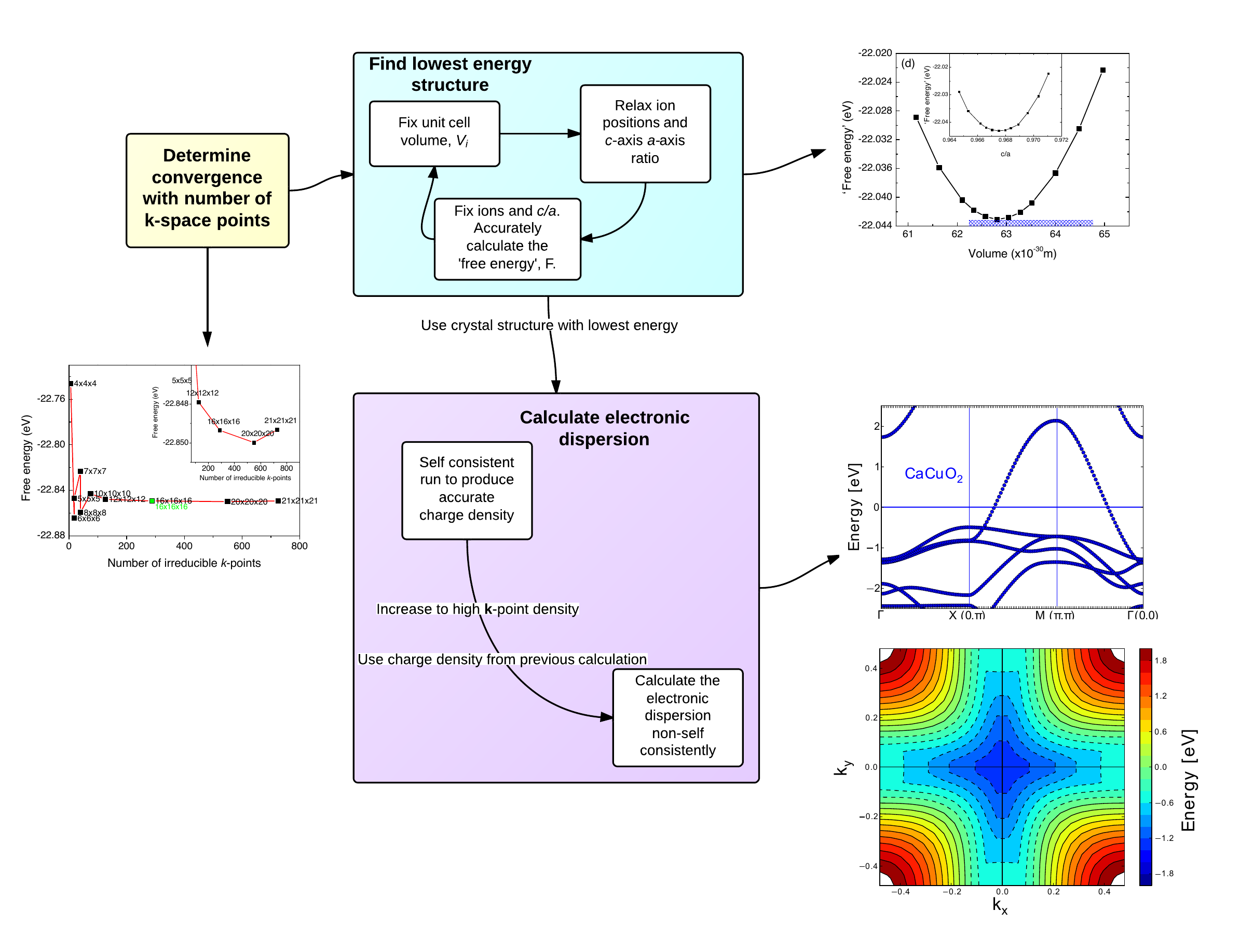}
	\caption[An illustration of the procedure followed to calculate the electronic structure with VASP.]{An illustration of the procedure followed to calculate the electronic structure.  The various calculation parameters that should be altered for each step can be found in the VASP manual \cite{vaspmanual}.  The inputs to the calculation are Projector Augmented Waves (PAW) pseudopotentials for Cu, O and A=\{Mg,Ca,Sr,Ba\}.  \added{Indicated in the flow-chart are typical data-plots obtained from each step.}}
	\label{fig:dftflowchart}
\end{figure}

We summarize the calculation procedure with a flow chart in \fig~\ref{fig:dftflowchart}.  

\begin{figure}
	\centering
		\includegraphics[width=0.75\textwidth]{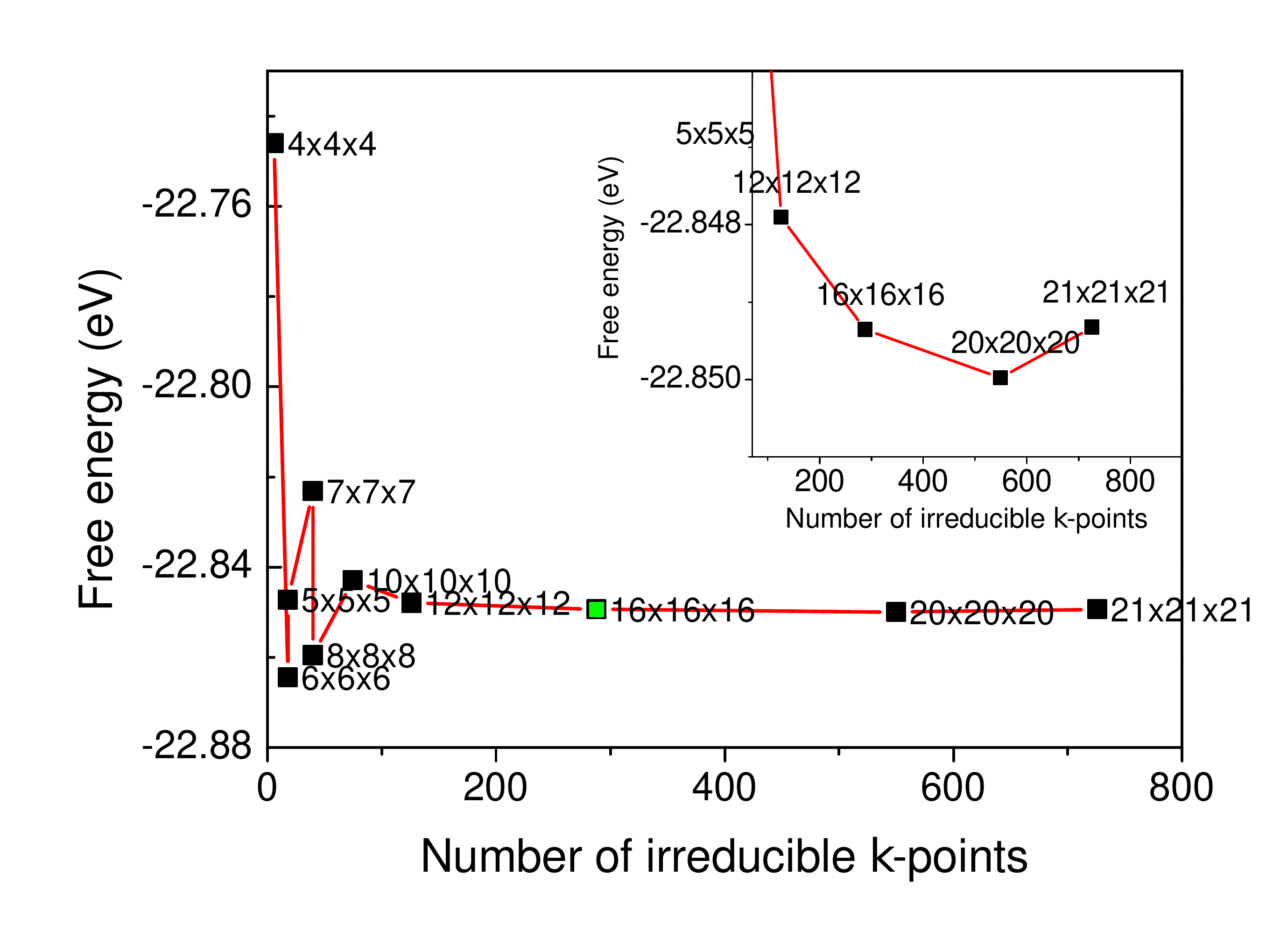}
	\caption[Calculated total energy of CaCuO$_2$ as a function of the density of $\kk$-space sampling.]{Calculated total energy of CaCuO$_2$ as the density of $\kk$-space sampling is increased.  A larger number of irreducible $\kk$-points returns more accurate calculations at the cost of increased computational time.  We find a 16x16x16 $\kk$-space mesh is sufficient for our system.  The green data point is from a `high precision' calculation.}
	\label{fig:fvskpoints}
\end{figure}

A critical step is to perform convergence tests on the total energy vs. $\kk$ space sampling of the Brillouin Zone (BZ) to determine a sufficient $\kk$ space sampling interval for accurate calculations. The results of this procedure are shown in \fig~\ref{fig:fvskpoints}.  From this convergence test we find that a $16\times 16\times 16$ $\kk$-space mesh is sufficient.  This mesh corresponds to a calculation over 288 irreducible\footnote{Or, `inequivalent', as in not related by the system's particular symmetry to other $\kk$-points.} $\kk$-points for a Monkhorst scheme. 

Before attempting any calculations of the electronic structure we must first determine the most stable crystal structure, as characterised by a minimum in the ``free energy'', $F$, of the material, $ F_{\textnormal{min}} $.  Note that we use the term ``free energy'' here to be consistent with the \replaced{nomenclature}{nomenculture} used by VASP, however, because the calculations are carried out at $T=0$ they really represent the ground-state internal energy, $U$. We can then compare with the experimentally determined cell volume and lattice parameters.  For CaCuO$_2$ the experimental values are $|a|=0.38556(6)$~nm and $|c|=0.31805(4)$~nm giving $\sfrac{c}{a}=0.8249$ and $V=4.728\times 10^{-2}$~nm$^3$ \cite{karpinski1994}. A more accurate way to determine $F$ in VASP is for a fixed unit cell volume.  We therefore calculate $F$ for various, fixed, unit cell volumes and locate the minimum.  

\begin{figure}
	\centering
		\includegraphics[width=0.45\textwidth]{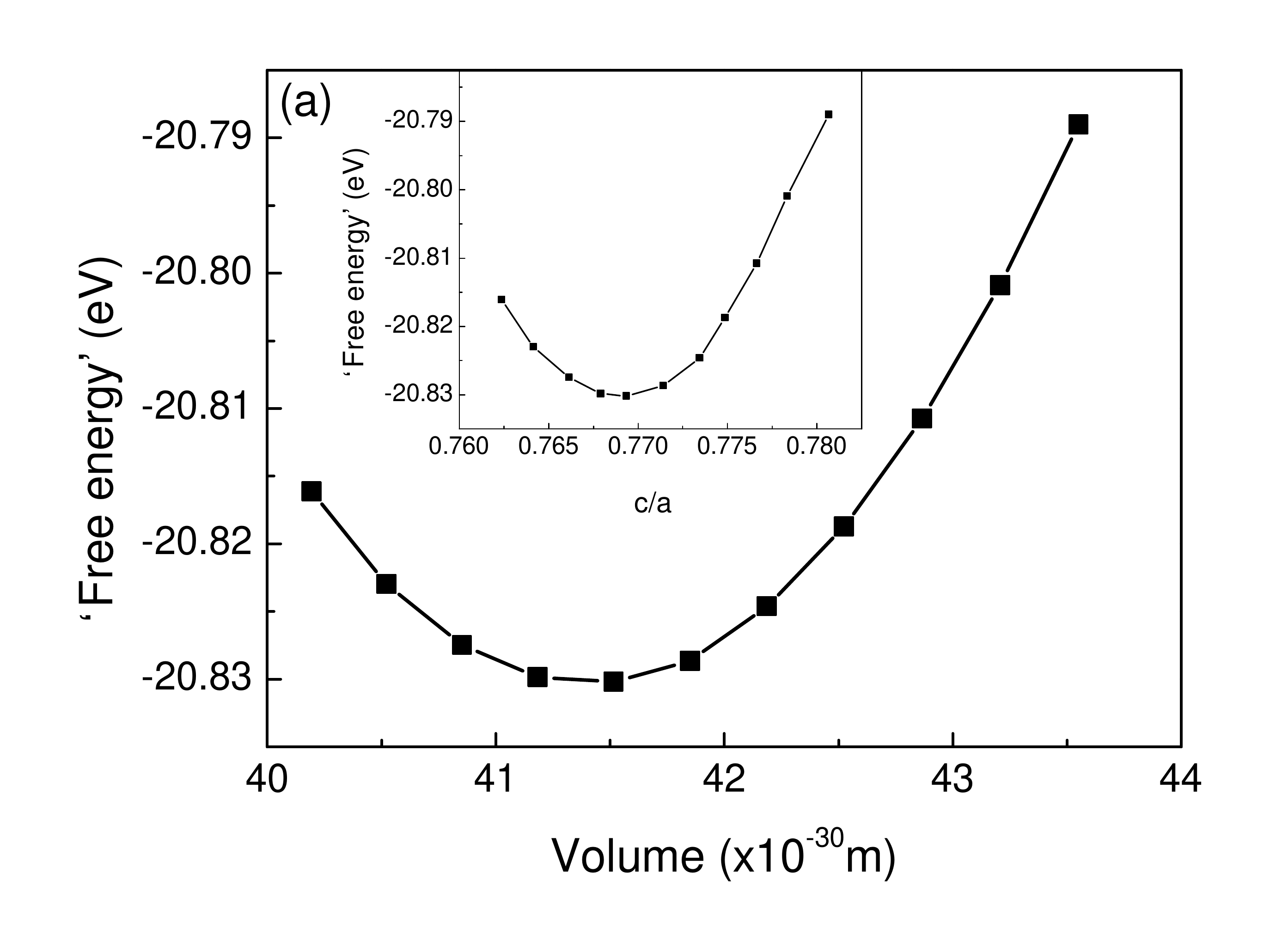}
		\includegraphics[width=0.42\textwidth]{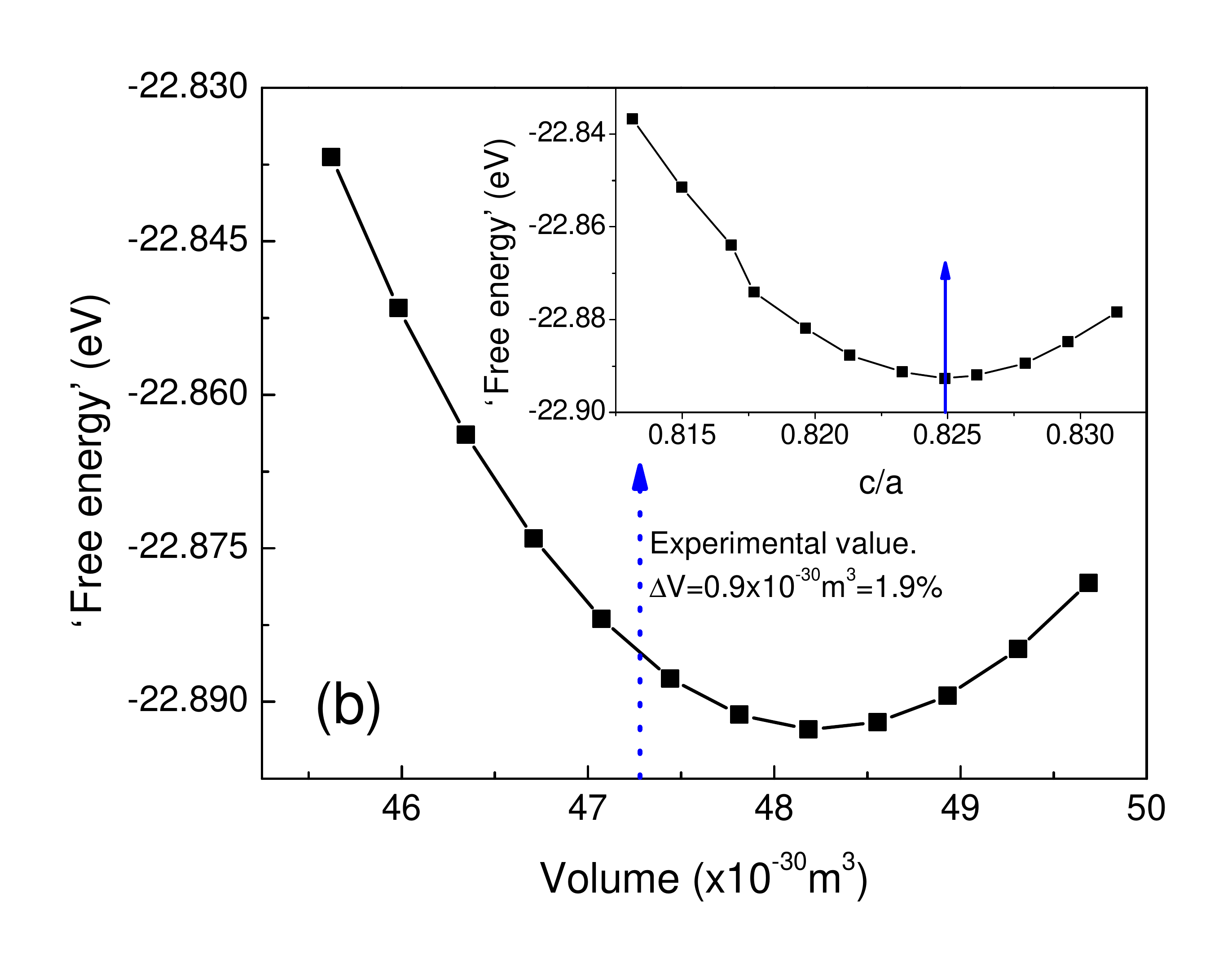}
		\includegraphics[width=0.45\textwidth]{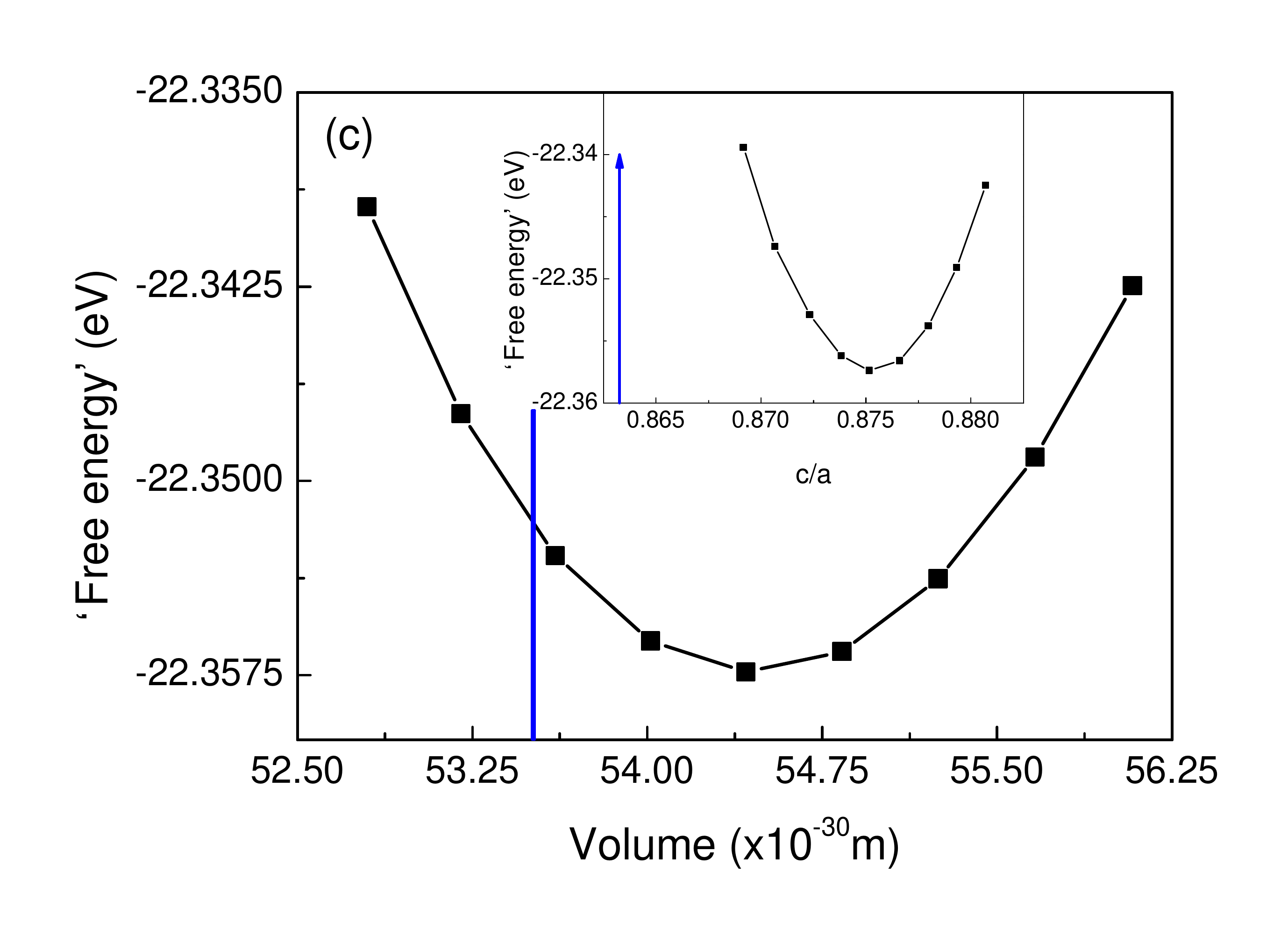}
		\includegraphics[width=0.42\textwidth]{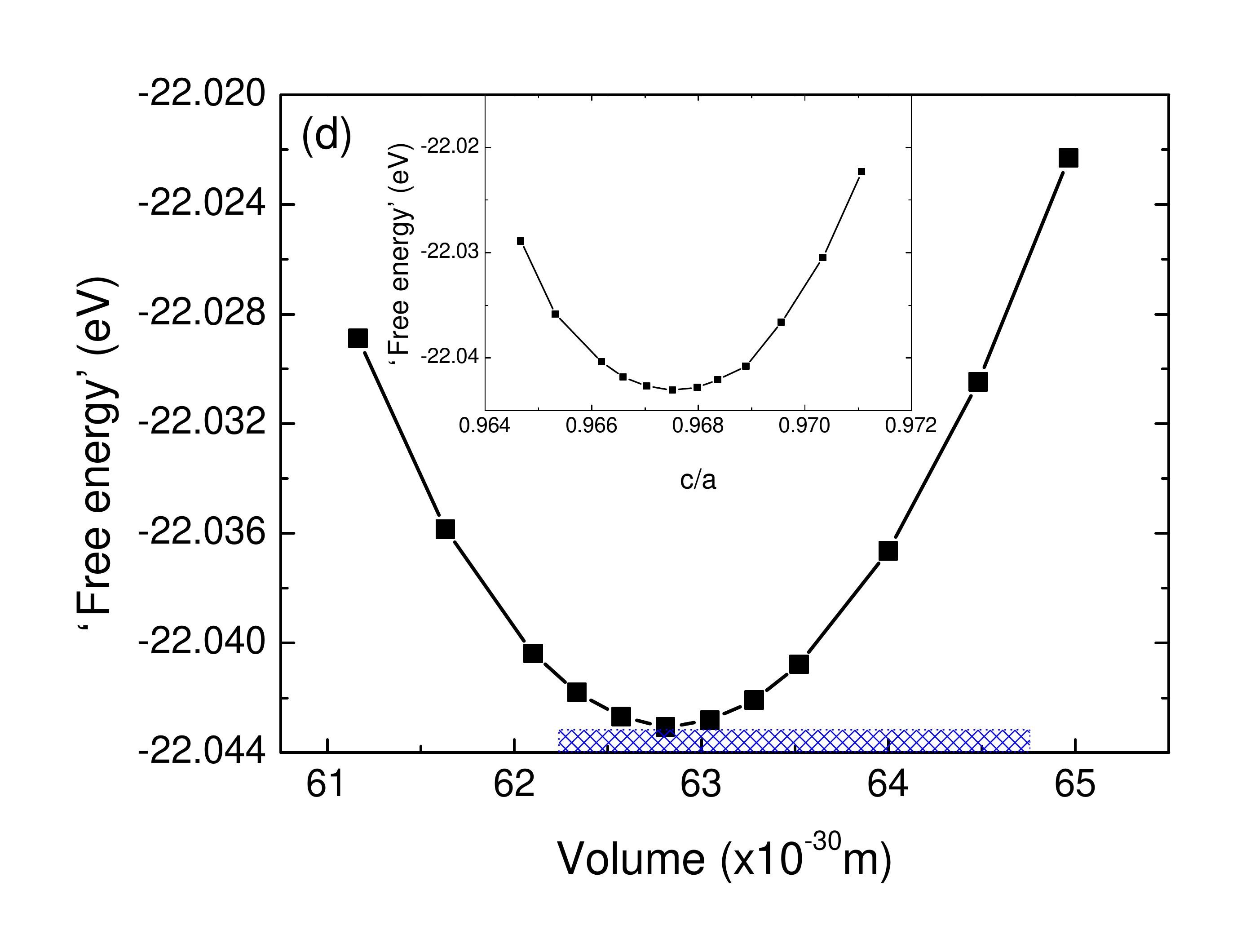}
	\caption[Calculated free energy vs. unit-cell volume for ACuO$ _{2} $ with A=Mg, Ca, Sr, Ba.]{The results of fixed-volume, ion relaxation calculations on ACuO$_2$ for (a) A=Mg, (b) A=Ca, (c) A=Sr and (d) A=Ba.  In the main panels the free energy is plotted against various fixed unit-cell volumes while insets show the corresponding $\sfrac{c}{a}$ lattice parameter ratios.  Annotated blue arrows/boxes represent reported experimental values of the unit-cell volume or lattice parameters.  These plots show that there is a satisfactory agreement between our DFT+LDA calculations and experimental results.}
	\label{fig:fvsv}
\end{figure}

For a fixed volume, the cell shape, i.e. $\sfrac{c}{a}$, and ion positions were first relaxed, for which VASP has sophisticated in-built algorithms, and then an accurate calculation of $F$ was performed without further relaxation.  The cell volume is then specified to be another value and $F$ calculated again. The results of this procedure are shown in \fig~\ref{fig:fvsv}(b). With this process we find $F_{min}$ occurs at $V=4.815\times 10^{-2}$~nm$^3$ for \cacuo which is 1.9\% higher than the experimentally determined unit cell volume\footnote{It is common for these DFT calculations to overestimate cell volumes by $\sim 1\%$.}.  On the other hand, the ratio $\sfrac{c}{a}$ is exactly the experimentally determined one at $F_{\textnormal{min}}$. 

These two results demonstrate good agreement between our VASP calculations and reality in the CaCuO$_2$.  

Similarly we have calculated the most stable unit cell parameters for A=Mg, Sr and Ba. These results are presented in \fig~\ref{fig:fvsv}. On these panels, the annotations in blue mark experimentally determined values of structural parameters where they are available. 

To our knowledge there are no reports that MgCuO$_2$ has been synthesized.  Other materials appear to not form in the simple 1:1:2 stoichiometry. For example, the infinite layer, stoichiometric BaCuO$_2$ has not been synthesized suggesting this compound is unstable - up to 30\% vacancies on the Cu site were reported by de Caro \etal \cite{decaro1999}, see also citations of this paper.  What \bacuo they could make had the lattice parameter $|c|\approx 0.42$~nm and $|a|=0.3906$~nm was assumed from lattice matching to the SrTiO$_3$ substrate \cite{decaro1999}.  The most stable crystal structure from our calculations has $|c|=0.3879$~nm and $|a|=0.4016$~nm which is a reversed $c$:$a$ ratio compared with these experimental results.  
Super-lattices of intercalated Ba-, and Ca- infinite layer compounds have been reported to be superconducting \cite{decaro1999, balestrino2003} although even numbers of Ba- unit-cells are needed \cite{decaro1999}.

In general, the comparison in \fig~\ref{fig:fvsv} reveals a good correspondence between our LDA calculations and experimental results.




\subsection{Initial band structure calculations}

Once the crystal structure with $F=F_{\textnormal{min}}$ has been found, we can use this structure for a precise calculation of the \textit{electronic} structure.  VASP is run with a high density of $\kk$-points ($24\times 24\times 24$ giving 936 irreducible $\kk$-points in most cases) to accurately calculate the electronic dispersion, $\ekm = E(\kk)-E_F$.  We wrote Python code to recover the $\kk$-points where the energy was not calculated because of their symmetry with another point. The code can be thought to `unfold' the symmetries in order to recover the full dispersion for visualisation. 

\begin{figure}
	\centering
		\includegraphics[width=0.535\textwidth]{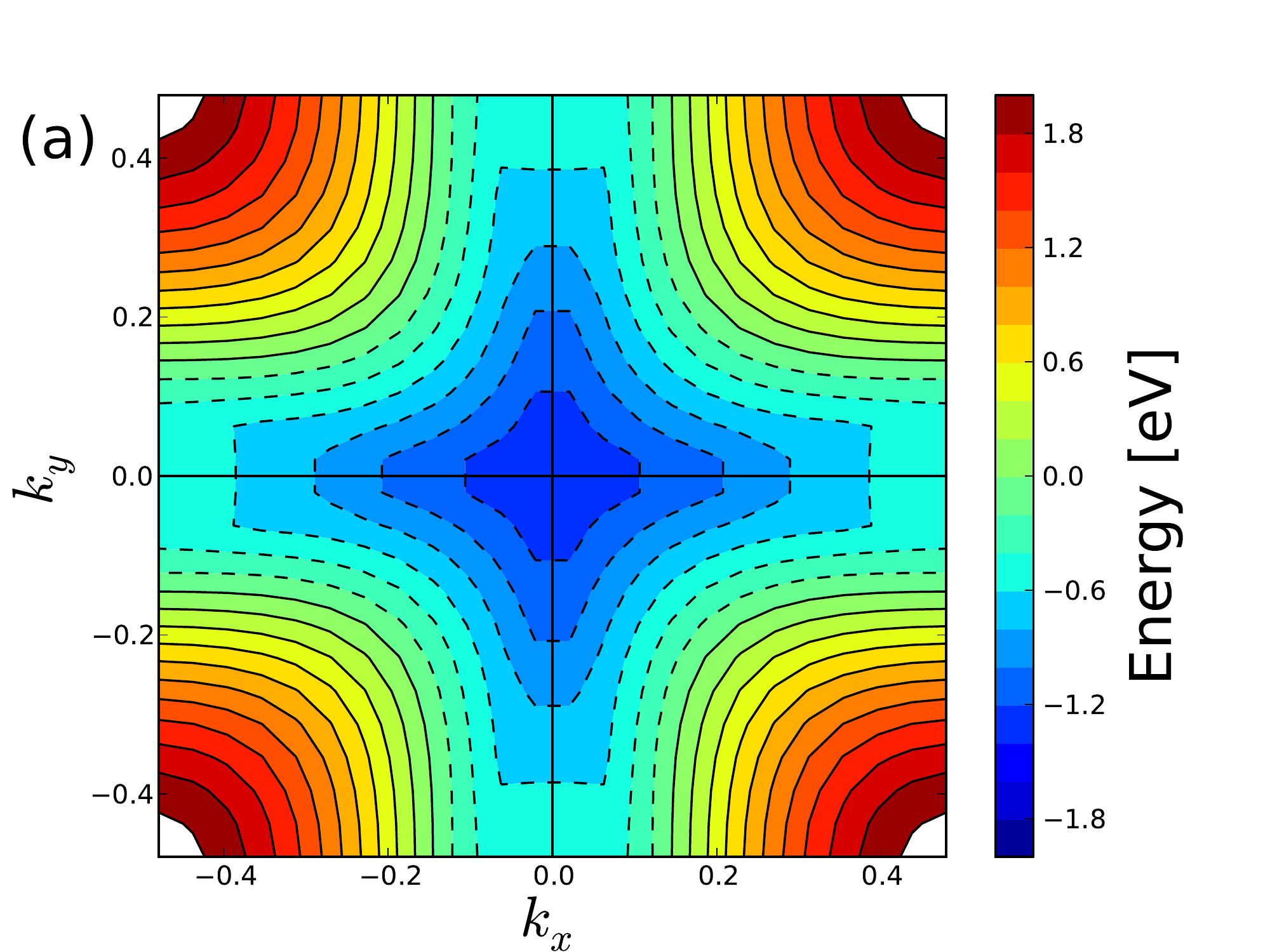}
		\includegraphics[width=0.535\textwidth]{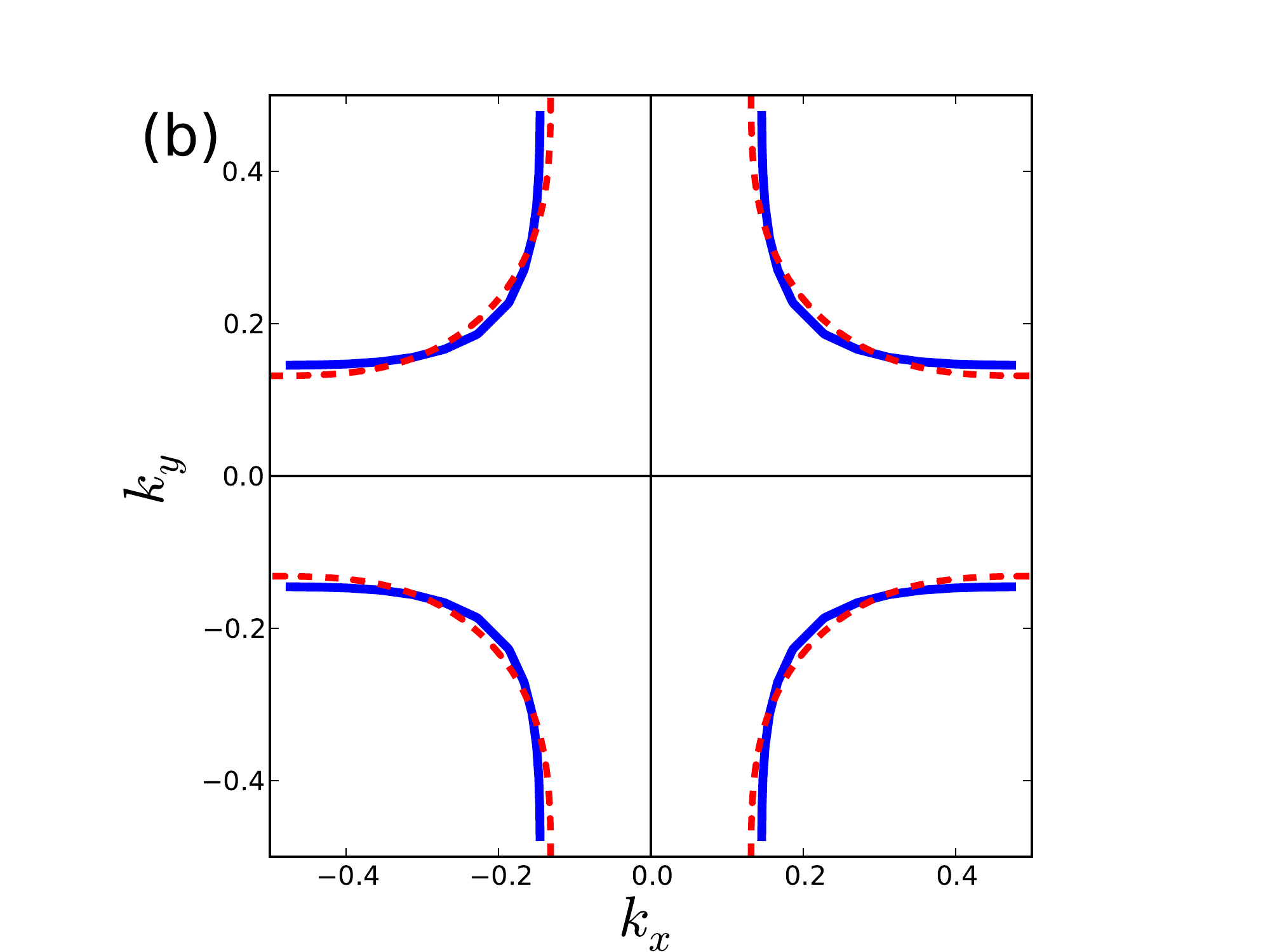}
		\includegraphics[width=0.535\textwidth]{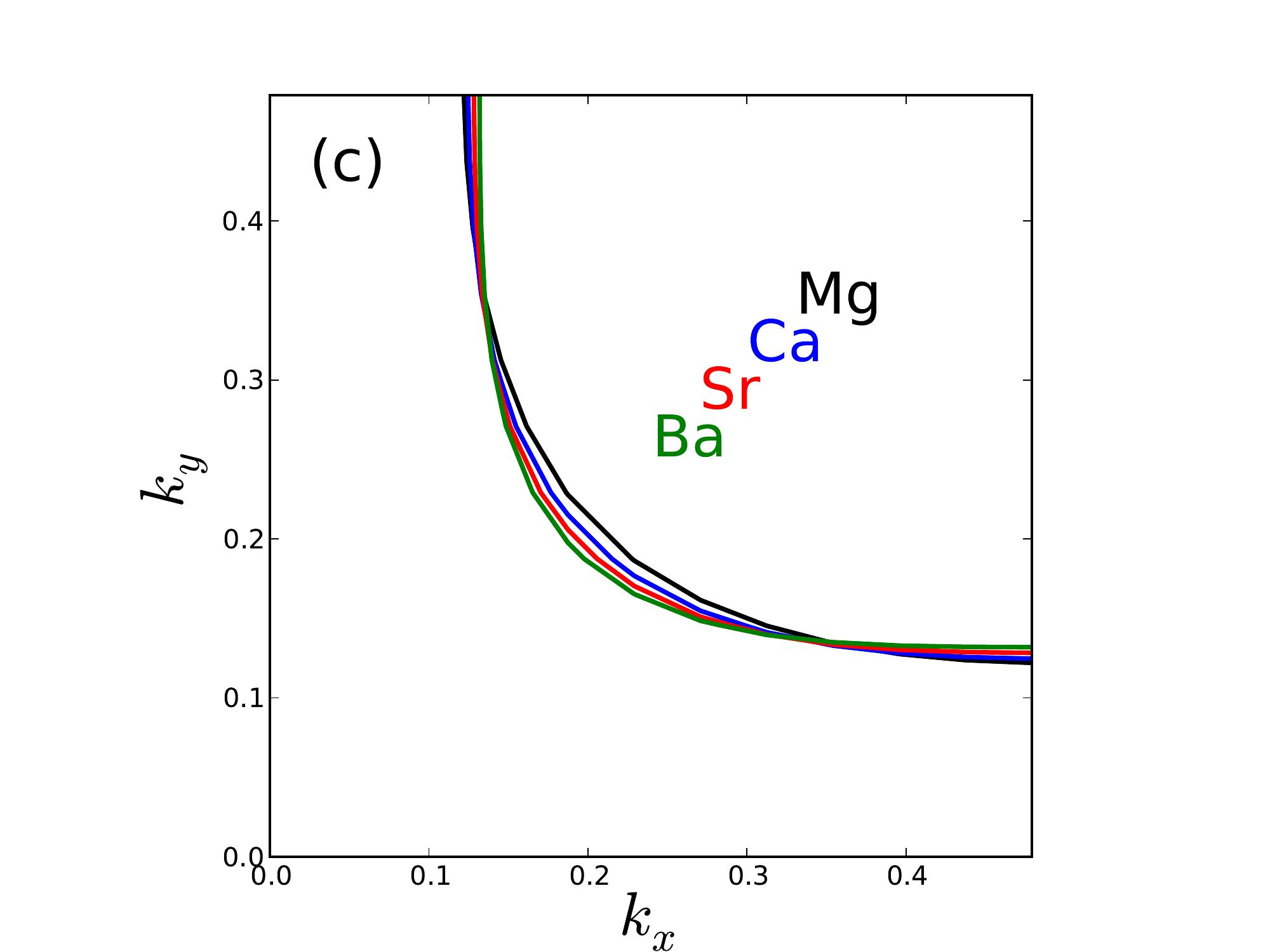}
	\caption[Calculated electronic dispersion of CaCuO$ _{2} $.]{\label{fig:cacuo2contourfermi} (a) The dispersion of the band that crosses $E_F$ in CaCuO$_2$ at $\kk_z=0$.  Energy is represented by colour from dark red at $+2$ eV to dark blue $-2$ eV.  Solid lines separate positive \ek whilst dashed lines separate negative $\ekm$.  Note the saddle point in \ek at the anti-node and that $\partial_{\kk}\ekm = 0$ results in a peak in the density of states \cite{ashcroftmermin}.  (b) $\ekm=0$ for \cacuo shown as a solid blue line and for comparison $\ekm =0$ derived from ARPES measurements \cite{kaminski2006, storey2007} on Bi$_2$Sr$_2$CaCu$_2$O$_{8+\delta}$ as the dashed red curve. (c) $\ekm=0$ for ACuO$_2$ with A=\{Mg,Ca,Sr,Ba\} indicated in the panel.  Here $\kk_z=0.27$ so that all Fermi-contours enclose half the Brillouin zone. }
\end{figure}

\fig~\ref{fig:cacuo2contourfermi} shows an example of such a calculation.  For CaCuO$_2$ with $\kk_z=0$, we plot the full dispersion of the band which crosses $E_F$ in panel (a) and $\ekm=0$ in panel (b). For comparison, we also plot on this panel the same $\ekm=0$ as parameterised from ARPES measurements on Bi2212 \cite{kaminski2006, storey2007} (red dashed curves). We note that there is a $\kk_z$ dispersion to this band that becomes more pronounced for small ion-size. 
By the Luttinger rule, the area enclosed by the Fermi contour should equal $(1+p)$ where $p$ is the hole concentration.  So, in \fig~\ref{fig:cacuo2contourfermi}, the area enclosed by our computed blue curve should be half of the area of the Brillouin zone. It is within the uncertainties from finite sampling of the Brillouin Zone and after taking into consideration the $\kk_z$ dispersion (which means one is measuring the volume enclosed by the Fermi-surface). 

\begin{figure}
	\centering
		\includegraphics[width=0.45\textwidth]{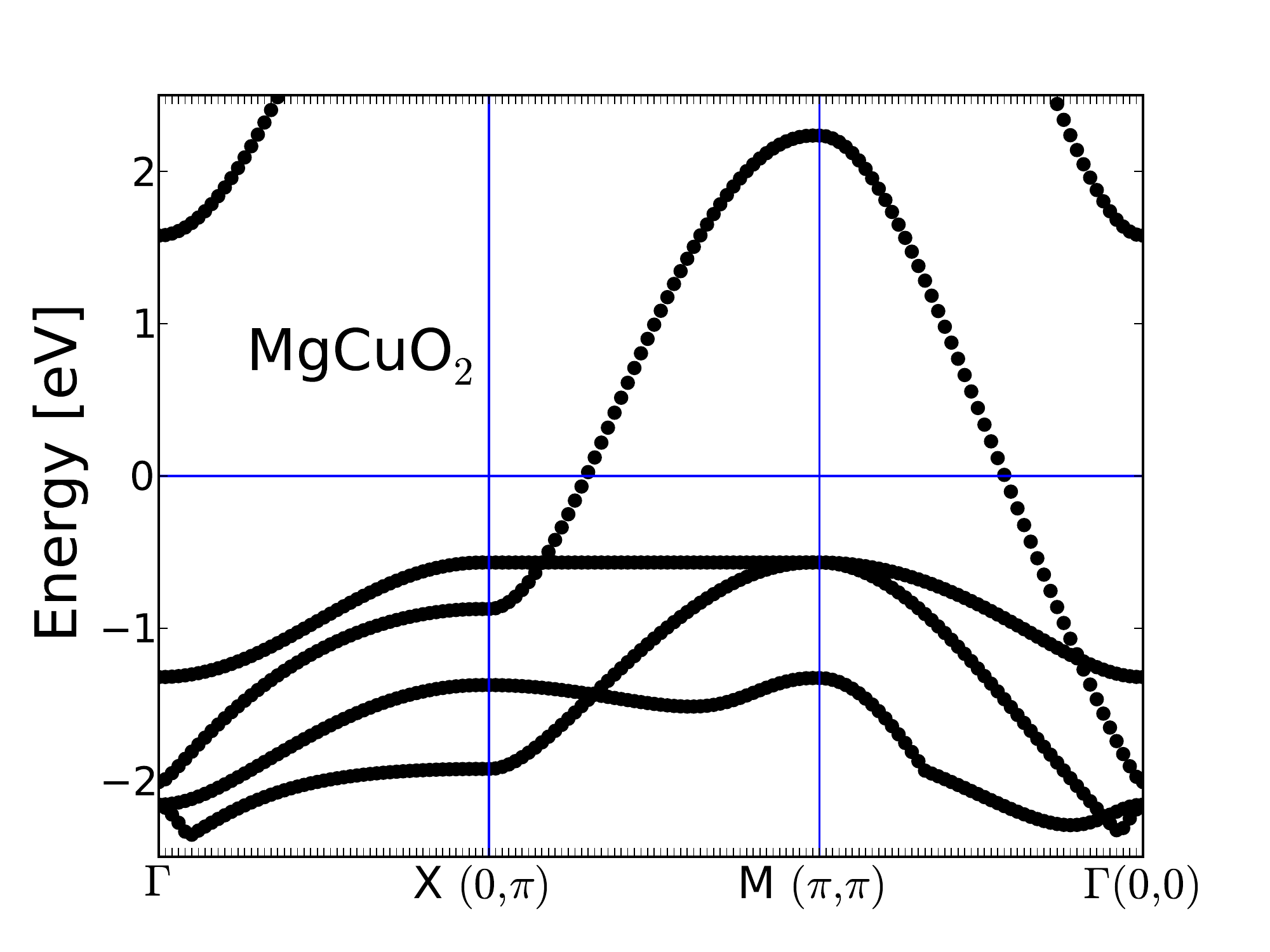}
		\includegraphics[width=0.45\textwidth]{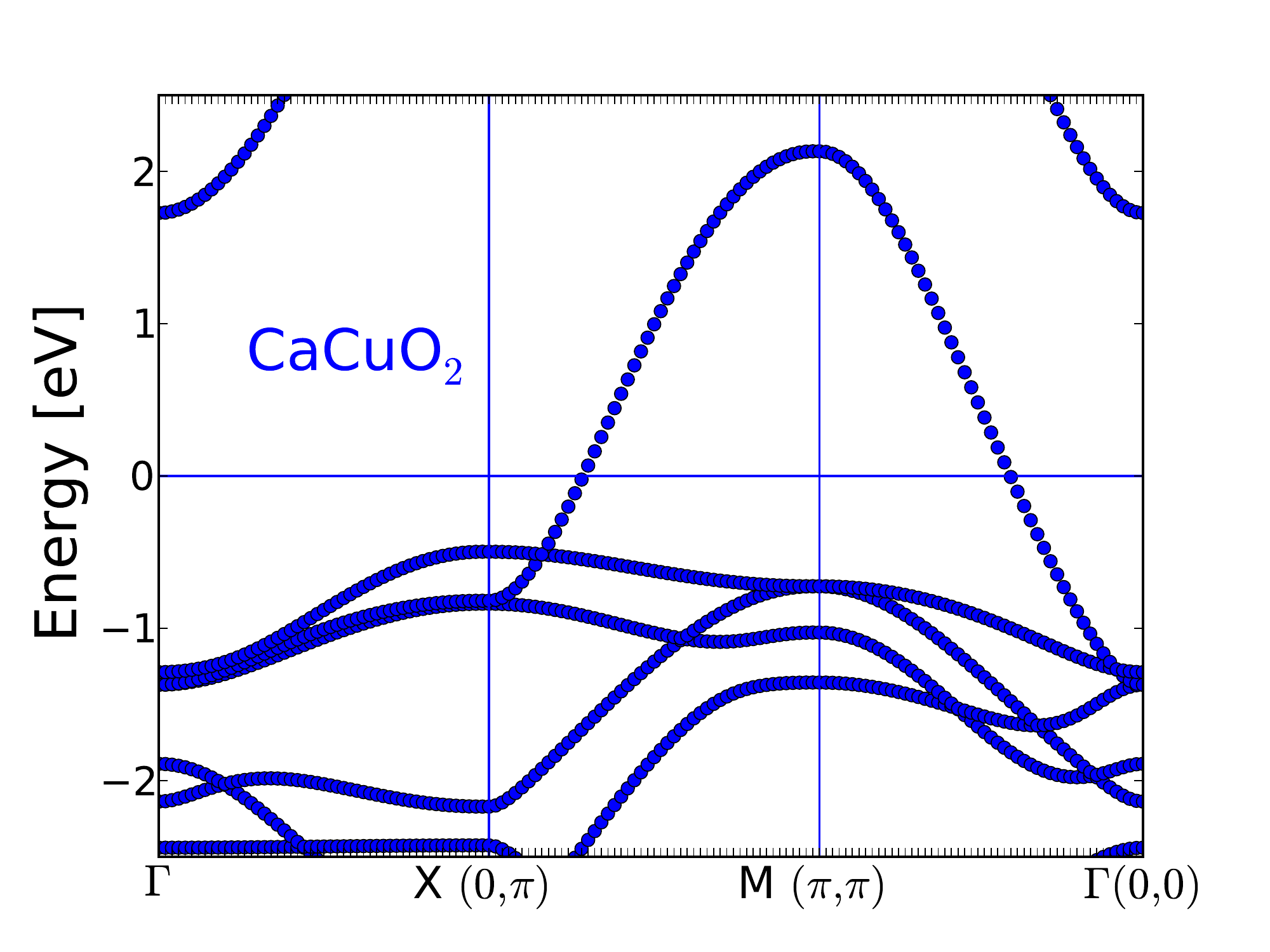}
		\includegraphics[width=0.45\textwidth]{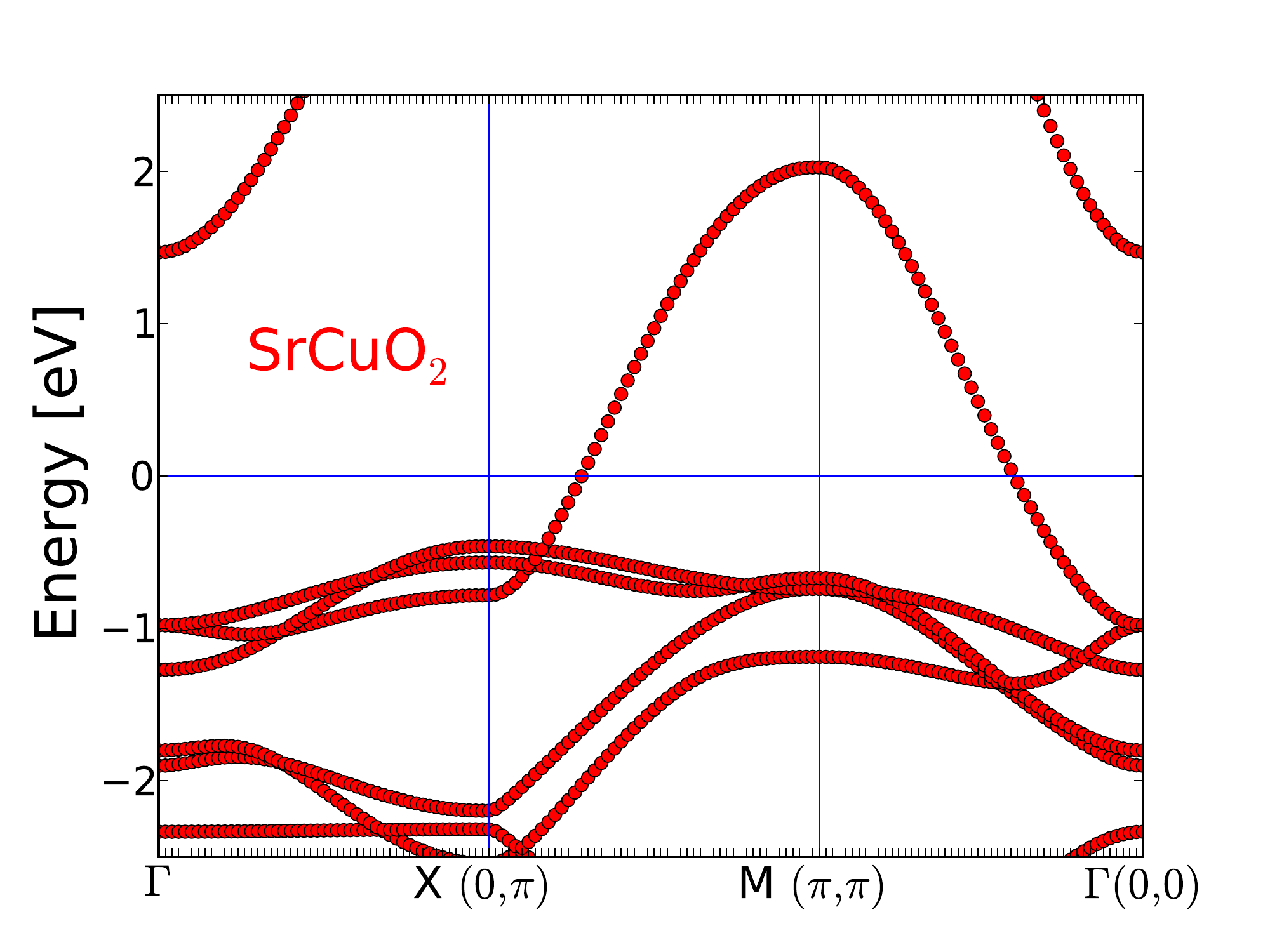}
		\includegraphics[width=0.45\textwidth]{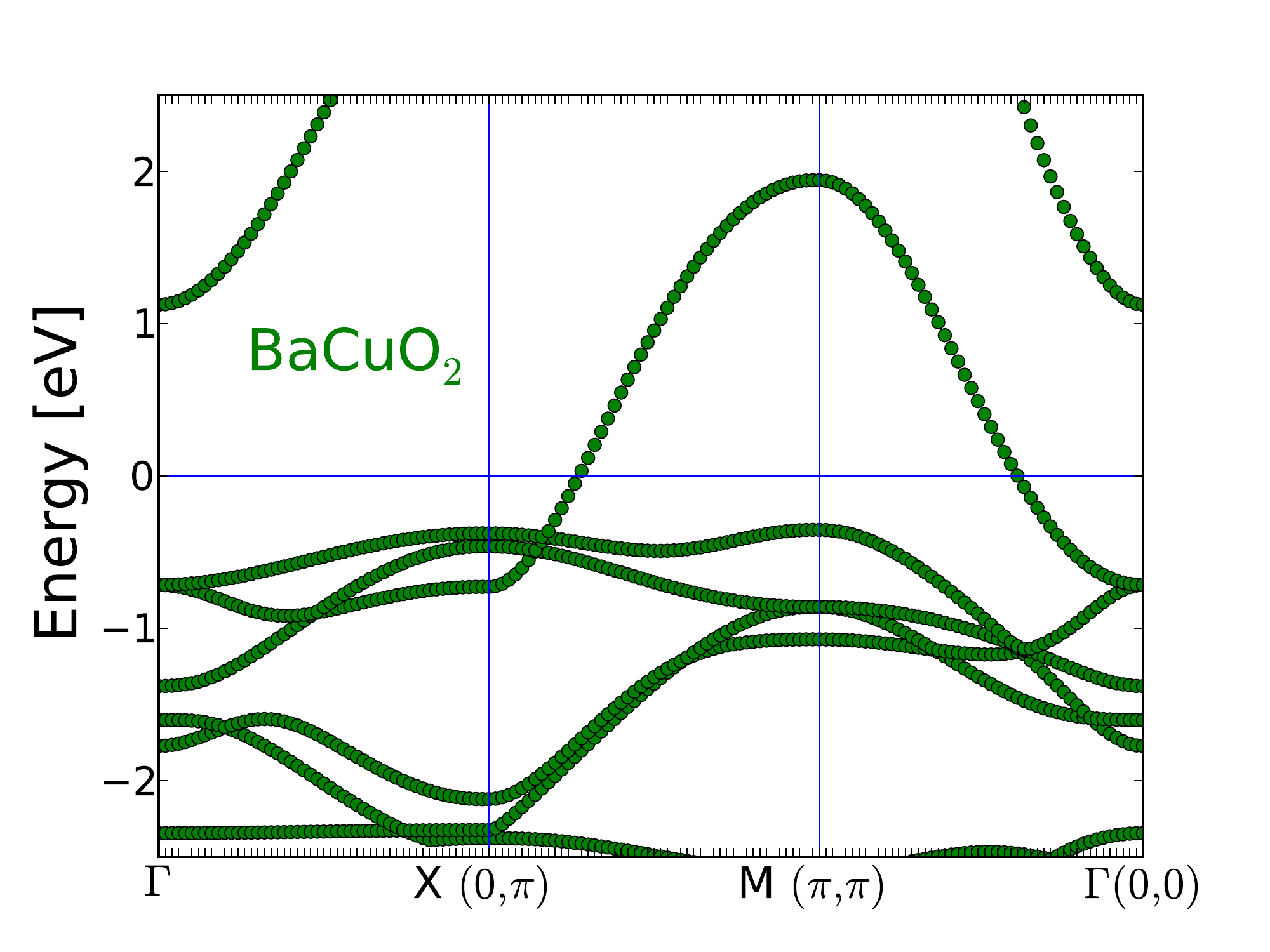}
		\includegraphics[width=0.45\textwidth]{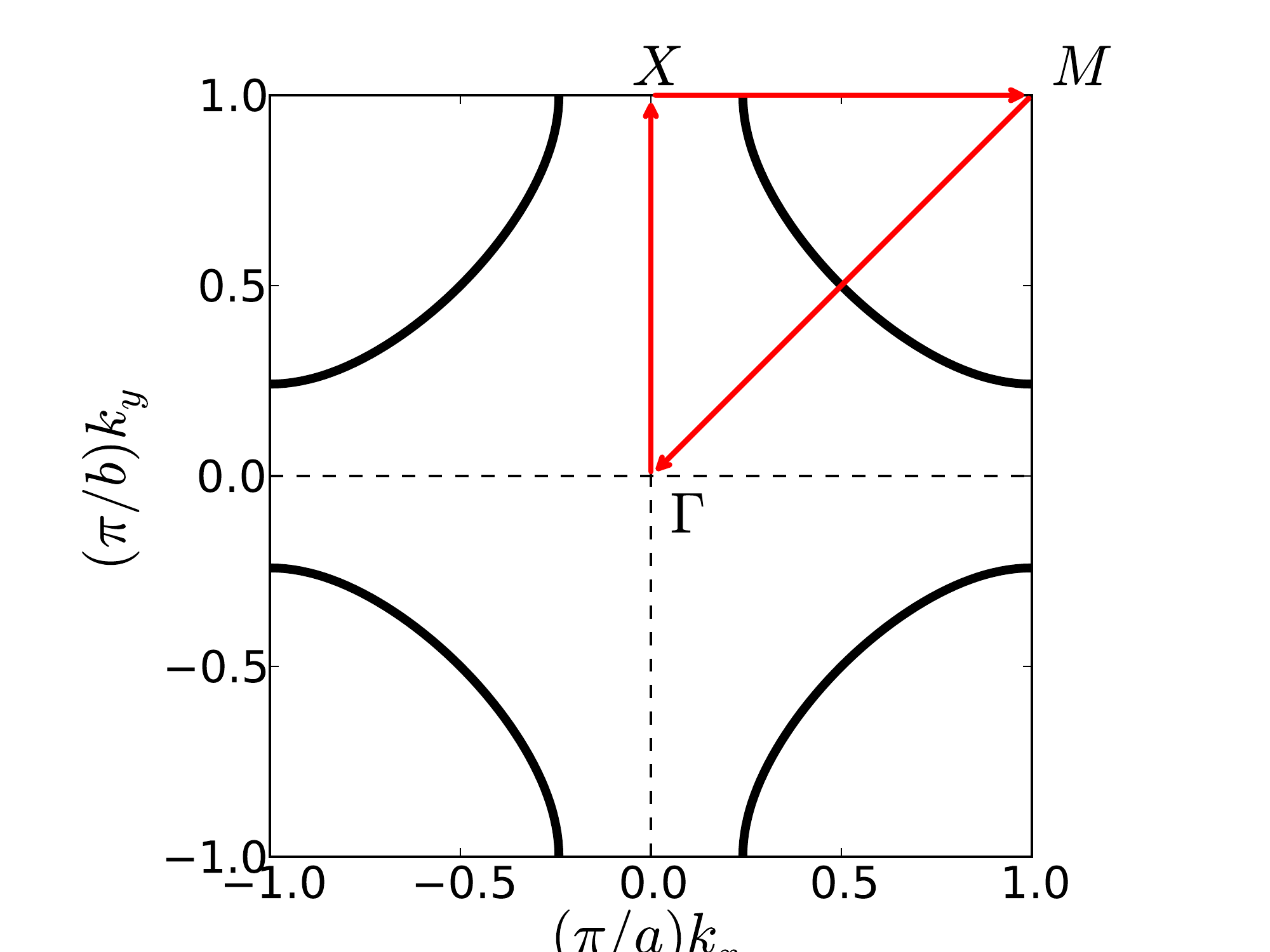}
	\caption[Calculated band structure of all ACuO$ _{2} $.]{The calculated electronic dispersion of for each ACuO$_2$ along certain paths in the Brillouin zone as illustrated in the lowest panel. \added{The energy is relative to the Fermi-energy, $ E_F $.} }
	\label{fig:bandstructure}
\end{figure}

It is also possible to calculate the electronic dispersion along specified $\kk$ directions of high symmetry, for example $\kk=(0,0)$ to $(0,\pi)$ to $(\pi,\pi)$ to $(0,0)$ for $\kk_z=0$ is a commonly chosen path and is illustrated in the lowest panel of \fig~\ref{fig:bandstructure}. In other notation this path is $\Gamma$-$X$-$M$-$\Gamma$. 

To calculate the dispersion, $\ekm$, for such restricted regions of the Brillouin zone it is \replaced{essential}{necessary} to perform the calculation in a different way, involving a non-self-consistent calculation where the $n(\rr)$ cannot be modified.  The optimal procedure is described in the VASP manual \cite{vaspmanual}.  

In \fig~\ref{fig:bandstructure} we plot \ek along the path discussed for each ACuO$_2$. With respect to \cacuo the resulting band structure shows three nearly degenerate bands at $\Gamma$ with $\epsilon = -1.3$~eV.  By $\kk=(0,\pi)$ only two are degenerate with one higher in energy.  Between $\kk=(0,\pi)$ and $\kk=(\pi,\pi)$ the degeneracy is completely lifted with a strongly dispersive band crossing the Fermi-level close to $\kk=(\sfrac{\pi}{4},\pi)$ and $\kk=(\sfrac{\pi}{2},\sfrac{\pi}{2})$ (compare these $\kk$ values with the Fermi-contour plotted in \fig~\ref{fig:cacuo2contourfermi}).  From other cuprates we know that undoped cuprates have a charge transfer gap of approximately 2~eV \cite{lee2006}, in contrast to these calculations showing the undoped \cacuo is a metal.  This same metallic band is seen in MgCuO$ _{2} $, SrCuO$ _{2} $ and BaCuO$ _{2} $.

For \added{ease of comparison between the four materials}, in the upper panel of \fig~\ref{fig:bandstructure} we plot the band structures of MgCuO$ _{2} $, CaCuO$ _{2} $, \srcuo and \bacuo together. 

\begin{figure}
	\centering
		\includegraphics[width=0.735\textwidth]{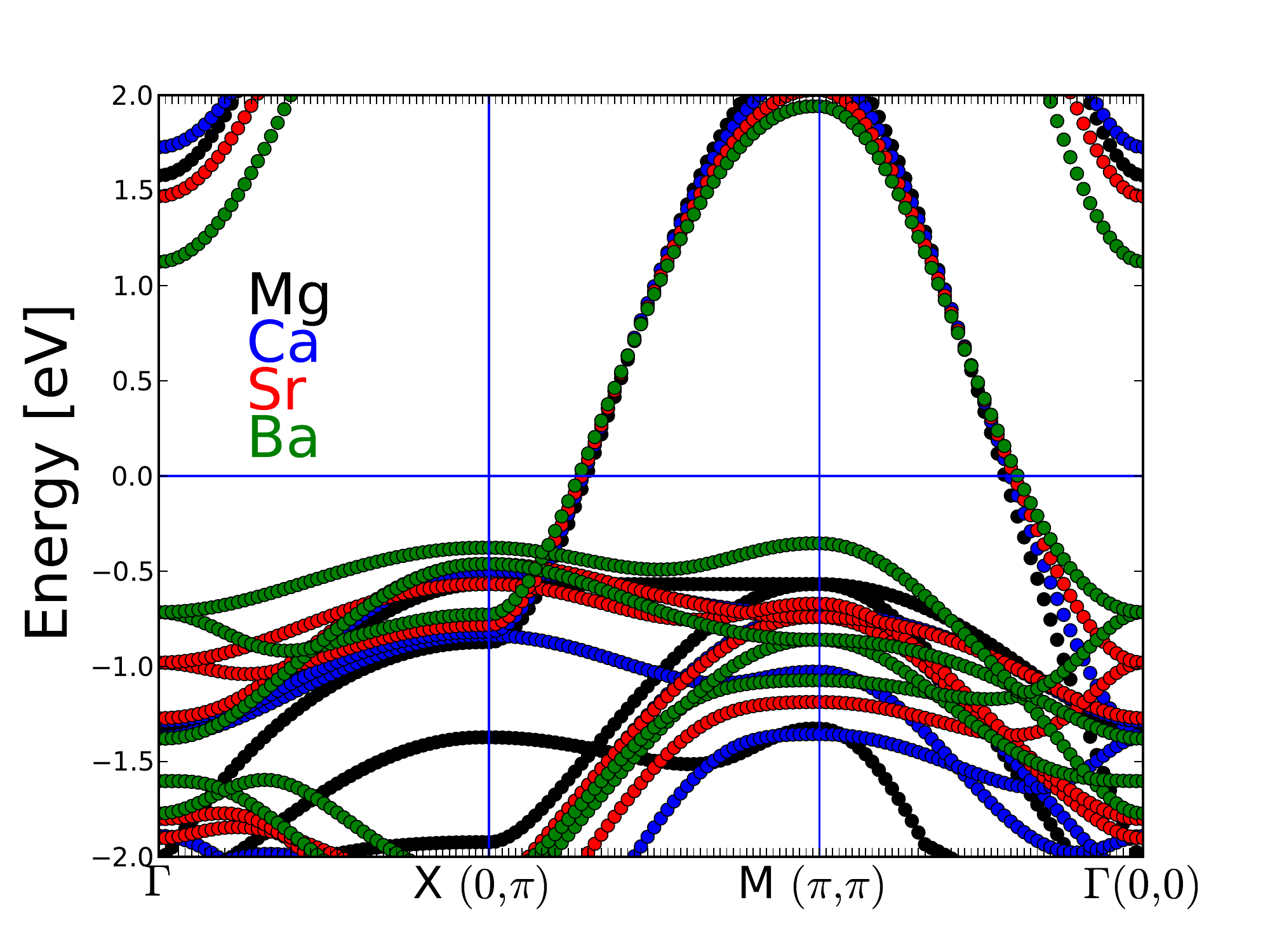}
		\includegraphics[width=0.735\textwidth]{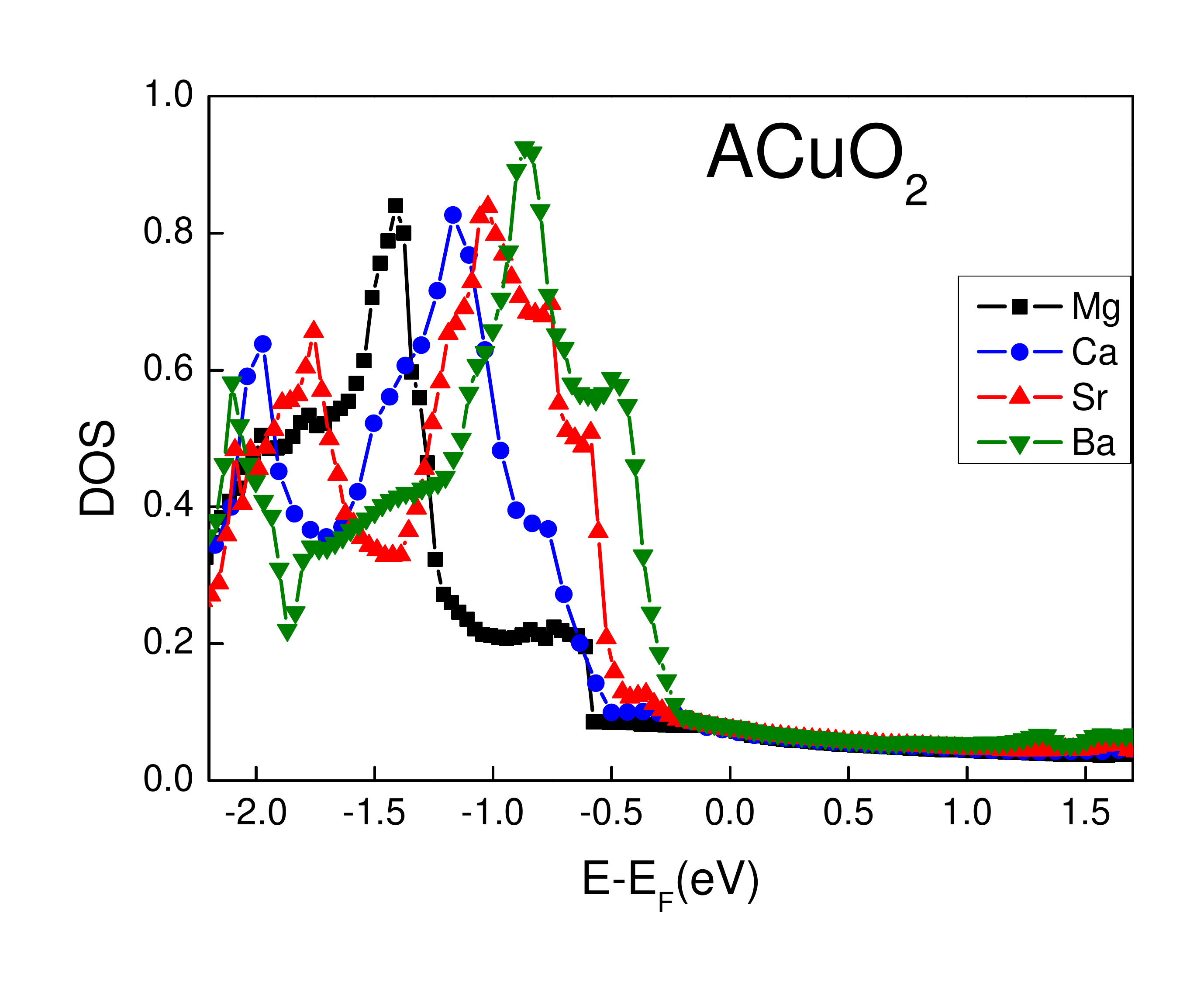}
	\caption[Band structure and density of states for all ACuO$ _{2} $.]{\label{fig:dosalld} The top panel shows the calculated electronic dispersion around the Fermi-level for (undoped) ACuO$_2$ where A=\{Mg,Ca,Sr,Ba\} are colour differentiated in the plot, as annotated.  Experimentally we know these materials (where synthesisable) are insulators with long range antiferromagnetic ordering of the Cu $d_{x^2-y^2}$ electron spins, however the band crossing the Fermi-level close to $\kk=(\sfrac{\pi}{4},\pi,0)$ and $\kk=(\sfrac{\pi}{2},\sfrac{\pi}{2},0)$ shows that these calculations predict our material to be a metal. This is also evident from the density of states shown in the lower panel.  The strong electronic correlations involving $d$ orbitals requires some corrections are made to the DFT method with the LDA.}
\end{figure}

With an accurate calculation of \ek it is straight forward to calculate the electronic density of states (DOS).  The results are plotted in the lower panel of \fig~\ref{fig:dosalld} and indeed show a finite DOS at $E_F$ confirming that we have calculated a metallic ground state.  Nevertheless, the calculated evolution of the DOS with ion-size shows a vHs like feature moving progressively closer to $E_F$. The vHs feature is a result of the weakly dispersive \ek around $X$ clearly visible in the upper panel of \fig~\ref{fig:dosalld}.


There are several reasons why a metallic ground state is to be expected from these LDA calculations. \added{The }LDA often under-estimates band gaps. Furthermore, in the cuprates the Cu \dxy orbital (hybridised with the O $p_{\sigma}$ orbitals) is the closest occupied band to $E_F$ as shown in \fig~\ref{fig:electronicconfig}.  We also know that strong on-site Coloumb interaction leads to long-range antiferromagnetic ordering of the Cu \dxy electron moments in the undoped cuprates.  More generally, the overlap of Cu \dxy and O $p_{\sigma}$ orbitals means that cuprates are strongly-correlated materials.  As such we cannot expect a simple LDA calculation to calculate a faithful band structure for ACuO$_2$.  One must at least perform a calculation that distinguishes electron spins and such a scheme is called LSDA.  Because undoped cuprates have strong electronic correlations it will also be necessary to add a further correction to our calculations. 

However, the dispersion reflects the rigidly shifted band structure observed using ARPES at finite doping where the strong-correlations are screened out by mobile carriers. Further, the saddle-point vHs is known to reside below $E_F$ and is crossed in the overdoped region leading to a change in the Fermi-surface topology \cite{storey2007}.  So these calculations do reveal the dispersion features that are known to exist when correlations are suppressed and we may therefore assume that the systematic changes with ion-size do reflect real band structure evolution.

\section{Discussion}

At the start of this chapter we discussed the correspondence between the results of our DFT calculations and reality.  We concluded that our calculations satisfactorily reproduced experimental lattice parameters where the materials have been reported to be synthesized.  


The calculated evolution of the DOS with ion-size provides a validation of the hypothesis that increase of \tcmax with larger ion size\footnote{Or conversely, the decrease of \tcmax with internal pressure} is a result an enhanced $ \dosm $. There are two caveats that must be mentioned;
\begin{enumerate}
	\item The doping level, $p$:  If we are to compare ion-size effects on \tcmax we should compare ACuO$_2$ at optimal doping. Our calculations however have been performed on undoped ACuO$_2$ and the DOS and band structure evolution with doping is not clear. A rigid-band approximation, where $E_F$ is modified by doping but not the electronic dispersion $E(\kk)$, is often used in the literature. However we note that some recent work has called into question the validity of this approximation for the cuprates \cite{lin2006}.  It is therefore desirable to repeat these calculations at different $p$. 
	\item  Is ACuO$_2$ a representative cuprate? The infinite layer ACuO$_2$ represents the fundamental component of a cuprate superconductor.  However the apical oxygen and other metal-oxide layers clearly play some role in the superconducting characteristics.  Therefore it is desirable to repeat these calculations on a more complicated system and suggest the LnA$_2$Cu$_3$O$_7$ system, where Ln is a member of the Lanthanide series and A=\{Ba,Sr\}, as it supports a wide range of ion-size variations.
\end{enumerate}




There is considerable scope for further studies in this direction.  With ACuO$_2$ we have a simple system for which these types of calculations can be performed accurately and relatively quickly.  There is a large range of ion-size that A can take between Mg and Ra.  On the other hand, there is not the same quality of experimental evidence to compare with as for e.g. Ln123 or the Bi-based compounds.  Furthermore, it is quite possible \mgcuo and \racuo cannot be synthesised in the real world. It would be interesting though to employ high-pressure/high-temperature synthesis in attempts to stabilise at least the Mg and Ba members. 

We identify two key avenues of further study;  
\begin{itemize}
	\item Vary the electronic doping state.  The easily accessible states are $p=(\sfrac{1}{8})n$ for an integer $n$ (because only integer numbers of electrons can be added/removed from the `unit cell' in VASP).  For intermediate doping states the number of CaCuO$_2$ groups in the `unit cell' must be increased.
	\item Vary the external pressure.  This would provide an interesting comparison with the ion-size variation calculations \added{and could be effected by performing the DFT calculations on a unit-cell volume restricted to a value smaller (for positive external pressure) or larger (for negative external pressure) from the calculated $ F_{\textnormal{min}} $ value shown in \fig~\ref{fig:fvsv}}. 
\end{itemize}

A similar breadth of calculations could be done for Ln(Ba,Sr)$_2$Cu$_3$O$_7$.  In this case we would be able to more directly compare the results of the computational studies with the experimental studies.  A DFT study of YBa$_2$Cu$_3$O$_7$ and YSr$_2$Cu$_3$O$_7$ has been carried out using VASP by Khosroabadi \etal \cite{khosroabadi2004} with a purpose of exploring the difference between ``mechanical'' pressure (external pressure) and chemical pressure (i.e. `internal', or ion-size) - as we hope to do.  The range of mechanical pressures they use is $ -15 $ to 15~GPa in 5~GPa intervals.   

We firstly note that their study does not explore ion-size variations as systematically as could be hoped.  Intermediate Ba:Sr concentrations would require a larger unit cell for the computations which would make them significantly more time-consuming.  With this in mind LaBa$_2$Cu$_3$O$_7$, LuBa$_2$Cu$_3$O$_7$ and YSr$_2$Cu$_3$O$_7$ (and perhaps YBaSrCu$_3$O$_7$) represents a better set of materials for exploring the ion-size effect in the Ln123 family of materials. 

Their calculations show quite distinct mechanical pressure dependences of the YBCO (\ybco or Y123) and YSCO structures. The pressure-induced charge transfer (PICT) is qualitatively similar for both YBCO and YSCO, but less dramatic in YSCO attributable to a less compressible $z$-axis.  Furthermore, under chemical pressure (smaller ion-size) they calculate an \emph{increase} in the apical oxygen bond length, which agrees with the experimental situation summarised in \refsec~\ref{sec:ln123systematics}, and in particular \fig~\ref{fig:ln123structure}, in distinction to the shorter apical bond length resulting from mechanical pressure.  Thus, they identify the Cu(2)-O(1) (the apical oxygen bond length) as the primary difference between mechanical and chemical pressures and the cause of their opposite effect on \tc.  

As mentioned in \refsec~\ref{sec:apical}, the apical oxygen likely has an important \cite{ohta1991, pavarini2001, weber2011, sakakibara2012, tallon2012}, but not fully understood, role in the electronic properties of the cuprates.  Indeed, the apical oxygen bond length is a parameter in the composite bond valence sum $V_+$ that correlates with \tcmax for all known HTS cuprates, \fig~\ref{fig:bvs} \cite{tallon1990}.  For example, the authors of \cite{pavarini2001, sakakibara2012} discuss the role of the apical oxygen in terms of its influence on the electronic energy levels of states in the \cuo layer, that being in addition to its role in charge-transfer from the charge reservoir layer to the \cuo layer.  For a different perspective, Raghu \etal argue that the charge reservoir layers themselves are the important parameter for \tcmax \cite{raghu2012optimaltc}. In our closing discussion we suggest a possible unified understanding of the physical role of the apical oxygen and charge reservoir layer.

\section{Conclusions}

We performed DFT calculations on undoped ACuO$_2$ for A=\{Mg, Ca, Sr, Ba\} to investigate ion-size effects the electronic dispersion, $ \ekm $, in this model cuprate system.  Where these materials have been synthesised we found good agreement between our calculated structural parameters and the experimental ones.  Our calculations show a peak in the density of states $\sim 1$~eV below the Fermi-level that moves closer to the Fermi-level with larger A ion-size.  This is similar to the hypothesis discussed in \refsec~\ref{sec:vhshypothesis} and is consistent with an interpretation of the ion-size affecting \tcmax via the density of states.

\chapter{Two-magnon scattering}
\label{ch:twomag}
\subsubsection{Summary and Introduction}

In this chapter we seek to understand the opposite effects of internal and external pressure on \tcmax by (i) measuring $J$ while controlling the internal-strain through isovalent ion substitution in LnA$_2$Cu$_3$O$_6$ and (ii) comparing it to data in literature for the external pressure dependence of $J$ and $ \tcmaxm $.  We find no resolution, $J$ and \tcmax anti-correlate with internal pressure as the implicit variable and correlate with external pressure as the implicit variable. 

Many of the systematics of Ln123 have been explored, but until now not the nearest-neighbour, antiferromagnetic, superexchange energy, $J$.  This is an important energy because it is widely believed that superconductivity in the cuprates is caused by a magnetic interaction. Many theoretical arguments \cite{bickers1987, moriya2003, uemura2009, scalapino2012} favour a pairing mechanism in the cuprates of magnetic origin.  For a long time it was not clear that magnetic correlations were present across the entire superconducting phase diagram but recently damped spin waves (paramagnons) have been detected using Resonant Inelastic X-ray Scattering, exhibiting a similar dispersion for many different cuprates across a broad doping range \cite{letacon2011}. In this case the energy scale of \(\omega_B\) is $J$ \cite{letacon2011}.  Even more recently, beautiful Raman data of Li \etal shows a correlation, or in their words a `feedback effect', between magnetic fluctuations as detected by two-magnon scattering and $\scgapm$ as measured by electronic Raman scattering in HgBa$ _{2} $CuO$ _{4+\delta} $ \cite{li2012}. $J$ can be deduced from Raman spectroscopy measurements of the two-magnon scattering. In the following we present our two-magnon scattering studies of the model system LnA$ _{2} $Cu$ _{3} $O$ _{y} $ as ion size is varied. 

For a full introduction to the techniques and theory used in this chapter, see \refsec~\ref{sec:twomagtheory} and \ref{sec:twomagmeasurements}.

\section{Two-magnon scattering in LnBa$_2$Cu$_3$O$_6$}

\begin{figure}
	\centering
		\includegraphics[width=0.75\textwidth]{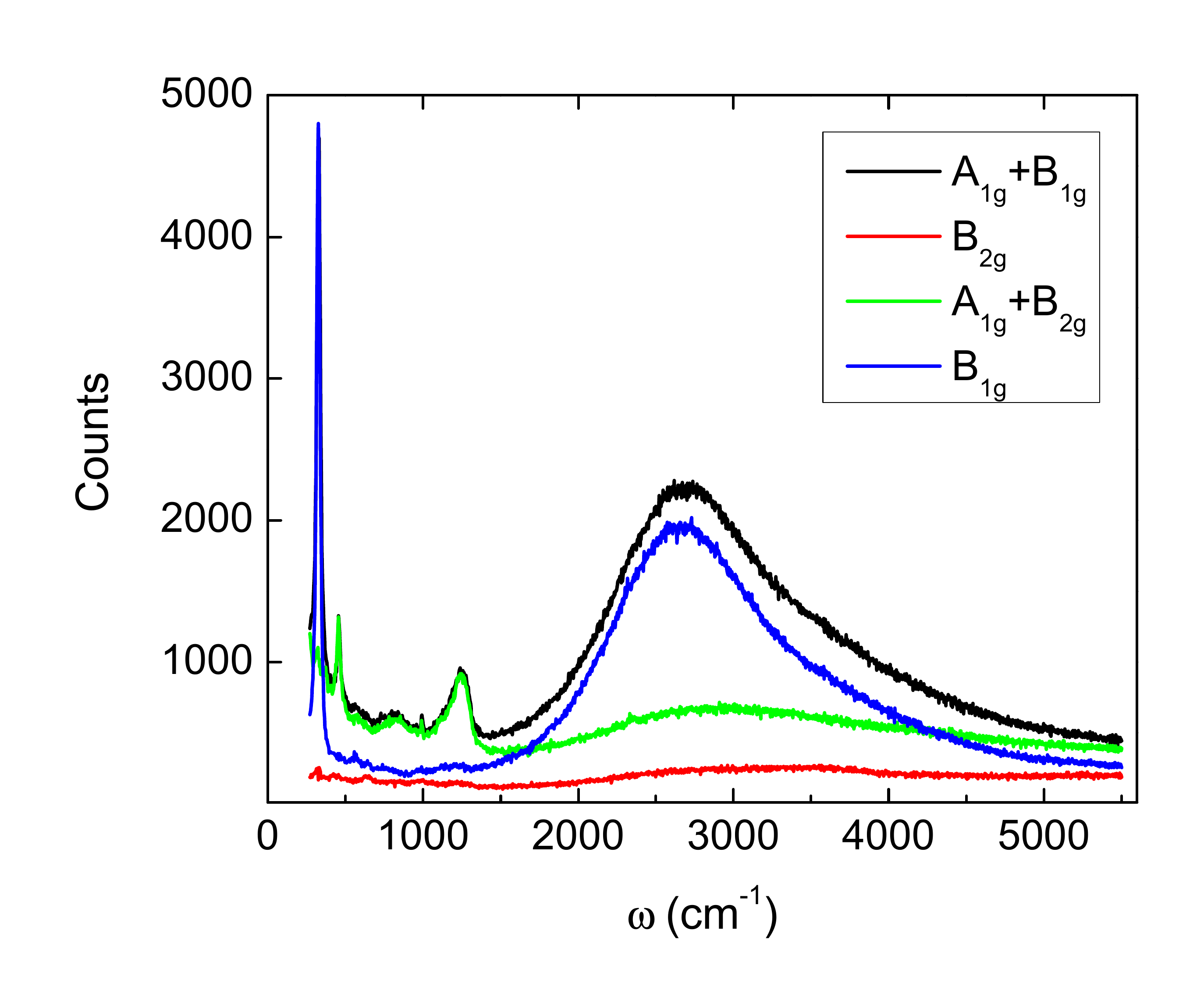}
	\caption[Representative raw Raman spectra from a Ln123 single crystal in all four scattering geometries.]{Raw Raman spectra from single-crystal SmBa$_2$Cu$_3$O$_6$ in $ \aogbogm $, $ \btgm $, \aogbtg and \bog symmetries. Possible small `leakage' \bog and \aogbog signal is observed in the \btg channel as a result of imperfect polarisers and alignment. Experimental set up; $514.5$~nm laser line at power $\sim 1$~mW, focused to a spot of size $\sim\!1$ $\mu$m with a x100 objective (NA=0.90). A 300 lines/mm diffraction grating is used to capture the entire spectrum in a single frame.}
	\label{fig:sm123allsymm}
\end{figure}

The raw Raman spectra from single-crystal SmBa$_2$Cu$_3$O$_6$ in $ \aogbogm $, $ \btgm $, \aogbtg and \bog symmetries are plotted in \fig~\ref{fig:sm123allsymm}. The broad, asymmetric peak around 2600~\cm in $ \bogm $, \aog (to a lesser extent - see \refsec~\ref{sec:a1g}) but not in \btg spectra is the result of two-magnon scattering. The position of its maximum intensity is linearly related\footnote{The estimation of $J$ can also be made from a more elaborate fitting procedure \cite{singh1989}.} to $J$; $\omega_{\textnormal{max}}\approx 3.2J$ \cite{chubukov1995}. 

\added{Also visible from these spectra are phonon modes below $ \sim 1000 $~$ \cmm $, especially the \bog phonon mode at $ \approx 330$~\cm that is due to out-of-phase vibration of the O(2) and O(3) ions in the \cuo layer \cite{limonov2000}, and a two-phonon mode at $ \approx 1300 $~\cm \cite{sugai2003}.} 

These data, and those reported below, were taken with the $514.5$~nm laser line at power $\sim 1$ mW, focused to a spot of size $\sim\!1$~$\mu$m with a x50 (NA=0.75) or x100 objective (NA=0.90). A 300 lines/mm diffraction grating is used to capture the entire spectrum in a single frame of the CCD.

\begin{figure}
		\centering
		\includegraphics[width=0.60\textwidth]{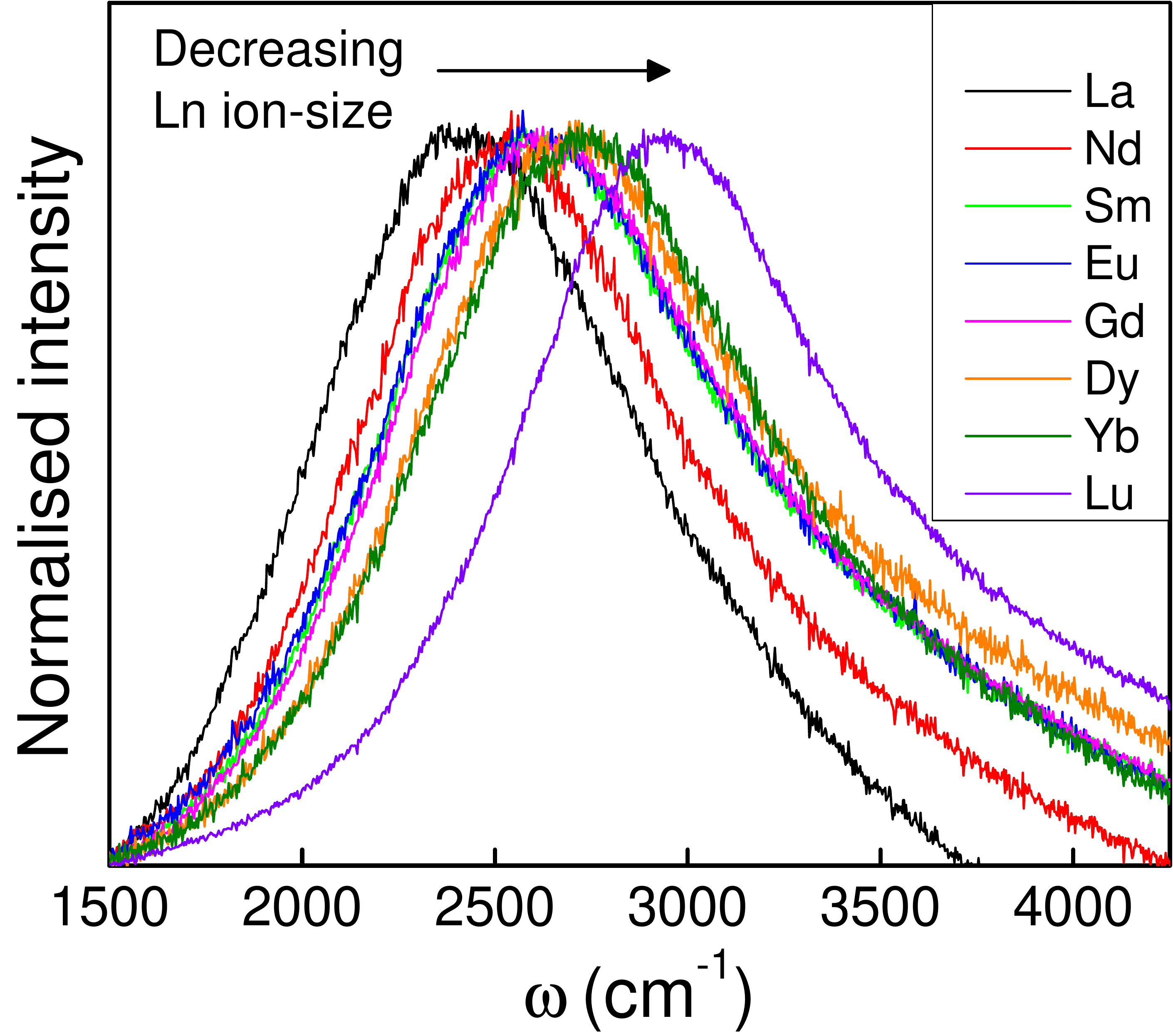} 
	\caption[The \bog two-magnon scattering peak for all Ln123.]{\label{fig:ln123b1g} Raman spectra of LnBa$_2$Cu$_3$O$_6$ showing a systematic increase in the \bog two-magnon peak position as the Ln ion-size is decreased (each Ln is indicated in the legend).  Data have had a constant fluorescence contribution subtracted and have been normalized by the peak intensity. }
\end{figure}

In \fig~\ref{fig:ln123b1g} we plot normalised Raman spectra for the entire LnBa$_2$Cu$_3$O$_6$ series, focusing on just the $\omega$-range of two-magnon scattering.  Quite nicely, we see a systematic increase in the two-magnon peak frequency as the Ln ion size decreases.  This is almost entirely because of a variation in the structural parameters: smaller Ln ion-size results in shorter in-plane bond lengths, particularly on the \cuo layers \cite{guillaume1994}, and thus a shorter superexchange path-length. This is the ``internal pressure'' effect \added{which can also be seen altering the phonon mode frequencies, see for example \fig~\ref{fig:peakpos}}.  $J$ is extremely sensitive to the superexchange path-length in the transition metal oxides \cite{massey1990} and to a lesser extent in the cuprates \cite{aronson1991}.  The shorter in-plane bond lengths result in increased overlap between Cu(2) \dxy and O(2,3) $p_{\sigma}$ orbitals that increases the superexchange energy.  

This previous statement can be illustrated by considering an expression for $J$ based on the two-band Hubbard model; $J\approx{4t_{pd}^4}\Delta^{-3}$ \cite{aronson1991, lee2006} where $\Delta$ is the charge-transfer gap energy and $t_{pd}$ the hopping integral between Cu $d$ and O $p$ orbitals.  $t_{pd}$ is itself inversely proportional to the superexchange path-length, $t_{pd}\sim d^{-n}$ with $2.5 < n < 3.0$ \cite{aronson1991}, so within this framework $J$ is shown to be strongly dependent on the superexchange path length.

\begin{figure}
		\centering
		\includegraphics[width=0.60\textwidth]{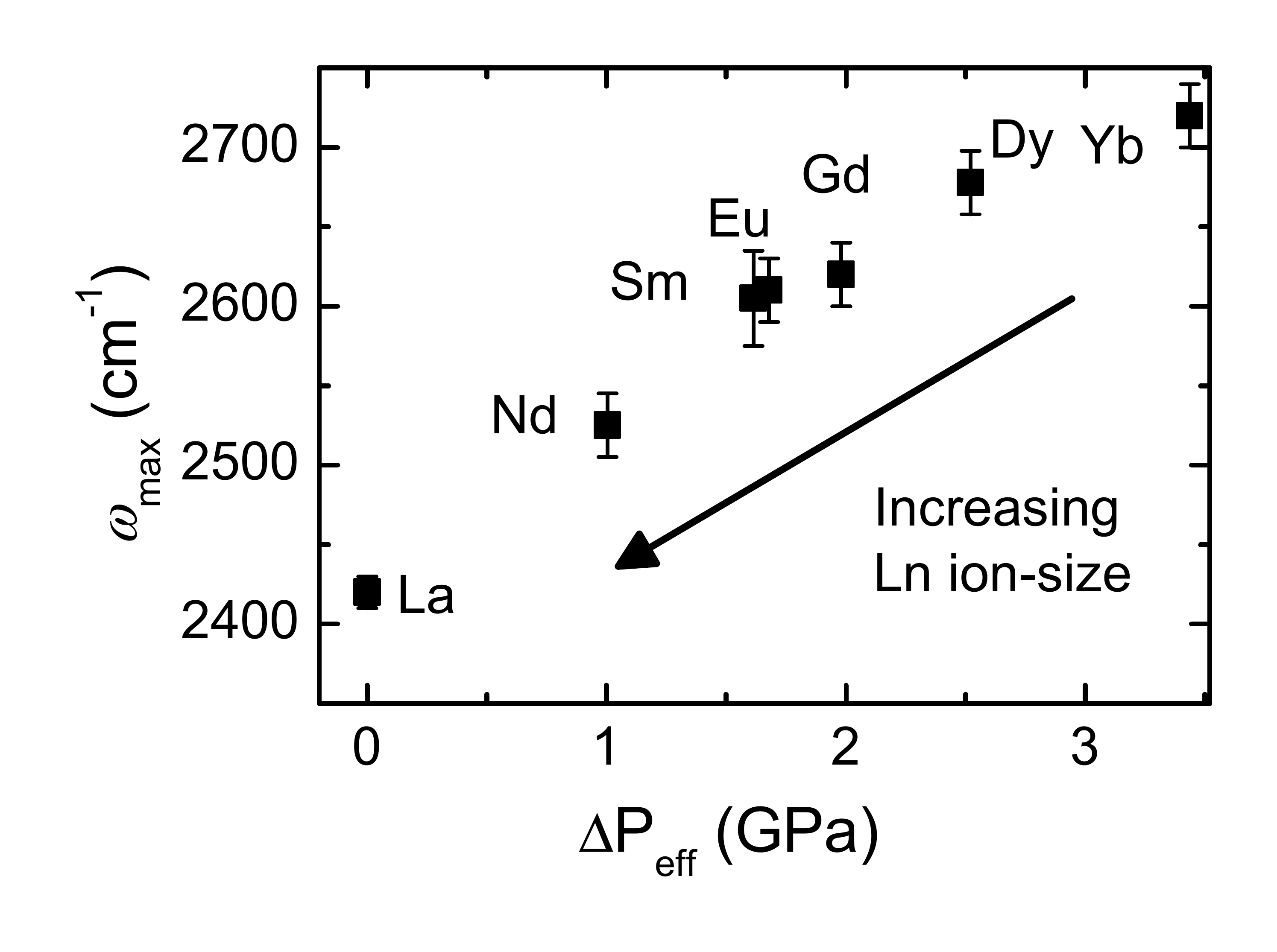} 
	\caption[The position of the two-magnon peak as a function of internal pressure.]{\label{fig:wvspeff} The \bog Raman two-magnon peak position, $\omega_{\textnormal{max}}$, for each Ln123 sample plotted against the effective internal pressure using $\Delta P_{\textnormal{eff}} = - B.\Delta V/V_0$ where $B= 78.1$ GPa is the bulk modulus for deoxygenated Y123 \cite{suenaga1991} and $\Delta V = V-V_0$ is referenced to La123.}
\end{figure}

We can estimate the effective internal pressure, $\Delta P_{e\!f\!f}$, from the change in unit cell volume, \(\Delta V = V- V_0\) where $V_0$ is referenced to La123, using $\Delta P_{e\!f\!f}=-B.\Delta V/V_0$. $B=78.1$ GPa is the bulk modulus, determined for oxygen deficient Y123 by Suenaga \etal \cite{suenaga1991}. In \fig~\ref{fig:wvspeff} we plot \(\omega_{max}\) against $\Delta P_{e\!f\!f}$.  Again, the observed positive gradient can be understood as an increased exchange interaction arising from increased overlap between Cu(2) $d$ and O(2,3) $p$ orbitals.

If $J$ is dependent on structural parameters, then we can quantify the shift in $J$ with Ln size using an area Gr\"{u}neisen parameter\footnote{It has been suggested from uniaxial high pressure studies and substrate strained-lattice studies that \cuo plane compression is the critical parameter \cite{schillingchapter}.}, 

\begin{equation}
\gamma_A=-\frac{\ln(J/J_{0})}{\ln(A/A_0)}
\label{eq:gruneisen}
\end{equation}

\noindent where $A$ is the area of the basal (\cuo) plane. Note the Gr\"{u}neisen parameter is unit-less and so is not sensitive to the constant relating the maximum two-magnon peak frequency, $\omega _{max}$ to the anti-ferromagnetic exchange constant $J$. 

\begin{figure}[tb]
	\centering
		\includegraphics[width=0.66\textwidth]{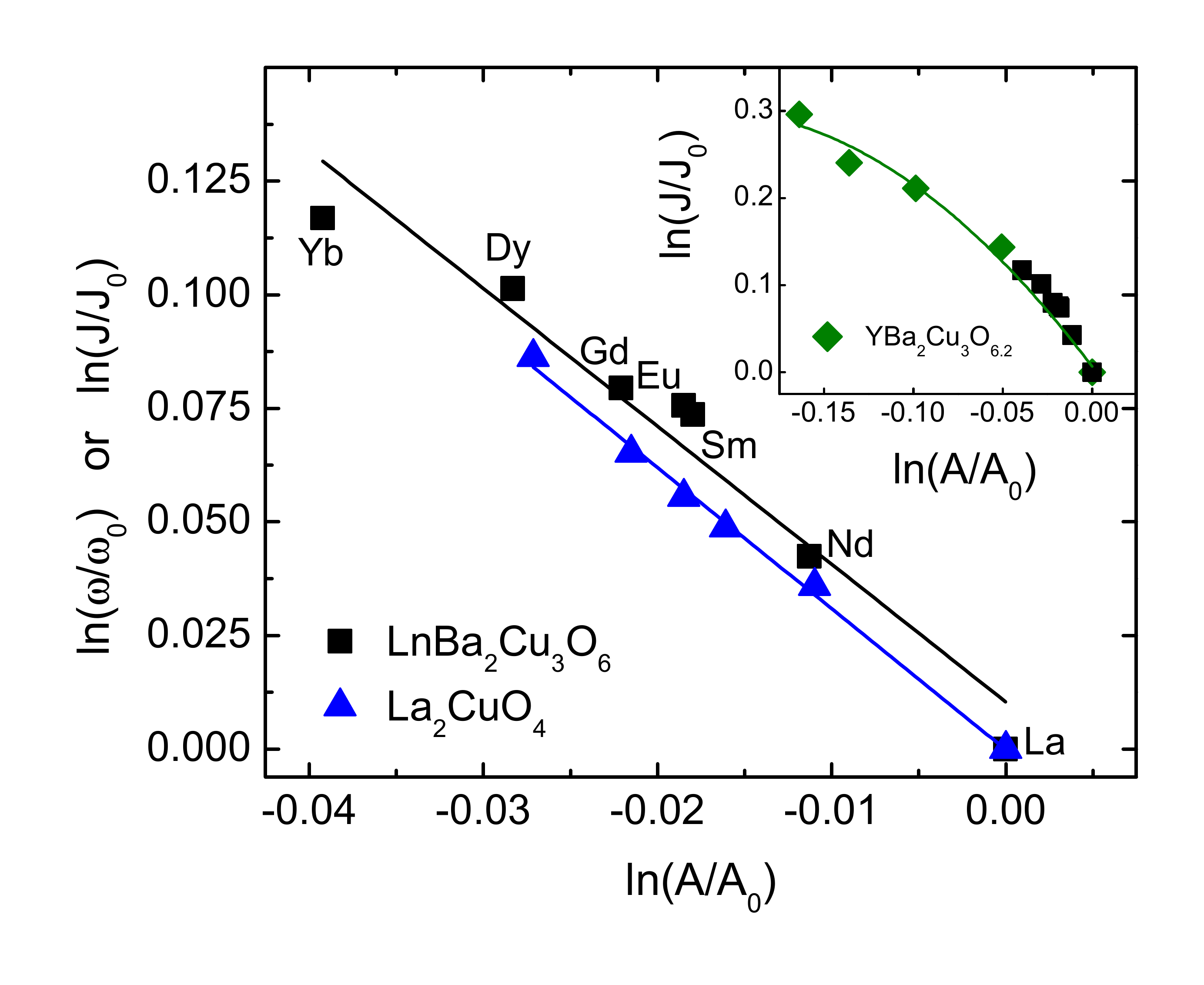}
	\caption[Gr\"{u}neisen scaling analysis of $J$ for internal and external pressure.]{To determine the Gr\"{u}neisen parameter for our Ln123 system and for external pressure data on La214 from \cite{aronson1991}, we plot the natural log of the \cuo basal plane area, $A$, against the natural log of $J$ where both values are referenced to La123 (for Ln123) or their value at atmospheric pressure (for La214).  We find $\gamma_A=3.0\pm 0.6$ for Ln123 and $\gamma _A = 3.1 \pm 0.1$ for La214.  This shows the effect of external pressure on $J$ is consistent with that of internal pressure. Inset: Data for YBa$_2$Cu$_3$O$_{6.2}$ from \cite{maksimov1994}. The dotted line is a guide to the eye.  Again, the effect of external pressure on $J$ appears consistent with that of internal pressure.}
	\label{fig:gruneisen}
\end{figure}

The area dependence of $J$ is plotted in log scale in \fig~\ref{fig:gruneisen}, where $A$ is calculated from Guillaume \textit{et al.} \cite{guillaume1994} and the zero subscript refers to La as reference.  From this plot we find $\gamma_A=3.0\pm 0.6$ where the large uncertainty reflects the value.  For example, we obtain \(\gamma_A=3.7\) if Yb123 is \textit{not} included in the fitting whereas we obtain $\gamma_A=2.5$ if we instead do not include La123 in the fitting. It is more likely, however, that $\gamma_A$ is not constant across the range, as can be seen by the non-linearity of the plot and as indicated by external pressure studies.

		\subsection{Comparison of external pressure and ion-size effect on $J$}
		\label{sec:jintandextp}
		
\begin{figure}
\centering
\includegraphics[width=0.75\columnwidth]{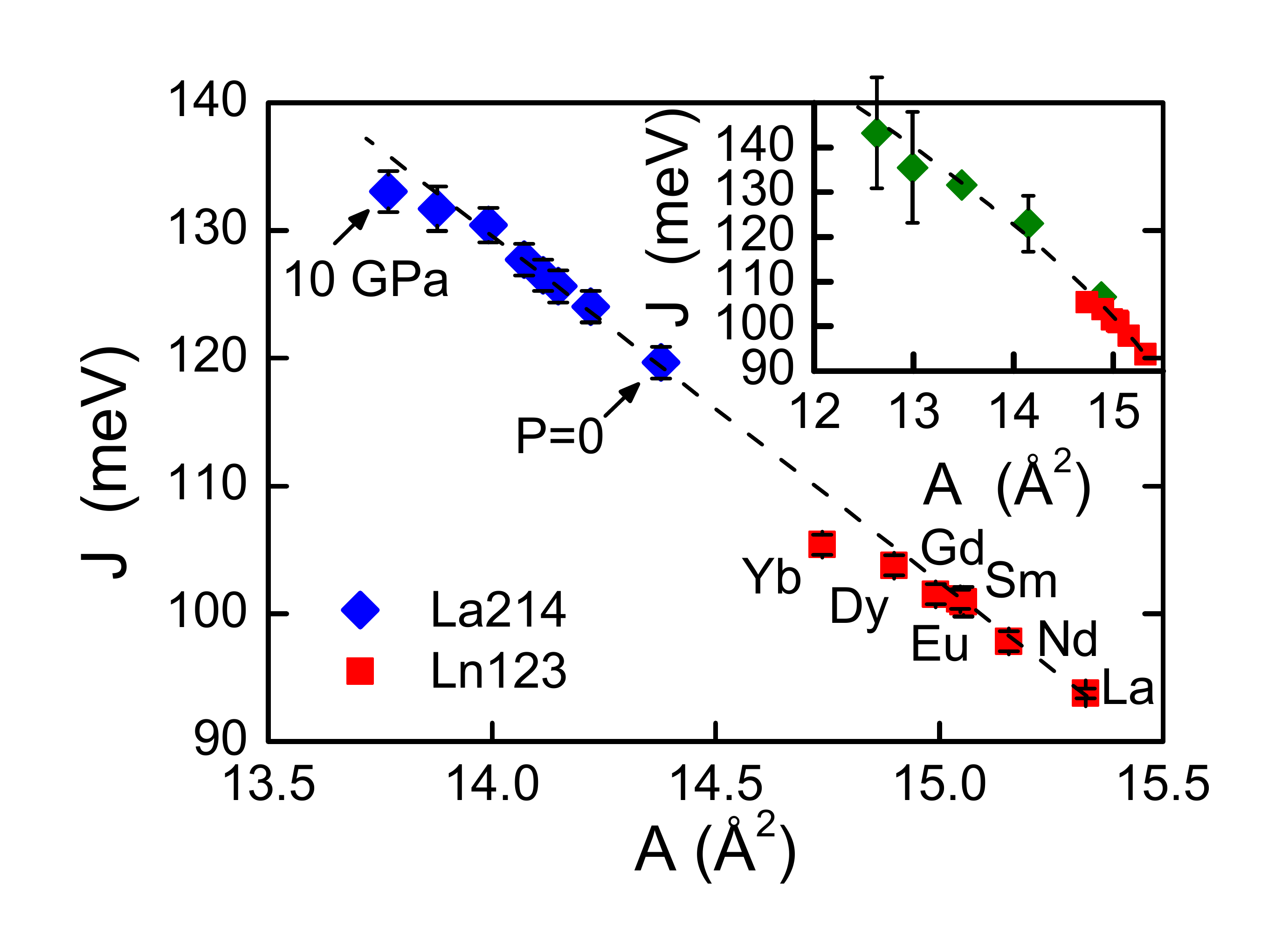}
\caption[Comparison of the effect of internal and external pressure on $J$.]{\label{fig:jintandextp}The basal area dependence of $J$ determined from our two-magnon scattering studies on LnBa$_2$Cu$_3$O$_6$ (red squares) where internal pressure (or ion size) is the implicit variable. This is compared with the same plot for La$_2$CuO$_4$ where external pressure (0 to 10 GPa) is the implicit variable \cite{aronson1991}. Remarkably a single behaviour for $J(A)$ is preserved across a 50\% change in $J$ irrespective of whether the effective pressure is internal or external. \deleted{We find the Gr\"{u}neisen parameters $\gamma_A=3.0\pm0.6$ for Ln123 and $\gamma_A=3.1\pm0.1$ for La214.  }Inset: $J$ versus $A$ for pressure-dependent two-magnon scattering data for YBa$_2$Cu$_3$O$_{6.2}$ to 80 GPa \cite{maksimov1994} (green diamonds) compared with our ion-size-dependent data for LnBa$_2$Cu$_3$O$_6$ (red squares). Again external and internal pressures appear to have quantitatively similar effects on $J$. The dashed lines are guides to the eye.}
\end{figure}

It is also instructive to plot the dependence of $J$ on absolute \cuo area, $A$.  These data are plotted in \fig~\ref{fig:jintandextp}. Plotting the data in this way allows us to readily compare the effect of internal and external pressure on $J$.  Thus in \fig~\ref{fig:jintandextp} we also show the effect of external pressure on $J$ in the related system La$_2$CuO$_4$ \cite{aronson1991} (blue diamonds) ranging from ambient pressure to 10 GPa, as annotated. 

Remarkably, the dependence of $J$ on $A$ is preserved across the entire range including two quite structurally disparate cuprates, irrespective of whether the pressure is {\it internal} or {\it external}. Collectively these cover a 50\% increase in the magnitude of $J$ arising from simple structural compression.  

We also plot these data of Aronson \etal on a log-log scale to further compare with the ion-size data, \fig~\ref{fig:gruneisen}. Using \eq~\ref{eq:gruneisen} we determine $\gamma _A = 3.1 \pm 0.1$ for La$_2$CuO$_4$ under external pressure (in this case the zero subscript in \eq~\ref{eq:gruneisen} refers to the value of $J$ and $A$ at atmospheric pressure). Within uncertainties, this value agrees with what we find for the internal pressure ion-size effect for Ln123. 

The only published data for two-magnon scattering in Y123 under external pressure is for YBa$_2$Cu$_3$O$_{6.2}$ \cite{maksimov1994} and this is plotted by the green diamonds in the inset to \fig~\ref{fig:jintandextp} along with our data for Ln123 (red squares) as well as in the inset to \fig~\ref{fig:gruneisen}. The residual 0.2 oxygens in the chain layer introduce uncertainties around possible pressure-induced charge transfer and we are therefore cautious in the use of this data. Nonetheless, they reveal a trend which is fully consistent with the ion-size effect for Ln123.  

Furthermore, Kawada \etal have shown that the pressure dependence of $J$ for a variety of cuprate-oxides is very similar \cite{kawada1998}.  This means we can be even more confident that internal and external pressure have the same effect on $J$ in the cuprates.

In an attempt to improve on the available data for two-magnon scattering under external pressure we undertook Raman scattering studies using a Diamond Anvil Cell to achieve high pressures. Although we were able to measure the temperature dependence of the \replaced{$150$ $ \cmm $, $310$ \cm and $475$ \cm phonon modes, which have already been extensively studied \cite{goncharov2003}}{\bog phonon mode}, we could not detect two-magnon scattering from our sample in the Diamond Anvil Cell.  These are tricky measurements. See \refsec~\ref{sec:dac} for more details.

Nevertheless, from \fig~\ref{fig:gruneisen} and \fig~\ref{fig:jintandextp} we can conclude that external pressure has a quantitatively similar effect to internal pressure on $J$ in the cuprates.  This is an important result. 

Our conclusion that external pressure has a similar effect to internal pressure on $J$ in the cuprates contrasts with what Aronson \etal find comparing their data to Cooper \cite{cooper1990}.

\section{Is $J$ related to $T_c$?}

Our motivation for these two-magnon measurements was that it is widely believed that a magnetic interaction between electrons is the pairing mechanism leading to Cooper pairs and superconductivity.  To leading order the energy scale of these magnetic interactions, $\omega_{\textnormal{B}}$, is set by $J$ \cite{letacon2011} and so we can use these Raman measurements to characterise magnetic interactions as a function of ion size.  In a fuller treatment of magnetic interactions however we must also consider longer-range interactions, see \refsec~\ref{sec:pairingmechanismdiscussion}.  Recalling the weak-coupling, $d$-wave BCS equation for \tc \ref{eq:bcsdwaveweak},

\[
k_\textnormal{B} T_c^{\textnormal{mf}} = 0.935 \hbar \omega_{\textnormal{B}} \exp \left[- 1/(N(E_F) V) \right]
\]

\noindent we might expect our systematic shift decrease of $J$ with ion-size to be reflected in the value of $ \tcm $.  Even when \tc is determined by solving the $ \kk $-dependent gap equation it will be\deleted{r} determined to leading order by $ J $.

\begin{figure}
\includegraphics[width=0.90\columnwidth]{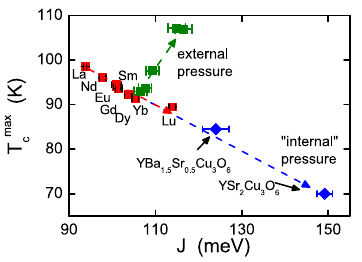}
\caption[\tcmax plotted against $J$ for Ln(Ba,Sr)$ _{2} $Cu$ _{3} $O$ _{y} $ under internal and external pressure.]{\label{fig:jvstc} $T_c^{\textnormal{max}}$ is plotted versus $J$ for single crystals of LnBa$_2$Cu$_3$O$_6$ (red squares) and for YBa$_{2-x}$Sr$_x$Cu$_3$O$_6$ (blue diamonds) with $x = 0.5$ and $2.0$. This reveals a systematic anticorrelation of $T_c$ with $J$ where ion size or ``internal pressure'' is the implicit variable. Greens squares show $T_c^{\textnormal{max}}$ versus $J$ under external pressure, revealing a behaviour which is orthogonal to that for internal pressure.  
}
\end{figure}

We thus plot in \fig~\ref{fig:jvstc} $T_c^{\textnormal{max}}$ versus $J$ for the Ln123 single-crystal series (red squares). We have used $T_c^{\textnormal{max}} = 98.5$ K for La123 \cite{lindemer1994} and $T_c^{\textnormal{max}} =96$ K for Nd123 \cite{veal1989} as these are the highest reported values of $T_c^{\textnormal{max}}$ in these compounds (where Ln occupation of the Ba site is minimised). Contrary to expectation from \eq~\ref{eq:bcsdwaveweak}, $T_c^{\textnormal{max}}$ anticorrelates with $J$ when ion-size or ``internal pressure'' is the implicit variable.

\begin{figure}[tb]
\includegraphics[width=0.50\textwidth]{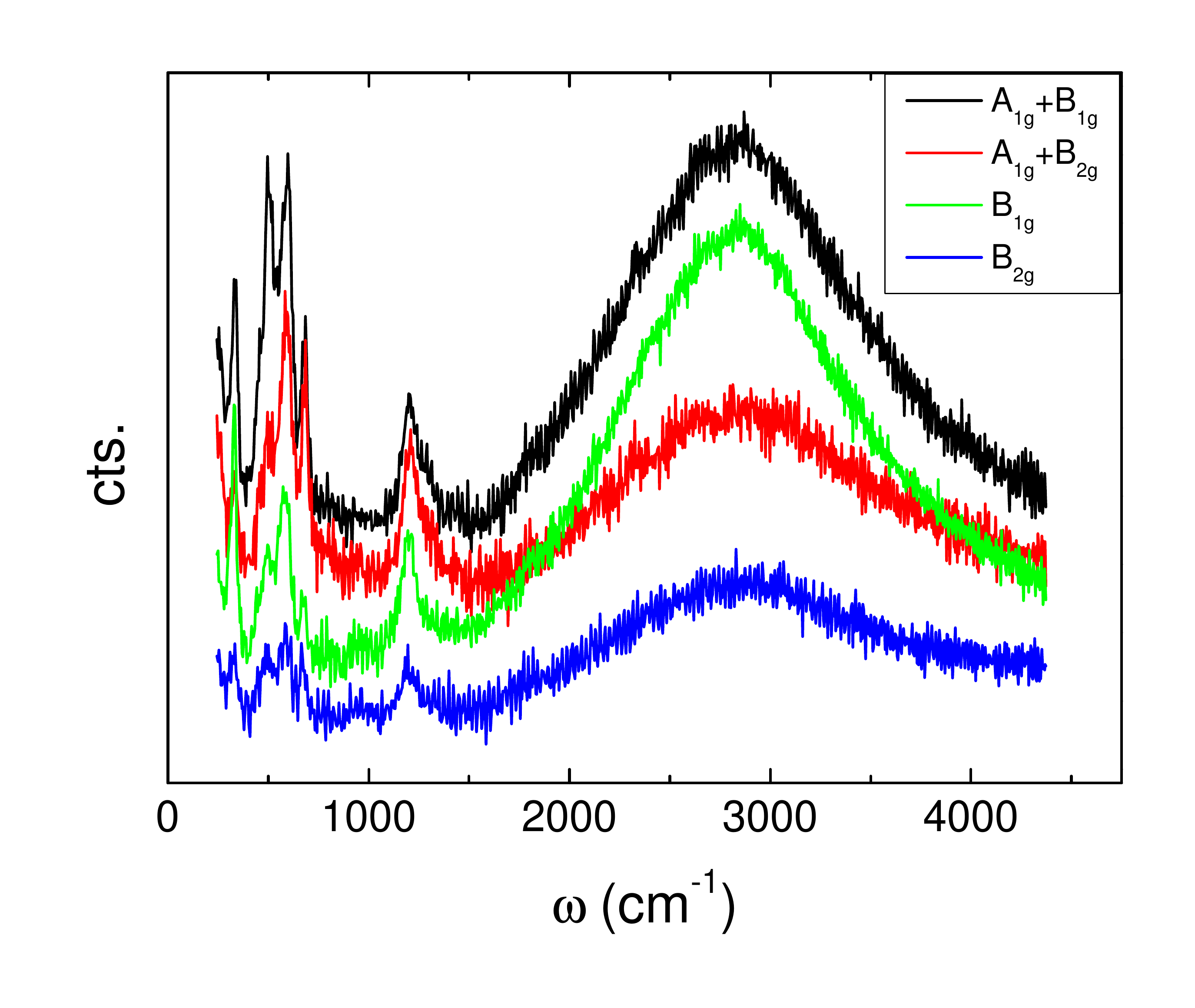}
\includegraphics[width=0.55\textwidth]{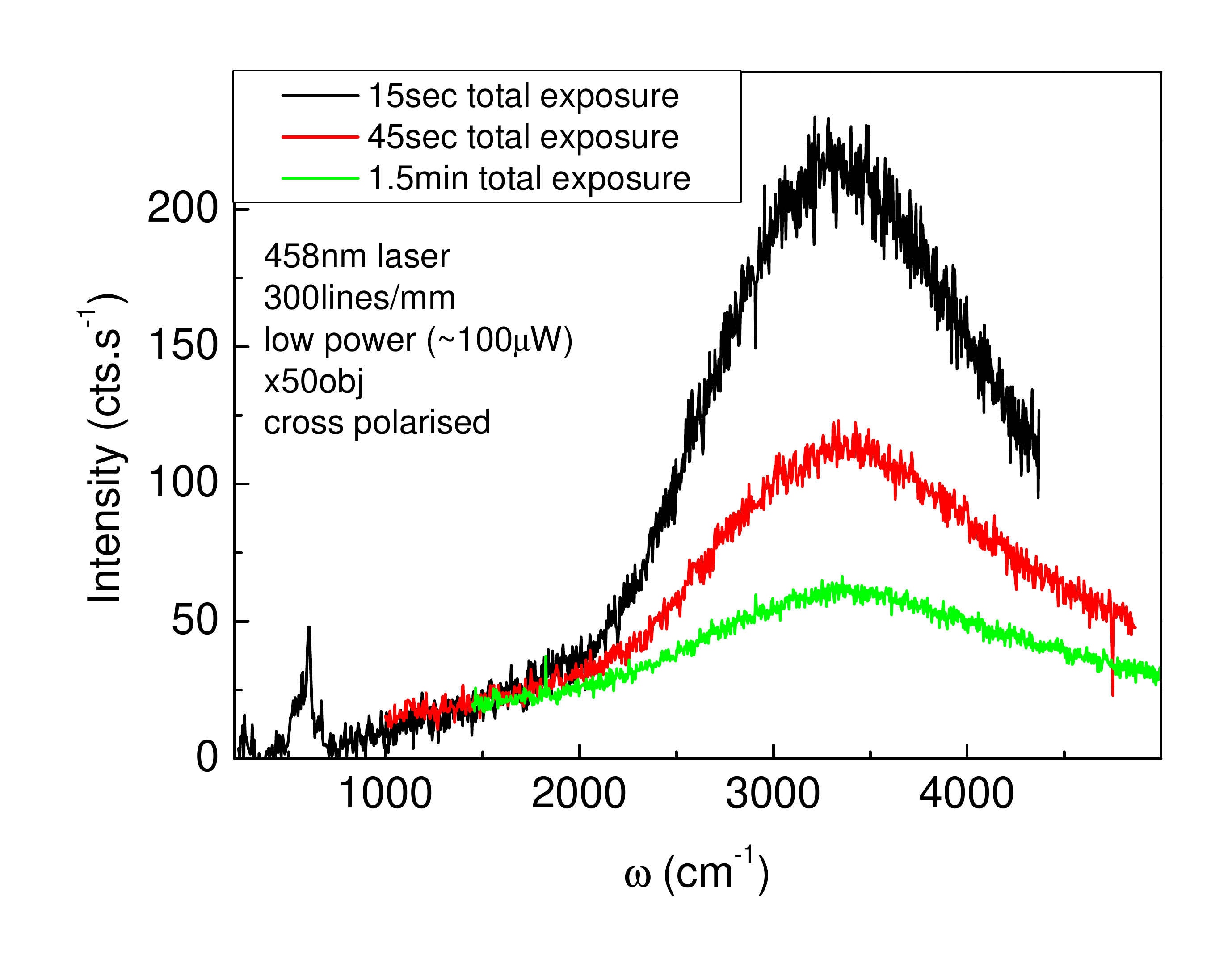}
\caption[Raw two-magnon scattering data for YBa$ _{1.5} $Sr$ _{0.5} $Cu$ _{3} $O$ _{6} $ thin films and for YSr$ _{2} $Cu$ _{3} $O$ _{6} $.]{\label{fig:yscoraw}(Left) Raw Raman scattering for epitaxial thin films of YBa$_{1.5}$Sr$_{0.5}$Cu$_3$O$_6$ made using the MOD-spin coating method on RABiTs substrates, \refsec~\ref{sec:modprocess}. (Right) Raw micro-Raman scattering (x50 microscope objective) from a single crystallite in a polycrystalline pellet of YSr$_{2}$Cu$_3$O$_6$. Because it is not possible to check \bog and \btg sample geometries on the single crystallite, we cannot be as confident this is true two-magnon scattering.  However, the peak does have the expected asymmetry, the characteristic disorder-mode at $\sim 600$~\cm and sharply reduces in intensity with longer exposure time - probably due to laser heating of the grain.}
\end{figure}

To push out to higher $J$ values, we repeated the Raman measurements on a $c$-axis-aligned thin film\footnote{See \refsec~\ref{sec:modprocess} for details of our thin-film synthesis process.} of YBa$_{1.5}$Sr$_{0.5}$Cu$_{3}$O$_6$ and on individual grains of polycrystalline YSr$_2$Cu$_3$O$_6$, prepared under high-pressure/high-temperature synthesis by Edi Gilioli \cite{gilioli2000}. The data is plotted by the blue diamonds in \fig~\ref{fig:jvstc}~ and the raw data can be seen in \fig~\ref{fig:yscoraw}. 

Significantly, the same anticorrelation between $T_c^{\textnormal{max}}$ and $J$ is preserved, and in this case now out to a more than 60\% increase in the value of $J$. This is a very large increase in $J$ and it is surprising indeed to not be reflected in the value of $T_c^{\textnormal{max}}$ if indeed magnetic interactions alone set the energy scale for pairing.

\label{sec:pressuretc}
Furthermore, to this plot we add data summarising the effect of external pressure on $T_c^{\textnormal{max}}$ and $J$ in YBa$_2$Cu$_3$O$_7$ (green squares: for 1 bar and 1.7, 4.5, 14.5 and 16.8 GPa).  The shift in $J$ with pressure is taken from the inset to \fig~\ref{fig:jintandextp} and the values of $T_c^{\textnormal{max}}$ at elevated pressures are from references \cite{mcelfresh1988, sadewasser2000}.  \replaced{We discuss the analysis to obtain these data points in more detail in the following paragraphs.}{ - there are many things to consider when doing analysis and these are discussed in \refappendix~\ref{sec:pressuretc}.}

Under ambient pressure \tcmax for YBa$_2$Cu$_3$O$_y$ is close to 93 K. To find \tcmax at higher pressures it is important to note that the application of pressure has two effects: (i) to raise the magnitude of \tcmax and (ii) pressure-induced charge transfer which increases the hole doping state \added{as was discussed in \refsec~\ref{sec:extpressuretheory} and summarised in \fig~\ref{fig:pict}}. Thus to determine \tcmax at elevated pressure one must investigate underdoped YBa$_2$Cu$_3$O$_y$. The closer is the system to optimal doping (on the underdoped side) the lower the pressure needed to attain optimal doping and \tcmax rises little above its ambient pressure value. But as doping is decreased, higher pressures are required to reach optimum doping and \tcmax is raised further. This applies until the 60 K plateau is attained around $p \approx  0.125$, when this pattern is broken and very much lower values of \tcmax are then encountered \cite{sadewasser2000}. Unfortunately no report has yet been presented of these systematics in small increments of doping. This is a gap that needs to be filled. However there are enough reports at several doping states to confirm the pattern. We then have the following data: ($P = 1$ bar, \tcmax= 93 K); ($P = 1.7$ GPa, \tcmax$ = 93.7$ K \cite{sadewasser2000});($P = 4.5$ GPa, \tcmax$ = 97.6$ K \cite{mori1991}); ($P = 14.5$ GPa, \tcmax$ = 107.2$ K \cite{mcelfresh1988}); and ($P = 16.8$ GPa, \tcmax$ = 107$ K \cite{sadewasser2000}).

For each of these pressures the value of $J$ is determined for YBa$_2$Cu$_3$O$_y$ from the inset to \fig~\ref{fig:jintandextp}, interpolating on a simple proportional basis. We obtain: ($P = 1$ bar, $J = 106.5$ meV); ($P = 1.7$ GPa, $J = 107.6$ meV); ($P = 4.5$ GPa, $J = 109.3$ meV); ($P = 14.5$ GPa, $J = 114.9$ meV); and ($P = 16.8$ GPa, $J = 116.5$ meV). From these, the values of \tcmax are plotted versus $J$ (with $P$ as the implicit variable) by the green squares in \fig~\ref{fig:jvstc}.

\fig~\ref{fig:jvstc} now reveals that the shift in data under external pressure is orthogonal to that found under internal pressure. This, again, reflects our central paradox and shows that the observed shifts in $J$ simply do not correlate with $T_c^{\textnormal{max}}$. 
 
This is the second main result from this chapter.  

\section{Discussion}
We have two main results to discuss; (i) external pressure and internal pressure have quantitatively similar effects on $J$ and (ii) we find $J$ to anti-correlate with \tcmax where internal pressure is the implicit variable, whereas a correlation is observed when external pressure is the implicit variable.  

There are several reasons why the observed shift of $J$ with ion size in \fig~\ref{fig:ln123b1g} is not a doping effect:
\begin{enumerate}
	\item For low oxygen concentration in the chain layer of Ln123, \(\delta \! \approx \!1\), the CuO$_2$ planes are undoped and, in fact, insensitive to small changes in the oxygen concentration \cite{tallon1995}. 
	\item Recent measurements of the pressure dependence of the thermopower of deoxygenated Y$_{1-x}$Ca$_x$Ba$_2$Cu$_3$O$_6$ show no evidence of pressure-induced charge transfer \cite{argenijevic}.  Typically when $\delta<1$ the application of pressure results in a strong decrease in the thermopower \cite{zhou1996} reflecting an increase in doping state \cite{oct}. However when the samples are near fully deoxygenated ($\delta\approx1$) the thermopower becomes pressure independent. 
	\item If $\delta\!<\!1$ then for a given \(\delta\) the doping on the CuO$_2$ planes increases with decreasing ion size in the Ln123 system \cite{samoylenkov1997}.  Given that $J$ decreases as doping is increased \cite{sugai2000, sugai2003}, this would have the effect of $J$ decreasing with decreasing ion size, the opposite to what we measure.
\end{enumerate}


Our results appear to differ from Ofer \etal \cite{ofer2006}. They correlate \tcmax with the Ne\'{e}l temperature, $ T_N $, in the (La$_{1-x}$Ca$_x$)(Ba$_{1.75-x}$La$_{0.25+x}$)Cu$_3$O$_y$ system (CLBLCO !). 
Using a model, $J\propto T_N\ln(\alpha)$, they relate \tn to $J$.  Here $\alpha$ is a measure of magnetic anisotropy where $\alpha=0$ is the 2D limit.  After performing a scaling analysis that utilises estimates of $\alpha$ from temperature-dependent local-magnetisation data, they are able to collapse the phase diagrams for $x=0.1, 0.2, 0.3, $ and $0.4$ onto a common phase diagram. From this they argue that the energy scale $J$ controls $ T_N $, the spin-glass transition temperature and $ \tcm $.  In other words, they argue $J$ and \tcmax correlate.

Their work is very interesting. However, \tn (what they directly measure) is not directly related to $J$, thus requiring a model to estimate the anisotropy as mentioned above. Moreover, this complex co-substituted sexenary system has large \added{Nuclear Quadrupole Resonance} (NQR) linewidths reflecting a high degree of disorder. It is our view that two-magnon Raman scattering in our model Ln123 system more direct and reliable in its implications.

On the other hand, several groups that have also observed $J$ does not correlate with \tcmax \cite{tassini2008, rullieralbenque2008, huckner2011, li2012feedback} although none have shown, as we have, a systematic anti-correlation between $J$ and \tcmax with internal pressure is the implicit variable.

\subsection{Interpretations}
\label{sec:twomaginterpretation}

Now there are several interpretations of the internal pressure data presented in \fig~\ref{fig:jvstc} and this possible anti-correlation between \tcmax and $J$;
\begin{enumerate}
  \item \label{sec:jvsdoping} We must first note that we are of course comparing two different doping states. $J$ measured at $p=0$ and \tcmax at optimal doping (possibly $p=0.16$).  Comprehensive studies of Sugai \etal \cite{sugai2003} show a roughly linear decrease of $J$ with doping as well as a suppression of two-magnon scattering intensity. Perhaps the doping dependence of $J$ for La123 is stronger than say for Lu123 so that, at between zero doping and optimal doping they have the same $J$ and by optimal doping the ordering of $J$ has reversed so that \tcmax indeed increases with $J$.  This hypothesis is sketched immediately below;  See \refsec~\ref{sec:jvsdoping} for details of our measurements investigating this possibility.
\begin{center}
\includegraphics[width=0.22\textwidth]{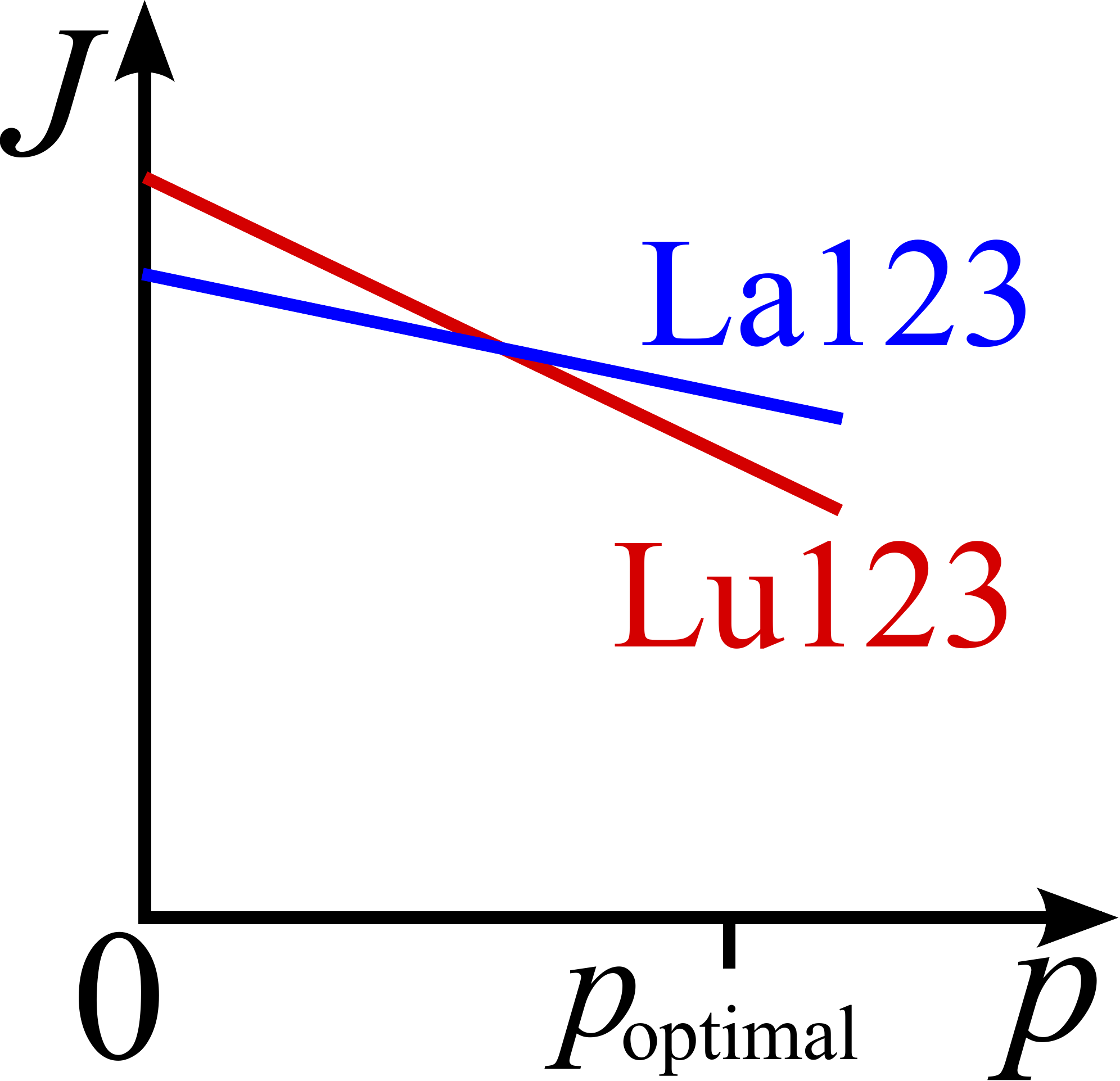}
\end{center}

On one level, if $J$ is primarily dependent on the superexchange path-length, as evidence suggests \refsec~\ref{sec:jintandextp} and \cite{chandramouli2003}, it is unlikely that this is the case.  There is no such `cross-over' behaviour in the superexchange path-length with doping in Ln123 (see \refsec~\ref{sec:ln123systematics}). 
Nevertheless this is an important consideration so we set out to track the doping dependence of the two-magnon scattering in Nd123, Eu123 and Yb123 single crystals.  

Our results are unexpected.  Representative raw spectra showing possible two-magnon scattering at different doping levels is shown in the left panel of \fig~\ref{fig:jvspnd123} for Nd123.  The possible two-magnon peak at finite doping does not uniformly lower in energy as expected nor does its intensity dramatically weaken at higher doping levels (as indicated by \tc).  Furthermore the profile of the peak at finite doping is different to the undoped peak profile. 

\begin{figure}[htb]
	\centering
		\includegraphics[width=0.45\textwidth]{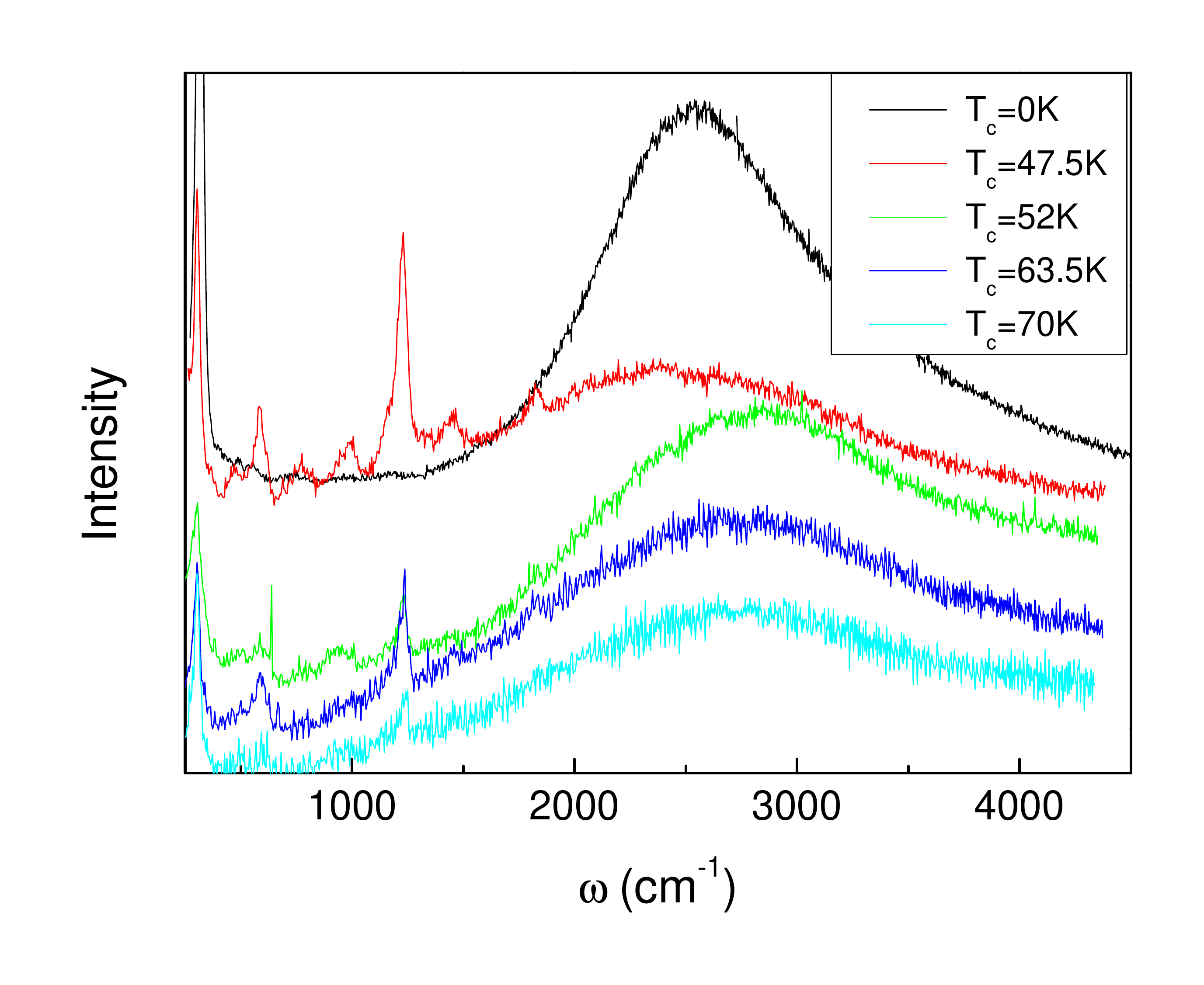}
		\includegraphics[width=0.45\textwidth]{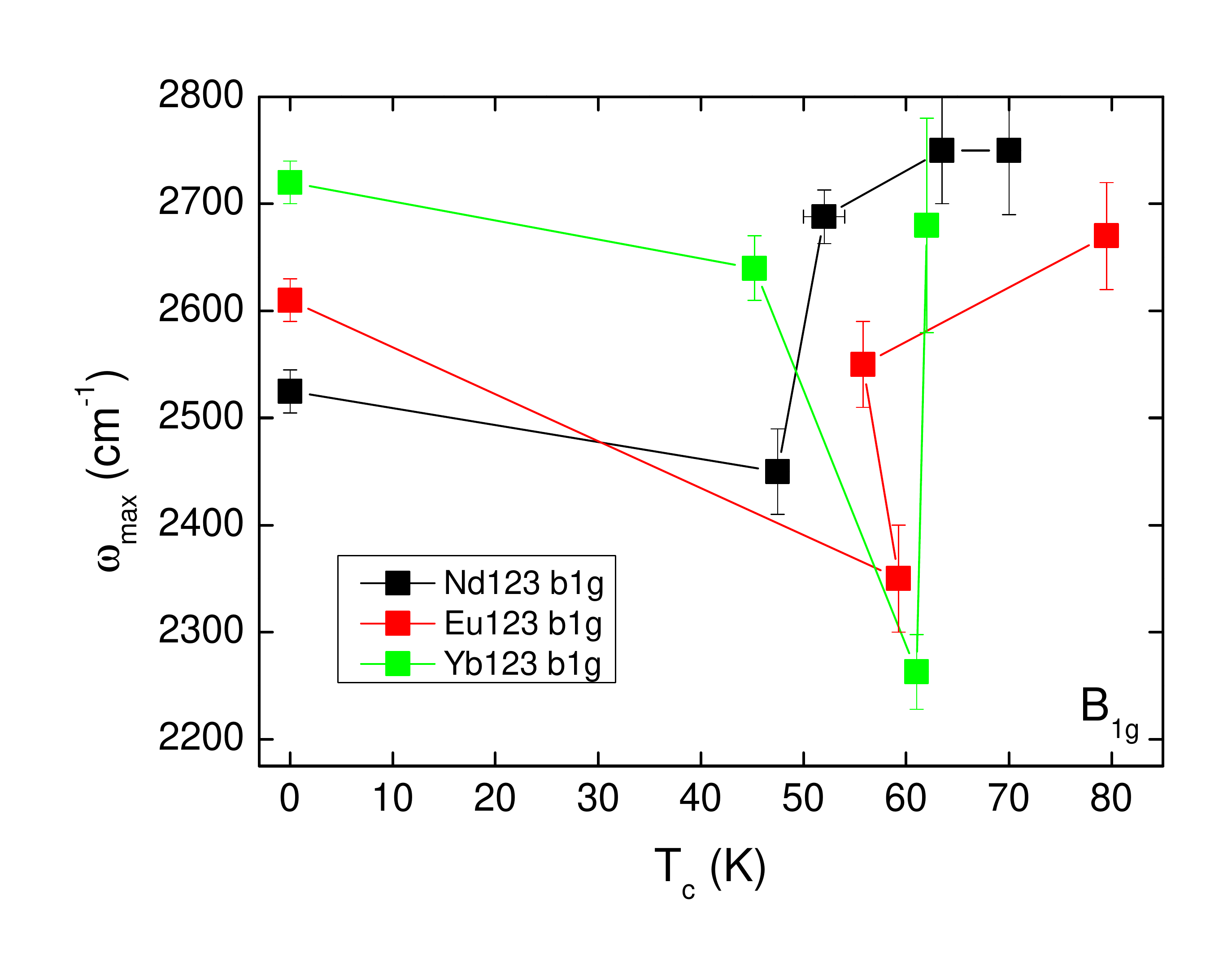}
	\caption[Possible doping dependence of $J$ for Ln123 with Ln=Nd, Eu and Yb.]{\label{fig:jvspnd123} (Left) Doping dependent \bog Raman spectra of Nd123.  The 458~nm laser line is used for all spectra except the undoped Nd123 spectrum (top). Note that once the crystal loses tetragonal symmetry with doping (due to a preferential direction of oxygen ordering on the chains) it is now no longer pure \bog and instead the lower \ag symmetry. (Right) The doping dependence of the supposed \bog two-magnon peak positions for Nd123, Eu123 and Yb123.  It appears possible that there is a cross-over in the magnitude of $J$ with doping between Nd123 to Yb123.  However, it is also not expected that $J$ \emph{increases} with doping and together with a strangely shaped `two-magnon scattering peak' shown in the left panel, suggests these rather erratic data may not be actual two-magnon scattering.}
\end{figure}

If we nevertheless plot the position of the `two-magnon' peak maximum intensity, $\omega_{\textnormal{max}}$, for \bog symmetry as a function doping for Nd123, Eu123, Yb123 we obtain the right-panel in \fig~\ref{fig:jvspnd123}. From these data, it would appear possible that there is a cross-over in the magnitude of $J$ with doping between Nd123 to Yb123, although there is considerable scatter. Because of this scatter and some issues discussed above, I am very cautious about the validity of these estimates of $ J $.  

%

However, we must also cite the recent, puzzling, Resonant Inelastic X-ray Scattering (RIXS) study of le Tacon \etal \cite{letacon2011}.  This study shows, in contradiction to the Raman results \cite{sugai2003},  paramagnons with spectral weight similar to the undoped cuprates and an essentially doping independent $J$.  To our knowledge this surprising RIXS result, and its discrepancy with the Raman results, is not well understood.  If we are to believe $J$ is in fact (essentially) doping independent from $p=0$ to $p>0.16$ - and indeed Nd123O6, Nd123O7, Y123O6.6, Y123O7 are materials used in le Tacon's study - it only further validates our anti-correlation between \tcmax and $J$ when ion-size is the implicit variable.

  \item $J$ might set the energy scale of an electronic order that competes with superconductivity.  The most likely candidate here is the pseudogap as there is evidence from inelastic neutron scattering \cite{storey2008pggroundstate} and specific heat data \cite{loram2001} that the pseudogap energy scale is set by $J$.  However, when we consider the effect of external pressure on $J$ and \tc shown in \fig~\ref{fig:jvstc}, this interpretation becomes less likely. For example, if we took the view that the suppression of \tc with ion-size was due to an enhancement of the pseudogap phase (or, similarly a Spin Density Wave or stripe phase) from an increasing $J$, we would need to explain why both \tc and $J$ increase under external pressure. 
  
  \item Perhaps $J$ does set energy scale for $\omega_{B}$, but other physical properties are affected by ion substitution which have a larger effect on $ \tcm $.  We consider this the mostly likely interpretation of \fig~\ref{fig:jvstc}.  Specifically, we consider the electronic density of states (\refsec~\ref{sec:lts}) and dielectric properties (\refsec~\ref{sec:pairingmechanismdiscussion}) in the following chapters.
  
  \item Disorder due to our ion substitutions could be responsible for the suppression of \tc and so we are not comparing `clean samples'.  We consider this unlikely for two reasons. (i) La123 and Nd123 are difficult to synthesize without disorder, but have a \tcmax higher than the most pure Y123\footnote{$T_c=94.5$~K - a thin film grown by the coated-conductor team at IRL.}.  (ii) With our Bi2201 studies we found \tc correlated more closely with ion size than disorder.  Nevertheless, as disorder is an oft-cited explanation for unconventional results, to test this possibility we carried out extensive \musr experiments.  This is the topic of the next chapter.
\end{enumerate}

We note that \bog two-magnon scattering only probes nearest-neighbour magnetic interactions \cite{singh1989} while recent RIXS studies \cite{guarise2010} reveal the presence of extended interactions involving next-nearest-neighbour, and higher, hopping integrals, $t'$  and $t''$ \cite{pavarini2001}. The associated additional exchange interaction is about half the magnitude of the changes that we have imposed by ion-size variation. It is remotely possible that inclusion of extended interactions might reverse the systematic anti-correlation between \tcmax and $J$ shown in \fig~\ref{fig:jvstc}. However, it is our view that variations in $t'$ and $t''$  will have a stronger influence via \dos by distorting the Fermi surface and shifting the vHs, presumably in a similar way to that found in the ACuO$_2$ system we studied using DFT.

\subsection{Foray into the literature for low-\tc superconductors}
\label{sec:lts}
Given the anti-correlation between $J$ and \tcmax across a wide range in our model cuprate system where internal pressure is the implicit variable, it is interesting to compare with the analogous energy scales in the low-temperature superconductors: the Debye temperature\footnote{Superconductivity in these materials is described by the BCS(+Eliashberg) theory with phonon-mediated pairing.} \cite{mcmillan1968, surma1983}, $\Theta_D$, and \tc. In the McMillan formula  \tc and the Debye temperature are related by \cite{mcmillan1968}

\begin{equation}
T_c=\frac{\Theta_D}{1.45} \exp \left[ - \frac{1.04(1+\lambda)}{\lambda-\mu^*(1+0.62\lambda)} \right]
\label{eq:mcmillan}
\end{equation}

\noindent $\mu^*$ accounts the screened, or `renormalised', Coulomb repulsion between electrons\footnote{A value difficult to experimentally measure\ldots}.  \newline
\noindent $\lambda=2\int^{\infty}_0{\Omega^{-1}\alpha^2F(\Omega)\dd\Omega}$ is the electron-phonon coupling constant which is the integral of the phonon density of states, $F(\Omega)$, weighted by the square of the electron-phonon coupling matrix, $\alpha$.  There are various modifications to \eq~\ref{eq:mcmillan} found in the literature \cite{surma1983}.

\begin{figure}
	\centering
		\includegraphics[width=0.5\textwidth]{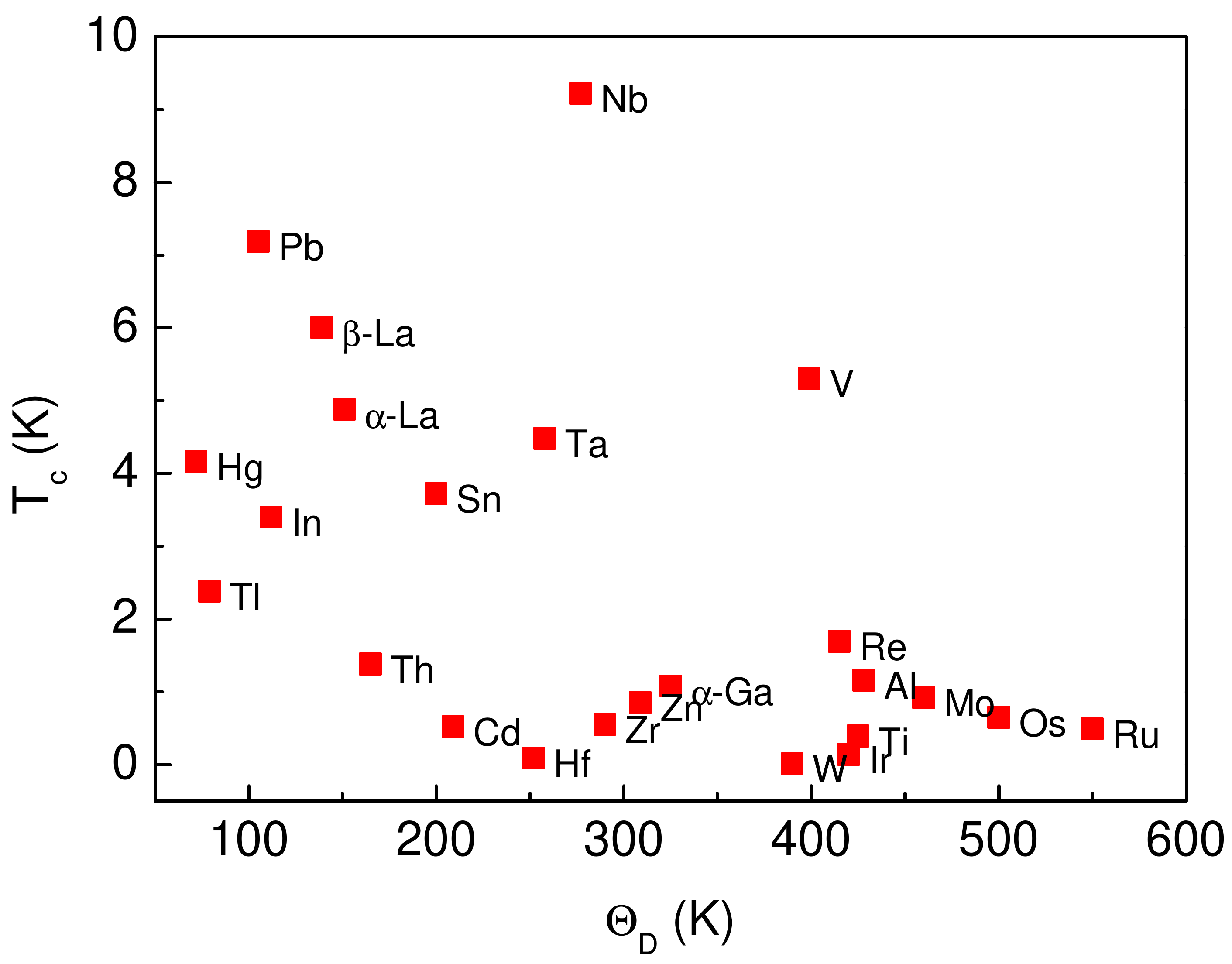}
		\includegraphics[width=0.5\textwidth]{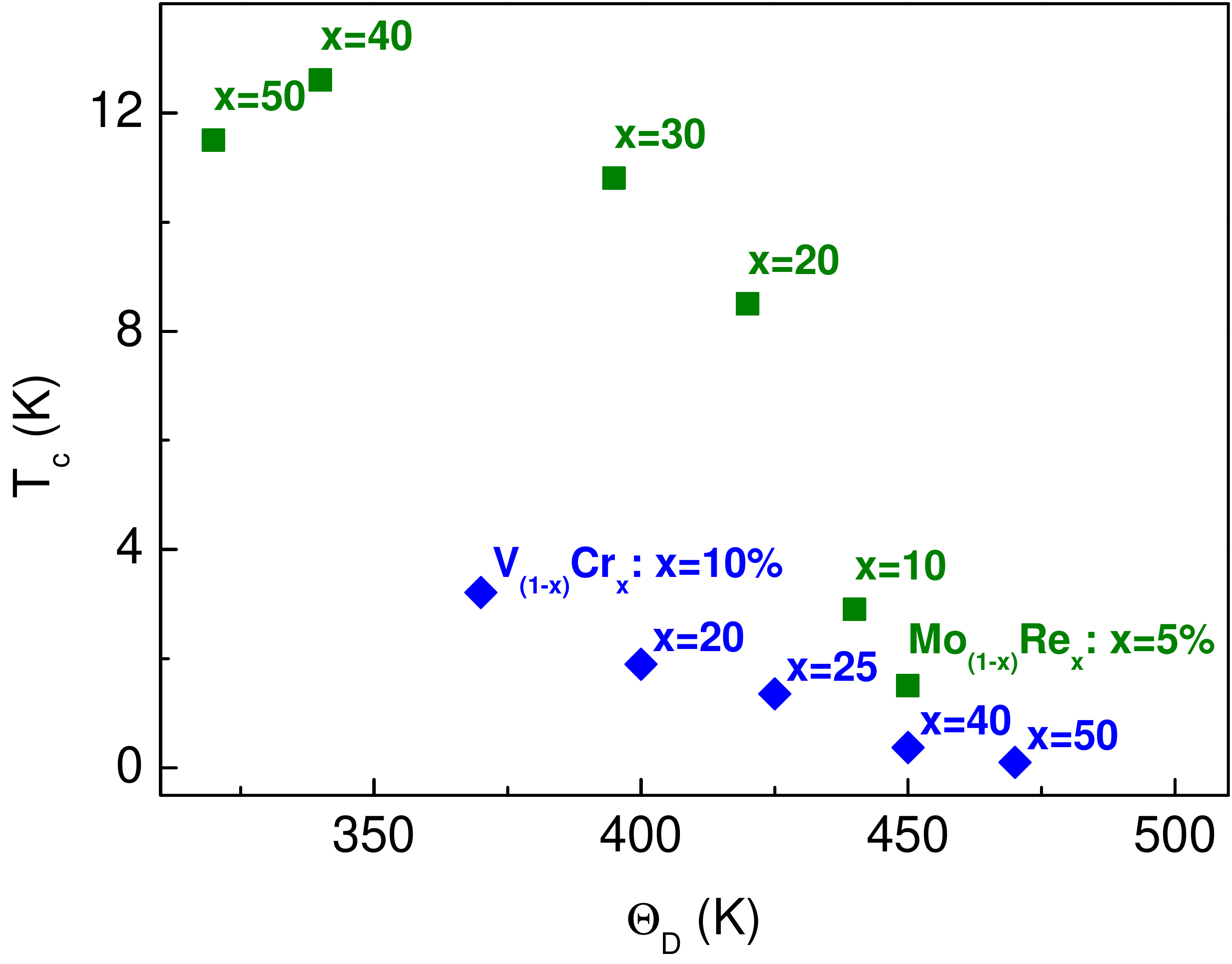}
	\caption[\tc plotted against the Debye temperature for superconducting elements and binary alloys.]{\label{fig:tcdebye} \tc plotted against $\Theta_D$ for (Top) the elements known to be superconducting at ambient pressure (in fact, the second lightest element known to be superconducting at ambient pressure, Be, with $\Theta_D = 1390$~K and $T_c = 0.026$~K, is not plotted) and (Bottom) alloy systems.  The Debye temperature is a prefactor in the McMillan formula for $ \tcm $, \eq~\ref{eq:mcmillan}, and so such a consistent anti-correlation between the two is noteworthy.  The electron-phonon coupling constant is the most important additional parameter needed to describe the \tc. Data is from \cite{mcmillan1968, poolehandbook}.}
\end{figure}

The results are interesting.  As shown by \fig~\ref{fig:tcdebye}, \tc and $\Theta_D$ in general also anti-correlate\footnote{\replaced{I}{We} use the notation \tc rather than \tcmax as, in general, the \tc of LTS materials is not optimised through doping as it is in the cuprates (and pnictides).  Hence while \tcmax is the material-specific value of interest for the cuprates, \tc is the material-specific value of interest for LTS.}, even in alloy systems.

Indeed, it is the electron-phonon coupling constant, $\lambda$, that has been found the most important additional parameter needed to describe the \tc values observed \cite{surma1983} (although an anti-correlation \tc and $\Theta_D$ over such a wide range is noteworthy). 

Given \tc and $\Theta_D$ anti-correlate in LTS materials, should we be surprised that $J$ and \tcmax anticorrelate in our model LnA$_2$Cu$_3$O$_6$ system if there is a magnetic pairing mechanism?  Perhaps not, but internal and external pressure have opposite effects on $ \tcm $, while we found they have quantitatively the same effect on $J$.  If we look at the analogous situation for some LTS materials we do not find this opposite behaviour.  For example, turning to the Mo-Re alloy system: $\frac{\partial T_c}{\partial P} < 0$ \cite{smith1975} while we can estimate the shift in Debye temperature with pressure by $\frac{\partial \Theta_D}{\partial P} \approx 3\frac{\Theta_D}{B} >0$ where $B=230$ GPa is the bulk modulus.  Thus, under external pressure, $\Theta_D$ increases as $T_c$ decreases similarly to the effect of alloying (although the gradient is greater with external pressure as the implicit variable). In other words, we do not see the opposite behaviour between internal and external pressure as we do with HTS cuprates.

\section{Summary}
In summary, we have sought to understand the opposite effects of internal and external pressure on \tcmax by (i) measuring $J$ while controlling the internal-strain through isovalent ion substitution in LnA$_2$Cu$_3$O$_6$ and (ii) comparing it to data in literature for the external pressure dependence of $J$ and $ \tcmaxm $.  We find no resolution: $J$ and \tcmax anti-correlate with internal pressure as the implicit variable and correlate with external pressure as the implicit variable. It is therefore most probable some other physical property plays a more dominant role in setting the value of $ \tcmaxm $.  In the following chapters we explore further the various interpretations and consequences for the results presented above.

\section{Appendix: Interesting extensions - \aog and \btg spectra}

\label{sec:a1g}

\begin{figure}[htb]
	\centering
		\includegraphics[width=0.45\textwidth]{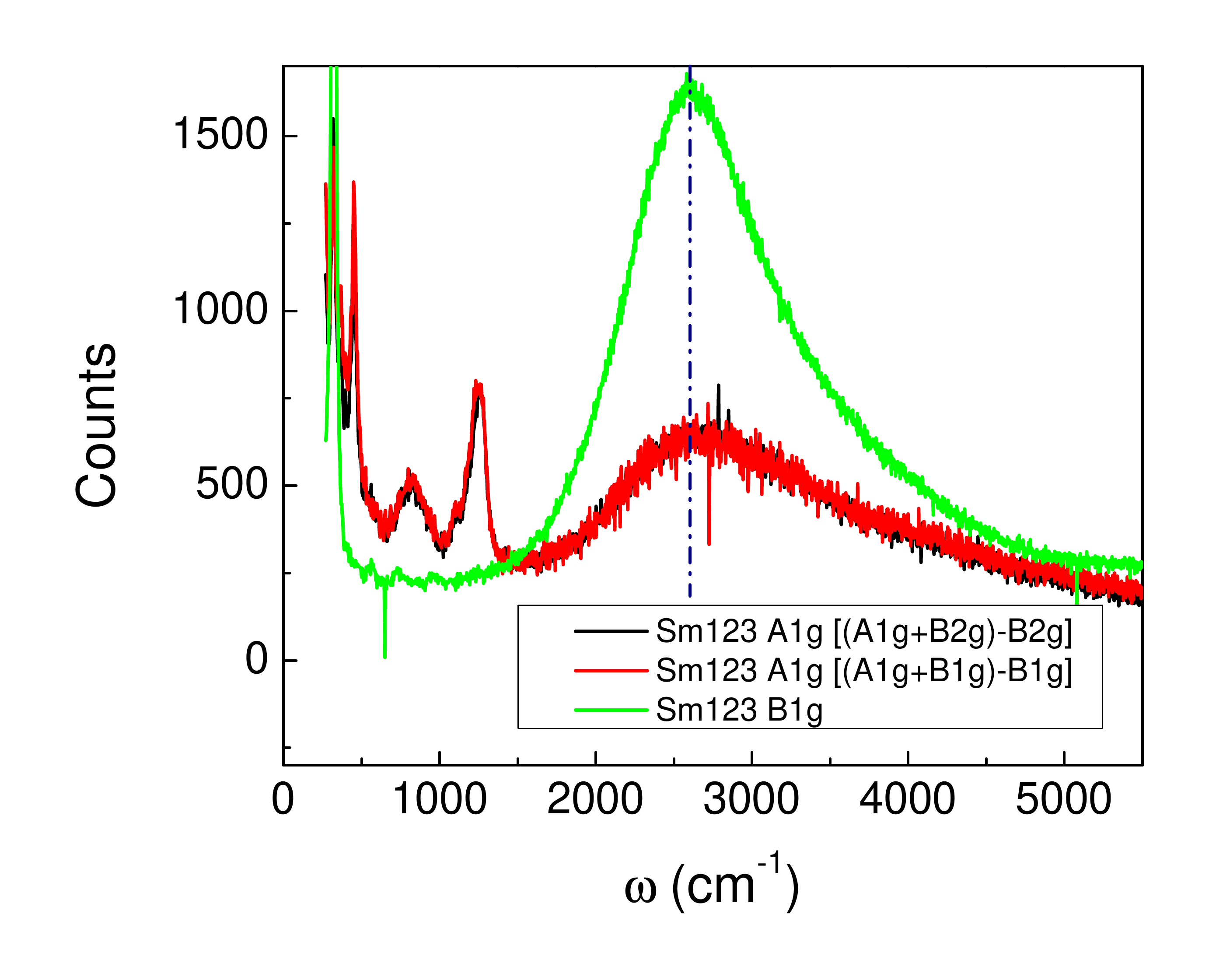}
		\includegraphics[width=0.45\textwidth]{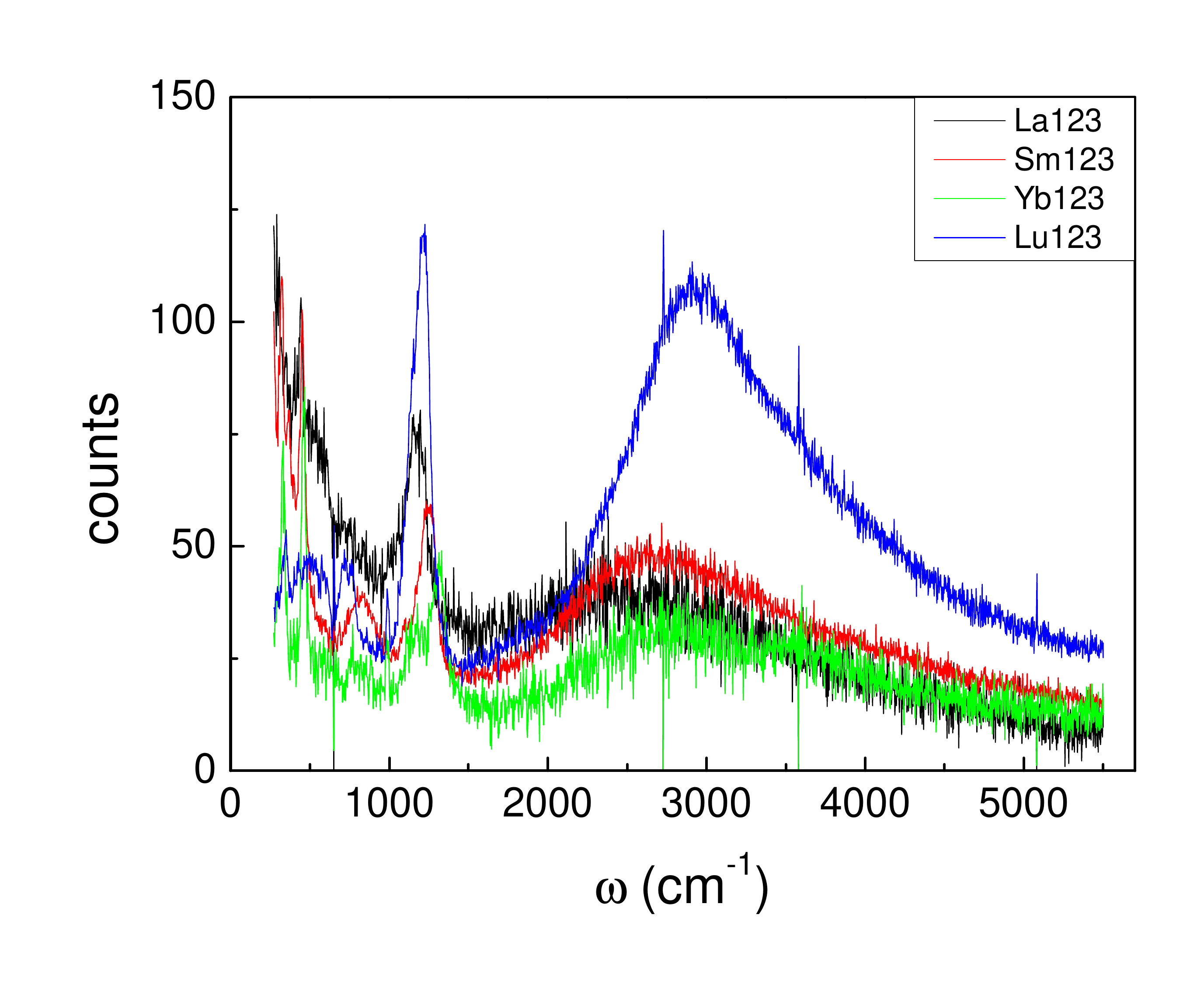}
	\caption[\aog Raman two-magnon scattering spectra of Ln123.]{\label{fig:a1g} (Left) SmBa$_2$Cu$_3$O$_6$ \aog spectra determined two different ways; firstly by subtracting \bog spectra from \aogbog spectra, secondly from subtracting the \btg spectra from the \aogbtg spectra. Note the two overlap almost perfectly - this nice overlap is seen in most, but not all of the Ln123 family.  Also shown, for comparison, is the \bog spectrum and a line marking the estimated peak maxima, \wmax (which coincides for \bog and \aog spectra). (Right) \aog spectra for the Ln123 series.}
\end{figure}

As mentioned in the theoretical introduction (\ref{sec:cupratetwomag}), the cuprates have the non-conventional feature of showing two-magnon Raman scattering in the \aog scattering geometry.  An example of this is shown in \fig~\ref{fig:a1g} for Sm123 (left-panel) and for the entire Ln123 series \added{(right-panel) where the \aog is determined by subtracting the \bog spectrum from the \aogbog spectrum or the \btg spectrum from the \aogbtg spectrum. For all spectra plotted in \fig~\ref{fig:a1g} we} find \aogbog - \bog = \aogbtg - \btg.  

\bog Raman two-magnon scattering contains information of the nearest-neighbour superexchange energy, $J$, and so it would be nice to know if something in the Raman-spectra that tells us about further-neighbour interactions. For example, perhaps \aog \added{spectra} may contain information about longer-range interactions? One published result suggests this may be possible.  It is the early work of Singh \etal \cite{singh1989} (a great read).  They argue \aog two-magnon scattering (and \btg scattering) arise from diagonal nearest-neighbour (next-nearest-neighbour) excitations (represented in the $t$-$J$ model by including the $t'$ hopping parameter).  These are Raman active due to quantum fluctuations. However, they only use their model of \aog (and $ \btgm $) scattering as a further estimate of $J$. On the other hand, an exact numerical treatment of the 2D Hubbard model by Tohyama \etal \cite{tohyama2002} finds \aog and \btg scattering to be a consequence of treating fully both spin and charge degrees of freedom. 

It may be fruitful to pursue this line of investigation further. As I understand, one takes the Hubbard Hamiltonian on a 2-D square-lattice with nearest neighbour hopping only and then project out the spin part of the Hamiltonian \cite{singh1989, chubukov1995, freitas2000}.  Doing this one gets the Heisenberg Hamiltonian which can then be solved to find the Raman response using the Loudon-Flurey Hamiltonian to describe the interaction of light with the spins.  Thus, a possible approach to this question is to add a next-nearest-neighbour interaction on our 2-D square lattice (i.e. spins along a diagonal interact via a hopping integral, usually denoted $t'$ in a Hubbard Hamiltonian). Perhaps it would then be possible to calculate what effect this has on the Raman-response from two-magnon scattering following a similar procedure?

In any case, elucidating the interplay between longer-range interactions and two-magnon scattering would be useful for studying other materials as well. 

\begin{figure}[htb]
	\centering
		\includegraphics[width=0.50\textwidth]{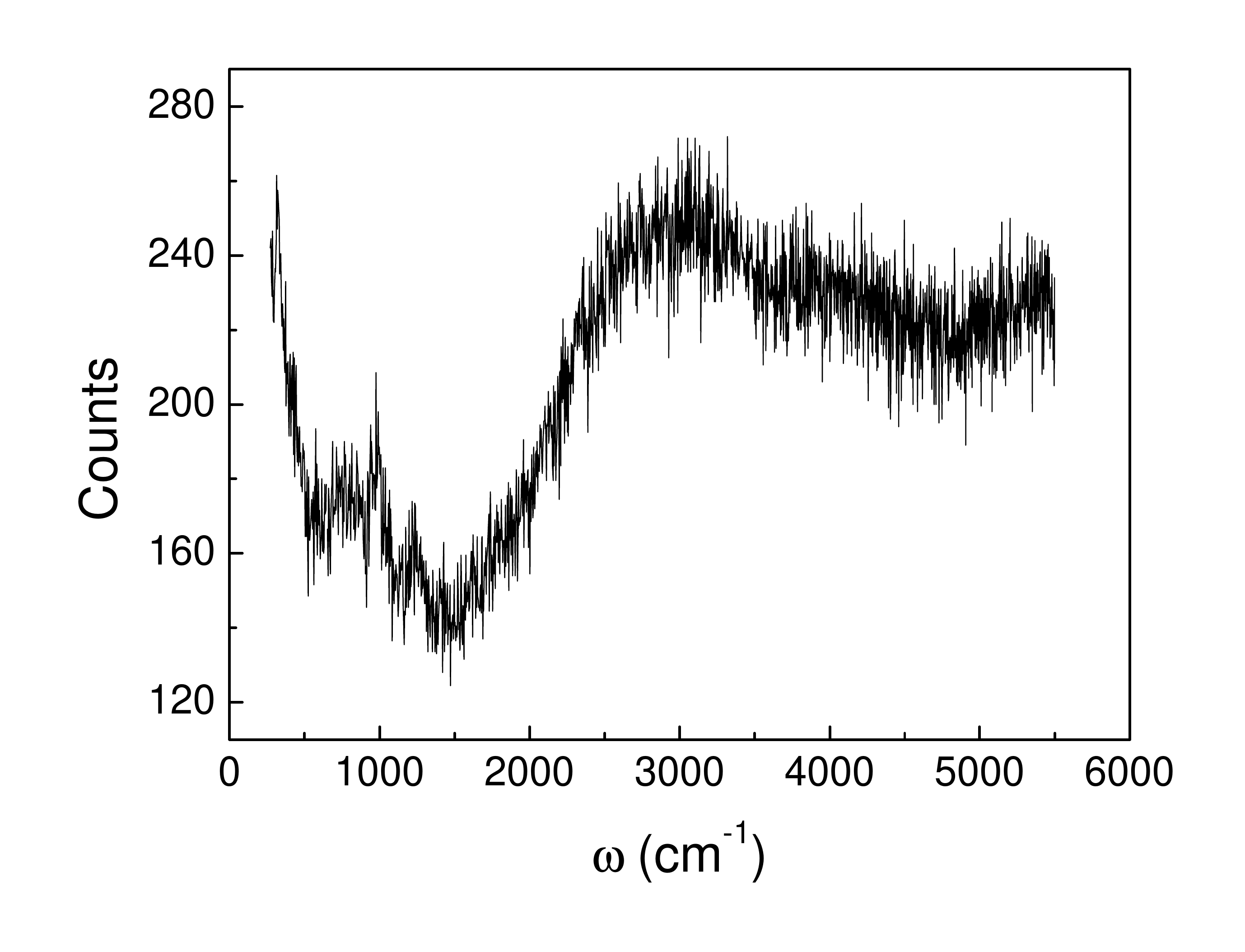}
		\includegraphics[width=0.45\textwidth]{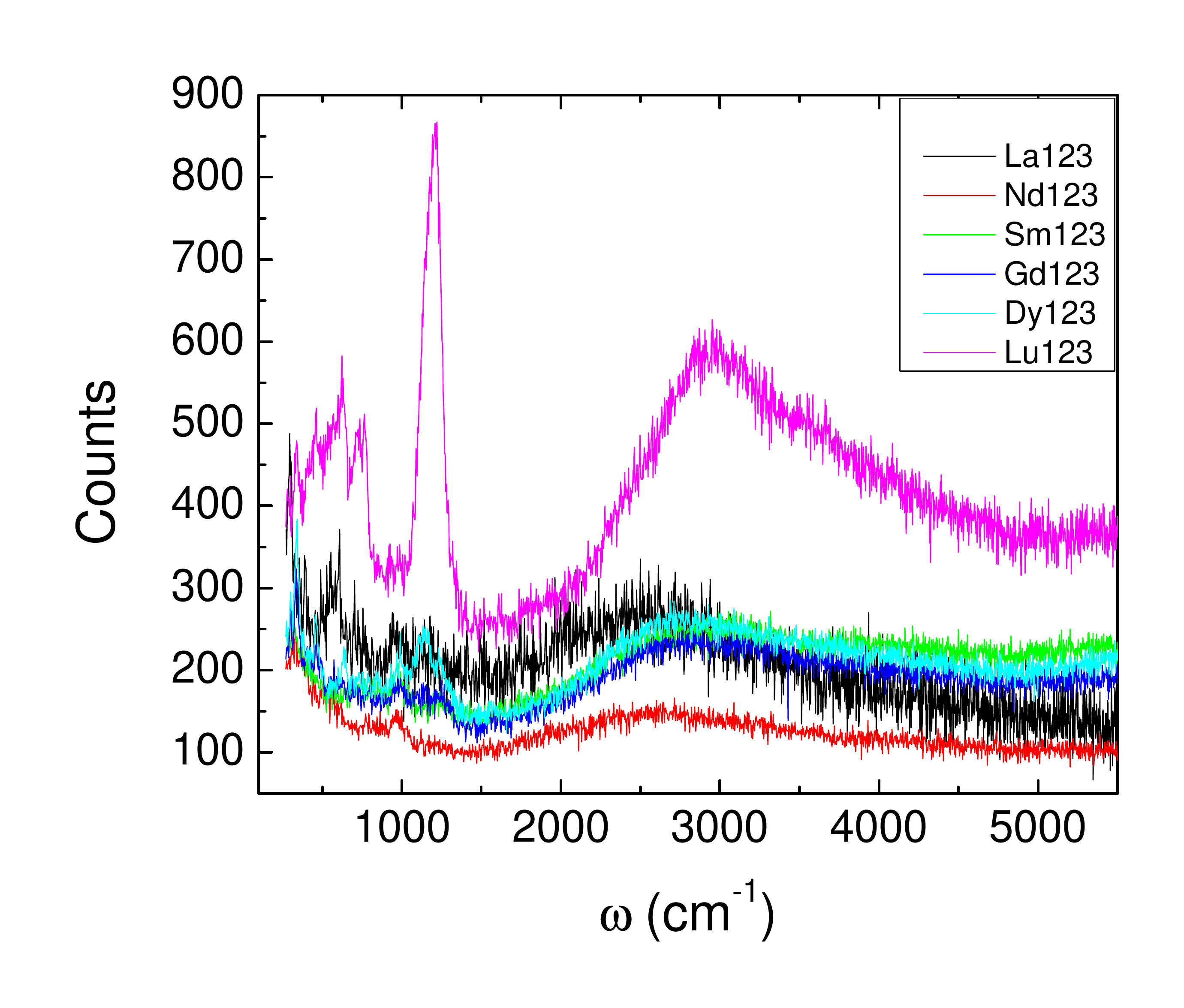}
	\caption[\btg Raman spectra of Ln123.]{\label{fig:b2g} (Left) SmBa$_2$Cu$_3$O$_6$ \btg spectrum. (Right) \btg spectra for the Ln123 series.}
\end{figure}

There is also an interesting `edge' feature in the \btg spectra at $\omega \sim 2500$~\cm which is shown in \fig~\ref{fig:b2g}. At first we suspected this was a fluorescence signal because it disappears when switching from the 458~nm to the 514.5~nm laser.  There is also no reported explanation of this feature, again leading us to believe it is not interesting. However from personal correspondence with Prof. Sugai it appears this feature may contain interesting information about the sample.  A paper on this feature may be published soon by his group\footnote{Unfortunately he did not reveal anything else!}.  We also see this `edge' feature at finite doping levels. The data in \fig~\ref{fig:b2g} show this `edge' moves to higher $\omega$ as the ion size decreases for undoped samples.  It is difficult to say what this means given the origin of this Raman feature is not known, but it certainly would be an interesting aspect to look into further.

\chapter{Muon Spin Rotation and the superfluid density}
\label{sec:musrchapter}
\label{ch:musr}
\subsubsection{Summary}

From our superfluid density measurements of \ybasr by \musr we find the nice result that this is a boring, well-behaved material!  The superfluid densities, $ \nsm $, are consistent with those previously reported for pure and Ca-doped \ybco \cite{uemura1989, tallon2003superfluid}.  These data are plotted in \fig~\ref{fig:uemura}.  Furthermore the suppression of \ns due to Zn doping is fully consistent with that found for \ybco \cite{bernhard1996} as shown in \fig~\ref{fig:sigmanormzn}.  The immediate conclusion is that disorder does not play a significant role in \ybasr as compared with Y123.  Consequently, the lower \tcmax seen in this compound, and those with lower Sr content, is a real `ion-size effect.'  Its systematic variation should therefore provide important insights into the SC mechanism and ultimately to the prediction of the magnitude of $ \tcm $.

\subsubsection{Motivation}
In such complicated, multi-ion compounds as the cuprates, the effect of disorder must always be considered. There may be deviations from perfect stoichiometry, e.g. say Y$_{0.88}$Ba$_{2.10}$Cu$_{3.2}$O$_{6.96}$ rather than YBa$_{2}$Cu$_3$O$_7$, or partial substitution of an ion, e.g. 

\noindent Y$_{0.8}$Ca$_{0.2}$Ba$_{2}$Cu$_3$O$_7$. Both are types of compositional disorder and belong to the art of the materials scientist, who may also have an eye on opportunities for flux pinning. A good real example of such is the Y$_{1-x}$Ca$_x$Ba$_{2-x}$La$_{x}$Cu$_4$O$_8$ cuprate, where the NMR line-widths broaden significantly as $x$ is increased \cite{williams1998nmr}. There may also be distortions of unit cell symmetry, bond lengths or bond angles which may be caused by ion-size mismatch. An instructive attempt to classify types of disorder in the cuprates and quantify their effects was made by Eisaki \etal \cite{eisaki2004}.  

There remains the possibility that the decreasing \tcmax with `internal pressure' in LnA$_2$Cu$_3$O$_y$ is primarily a disorder effect, as is the orthodox view of the situation for Bi2201 (Chapter~\ref{ch:bi2201}).  Alternatively, the suppression of \tc is possibly the result of the growth of a competing stripe-ordered phase e.g. because of the larger $J$ \cite{tranquada1997, wolf2004}.   We consider both scenarios  unlikely, however this can and should be tested. 

Therefore we seek to ascertain whether the suppression of \tcmax in our model LnA$_2$Cu$_3$O$_y$ system is a disorder effect.  The superfluid density is very sensitive to disorder in $d$-wave superconductors (e.g. see \fig~2 of reference \cite{bernhard1996}), as well as to stripe-order, making it an ideal quantity to measure to test this hypothesis. If \tcmax were reduced because of disorder one would see a concurrent, and larger, reduction in the superfluid density. 


%


\section{Experiment}
\label{sec:musrexperiment}
\begin{figure}[t]
	\centering
		\includegraphics[width=0.65\textwidth]{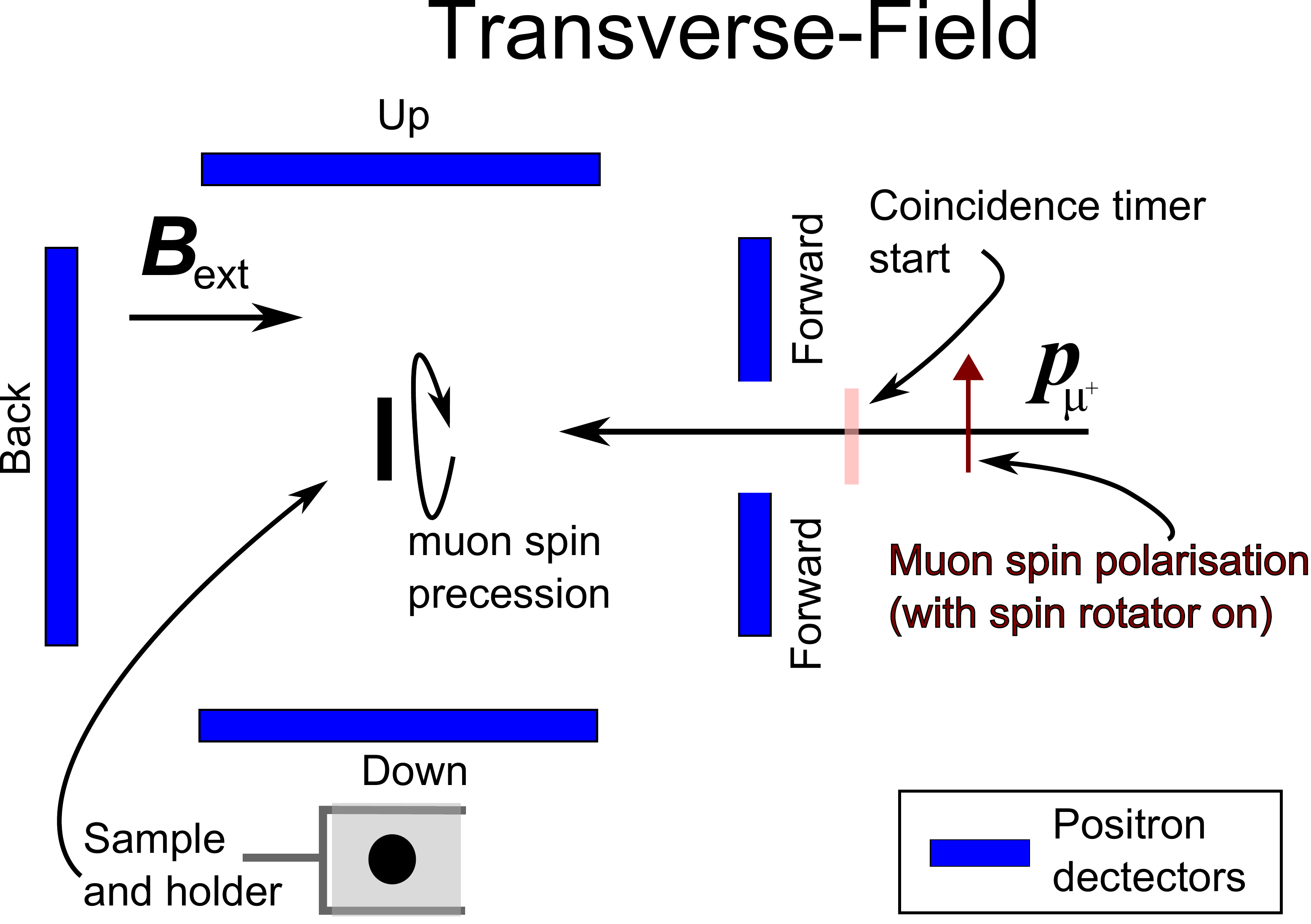}
	\caption[A schematic diagram a \musr experiment.]{\label{fig:musrsetup2} A schematic diagram of the experimental set-up at the GPS beamline (used to make our \musr measurements).  Normally, the spin of the muon is anti-parallel to its momentum, however, GPS is equipped with spin-rotating magnets which flip the muon spin near perpendicular to its momentum - as annotated. A magnetic field up to 6000 G ($0.6$ T) at the sample can then be applied, around which the muons will precess. In this geometry, to make a TF measurement we would measure the difference in the number positron detections resulting from muon decay after a given time $t$ between the Up and Down detectors.  $t=0$ is when a muon is detected passing through the coincidence timer in the Front positron detector.  At the bottom a sample (black circle) is sketched taped (light grey square) to a silver holder.}
\end{figure}

\added{Recall that the basic principles of the \musr technique were discussed in \refsec~\ref{sec:musr} and the superfluid density in \refsec~\ref{sec:sfintro}.} The experiment was performed on the GPS beam line at the PSI muon source (Villigen, Switzerland).  Because there is a muon-spin rotator magnet installed on this beam line, it is capable of both transverse and longitudinal geometries, see \fig~\ref{fig:musrsetup2} (which is a reproduction of \fig~\ref{fig:musrsetup}). 

A coincidence counting method is used.  An incoming muon starts the timer.  If a positron is detected within $10$ $\mu$s, the event and where it was detected, is recorded.  If instead a second incoming muon is detected within $10$ $\mu$s, then the next two positron detections must be discarded. Otherwise it could not be determined from which muon the detected positron came and hence how long those muons spent precessing in their $\mathbf{B}_{\textnormal{loc}}$ before decaying.  The remarkably intense GPS line has this luxurious problem and so we use shutters to reduce the rate of muons incident on the sample.  

Our samples are mounted between the prongs of a silver holder, held in place by thin tape. This is sketched on the bottom of \fig~\ref{fig:musrsetup2} where the black circle represents the sample.  The tape is thin enough that the muons have too much energy to stop inside it.  The sample is cooled by a He-flow cryostat and the temperature monitored by a nearby thermocouple.  

\begin{figure}[]
	\centering
		\includegraphics[width=0.90\textwidth]{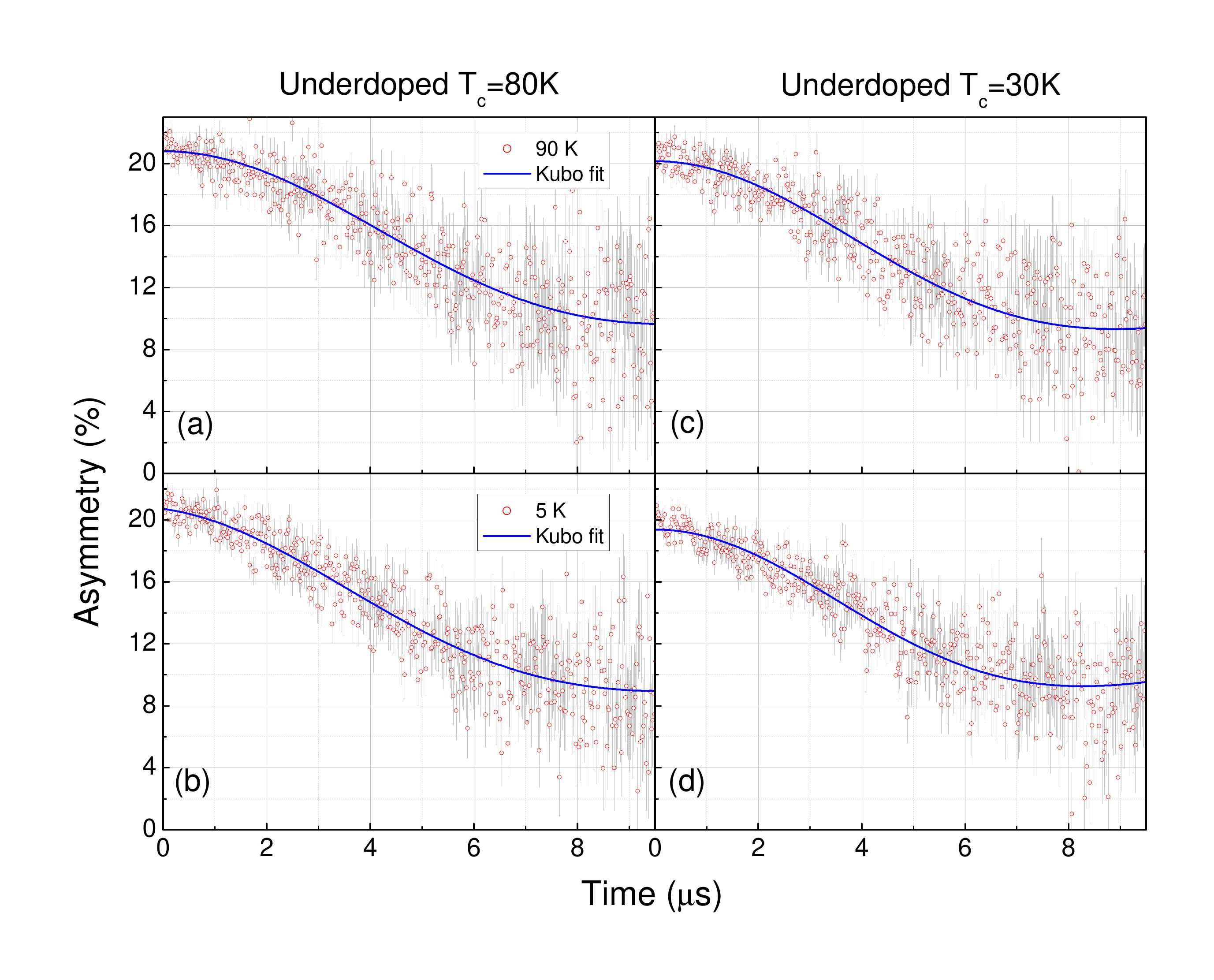}
	\caption[Typical Zero-Field \musr data.]{Typical Zero-Field \musr data (red circles with grey error bars) on YBaSrCu$ _{3} $O$ _{7-\delta} $; In the first column are data for UD80K ((a) and (b)), in the second column data for UD30K ((c) and (d)).  In the first row are data taken at 90~K, in the second row data are taken at 5~K. Each data set can be fitted well to the Kubo-Toyabe function (\eq~\ref{eq:kubo}), as shown by the blue lines. In addition, the initial asymmetry is close to the calibrated value, $A_0\approx 21$\%, and there is little change between ZF data at $T=5$~K and $T=90$~K. We conclude that these samples are do not have significant internal magnetic fields.}
	\label{fig:kubo}
\end{figure}

In zero-field (ZF), relaxation of the muon polarisation results purely from local magnetic fields. Omni-present nuclear magnetic moments are generally weak and isotropic in which case a ZF asymmetry plot is well described by the Kubo-Toyabe function, \eq~\ref{eq:kubo} \cite{kubo1981}.  An example of such is shown in \fig~\ref{fig:kubo}.  We perform zero-field \musr measurements to determine if there is additional static, or dynamic\footnote{With respect to the time scale relative to the muon - which is between $\sim$ns at GPS (instrumental resolution limited) and $\sim 10$~$\mu$s.}, magnetism in our samples.  The signature of such magnetism is an initial exponential suppression of the asymmetry, or it may be so rapid (large field distribution, fast dynamical processes) that it is instead inferred by a low initial asymmetry, which should be $A_0\approx0.21$ from our calibration on Ag foil.  ZF measurements are performed in longitudinal geometry which at the GPS beamline means measuring with the up/down detectors if the muon spin rotator is on \fig~\ref{fig:musrsetup2}.

Generally, we fit TF data using a single-frequency cosine oscillation modulated by a simple Gaussian dephasing, \eq~\ref{eq:gaussian}, discussed in \refsec~\ref{sec:musr}.  The fits are excellent and have typically cumulated error of $\chi^2\approx 1.0$.

As will be discussed below, the YSCO TF data were better fitted to either a Lorentzian decay envelope,
\begin{equation}
P_X(t)=A \exp \left( - \Lambda t \right) \cos(\gamma_{\mu}\BB_{\textnormal{loc}}t+\delta)
\end{equation}

\noindent or to the sum of two Gaussian components.

To perform the fits we use the freely-available WiMDA software package written by Francis Pratt \cite{wimda}.

\subsection{Materials studied}
We plan to study NdBa$_2$Cu$_3$O$_y$ (Nd123), \ybasr and \ysco. These, along with previous measurements on YBa$_2$Cu$_3$O$_y$ \cite{tallon1995, bernhard1995, bernhard1996, niedermayer1998}, span the internal pressure range in our model Ln(Ba,Sr)$_2$Cu$_3$O$_y$ system.  With our beam-time allotment we had time for \musr measurements of six \ybasr samples and three \ysco samples.  They are listed, along with a short reference code, in \tab~\ref{tab:musrsamples}.   

The polycrystalline \ysco samples were grown by our collaborator Dr. Gilioli at IMEM in Parma, Italy. Dr. Gilioli has refined the technique of \ysco synthesis which involves specialised high-pressure, high-temperature equipment \cite{gilioli2000}.  The first sample was an as-prepared, overdoped $T_c=60$~K sample and is labelled Edi35. The second, with $T_c=66$~K, had been slightly de-oxygenated by annealing in 10~bar O$_2$ while slow cooling from 400\degc to 300\degc and is labelled Edi22. Unfortunately, the \ysco were found to contain a considerable fraction of magnetic impurities (see the ZF measurements section \ref{sec:zfmusrysco}).  In an attempt to dissolve the KCl out of the pellet, which we thought may be the magnetic impurity, we ground pellets of the $T_c=66$~K sample (Edi22) to a powder in dry Methanol (KCl is soluble in Methanol).  We then poured the mixture onto ``Kim Wipe'' tissue paper, flushed with more Methanol, and then evaporated excess Methanol with a hair drier.  The \ysco recovered from this process would have spent 10 minutes in Methanol.  The powder, which is named \textbf{Edi22w}, was then immediately re-measured with muons.

Unfortunately, there was insufficient time to take measurements on Nd123.  Instead Nd123 and further \ysco measurements are the subject of a new \musr proposal, now awarded at the time of writing.

\begin{table}
	\centering
		\begin{tabular}{lclll}  \toprule
		 Code & Material formula & \tc(K) & $\rttep$($\mu$V.K$^{-1}$) & $p$ \\ \midrule
		 OD80K & YBaSrCu$_3$O$_{7}$ & 80 & -1.0 &  0.184 \\ 
		 OP84K & YBaSrCu$_3$O$_{6.91}$ & 83.5 & 1.5 &  0.160 \\ 
		 UD80K & YBaSrCu$_3$O$_{6.88}$ & 80 & 2.8 &  0.135 \\ 
		 UD30K & YBaSrCu$_3$O$_{6.67}$ & 30 & 20.0 &  0.075 \\ 
		 1\% Zn OP77K & YBaSrCu$_{2.98}$Zn$_{0.02}$O$_{6.98}$ & 77 & -0.55 &  0.167 \\ 
		 3\% Zn OD61K & YBaSrCu$_{2.94}$Zn$_{0.06}$O$_{6.99}$ & 61 & -0.50 &  0.160 \\ 
		 3\% Zn OP62K & YBaSrCu$_{2.94}$Zn$_{0.06}$O$_{6.94}$ & 62 & 1.50 &  0.160 \\ 
		 Edi35 & YSr$_2$Cu$_{3}$O$_{y}$ & 62 & -6.1 &  - \\ 
		 Edi22 & YSr$_2$Cu$_{3}$O$_{y}$ & 66 & -9.5* &  - \\  
		 Edi22w & YSr$_2$Cu$_{3}$O$_{y}$ & 64 & - &  - \\ \bottomrule
		 		  		  		 
		\end{tabular}
		\caption[Samples studied by $ \mu $SR.]{\label{tab:musrsamples} A table of the samples studied by \musr along with the code used to refer to them throughout this section. See \refsec~\ref{sec:annealingconditions} for more details on sample preparation and annealing. *We consider this measurement less reliable due to poor electrical and thermal contact between the sample and probe.}
\end{table}

\section{Results}

\subsection{Zero-field measurements}
\label{sec:zfmusr}
We perform zero-field (ZF) measurements ($\BB_{\textnormal{ext}}=0$ T) at low temperature ($T=5$ K) and high temperature ($T=90$ K) before the transverse-field measurements (TF).  These measurements are to check for additional magnetic order in our samples - the signal of which may be very small compared with the SC signal in a TF measurement. In \fig~\ref{fig:kubo} we plot $P_Z(t)$ (\eq~\ref{eq:asymm}\footnote{$Z$ is usually defined as the direction parallel to the muon spin polarisation.}) in ZF for UD80K (underdoped, $T_c=80$ K) and UD30K (heavily underdoped\footnote{This doping is only just above the spin-glass magnetic phase which forms at low temperatures and doping.}, $T_c=30$ K) \ybasr at both $T=90$ K (top panels)  and $T=5$ K (lower panels).   For high and low dopings we see very little change between ZF data at $T=5$ K and $T=90$ K.  This tells us \dbloc is the same at both temperatures, which in turn means no magnetic phase transition between these two temperatures. At 90 K \ybasr is a well-behaved paramagnet and so at 5 K in zero-field the only local magnetic fields are from, presumably, nuclear moments.  Furthermore, the initial asymmetry is high, $A_0\approx21$\%, and consistent with our calibration which means there is no rapid relaxation or dephasing of the muons from a broad field distribution or rapidly dynamic magnetic fields. 

There is small, $0.5$\%, reduction in asymmetry of the UD30K sample at 5 K compared with 90 K.  This suggests a small ($\sim2$\%) magnetic phase which rapidly dephases the muons.  In TF we also see a rapidly dephasing component, but it is gone by 8 K.

All $P_Z(t)$ in \fig~\ref{fig:kubo} for \ybasr fit well to a Kubo-Toyabe function (\eq~\ref{eq:kubo}) as shown by the solid blue lines. The extracted widths of the field distributions at 5 K are $\sigma = 0.18$ $\mu$s$^{-1}$, $0.14$ $\mu$s$^{-1}$ for underdoped and heavily-underdoped \ybasr respectively.  These dephasing rates are a consequence of the isotropic nuclear moments in the sample.

We conclude from these data that our \ybasr are free from an additional magnetically ordered phase which could have resulted from magnetic impurities or from an intrinsic property of \ybasr due to e.g. its larger super-exchange energy, $J$.  This conclusion is also supported by our bulk magnetisation measurements.

\subsubsection{YSCO}
\label{sec:zfmusrysco}
\begin{figure}[]
	\centering
		\includegraphics[width=0.85\textwidth]{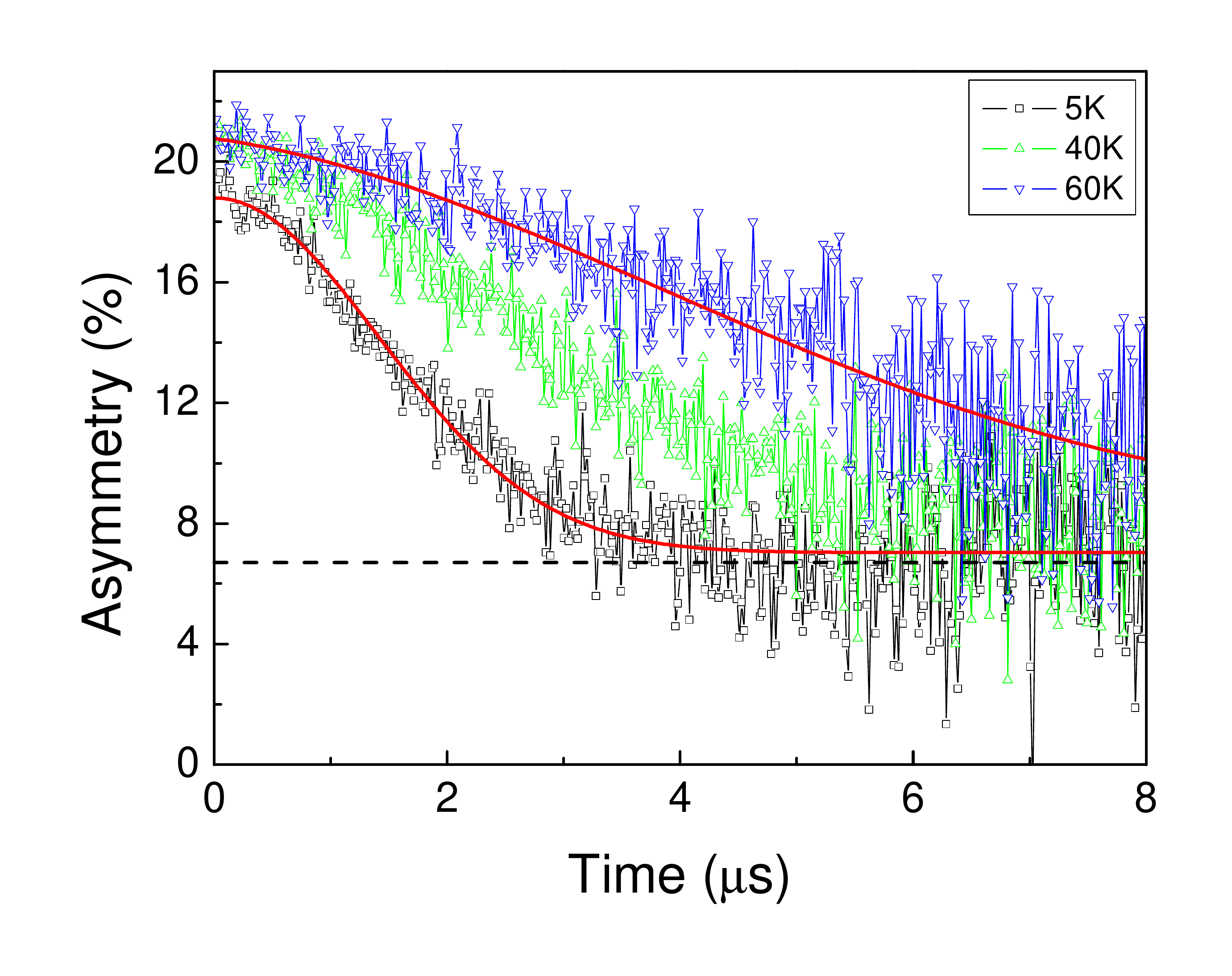}
	\caption[Representative zero-field \musr data for YSCO.]{Zero-field \musr data from Edi35 (as prepared \ysco, slightly overdoped, $T_c=62$~K) at three representative temperatures.  At $T\geq60$~K the data is well fitted by either the Kubo-Toyabe function (as shown by the solid red line for $\sigma = 0.19$~$\mu$s$^{-1}$) implying an isotropic Gaussian distribution of \bloc with $\left\langle \blocm \right\rangle=0$, or by $P_Z(t)/P_Z(0)=\exp(-\sigma^2t^2/2)+1/3$ with $\sigma=0.27$~$\mu$s$^{-1}$. For $T<20$~K however the data is more appropriately fitted by the latter expression, plotted red line, with $\sigma=0.71$~$\mu$s$^{-1}$. A dotted black line annotates 1/3 of the initial asymmetry value.  Note the lower asymmetry at $t=0$~$\mu$s at $T=5$~K.}
	\label{fig:yscozf}
\end{figure}

\added{In \fig~\ref{fig:yscozf} we show representative ZF data for our \ysco samples.} We find that the time dependence of the ZF data is more appropriately fitted to a Gaussian dephasing expression, $P_Z(t)/P_Z(0)=\exp(-\sigma^2t^2/2)+1/3$, rather than a Kubo-Toyabe function (\eq~\ref{eq:kubo}). While the distinction is not clear for $T\geq60$~K, it is clear for $T<60$~K.  Typical fitting values are $\sigma\approx 0.7$~$\mu$s$^{-1}$ (Gaussian fit) and $\sigma\approx 0.4$~$\mu$s$^{-1}$ (Kubo-Toyabe fit) for $T<40$~K, and $\sigma\approx 0.25$~$\mu$s$^{-1}$ (Gaussian fit) and $\sigma\approx 0.20$~$\mu$s$^{-1}$ (Kubo-Toyabe fit) for $T>40$~K. 

Also evident from these data is a significant `missing fraction' at low temperatures.  The initial asymmetry measured is $A_0\approx 19$\% whereas from calibration we expect $A_0=21$\%.  For Edi22w this drops to $A_0\approx16$\%. Furthermore, the initial (linear) decrease of $P_Z(t)$, most pronounced at low temperatures and for Edi22w, is not captured by the Kubo-Toyabe or Gaussian profiles. Subsequent magnetisation measurements reveal a magnetic transition at $T\approx 20$~K (the actual temperature is field dependent), in addition to the SC transition.  

In light of these complications, we quarantine the \ysco data for now and instead discuss it in its own separate section, \refsec~\ref{sec:musrysco}.

		\subsection{Transverse-field measurements}
		\label{sec:tfmusr}

Given the nice ZF data for \ybasr we expect the TF measurements to not throw up any surprises.  To conduct a TF measurement we always apply the field above \tc (typically at 90~K) so that the external magnetic field fully penetrates the sample before lowering the temperature (i.e. field-cool).  Unless otherwise stated, $\BB_{\textnormal{ext}}=0.1$~T.  

\begin{figure}
	\centering
		\includegraphics[width=0.52\textwidth]{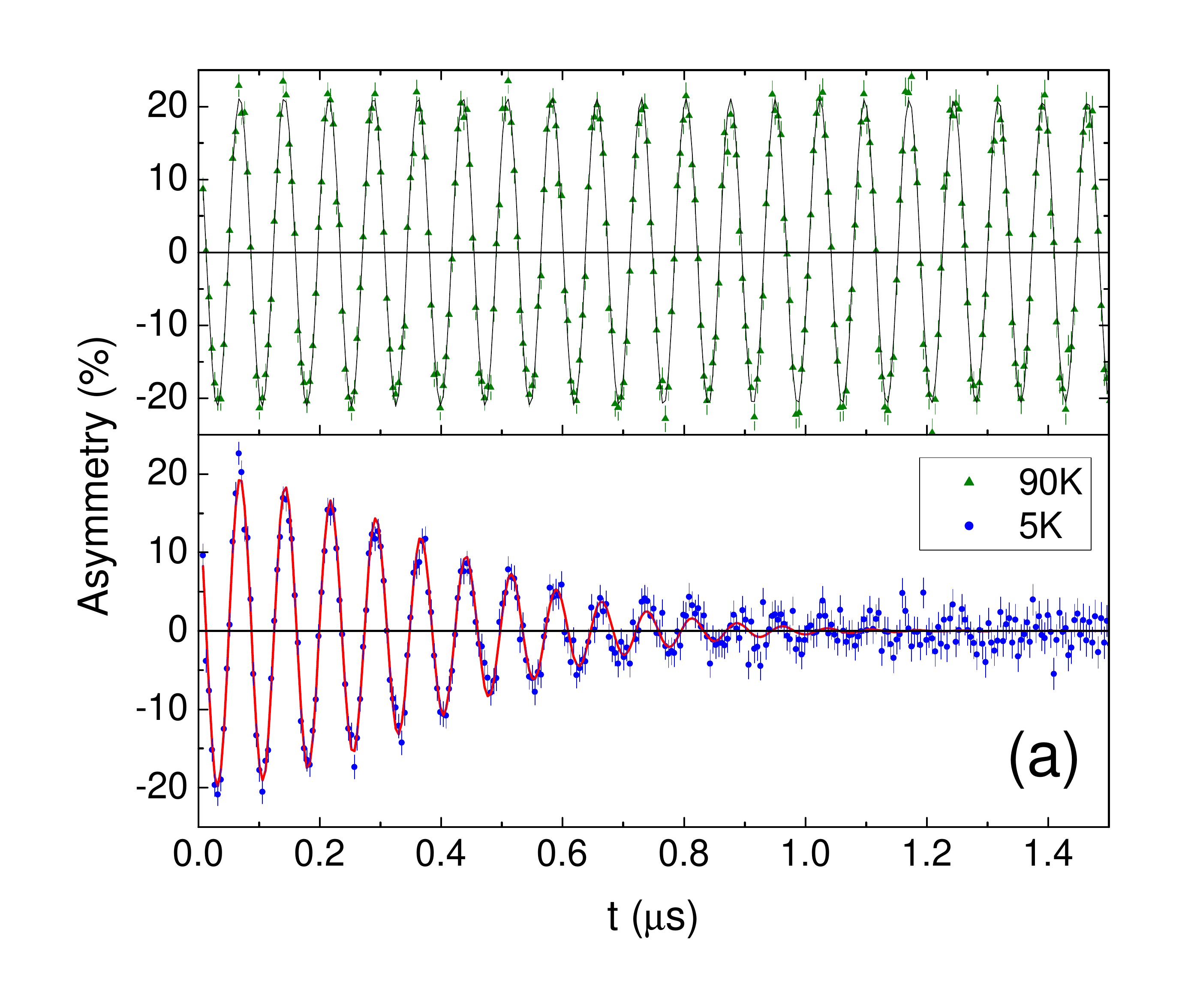} 
		\includegraphics[width=0.52\textwidth]{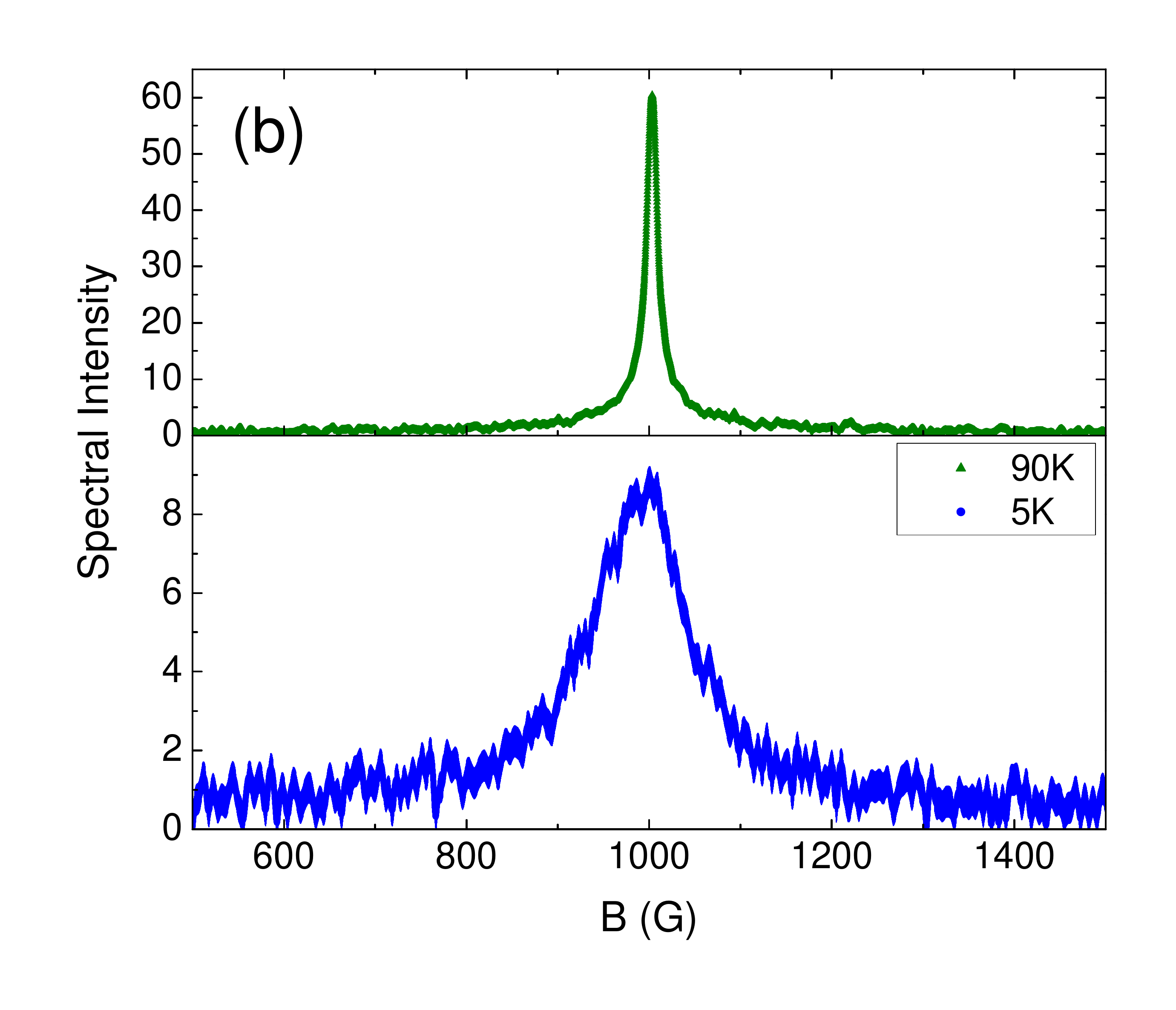} 
		\includegraphics[width=0.52\textwidth]{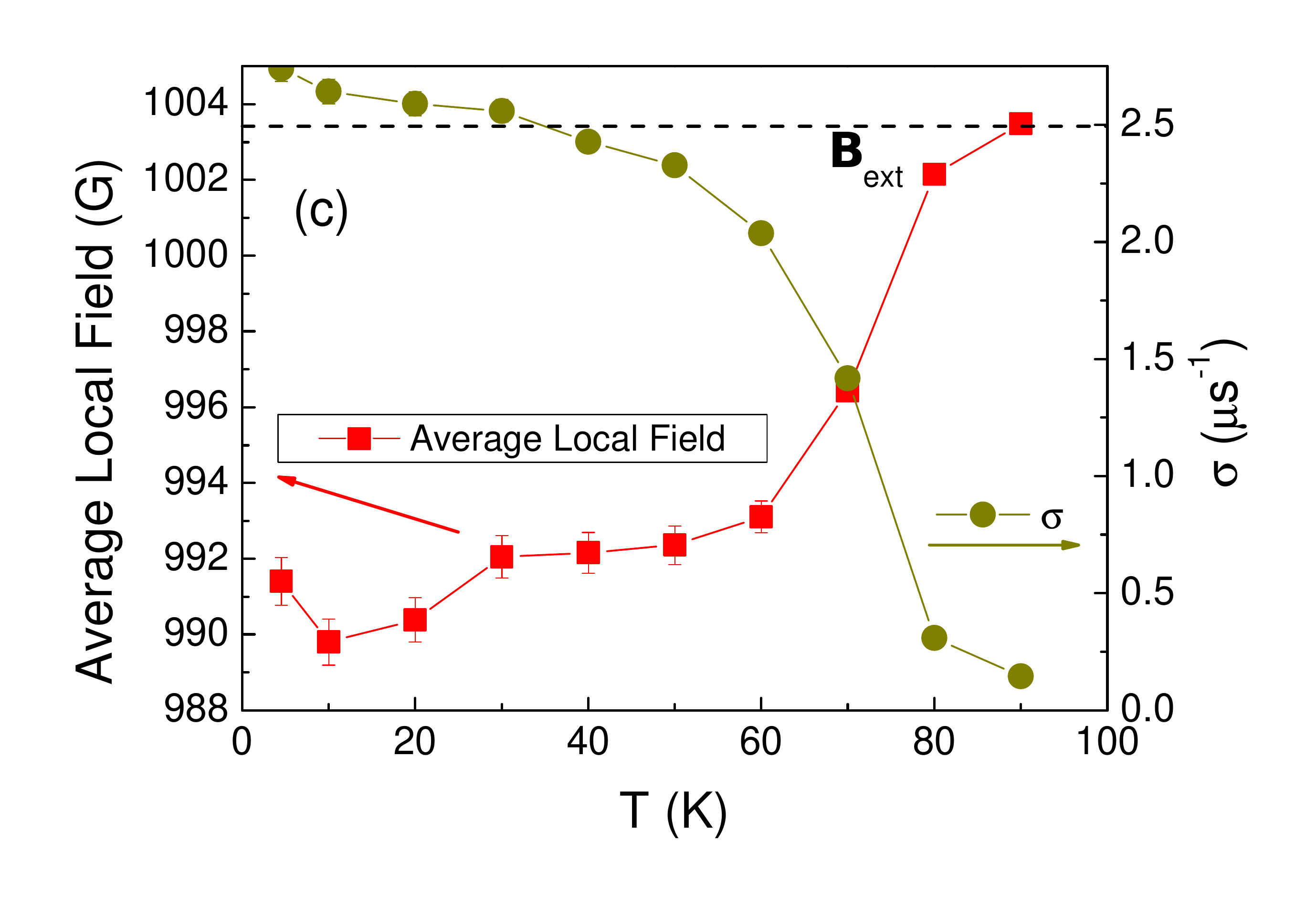}
	\caption[Representative raw data in transverse field mode for \ybasr above and below \tc.]{(a) Representative raw data in transverse field mode for \ybasr above ($T=90$ K) and below ($T=5$ K) \tc.  For clarity the data after $t=2$~$\mu$s is not shown. (b) Amplitude of the components of the Fourier transform of the data shown in (a), i.e. the local magnetic field distribution. (c) The temperature dependence of the average local field, $\left\langle \blocm \right\rangle$ - a result of diamagnetic screening of the external field.  Also shown for comparison is the temperature dependence of $\sigma$ as defined in \eq~\ref{eq:gaussian}.}
	\label{fig:tfraw}
\end{figure}

In \fig~\ref{fig:tfraw}a we plot representative TF-\musr data for \ybasr above and below $\tcm$.  Above \tc one sees the precession of the muon in \bloc=\bext with only a small dephasing due to inhomogeneity in $ \blocm $.  This is shown by the sinusoidal oscillation of $P_X(t)$ with a small decrease in amplitude over time.  Below \tc one sees the precession of the muon spin in \bloc < \bext with the amplitude of the oscillation decreasing as $\exp (-\sigma^2 t^2/2)$.  

\dbloc is determined by taking the Fourier transform of $P_X(t)$ and is plotted in \fig~\ref{fig:tfraw}b. The centre of the peak at $T=5$ K is $\blocm = 0.992$~T. As expected from the good fits of $P_X(t)$, we consistently see a nice Gaussian distribution of \bloc when looking at the Fourier transform.  \fig~\ref{fig:tfraw}c shows the mean of \dbloc is less than \bext below \tc and this is because of diamagnetic screening of \bext in the SC state.  

If there were also a non-SC, paramagnetic volume fraction within our sample\footnote{Recall that the ZF measurements rule out a magnetically ordered inclusions.} we would see remnant sinusoidal oscillation which would relax at the much slower rate observed in the $T>T_c$ TF data. Within uncertainties (which could be reduced by better statistics - i.e. longer counting times) we do not see an additional paramagnetic volume fraction.  Furthermore the initial asymmetry, $P_X(t=0)$ is close to $0.21$. Thus we conclude that our samples have close to 100\% superconducting volume fraction.

To make the discussion above more quantitative we now discuss the fits of the \\ \ybasr data to a Gaussian `relaxation' function \eq~\ref{eq:gaussian}.  Over all temperature ranges and for all \ybasr samples the fits are excellent. The `relaxation'\footnote{Technically it is a dephasing of the muon spins due to the vortex-lattice-induced field distributions rather than a relaxation, which would be caused by time-varying local magnetism.  This is being pedantic though and $\sigma$ is often happily called a relaxation rate anyway.} rate $\sigma$ is proportional to the field distribution and the superfluid density by \eq~\ref{eq:sfdensity}.  

\begin{figure}
	\centering
		\includegraphics[width=0.635\textwidth]{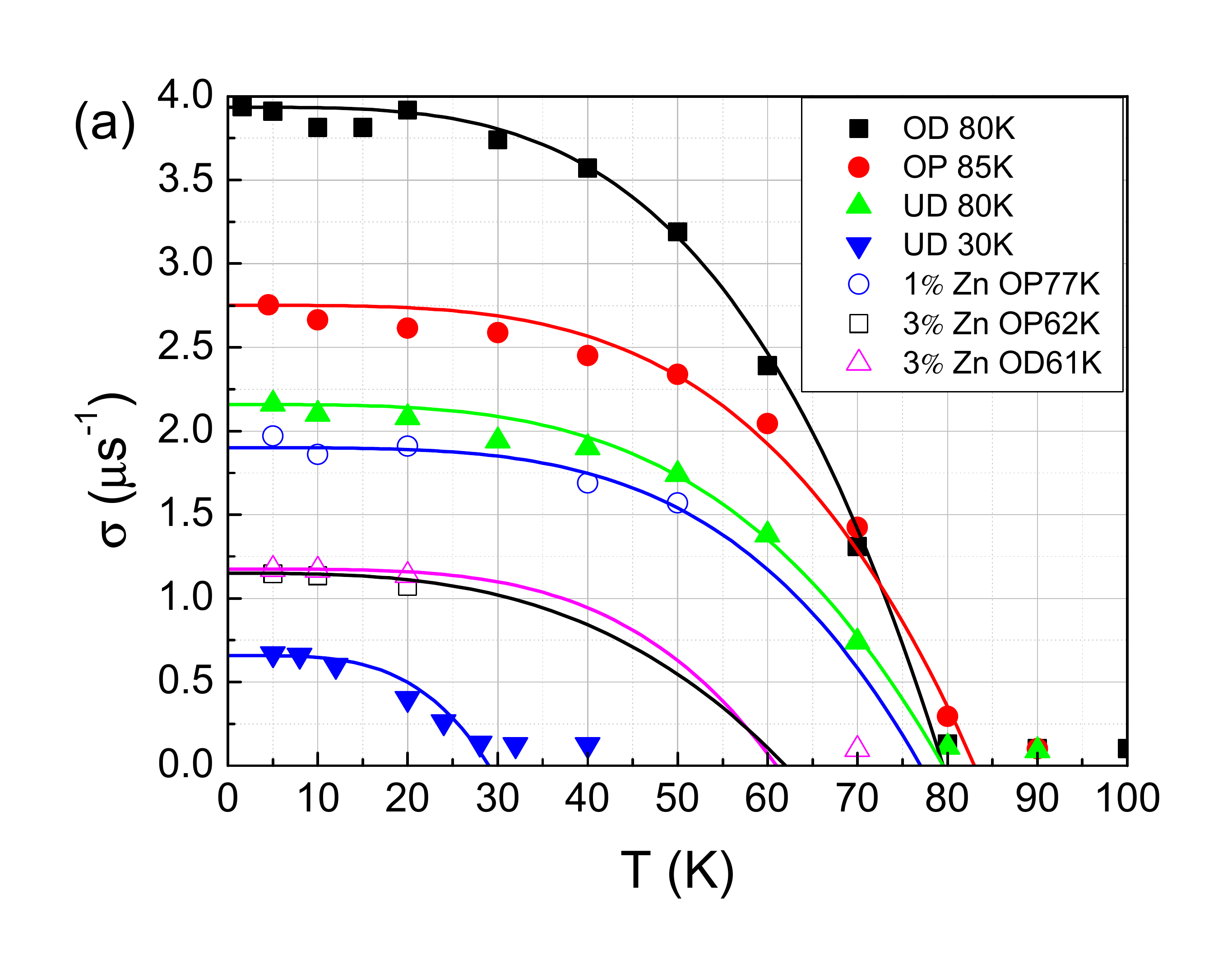} 
		\includegraphics[width=0.635\textwidth]{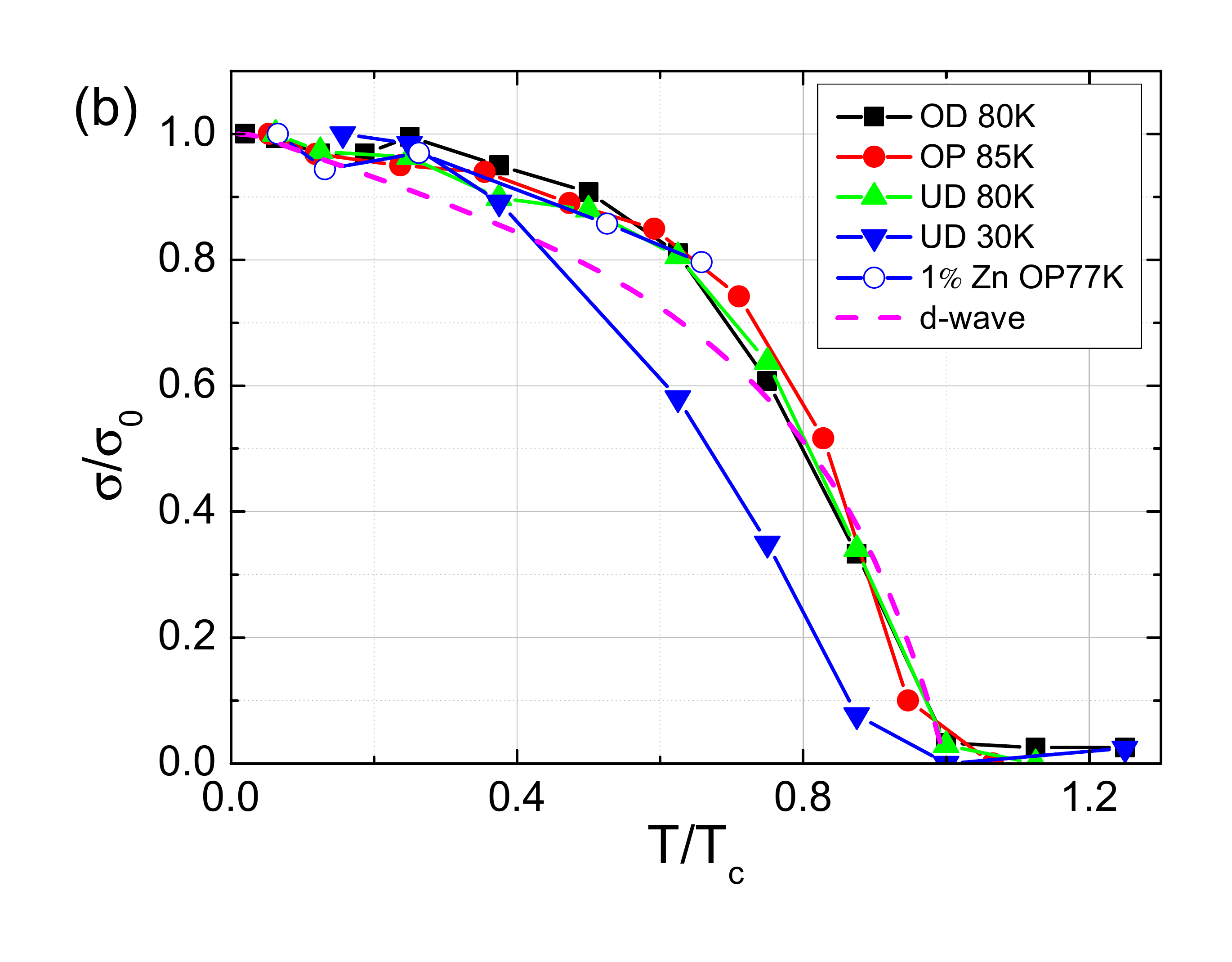}
	\caption[The TF-\musr relaxation rate for \ybasr samples studied.]{(a) The TF-\musr relaxation rate $\sigma$ versus temperature for various doping states for pure and Zn-doped YBaSrCu$ _{3} $O$ _{7-\delta} $.  ``OD'' refers to overdoped samples, ``OP'' to optimally doped and ``UD'' to underdoped. These letters are followed by the \tc of that doping state as measured from bulk susceptibility.  $\sigma(T)$ is determined from Gaussian fits (\eq~\ref{eq:gaussian}) to the raw $P_X(t)$ data - an example of which is shown in \fig~\ref{fig:tfraw}.  The slow relaxation rate above \tc of $\sigma \approx 0.1$~$\mu$s$^{-1}$ comes from the nuclear moments.  Solid curves are fits of the data with $\sigma=\sigma_0\left( 1 - t^{\alpha} \right)$ where $t=T/T_{c}$. $\alpha$ and \tc are free parameters, except in the case of Zn doping where the susceptibility derived \tc values are used.  The \tc determined from fitting these \musr measurements are generally lower than the bulk susceptibility values, but by no more than 2~K. $\alpha$ values range between $3.5$ and $3.7$. (b) Here the data is normalised to $T_c$ and to the lowest temperature $\sigma_{\textnormal{SC}}(T)$ value, $\sigma_{\textnormal{SC}}(0)$.  $\sigma_{\textnormal{SC}}(T)$ is taken to be $\sigma_{\textnormal{SC}}(T)^2=\sigma(T)^2 - \sigma(T>T_c)^2$. All except the most underdoped \ybasr scale well with $\sigma=\sigma_0\left( 1 - t^{\alpha} \right)$.  The faster suppression of \ssc with temperature for the most underdoped \ybasr is discussed in the text. The yellow dashed line is the theoretical temperature dependence of $\sigma$ for a $d$-wave superconducting gap \cite{won1994}. }
	\label{fig:sigmavstall}
\end{figure}

In \fig~\ref{fig:sigmavstall} we plot the relaxation rate versus temperature for all \ybasr samples we had time to measure before the muon beam crashed\footnote{Half-way through our allotted beam time the carbon target broke (three weeks before the Christmas shutdown period).  This section of the beam line is heavily radiation shielded by huge 1000+kg concrete `lego' blocks and these had to be removed before the C target could be replaced.  Needless to say this is no minor undertaking and the beam was not on before our time ran out. At the time of writing, more beam time had just been allotted following the next proposal round.}.  To improve the quality of the Gaussian fits, we perform a global fit for all temperatures and fields for a particular \ybasr composition which allows us to fix the phase $\delta$ and $A_0$ as these should not vary for a given sample and set-up (they are related only to the experimental set up if there are no additional magnetic phases).  

In addition we measure $\sigma$ with low ($\BB_{\textnormal{ext}}=0.03$~T) and high fields ($\BB_{\textnormal{ext}}=0.3$~T) at 5~K.  The purpose of these measurements is to quickly probe the vortex lattice properties, see Section $5.2$ of \cite{reotier1997} and references therein.  We would expect a dramatic increase in the symmetry of \dbloc going from low to high fields if the vortex lattice were to undergo a phase transition to a less ordered, or lower dimensional, state. That we see very little variation in \dbloc shows the vortex lattice is quite robust in these conditions - as it is for \ybco. In fact, a robust lattice gives us further confidence that $\sigma \propto \lambda^{-2}$. Correspondingly we find only a small variation in the calculated value of $\sigma$; for example on optimally-doped YBaSrCu$_3$O$_y$ we measure $\sigma = \{2.6\pm 0.1, 2.8\pm 0.1, 2.7 \pm 0.1 \}$~$\mu$s$^{-1}$ for $\BB_{\textnormal{ext}} = \{ 0.3, 0.1, 0.03\}$~T respectively. 

We now discuss the temperature dependence of $\sigma$ for YBaSrCu$_3$O$_y$ shown in \fig~\ref{fig:sigmavstall}. 

For $T \ll T_c$, a linear suppression of $\lambda^{-2}\propto \sigma$ with temperature is characteristic of a SC order parameter with nodes.  This is a result of small scattering vectors at low thermal excitation being able to connect regions of the Fermi-surface close to where the order parameter changes sign, for example across the nodes in \dxy symmetry \cite{sunmaki, reotier1997}. If there were significant impurity scattering, $\sigma$ would change little with temperature for $T\ll T_c$, much like for an $s$-wave superconductor, because the scattering potential of the impurities is enough to break Cooper-pairs at the nodes in analogy to the effect of a finite temperature \cite{sunmaki, bernhard1996}. We see a linear depression of $\sigma(T)$ and this indicates, again, good sample quality and the absence of disorder scattering.

Closer to \tc we expect $\sigma=\sigma_0\left( 1 - t^{\alpha} \right)$ where $t=T/T_{c}$.  We use this relation to fit our data and estimate $T_c$, fitting only data close to $T_c$. These fits are shown as solid lines in \fig~\ref{fig:sigmavstall}(a).  We generally find values \tc values $\sim1-2$K lower than those obtained from bulk susceptibility measurements, although in the latter case they are zero-field-cooled \tc estimates. Fitted values for $\alpha$ range between $3.5$ and $3.7$.  The paucity of data points close to \tc reflects that our primary interest was $\sigma$ in the low temperature limit.  

The two-fluid model has $ \alpha = 4 $ and our values are of this order.  Closer to \tc the Ginzberg-Landau theory requires $ \alpha =1 $ so that a single $ \alpha $ value in the fit, though common, is not rigorous. The cross over to linear suppression of \ns may well account for the lower \tc values that we recover from the fits.  The full theoretical $d$-wave temperature-dependence of \ns \cite{won1994} is shown by the dashed curve in \fig~\ref{fig:sigmavstall}(b) for $\tcm = \hscgapm/3.5k_B$. The match is not good in the mid-temperature range and this is typically found for polycrystalline samples, while single crystal microwave measurements recover the $d$-wave temperature dependence.  This failure is not well understood.  However, it is also clear that this simple $d$-wave $T$-dependence is not expected where there is a pseudogap present which does not have $d$-wave symmetry, wiping out only those states near the antinodes \cite{khasanov2008}.  This gives a stronger $T$-dependence with $\alpha$ rising above 1, as observed. 

Hence, the temperature dependence of \ssc is generally as we would expect.  The exception is the UD30K data which decreases more rapidly, as can be clearly seen in \fig~\ref{fig:sigmavstall}(b).  Note this is not merely the result of an incorrect \tc, but rather a different temperature dependence as can be seen by the poor fit to $\left( 1 - t^{\alpha} \right)$ in \fig~\ref{fig:sigmavstall}(a). To test the possibility of an inhomogeneous doping state the raw data were fitted again with multiple components. The poor resulting fit with additional components however discredits this possibility.  It is instead likely a consequence of a large pseudogap energy at this doping \cite{khasanov2008} or competing charge order, as observed at this doping state \cite{taillefer2009}. Many other physical properties show anomalous temperature dependence in this doping region, e.g. the strong pseudogapped $T$-dependence of $ S(T)/T $, where $ S(T) $ is the electronic entropy, becomes rather flat at this doping \cite{loram1994}. \added{In addition, the estimated $ p=0.075 $ of this sample places it close to the spin-glass state, see \refsec~\ref{sec:phasediagram}, and this could be expected to additionally contribute to the dephasing of muons.}


		\subsection{\ns and \tc}
		\label{sec:nstc}
\begin{figure}
	\centering
		\includegraphics[width=0.85\textwidth]{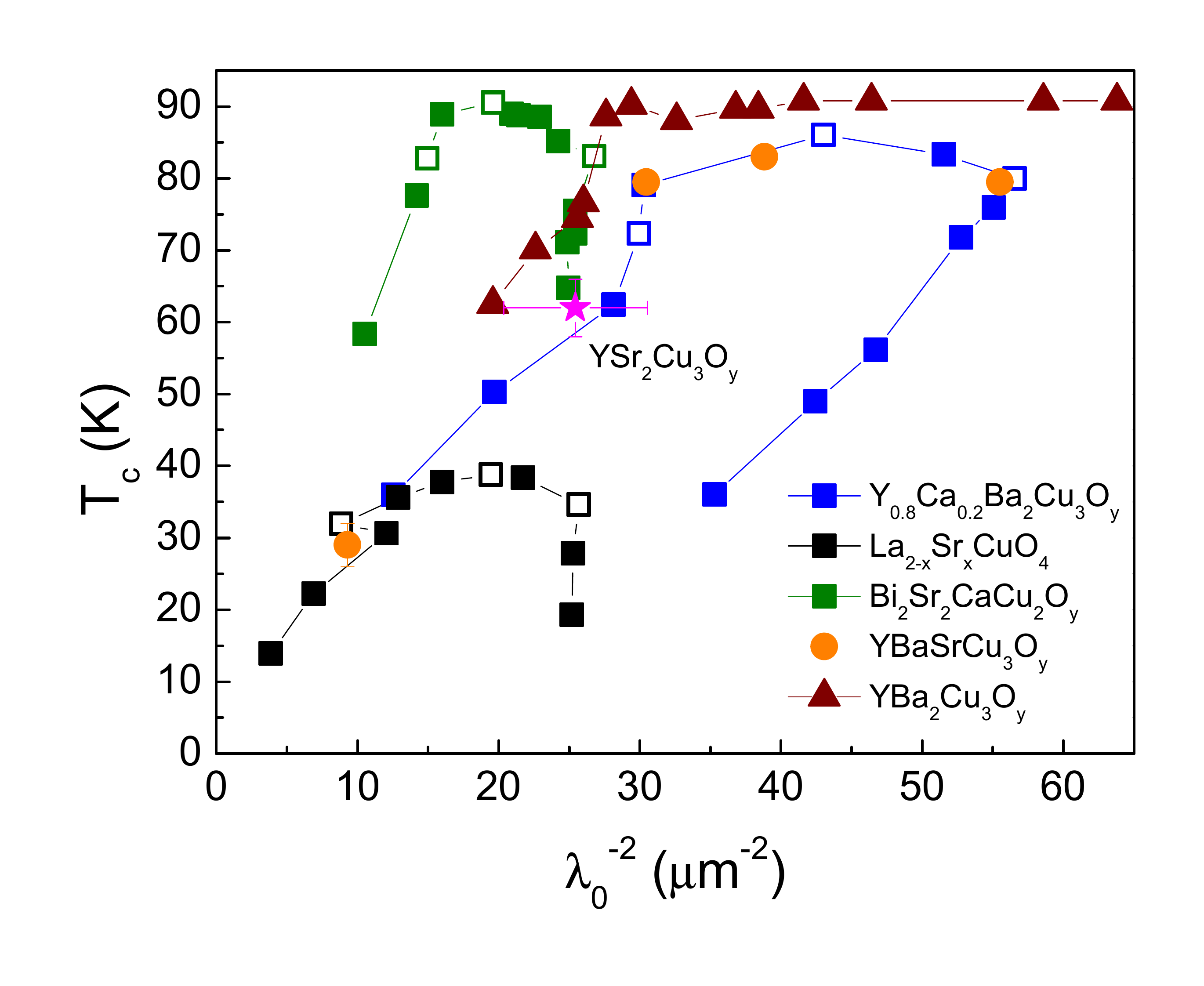}
	\caption[A `Uemura plot' of \tc vs. superfluid density.]{\label{fig:uemura} \tc plotted against the superfluid density, $\lambda_0^{-2}$, in the low temperature limit. The plot is adapted from \cite{uemura1989} and \cite{tallon2003superfluid}. Open symbols indicate the doping states $ p=0.125 $, $ 0.16 $ and $ 0.19 $. To these published data we add our own data from \ybasr~.  Increasing doping proceeds in a clockwise direction and gives rise to the `boomerang', although for pure \ybco contributions to $\lambda_0^{-2}$ from the chains lead to a plateau.  Presented in this way the \ybasr superfluid density is comparable with both pure and Ca-doped \ybco for similar doping and \tc values.  If Sr-substitution-induced disorder were significant we would have expected a suppression of $\lambda_0^{-2}$ relative to \ybco.}

\end{figure}

We now discuss the relation between the superfluid density in the low temperature limit of \ybasr and the well-characterised YBa$ _{2} $Cu$ _{3} $O$ _{7-\delta} $.  In \fig~\ref{fig:uemura} we reproduce the data presented in \fig~$1$ of \cite{tallon2003superfluid} showing \tc plotted against the superfluid density \ns in Y$_{0.8}$Ca$_{0.2}$Ba$_2$Cu$_3$O$_y$, Bi$_2$Sr$_2$CaCu$_2$O$_{8+\delta}$ and La$_{2-x}$Sr$_x$CuO$_4$. This is the so called `Uemura plot' \cite{uemura1989}, see \refsec~\ref{sec:sfintro} for a discussion of its features.  Although not included in \cite{tallon2003superfluid}, we add similar data for the pure \ybco compound from \cite{uemura1989}. Unlike the Ca-doped material, oxygen filling of the chains is required to optimally dope \ybco and these filled chains become superconducting themselves.  This contributes significantly to the superfluid density and leads to the `plateau' at high doping in this material where \ns doubles with almost no change in \tc \cite{bernhard1995musr}. 

To this published data we add our results on \ybasr presented above.  We use \eq~\ref{eq:sfdensity} to convert the measured relaxation rates $\sigma$ into \ns values\footnote{Pending thorough optical reflectometry and ellipsometry measurements we do not know the effective carrier mass $m^*$ seen in \eq~\ref{eq:superfluid}.} and take \tc to be the values determined from \musr with uncertainties estimated from their difference to the bulk susceptibility measurements. Note that the $\lambda$ here is really the $ab$-plane value $\lambda_{ab}$.

The data in \fig~\ref{fig:uemura} show that \ns for \ybasr is similar to \ns for Ca-doped Y123 of similar doping states. If \tc were suppressed by disorder there would be a characteristic more rapid suppression of \ns than $ \tcm $. For example, see \cite{sunmaki, tallon2003superfluid} or \fig~\ref{fig:sigmanormzn} where Zn substitution is seen to suppresses \ns twice as fast as $ \tcm $. This is not observed.  Consequently the lower \tc in \ybasr with respect to Y123 can be attributed to disorder no more so than the small decrease in \tcmax of Ca-doped Y123.

Our \ybasr data is also consistent with the pure Y123 of similar doping states with the UD30K data even falling on the so called `Uemura line' of \cite{uemura1989}.  The pure Y123 \ns does become larger than our most overdoped \ybasr sample, but presumably this is due to an additional chain contribution rather than a suppression of \ns in \ybasr (which may not be fully oxygen loaded, or be more susceptible to chain disorder \cite{ying2002} - we do not know enough about the effect of Sr substitution to say which).  

We conclude from these data that it is \textit{not} disorder, or a developed stripe phase, causing the 8K suppression of \tcmax upon Sr substitution for Ba\footnote{A weaker, more acceptable, conclusion would sound like; \textit{disorder does not have a significant effect on the superconducting properties of \ybasr as compared with \ybco}.}. 


The nuclear quadrupole resonant (NQR) line-widths in \ybasr have been measured by Ying \etal \cite{ying2002}.  They find a broadening of the line-widths with Sr substitution and a blue-shift of the Cu(2) peaks and red-shift of the Cu(1) peaks.  The peak frequency shifts are consistent with an increasing hole concentration with Sr substitution (the internal pressure effect).  The broadening line-widths are attributed to Sr substitution induced disorder\footnote{In particular, the authors conclude a Sr-substitution induced O(4) (chain oxygen) $b$-axis `buckling' leads to the observed suppression in \tc with Sr substitution.}. It is interesting that while these authors clearly see a pronounced increase in disorder in \ybasr, our \musr results show no appreciable suppression of \ns from this disorder. Note that NQR measures electric field gradients whereas muons do not couple to electric field gradients (they are spin 1/2 particles). That the superfluid density remains consistent with the pure Y123 material, which has sharp NQR line widths, shows the superconducting properties are less sensitive to disorder causing broad NQR line-widths on the Cu sites.  The same conclusion can be drawn from NMR \cite{williams1998nmr} and NQR \cite{williams2007nqr} studies of the co-substituted Y$_{1-x}$Ca$_{x}$Ba$_{2-x}$La$_{x}$Cu$_3$O$_8$ cuprate.

\subsection{YSr$_2$Cu$_3$O$_y$}
\label{sec:musrysco}
\begin{figure}
	\centering
		\includegraphics[width=0.850\textwidth]{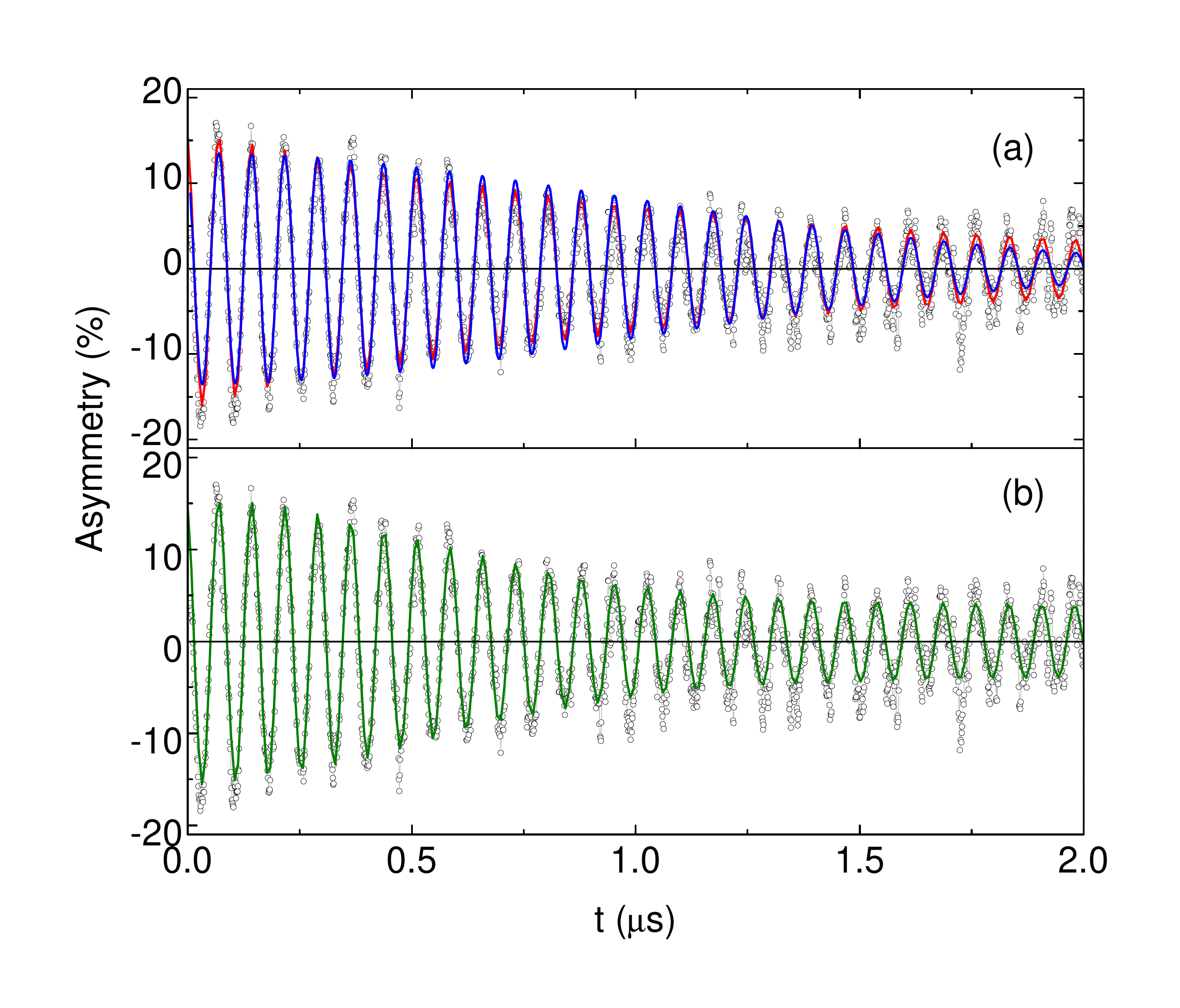}
	\caption[Representative TF-\musr data for \ysco at $ T=5 $~K.]{Representative TF asymmetry data on \ysco at $T=5$~K. Shown in (a) is the fit to a single Gaussian relaxation (blue line), \eq~\ref{eq:gaussian}, as well as to a single `Lorentzian' relaxation (red line).  In (b) the data is fitted to the sum of two Gaussians. Data are of the Methanol-washed, $T_c=66$~K powder \ysco sample.}
	\label{fig:tfrawysco}
\end{figure}

We also spent considerable time measuring the \ysco samples Edi35, Edi22 and Edi22w.  In \fig~\ref{fig:tfrawysco} we show representative TF asymmetry data on \ysco at $T=5$~K. To the data in \fig~\ref{fig:tfrawysco} we show a single (Gaussian) component fit in panel (a) and a two component fit in panel (b) and solid lines.  We have not presented these data with the \ybasr data as the \ysco were found to contain a considerable fraction of magnetic impurities (see the ZF measurements section \ref{sec:zfmusrysco}).  

In an attempt to partially account for this we fit the TF data to two Gaussians, \eq~\ref{eq:gaussian}: component 1 is slowly dephasing component while component 2 is more rapidly dephasing, i.e. $\sigma_2>\sigma_1$.  Also, from the raw TF data shown in \fig~\ref{fig:tfrawysco}(a), it is clear that a single Gaussian dephasing component is not sufficient, although a better fit to the decreasing oscillation amplitude can be obtained from the `Lorentzian' function; $\sigma(t)\sim\exp(-\Lambda t)$ - see \refsec~\ref{sec:musrexperiment}. 

%
%

We have two possible ways to proceed;
\begin{itemize}
	\item Use only a single dephasing component, i.e. a Lorentzian, to fit the decreasing amplitude of the oscillation. This is the simplest approach. However, the additional magnetism in the material contributes \dbloc and a single component will not distinguish between this and any flux-line lattice field broadening.  With this approach we will not gain any useful information about the superfluid density.
	\item Fit with more components, of which using only two is the simplest (e.g. the sum of two Gaussians or a Gaussian and Lorentzian), and then see if one of these components is consistent with what we expect from a superconducting phase. 
\end{itemize}
In an attempt to extract some useful information from the \ysco \musr data, we proceed with the latter.

The Gaussian function is chosen to keep the model simple and to introduce as few additional parameters as possible.  The free parameters in the fit are now $\{A^{(1)}$, $A^{(2)}$, $\sigma ^{(1)}$, $\sigma ^{(2)}$, $\blocm ^{(1)}$, $\blocm ^{(2)}$, $\delta ^{(1)}=\delta ^{(2)}$, $A_B\}$ where the superscript refers to component 1 or 2. The ratio $A^{(i)}/(A^{(1)}+A^{(2)})$ gives the volume fraction of the $i^{th}$ component. The `baseline' asymmetry, $A_B$, as a free parameter is typically small, $\approx 0.0$, for the two-component fits.  Thus, for the fit results displayed it has been set to $0.0$ to remove another free parameter.  In \fig~\ref{fig:tfrawysco}(b) we show such a two-component fit to data at $T=5$~K from the Methanol-washed $T_c=66$~K sample. Note that the more-rapidly decaying oscillations for $t<1.5$~$\mu$s are better reproduced.

\begin{figure}
	\centering
		\includegraphics[width=0.85\textwidth]{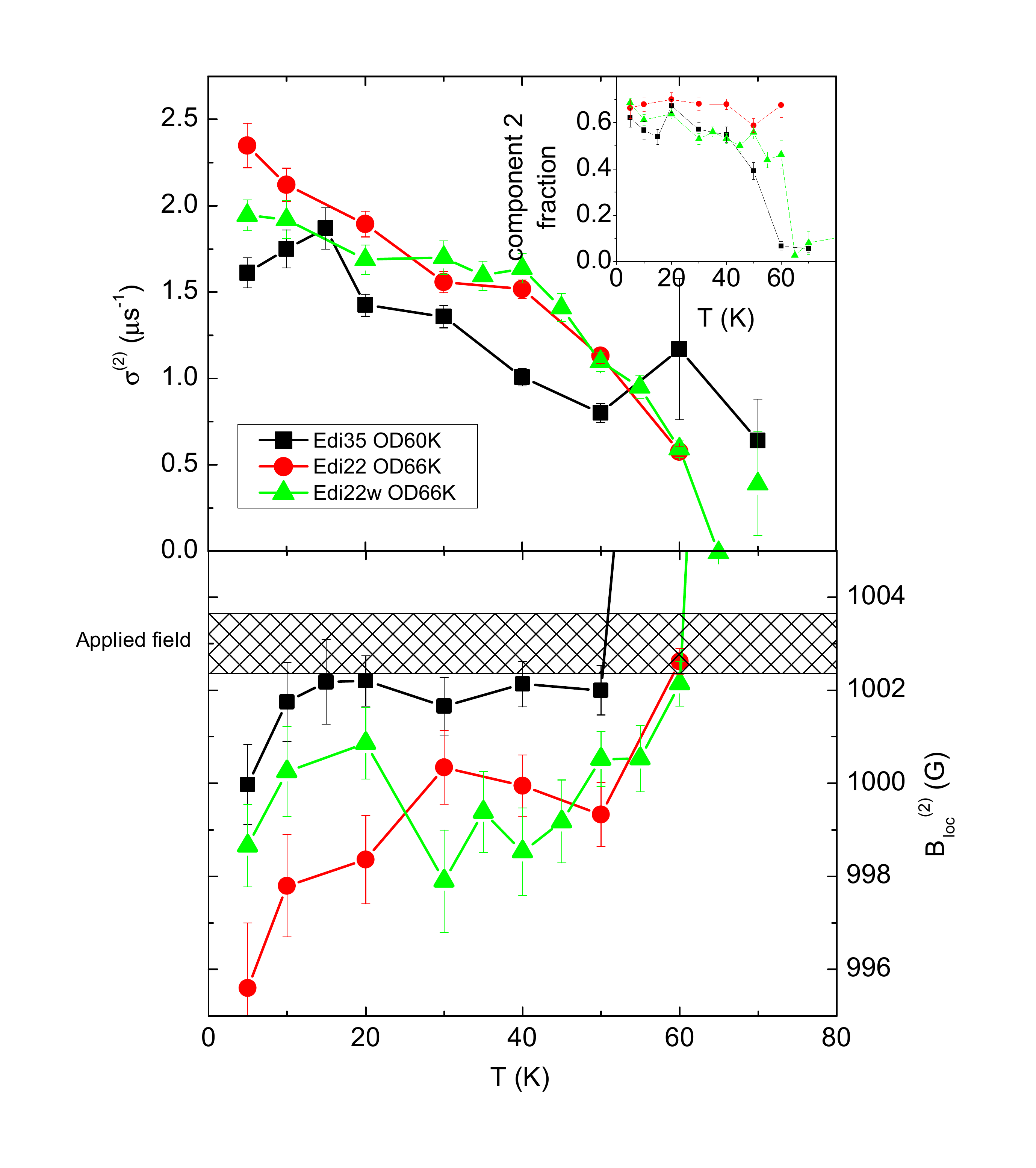}
	\caption[TF-\musr relaxation rate from a two-component fit for all \ysco samples studied.]{\label{fig:ysco2comp} Plotted in the top panel is the temperature dependence of relaxation rate, $\sigma ^{(2)}$, for the second component, for all three \ysco samples measured. Shown in the inset of this panel is the volume fraction of this second component, also as a function of temperature.  In the lower panel is the average local field, $\left\langle \blocm \right\rangle$, of the second component.  The externally applied field, \bext $= 1003\pm 1$ G, is annotated for comparison.}
\end{figure}

In \fig~\ref{fig:ysco2comp} the results of this two-component fitting approach for all TF-\musr YSCO data are presented.  Only parameters for component 2 are plotted because this component is possibly related to the superconducting volume fraction.

This approach is at least partially successful as shown by several encouraging features in \fig~\ref{fig:ysco2comp} for the Edi22w and Edi35 samples (green and black respectively); 
\begin{enumerate}
	\item The fraction of this phase goes to 0 at \tc (above \tc the errors for this component become very large and the TF data are reasonably fitted with one component only).
	\item $\blocm ^{(2)}<$\bext for $T<T_c$ which implies diamagnetic screening.
	\item The temperature dependence of $\sigma ^{(2)}$ is what we would expect for a SC phase with $T_c\approx60$~K, especially for Edi22w.
\end{enumerate}

The volume fraction of the second component is $\approx 0.6$. 

The relaxation of the second component is $\sigma^{(2)}(T\rightarrow 0)=(2.0\pm0.4)$~$\mu$s$^{-1}$ for three YSCO samples and so we \emph{tentatively} include this data point in \fig~\ref{fig:uemura}.  The data point has a low superfluid density when compared with optimally-doped or overdoped Y123.  Perhaps our samples are on the underdoped side of \tcmax (=70 K in the best samples) despite their high pressure O$_2$ anneals? This seems unlikely, especially given a thermopower of $-6$~$\mu$V.K$^{-1}$ in Edi35, but it could be due to significant O(5) occupation.  Alternatively, this may reflect a real superfluid suppression from disorder and/or from an enhanced pseudogap or stripe-phase (enhanced by large $J$). To investigate further requires samples free of magnetic impurity phases, which in turn requires considerable materials-chemistry effort (now underway), followed by characterisation of the material with other techniques and a range of \musr measurements from over- to underdoping.

		\subsection{Zn substitution in \ybasr}
		\label{sec:znmusr}
We turn now to our initial \ns data for Zn-doped YBaSrCu$_{3-z}$Zn$_z$O$_y$ with z=0.02 (``1\%'') and z=0.06 (``3\%'') shown in \fig~\ref{fig:sigmavstall}. The substitution of only a small amount of Zn on the \cuo layer rapidly lowers \tc as can be seen in \tab~\ref{tab:musrsamples}, but even more rapidly suppresses \ns (in the former case it is linear \cite{abrikosovgorkov} while in the latter case it is super-linear \cite{sunmaki}). 

\begin{figure}[t]
	\centering
		\includegraphics[width=0.75\textwidth]{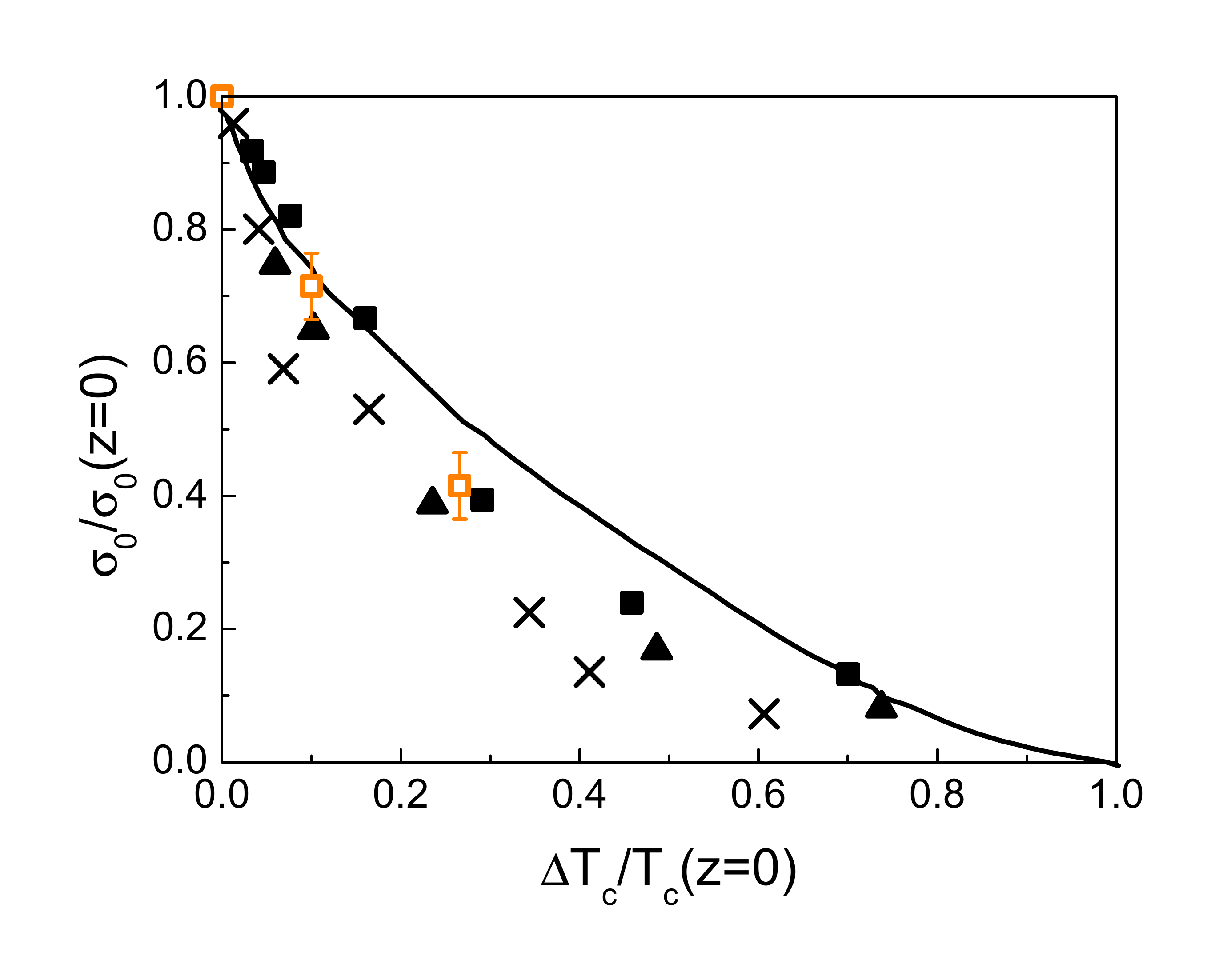}
	\caption[Normalised superfluid density plotted against relative decrease in \tc due to Zn doping in YBaSrCu$ _{3-z} $Zn$ _{z} $O$ _{y} $.]{The normalised superfluid density plotted as a function of the relative decrease in $T_c$.  The data are been normalised to the Zn free material values.  The plot is adapted from \cite{bernhard1996}.  Data for Y$_{0.8}$Ca$_{0.2}$Ba$_2$Cu$_{3-z}$Zn$_z$O$_y$ optimally-doped (black squares) and overdoped (black triangle) is from Bernhard \etal \cite{bernhard1996} and data for YBa$_2$Cu$_{3-z}$Zn$_z$O$_y$ near optimally doped (black crosses) is from Bucci \etal \cite{bucci1994}. The solid line show the predicted superfluid suppression for a system with \dxy order parameter symmetry due to isotropic scattering in the unitary limit \cite{sunmaki}. To this plot we add our data for YBaSrCu$_{3-z}$Zn$_z$O$_y$ optimally doped (open orange squares).  From these initial data \ybasr appears entirely consistent with Y123.}
	\label{fig:sigmanormzn}
\end{figure}

Here is a situation where disorder really does matter! 

Bernhard \etal have measured \ns for a range of Zn concentrations in overdoped, optimally-doped and underdoped Ca-doped Y123 \cite{bernhard1996}.  From these data, reproduced in \fig~\ref{fig:sigmanormzn}, they could conclude the SC order parameter has \dxy (or similar) symmetry.

To these published data we add our results from YBaSrCu$_{3-z}$Zn$_z$O$_y$ on overdoped and optimally-doped samples\footnote{Which, unfortunately, was all we had time to measure before the beam failed.}.  In taking the ratio of \ns at $z=0$ and $z>0$ we assume that $m^*$ changes negligibly with Zn substitution, which is most likely justified \cite{bernhard1996} but could be tested by reflectometry and ellipsometry measurements, for example. As was found in the previous section, \ybasr and \ybco behave in the same way, this time under Zn substitution. This verification is additionally important because it shows a consistency not only between \ybasr and Y123 at different oxygen concentrations, but also a reproducibility of that consistency between different `batches' of YBaSrCu$_{3-z}$Zn$_z$O$_y$.

These data also confirm \ybasr has a SC order parameter with \dxy symmetry and that Zn acts as a unitary scatterer.

\section{Conclusions}
We have used \musr to measure the superfluid density, $ \nsm $, of \ybasr and we find the nice result that this is a boring, well-behaved material.  \fig~\ref{fig:uemura} showed that the \ns are consistent with those previously reported for pure and Ca-doped \ybco \cite{uemura1989, tallon2003superfluid}. Furthermore, the suppression of \ns due to Zn doping shown in \fig~\ref{fig:sigmanormzn} is fully consistent with that found for \ybco \cite{bernhard1996} and confirms that \ybasr has a SC order parameter with \dxy symmetry and that Zn acts as a unitary scatterer.  The immediate conclusion is that disorder does not play a significant role in \ybasr as compared with Y123.  Consequently, the lower \tcmax seen in this compound, and those with lower Sr content, is a real `ion-size effect.'  

Our measurements of \ns by \musr for YSCO are less certain as it appears our sample contain additional magnetic impurities.  The values of \ns extracted from a two-component fit of the data revealed a comparatively low $ \nsm $.  This indicates that disorder may significantly suppress the superfluid density in the fully Sr substituted material. Further measurements on pure \ysco samples are desirable to confirm or refute this result.

\chapter{Gap estimates from Raman Spectroscopy}
\label{ch:ramangaps}
\subsubsection{Summary}
The purpose of the work presented in this chapter is to measure the spectral gap in \bog and \btg symmetry for Yb123, Eu123 and Nd123 single crystals at optimal doping.   To the best of our knowledge these measurements have not been done before. We find \bog spectral gap energies similar to Y123 but no clear systematic ion-size dependence which was the object of our study.  We were unable to unambiguously identify spectral gap features in \btg scattering geometry.

\section{Introduction and Motivation}
\label{sec:ramangapsintro}

In BCS theory, the superconducting energy gap, $\hscgapm$, is proportional to the mean-field transition temperature $\tcmfm$, $\scgapm =\alpha.k_BT_c^{(\textnormal{mf})}$.  For $d$-wave \hscgapk symmetry in the weak-coupling limit, $\alpha=4.28$ while for the strong-coupling limit the value is larger.  Observation of an energy gap in Raman spectra provided early estimates of $\alpha$ for the cuprates.  For example, the early work of Friedl \etal \cite{friedl1990} on Ln123 found $\alpha=4.95\pm0.10$, from which they concluded that the cuprates are strongly-coupled superconductors\footnote{At the time an $s$-wave symmetry \scgapk was popular, in which case the weak-coupling BCS result is $\alpha=3.53$.}.  Recent work by Guyard \etal \cite{guyard2008} on Hg1201 illustrates a symmetry and doping dependent value of $\alpha$; at the nodes $\alpha\approx 6.4$ independent of $p$, whilst at the anti-nodes $\alpha$ decreases from $\approx 12$  in the under-doped regime to $\approx 6.4$ in the over-doped regime.  These strange results highlight two important facts about the cuprates: (i) The difference between \tc and \tcmf  can be significant \cite{dubroka2011, tallon2011}, but is often overlooked.  (ii) Both the pseudogap and superconducting energy gap contribute to the energy of the spectral gap, $\specgapm$, observed by Raman spectroscopy \cite{tallonarXiv}, \refsec~\ref{sec:gaps}.  In addition, the pseudogap modifies the temperature dependence of the \specgap from that expected from a mean-field superconducting gap \cite{guyard2008}. 

There are two ways by which $\specgapm$ is manifest in Raman spectra. Firstly in the Electronic Raman Scattering (ERS) continuum and secondly, via the renormalisation of phonon energy and life-time\footnote{Or to put it another way, the renormalisation of the real and imaginary parts of the phonon self-energy, $\Sigma$, respectively.} \cite{zeyher1990, nicol1993}.

There have been many studies exploring the doping dependence of the \bog and \btg ERS features, e.g. \cite{chen1993, opel2000, venturini2002, hewitt2002, sugai2003, letacon2006, guyard2008breakpoint}, as well as recent measurements of the temperature dependence of these features \cite{guyard2008, guyard2008breakpoint, li2012}.  

An early study made measurements on Ln123 with the purpose of estimating \specgap \cite{friedl1990, thomsen1990}.  Here they made use of the renormalisation of phonon life-time to estimate $\specgapm$. In distinction to this work, they assume that \specgap is the same in these materials and instead use the variation in phonon life-time with Ln ion-size to provide more data points for a fit to the Zeyher-Zwicknagl model \cite{zeyher1990}.  

With our measurements however we are primarily interested in comparing \specgap \textit{between materials at a given doping state}, namely, between Yb123, Eu123 and Nd123 at optimal doping.  Optimal doping is chosen for two reasons; (i) it is a suitable doping state for inter-comparison between materials and (ii) $\tcm=\tcmaxm$ here. These samples were chosen because they span the ion-size range available and because polycrystalline samples of the same material can be synthesized in the lab \cite{williams1996}.  At best, these measurements can provide estimates of the pseudogap energy and symmetry \cite{loram1994, guyard2008}, the superconducting gap energy at the nodes and anti-nodes \cite{hewitt2002, guyard2008} and information on the quasi-particle dynamics \cite{opel2000, letacon2006} as the ion-size varies across our Ln123 series. 

\section{Experiment}

A description of the experimental set up for these variable temperature Raman spectroscopy measurements is given in \refsec~\ref{sec:ramanvtexptechniques}.

\subsection{Experimental considerations}

Optimally-doped Ln123 is distorted from a tetragonal unit cell into a slightly orthorhombic unit cell due to the formation of Cu-O chains along the $b$-axis direction.  This orthorhombic distortion leads to a mixing of the $ \bogm $, \aog and \btg  symmetries \cite{cardona1999}.  Because the distortion is small, the approximation of tetragonal symmetry is made giving the usual scattering geometries: $ \aogbogm $, \bog etc. \cite{kendziora1995, limonov2000, weber2000}.

Early studies found no significant ERS resonance effects between the 647~nm and 458~nm laser lines \cite{venturini2002}, that is, there is no significant change in the \bog spectra below \tc for incident laser wavelengths between 647~nm and 458~nm. However, more recent studies of le Tacon \etal show a resonant enhancement of the \bog gap and phonon modes \cite{letacon2006, li2013} that may be related to the electronic band structure via inter-band transitions \cite{li2013}.  We use the 514.5~nm laser line as this is the most common wavelength used by others in the field.

The temperature of the sample should be as close as possible to that measured by the thermocouple.  The two main reasons why this may not be the case are:
\begin{enumerate}
 \item The laser heats the sample.  To mitigate, the laser power is kept at less than $10$~mW - an experimentalist's `rule of thumb.'  In this regard we are also helped by the large laser-spot size, $\sim 1$~mm$^2$, reducing the intensity of light on the sample.
 \item There is poor thermal conductivity between the sample and cold finger.  A high surface area contact was made with silver paint or vacuum grease to ensure good thermal contact.
\end{enumerate}

Fortunately for these measurements, phonon thermal conductivity dominates over electronic thermal conductivity in the cuprates (at least in the normal state) \cite{poolehandbook} which means that the thermal conductivity remains large, and is in fact enhanced in the superconducting state due to the decrease in phonon scattering. From a simplified model of the experimental situation we estimate $\sim 1$~K laser heating. 

\begin{figure}
	\centering
		\includegraphics[width=0.50\textwidth]{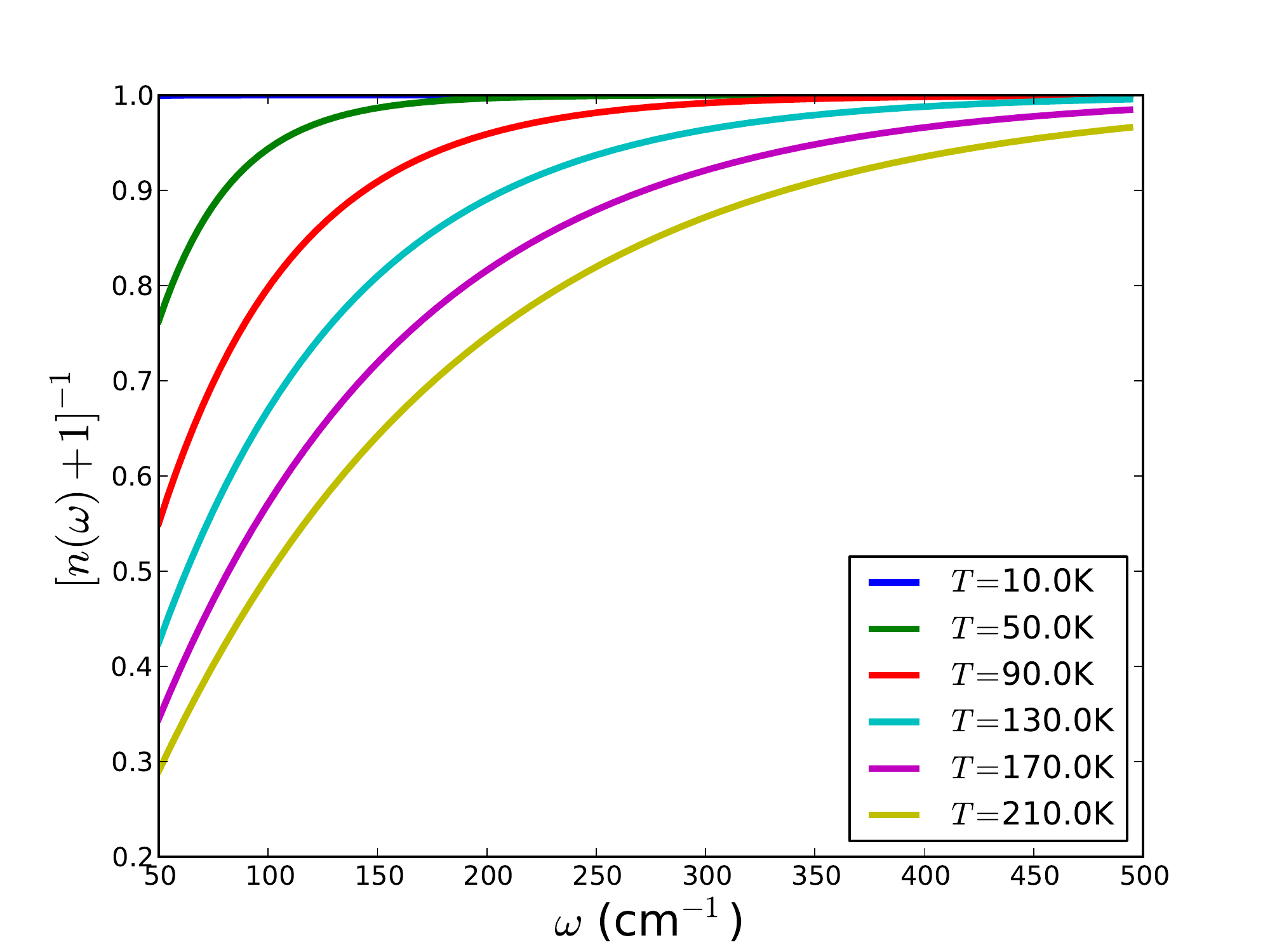}
	\caption[Examples of the Bose-Einstein correction factor at various temperatures.]{Examples of the correction factor used to scale spectra at various temperatures. $n(\omega,T)=[\exp(\frac{\hbar \omega}{k_B T})-1]^{-1}$ is the normal Bose-Einstein distribution.}
	\label{fig:bosescalingfactor}
\end{figure}

 The high-vacuum chamber is always handled using powder-free latex gloves.  Organic materials from one's hands will out-gas at high vacuum and subsequently condense on the sample's surface (because the sample is the coldest part of the chamber).  This will result in a spurious florescence signal which could possibly mask the weak electronic Raman scattering (ERS) from the sample. The chamber is continuously pumped on by a turbo pump which avoids the possibility in diffusion pump systems where oil-droplets back-diffuse and condense on the sample. 

Another issue in this regard is our use of vacuum grease to attach the sample to the cold-finger. Despite our most conscientious efforts, it is possible some of this grease found its way to the surface of our Eu123 sample.  We speculate this is the cause of a linear component to the Eu123 \bog and \btg spectra shown in  \fig~\ref{fig:eu123ersb1g} and \fig~\ref{fig:eu123ersb2g} respectively.

After subtraction of a suitable `dark count'\footnote{These are the counts recorded by the CCD with the shutters to the spectrometer closed.  It is calibrated at the start of each measurement.}, every spectrum is multiplied by the Bose-Einstein factor, $[n(\omega,T)+1]^{-1}$ where $ n(\omega,T)= \left[ \exp(\frac{\hbar \omega}{k_B T})-1 \right]^{-1}$ is the normal Bose-Einstein distribution (see Chapter 1 and 5 of \cite{weber2000}). \fig~\ref{fig:bosescalingfactor} shows the correction factor for the range of interest, $[50,700]$~$ \cmm $, for several representative temperatures.

\subsection{Raman-active phonon modes in Y123 structures}

There are six Raman-active phonon modes in the Y123 structure which will show up in addition to any electronic Raman scattering (ERS).  They are listed in \tab~\ref{tab:phononmodes}. The `names' given to them are the approximate energy shift at which they are observed.  For each mode there is a cartoon of the main ion displacements involved in the mode. \added{Also listed in the table are experimentally determined, mode-specific Gr\"{u}neisen parameters, $\gamma = -\dd\ln(\omega)/\dd\ln(V) $, from reference \cite{syassen1988}.  These Gr\"{u}neisen parameters are used to determine the dependence of the phonon mode energy on the volume of the unit-cell.} Assignment of the modes are documented, for example, here \cite{limonov2000}. The classification of the symmetry of each phonon mode is listed and is important with regard to which region of the Fermi-surface the phonon mode is coupled \cite{devereaux1994}.

\begin{center}
\LTcapwidth = \textwidth 
\begin{longtable}{lccc} \toprule
			Name & Symmetry & Gr\"{u}neisen parameter &  Ion displacements \endhead \midrule \midrule
			120~\cm  & $A_g$  &  -  & \includegraphics[width=0.225\textwidth]{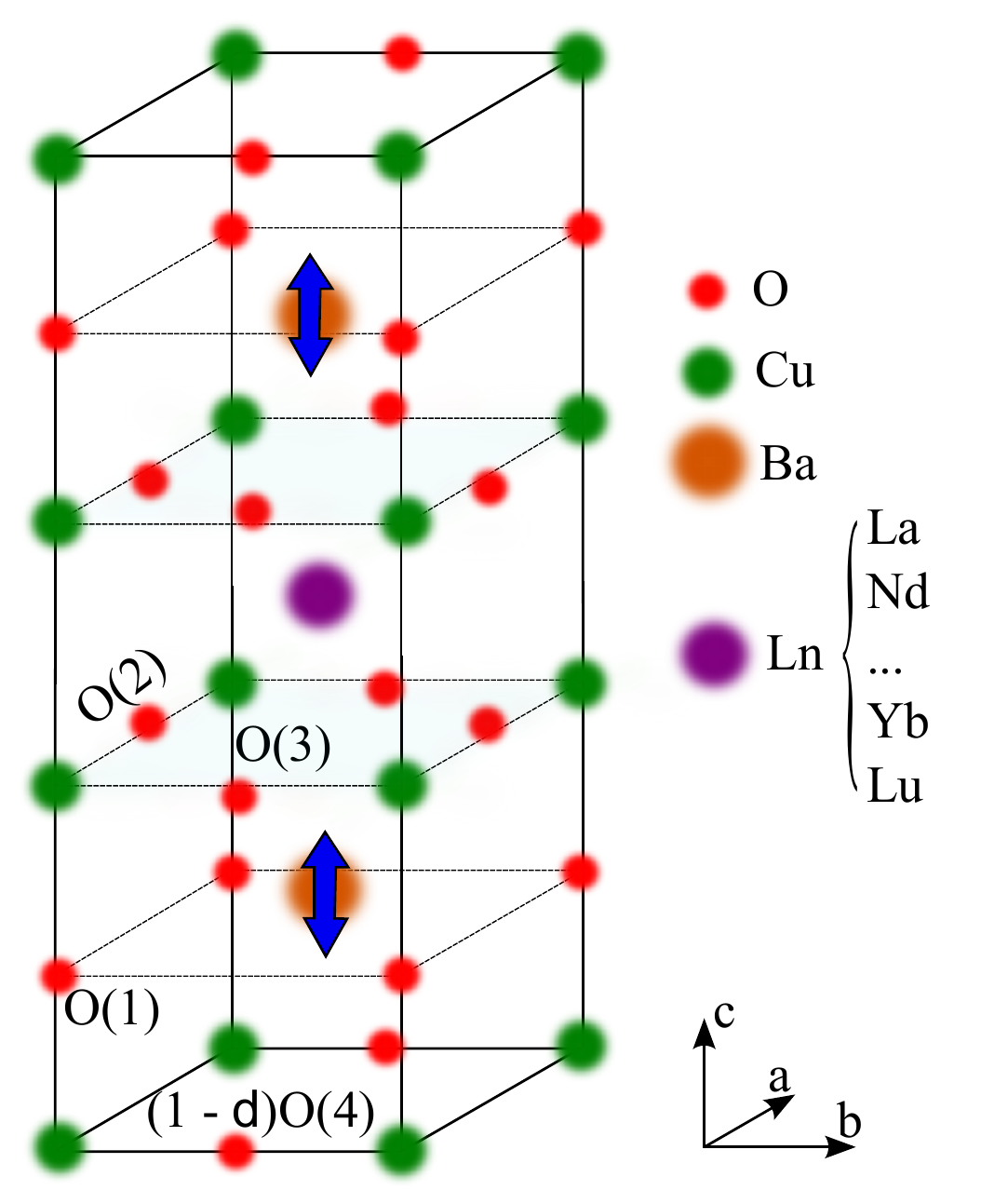} \\ \midrule
			150~\cm  & $A_g$  &  1.5  & \includegraphics[width=0.225\textwidth]{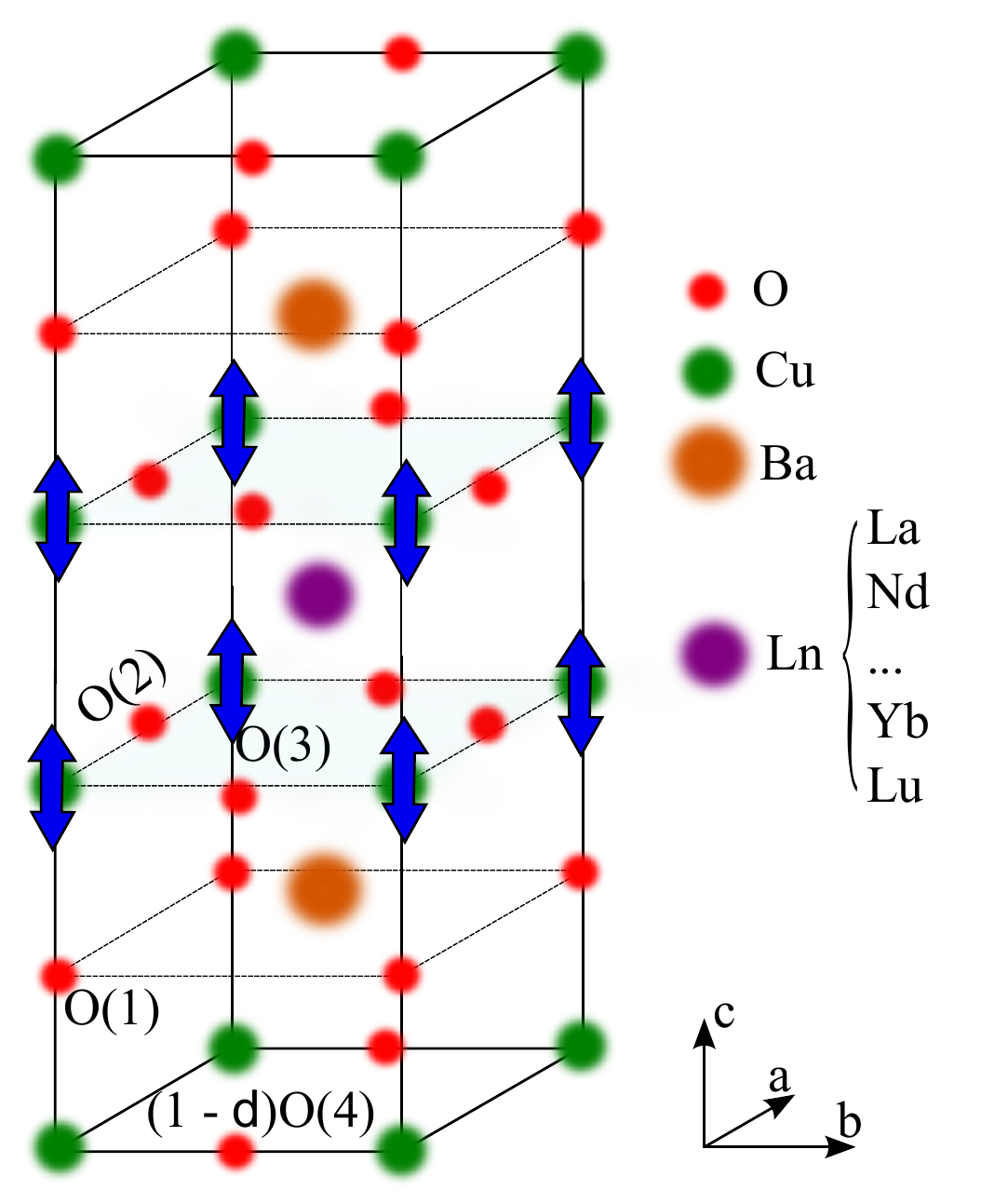} \\ \midrule
			330~\cm  & quasi-\bog  &  1.6   & \includegraphics[width=0.225\textwidth]{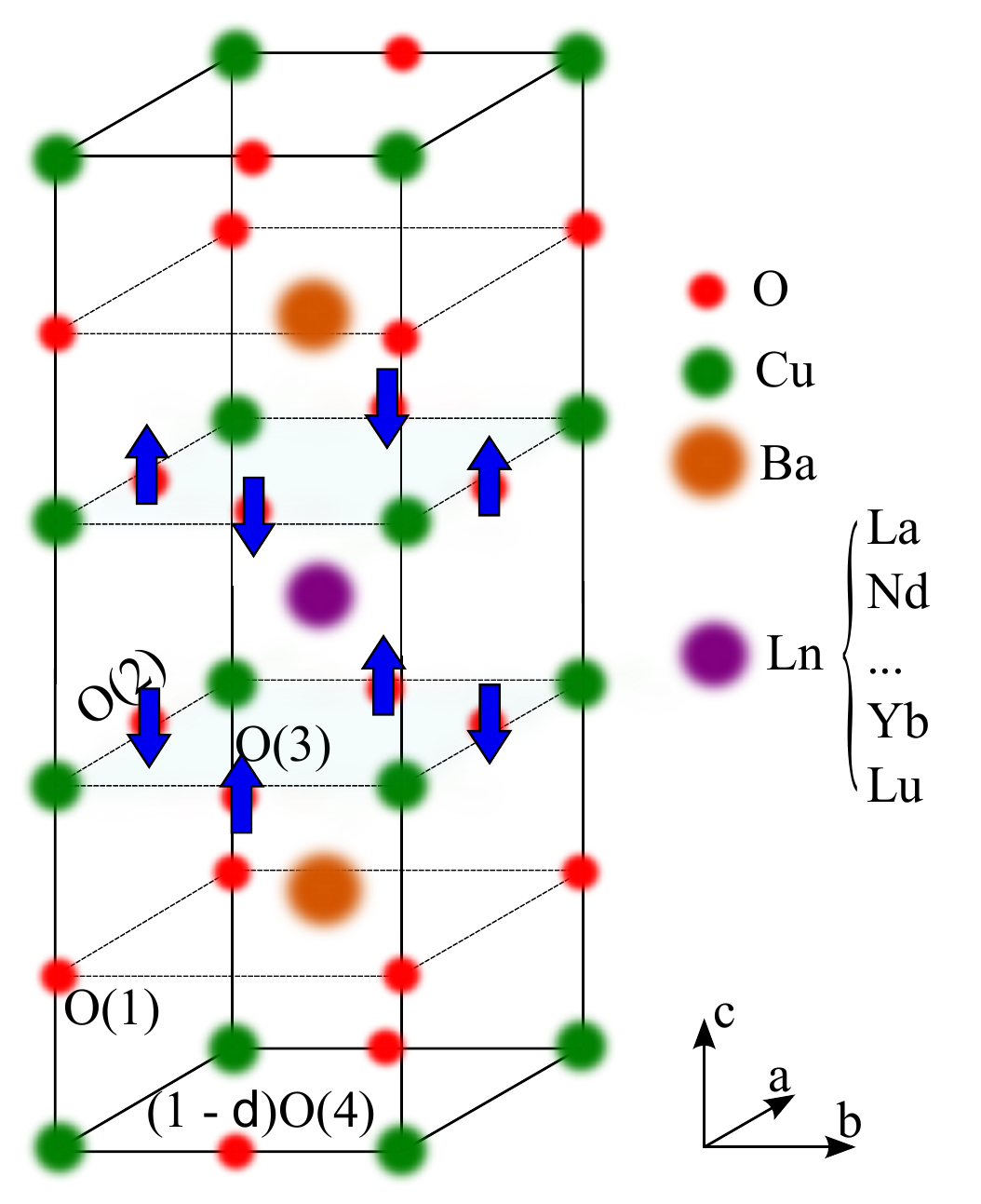} \\ \midrule
			430~\cm  & $A_g$  &  1.7  & \includegraphics[width=0.225\textwidth]{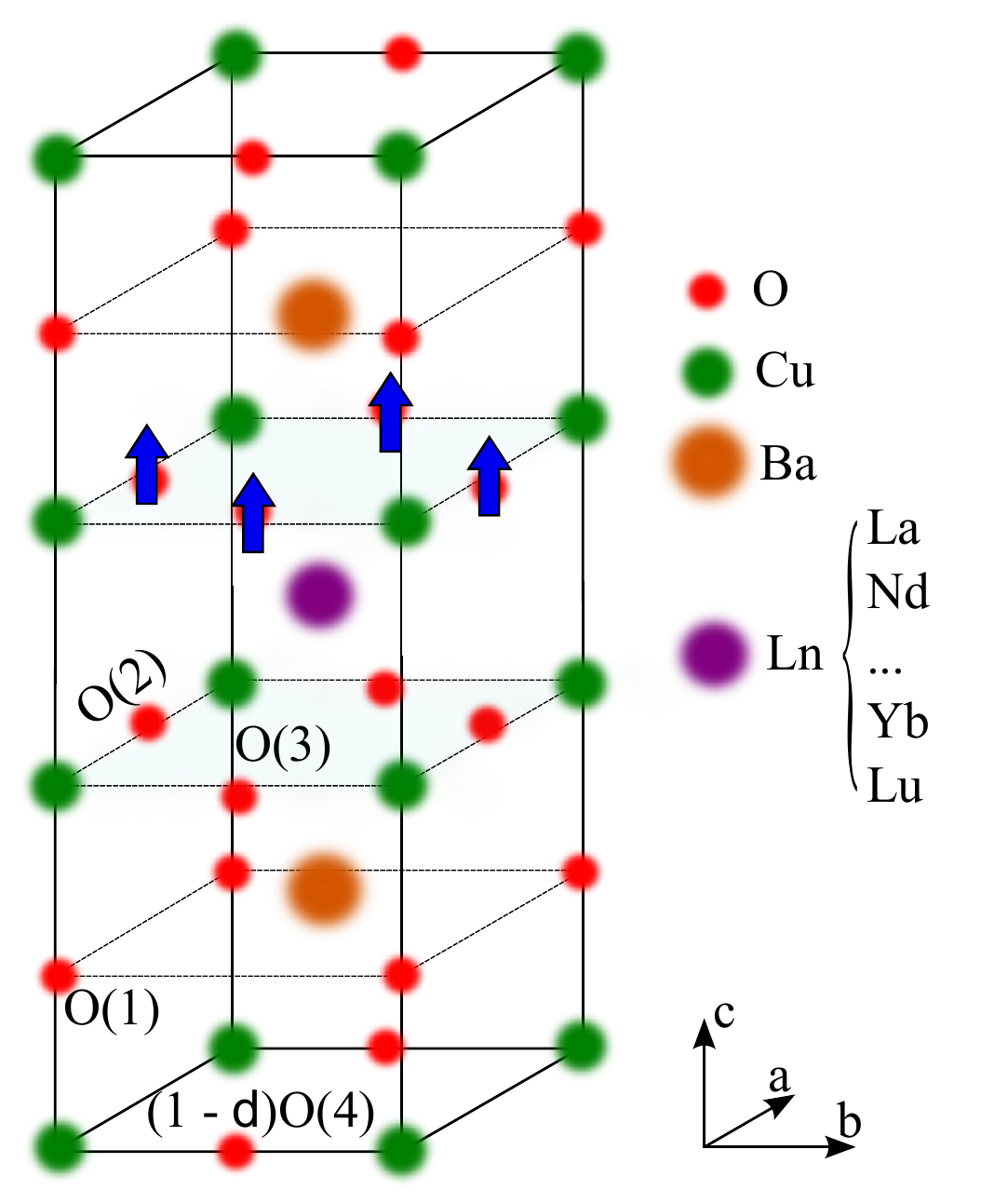} \\ \midrule
			500~\cm  & $A_g$  &  1.9  & \includegraphics[width=0.225\textwidth]{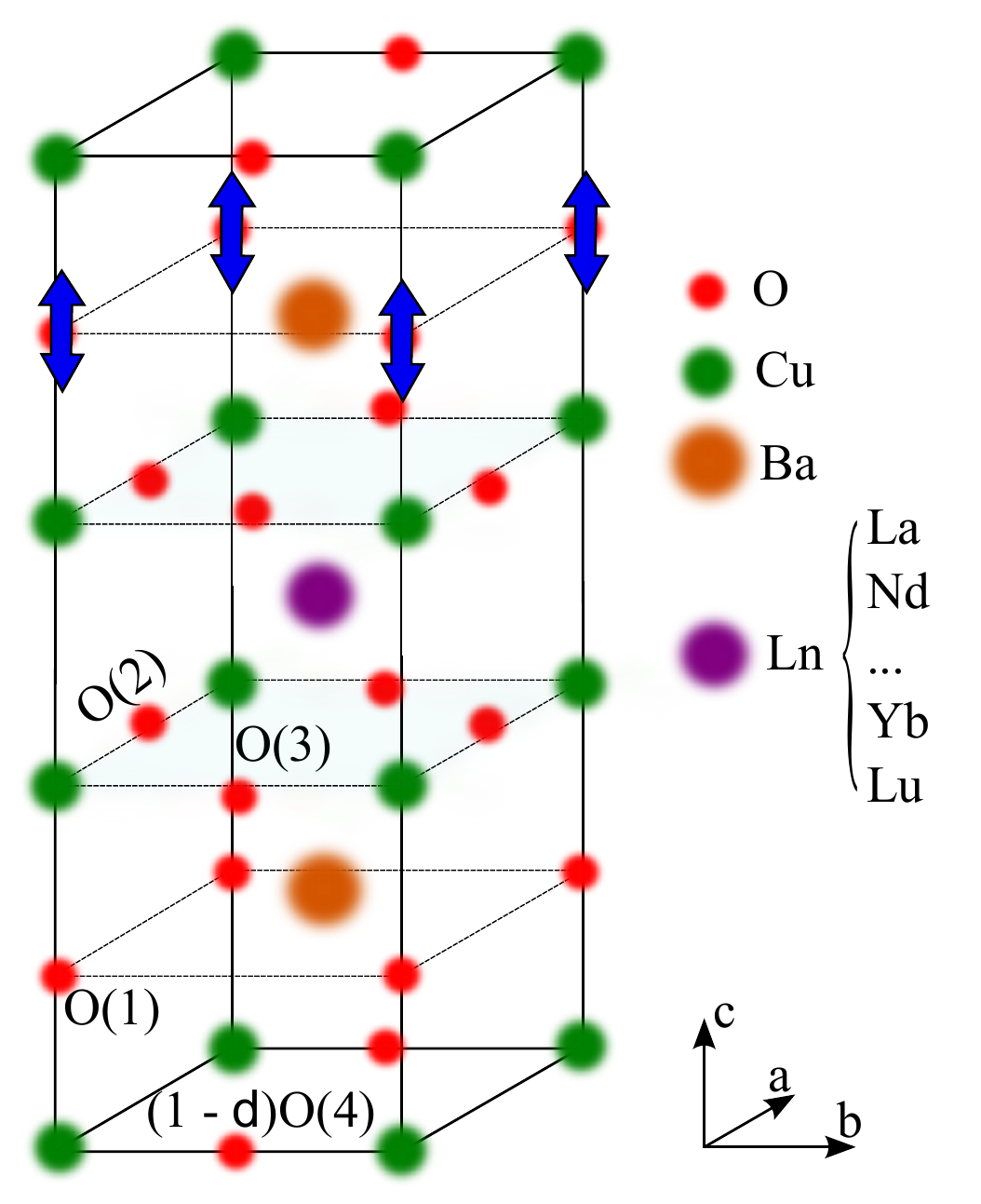} \\ \midrule
			580~\cm  & $A_g$  &  -  & \includegraphics[width=0.225\textwidth]{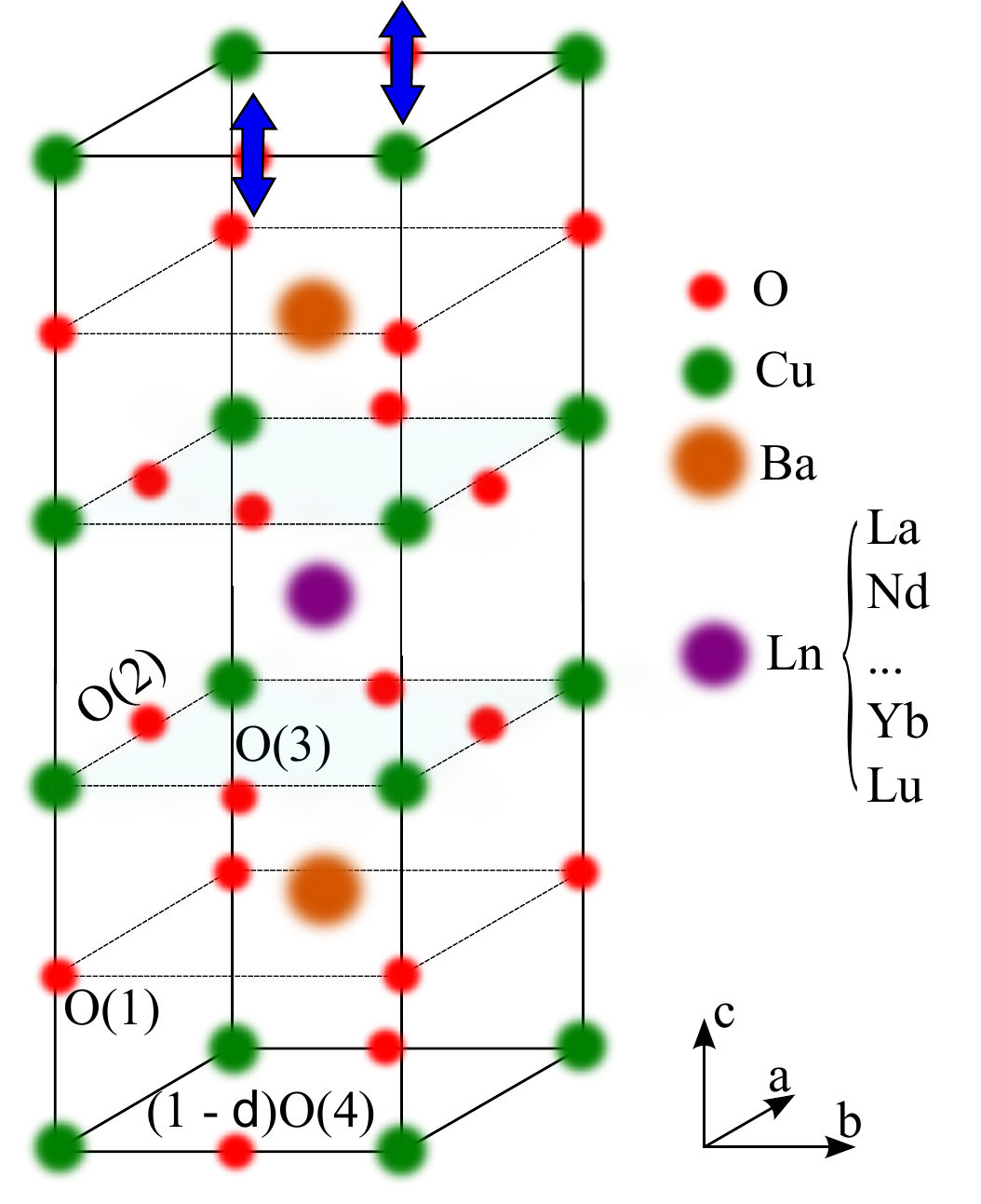} \\ \bottomrule
			
	\caption[The Raman-active phonon modes in Ln123.]{\label{tab:phononmodes}The Raman-active phonon modes in optimally-doped Ln123 named by their approximate energy in Y123.  For each mode a cartoon shows the main ionic displacements involved. The Gr\"{u}neisen parameters, $\gamma = -\dd\ln(\omega)/\dd\ln(V) $, are from \cite{syassen1988}.}
\end{longtable}
\end{center}

\subsection{Optimal doping}
Many physical properties of these materials are doping dependent and \specgap is no exception.  Measuring this doping dependence has been the focus of previous Raman studies.  Here however we are interested in Ln substitution effects and so we compare samples at the same doping state.  When measuring two-magnon scattering we compared undoped Ln123.  When measuring the superconducting energy gap we should compare \emph{optimally doped} Ln123.  

Having determined the annealing conditions for optimal doping on polycrystalline samples, \tab~\ref{tab:ln123annealingconditions}, we co-anneal Ln123 single crystals with polycrystalline pellets of the same material under the conditions required for optimal doping.  As a check, $\rttep$ is measured on the polycrystalline sample after the anneal to confirm that we indeed have optimally doped our polycrystalline (and therefore single crystal) samples\footnote{It has been known for $p$ resulting from some given annealing conditions to be dependent on the history of the sample.}.  The exception is Nd123 which we were not able to overdope.  Instead, we use the most fully oxygenated sample we could prepare\footnote{This involved high O$_2$ pressure, low temperature anneals - see \tab~\ref{tab:o7anneal}.}. 

The \tc values of the single crystals measured below are $T_c=90.5, 94.5$ and $88$~K for Yb123, Eu123 and Nd123 respectively.

\subsection{Data analysis}
We use the `Python' coding language and libraries to analyse much of the following data.

In order to characterize the phonon modes, we first subtract a linear background over a suitable range, which we determine for each peak or pair of peaks, before fitting the peaks themselves.  The peaks are fitted, individually or as pairs, to a Fano profile as specified in \eq~\ref{eq:fano}. Next the `minimize' package \cite{pythonlmfit, scipyoptimize} uses a least-squares minimization algorithm to determine the best fitting parameters within the physically sensible bounds that we specify.  Every fit is visually inspected, and rejected if necessary, before the calculated parameters are recorded. 

\label{sec:greensfitting}
A more sophisticated fitting procedure is outlined in \cite{limonov2000} and the earlier papers \cite{chen1993, bock1999, panfilov1999}.  Below we outline the function and parameters involved, but firstly note that the reason for using this alternative procedure is to obtain an alternate estimate of the ERS component of the spectra.  The fit parameters of the phonons are just as accurately obtained by the procedure described above. 

The entire spectrum is fitted to a coupled phonon (the 330~\cm mode) and electronic term plus five independent, Lorentzian line-shape phonons.  Each independent phonon is described by three parameters; an intensity $A_i$, HWHM $\Gamma_i$ and energy $\Omega_i$ (following similar notation in \cite{limonov2000});
\[
I_i(\omega)=A_i\frac{\Gamma_i}{\Gamma_i^2+(\omega-\Omega_i)^2}
\]
The coupled phonon and electronic term we take from \cite{limonov2000} which is based on a Green function operator to relate the electronic and phonon Raman tensor matrix elements to the imaginary part of the susceptibility tensor $\chi''(\omega)$ \cite{chen1993};
\begin{equation}
I(\omega,T)=A[1+n(\omega,T)]g(\omega)\left\{ 1+ \frac{g(\omega)}{1+g(\omega)}\left[ \frac{\left(S\frac{1+g(\omega)}{g(\omega)}+(\omega-\Omega)\right)^2}{\Gamma_0^2(1+g(\omega))^2+(\omega-\Omega)^2}-1 \right]  \right\}
\label{eq:limonov}
\end{equation}

\noindent This complicated looking expression has the following free parameters; intensity $A$, a scaled $\omega$-independent phonon-electron coupling term $S$, un-renormalised HWHM of the 330~\cm mode $\Gamma_0$, renormalised energy of the 330~\cm mode $\Omega$ and $g(\omega)=V^2\rho(\omega)/\Gamma_0$ where $V$ is the electron-phonon coupling and $\rho(\omega)$ is the imaginary part of the electronic response, see \refsec~\ref{sec:ramaners}.  After Limonov \etal we fit to the following form of $g(\omega)$ with five parameters $C_0$, $C_1$ and $C_2$, $\Gamma_e$ and $D$;
\[
g(\omega)=C_0+\frac{C_1\omega+C_2\omega^2}{\Gamma_e^2+(\omega-D)^2}
\]

\begin{figure}
	\centering
		\includegraphics[width=0.50\textwidth]{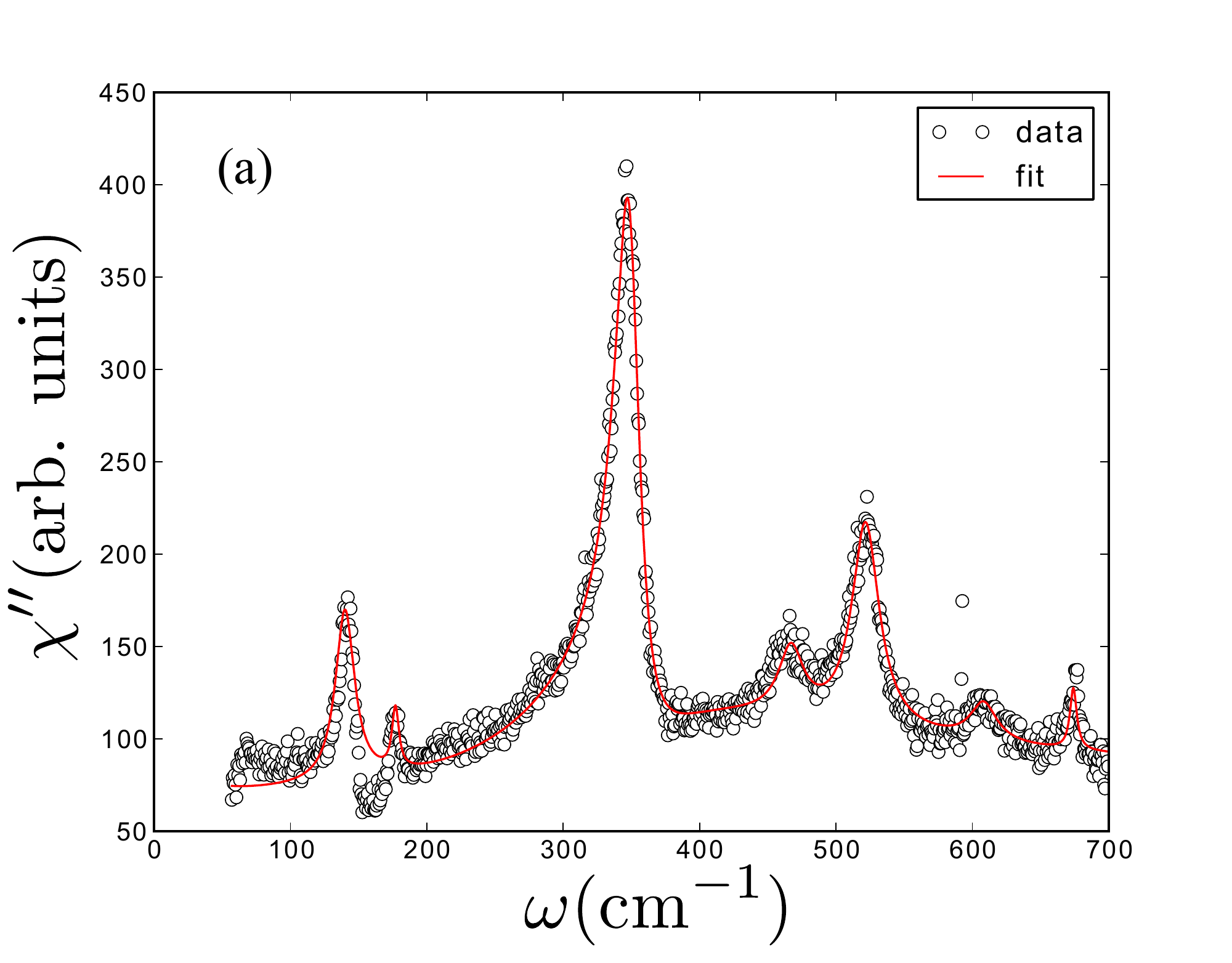}
	\caption[A typical fit to a Raman spectrum using a Greens function approach]{A typical fit using the procedure described in the text, \refsec~\ref{sec:greensfitting}.  The data is from Yb123 in \aogbog symmetry at $T=20$~K.}
	\label{fig:limonovfittingexample}
\end{figure}

We are interested in the frequency where $g(\omega)$ is maximum as it can be identified with the renormalisation peak associated with $\scgapm$.

In practice, fewer than 6 parameters are refined at a time. In general the independent phonon parameters are easily deduced and many other parameters have restrictive sensible bounds; e.g. $\Gamma_e \approx 200$ $\cmm$, $\Gamma_0\approx10$ $\cmm$, $D\approx500$ $\cmm$, $\Omega\approx330$ $\cmm$ and $S\approx - 6$.  A typical fit is shown in \fig~\ref{fig:limonovfittingexample}. The data is from Yb123 in \aogbog symmetry at $T=20$~K.

\section{Results}
\subsection{Instructive spectra}

\begin{figure}
 \includegraphics[width=1.0\textwidth]{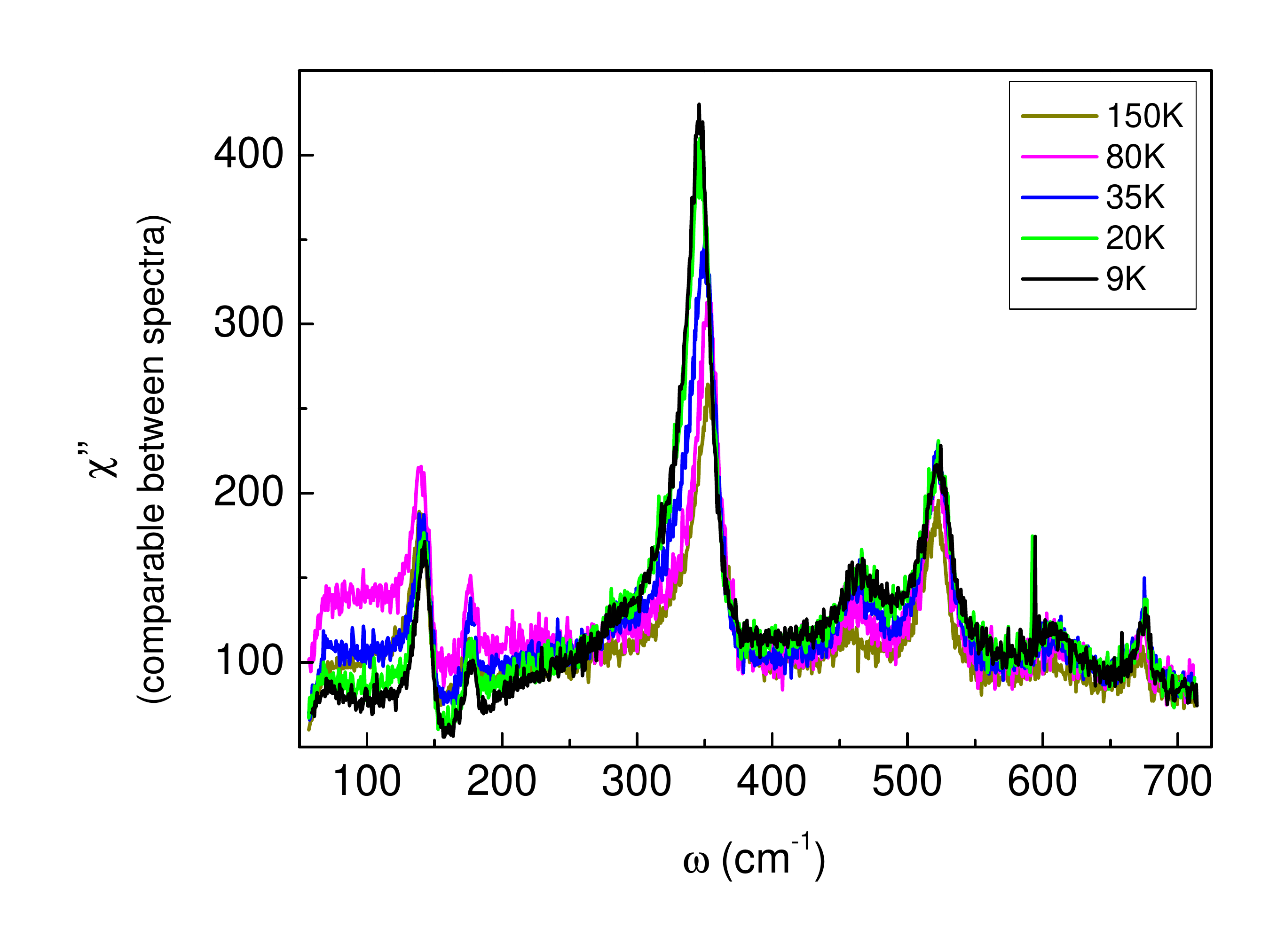}
	\caption[Temperature-dependent \aogbog Raman spectra of Yb123.]{\label{fig:yb123a1gb1graw}Temperature-dependent Raman spectra of optimally-doped Yb123 in \aogbog scattering geometry. The temperature of the cold finger is shown in the legend and each spectrum has been multiplied by the appropriate $[n(\omega,T)+1]^{-1}$ factor. We estimate that the sample temperature may be $\sim1$~K higher than the indicated temperature due to laser heating. Other than the temperature, all experimental parameters were kept constant so the intensity differences between spectra are due to temperature-dependent sample properties. These data show several interesting physical phenomenon including modification to the electronic scattering response (ERS) and phonon self-energy renormalisation due to the a gap in the electronic density of states (DOS).}
\end{figure}

In \fig~\ref{fig:yb123a1gb1graw} we plot spectra from optimally-doped Yb123 in \aogbog symmetry at various temperatures as indicated in the legend.  The only manipulation of these spectra has been multiplication by the appropriate Bose occupation factor, $[n(\omega,T)+1]^{-1}$. These data nicely demonstrate some really interesting physics;

\begin{enumerate}
 \item `Hardening'\footnote{Nomenclature used in the Raman community meaning increasing Raman shift, $\omega$, when an experimental parameter is varied e.g. an increase in energy of the feature.} of phonon modes with decreasing temperature due primarily to lattice contraction with temperature.
 

The temperature dependence of the $i^{\textnormal{th}}$ phonon mode's energy due to lattice contraction can be expressed as \cite{postmus1968, borer1971, menendez1984};
\begin{equation}
\omega_{i}(T)=\omega_{i}(0)\exp\left[ 3\gamma_i \int^T_0{\alpha(T')\dd T'}\right]
\label{eq:modenormaltempdependence}
\end{equation}
\noindent where $\gamma_i$ is the Gr\"{u}neisen parameter and $\alpha$ is the linear coefficient of thermal expansion.  $\alpha$ is anisotropic in Ln123 \cite{meingast1990}, so we replace $3\gamma_i \int^T_0{\alpha(T')\dd T'}$ in \eq~\ref{eq:modenormaltempdependence} with 
\[
\gamma_i \sum_{j=a,b,c}{ \int^T_0{\alpha_j(T')\dd T'}}
\] 

\begin{figure}[ht]
\centering
	\includegraphics[width=0.55\textwidth]{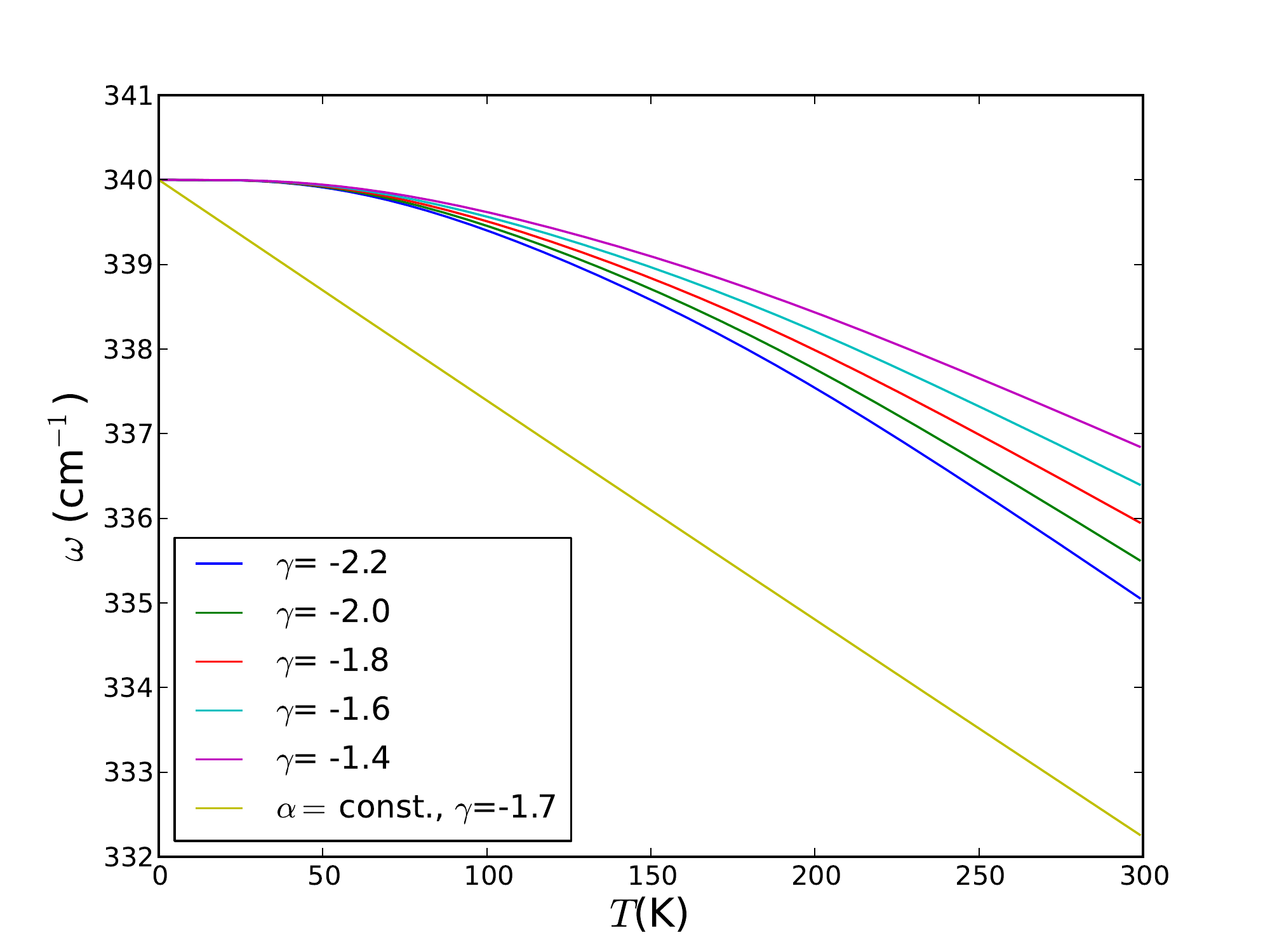}
	\caption[Calculated temperature dependence of the shift phonon frequency solely due to lattice contraction.]{\label{fig:340modetempdependence} Calculated temperature dependence of the shift in a $T=0$~K, 340~\cm phonon frequency solely due to lattice contraction \cite{postmus1968, borer1971, menendez1984}. This is calculated from \eq~\ref{eq:modenormaltempdependence} with the thermal expansion data of \cite{meingast1990} and for various sensible values of the Gr\"{u}neisen parameter, $\gamma$, as indicated in the legend \cite{syassen1988}.  For comparison the temperature dependence is also shown if thermal expansion is taken to be a temperature independent value, $\alpha=15.1\times 10^{-6}$~K$^{-1}$. Note that the Gr\"{u}neisen parameter determines the gradient of the curve.}
\end{figure}

\fig~\ref{fig:340modetempdependence} shows $\omega (T)$ calculated with $\omega (0)=340$~$ \cmm $, thermal expansion data from Meingast \etal \cite{meingast1990} and a range of experimentally and theoretically estimated values of $\gamma$ for the relevant Y123 modes: $\gamma = - 1.7 \pm 0.5$ \cite{syassen1988}.   

In \fig~\ref{fig:peakpos}, fits of this model temperature dependence to our data are shown as solid lines.  Mode-specific $\gamma$ were from Syassen \etal \cite{syassen1988} were used in the fitting. Only $T>T_c$ data points were included in the fitting, and although there are sometimes few data points to determine the fit from, generally\footnote{Occasionally it seemed appropriate to also fit $\gamma$ within the physically sensible range $[-1.0,-3.0]$ when there was sufficient data points above \tc.  As can be seen from \eq~\ref{eq:modenormaltempdependence} and shown in \fig~\ref{fig:340modetempdependence}, $\gamma$ determines the gradient of $\omega_{i(T)}$.} $\omega_{i(0)}$ is the only free parameter.   As can be seen in \fig~\ref{fig:peakpos}, the fits are generally satisfactory and serve to clearly show where $\omega_{i(T)}$ is renormalised by a gap in the electronic density of states, as discussed below.  


\item The asymmetric line shape of the peaks is due to the Fano effect and indicates coupling between phonons and the electronic continuum, \refsec~\ref{sec:fanolineshape}. 

We use the Fano line shape in the fitting routine noting that more complicated fitting routines give similar results \cite{limonov2000}.

\item `Softening' of some phonon modes with decreasing temperature - contrary to the behaviour described above.  The theory behind this effect is explained by Zeyher and Zwicknagl \cite{zeyher1990} in the case of an $s$-wave superconducting order parameter and generalised by Nicole, Jiang and Carbotte \cite{nicol1993} to the case more relevant for the cuprates, of a gap with nodes. An opening of a gap in the electronic density of states (e.g. due to the opening of a superconducting gap) renormalises the energy and life-time (as revealed by the HWHM) of a mode via electron-phonon coupling and via modified scattering channels.

In the simplest model, a particular mode will `anomalously' \emph{soften} if its unrenormalised energy is below \specgap while \emph{harden} if its unrenormalised energy is above $ \specgapm $.  This is illustrated in \fig~8 and 9 from Zeyher and Zwicknagl \cite{zeyher1990}. 


\item The conventional temperature dependence of the half-width-at-half-maximum (HWHM) of a phonon mode (which is related to the half-life of that phonon excitation) takes the form \cite{klemens1966, weber2000, limonov2000};
\begin{equation}
 \label{eq:anharmhwhmbroaden}
\Gamma{(\omega_p,T)} = \Gamma{(\omega_p,0)} \left[ 1+ 2n{(\omega_p/2,T)}\right] + \Gamma_{\textnormal{imp}}
\end{equation}

\noindent $\Gamma_{\textnormal{imp}}$ results from phonon scattering off impurities, similar to the residual resistivity in metals as $T\rightarrow 0$K, and $n(\omega,T)$ is the Bose-Einstein distribution function.  Physically this represents a phonon decaying into two phonons of opposite momenta, each with energy of $\omega_p/2$.  Fits to this function, with similar comments relating to the phonon energy fits, are shown as solid lines in \fig~\ref{fig:peakpos}.

The HWHM is also renormalised by the \specgap as shown in \fig~\ref{fig:nicolfig1}. This manifests as an increased HWHM when the energy of the mode is close to $\specgapm$. Physically it results from additional scattering opportunities available to the phonon, e.g. Cooper pair breaking.
   
\item Depression of the `background' at low temperature and low $\omega$.  The background is due to Raman scattering from the electronic continuum in the material (ERS) as opposed to phonon modes which have well-defined energies.  The reduction in intensity at low energies arises from the opening of a gap in the electronic density of states at temperature $T<T_c$.

\end{enumerate}


\subsection{\specgap estimates from phonon shifts}

\fig~\ref{fig:peakpos} shows the energy and half-width-at-half-maximum (HWHM) as a function of temperature for various phonon modes.  These parameters were determined from fitting to a Fano line-shape with a linear background.

In the simplest model the renormalised energy of a particular mode will be lower if its unrenormalised energy is below $\scgapm$ but larger if its unrenormalised energy is above $\scgapm$.  This is illustrated in \fig~8 and 9 from Zeyher and Zwicknagl \cite{zeyher1990}.  In our data we clearly see renormalisation of both the energy and HWHM of phonon modes. There are however several further considerations which we now discuss.  

We will actually be measuring some combination of the superconducting gap and pseudogap, namely $\specgapm$.  Again, we note that \specgap is a \emph{pair-breaking} energy. The \specgap energy obtained will depend on the symmetry of the particular mode.  For example, calculations by Devereaux \etal \cite{devereaux1994} show the different responses of an \aog and \bog symmetry phonon to a spectral gap with nodes.  In these calculations the transition from renormalised hardening to softening for \aog phonons broadens but remains centred close to $\scgapm$. The range of \aog phonon energies whose HWHM are renormalised by the energy gap increases, and the energy of maximum broadening decreases with respect to \bog phonons.  Physically, these \aog phonons are providing pair-breaking excitations in regions of the Fermi-surface where \scgapk vanishes at the nodes.

Impurity scattering will further broaden out in energy the transition between renormalised and unrenormalised phonon excitations. The main concern for us however is that impurity scattering also alters the magnitude of the relative phonon shifts that would have allowed a more accurate estimate of \specgap, as shown in \fig~8 and 9 from Zeyher and Zwicknagl \cite{zeyher1990}.  

However, the chief consideration, comes from the extension from the $s$-wave to $d$-wave gap model described by Nicol \etal \cite{nicol1993}.  Their findings are more applicable to our materials as they consider the effects of anisotropic gaps with nodes rather than an isotropic $s$-wave gap. 

\begin{figure}[tb]
	\centering
		\includegraphics[width=0.50\textwidth]{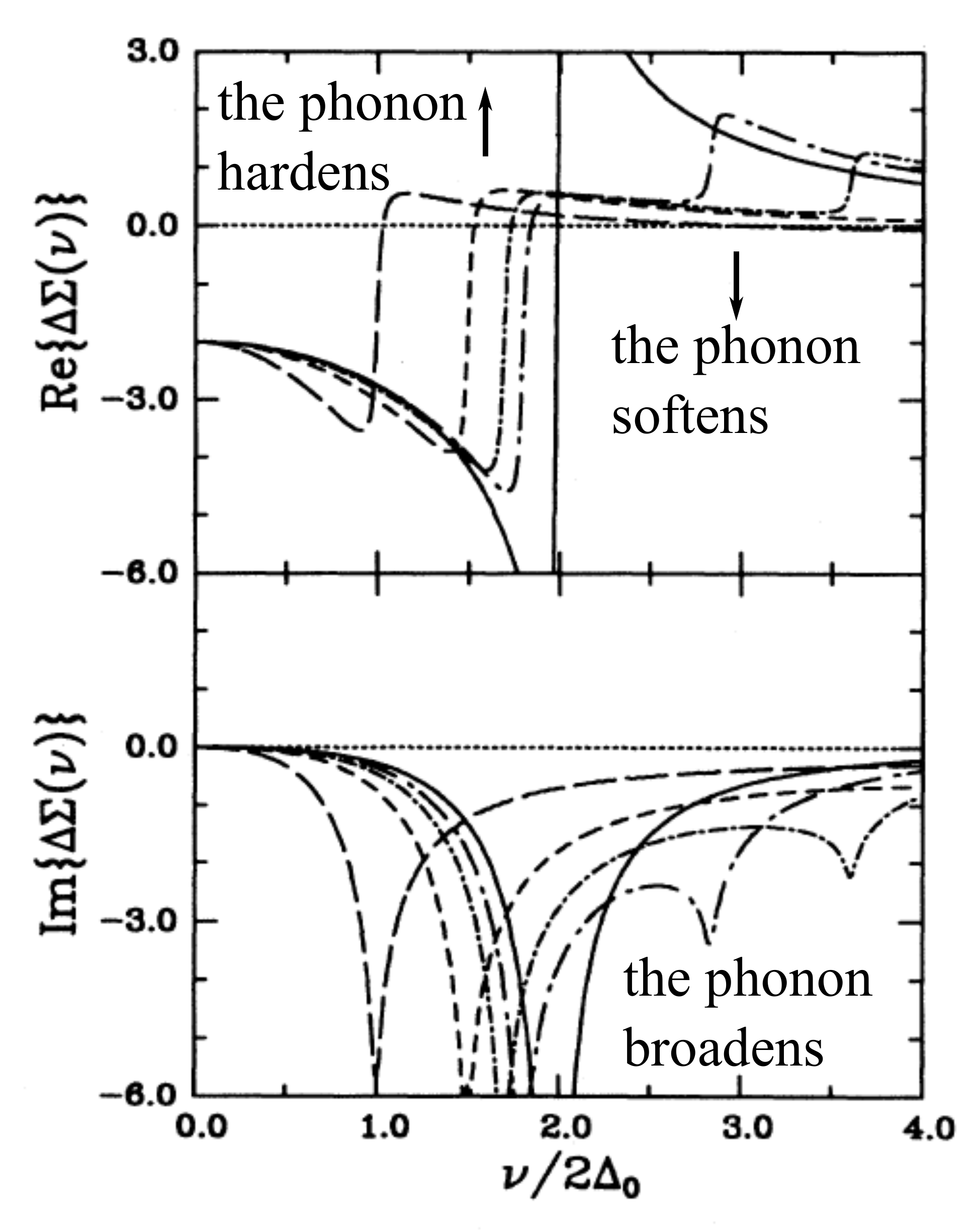}
	\caption[Reproduction of Fig.~1 from Nicol \etal showing the renormalisation of phonon energy and lifetime in the presence of a $ \specgapm $.]{\fig~1 from \cite{nicol1993} supplemented with our annotations. The figure presents the results of calculations of the phonon self-energy, $\Delta \Sigma(\upsilon)$, as a function of energy $\upsilon$ scaled by the spectral gap in the low-temperature limit, $2\Delta_0$. As annotated the real part of the self-energy results in a shift (renormalisation) of the phonon energy, whilst the imaginary part relates to the width (life-time or scattering rate) of the phonon mode. The solid line corresponds to $\bar{\mu}\equiv \mu/2t=0$ where $\mu$ is the chemical potential and $t$ the hopping integral in the Hubbard model, which is extremely sensitive to the super-exchange path-length. The long dashed curve is $\bar{\mu}=-1$ and the other curves are for intermediate values.}
	\label{fig:nicolfig1}
\end{figure}

\begin{figure}
	\centering
		\includegraphics[width=0.45\textwidth]{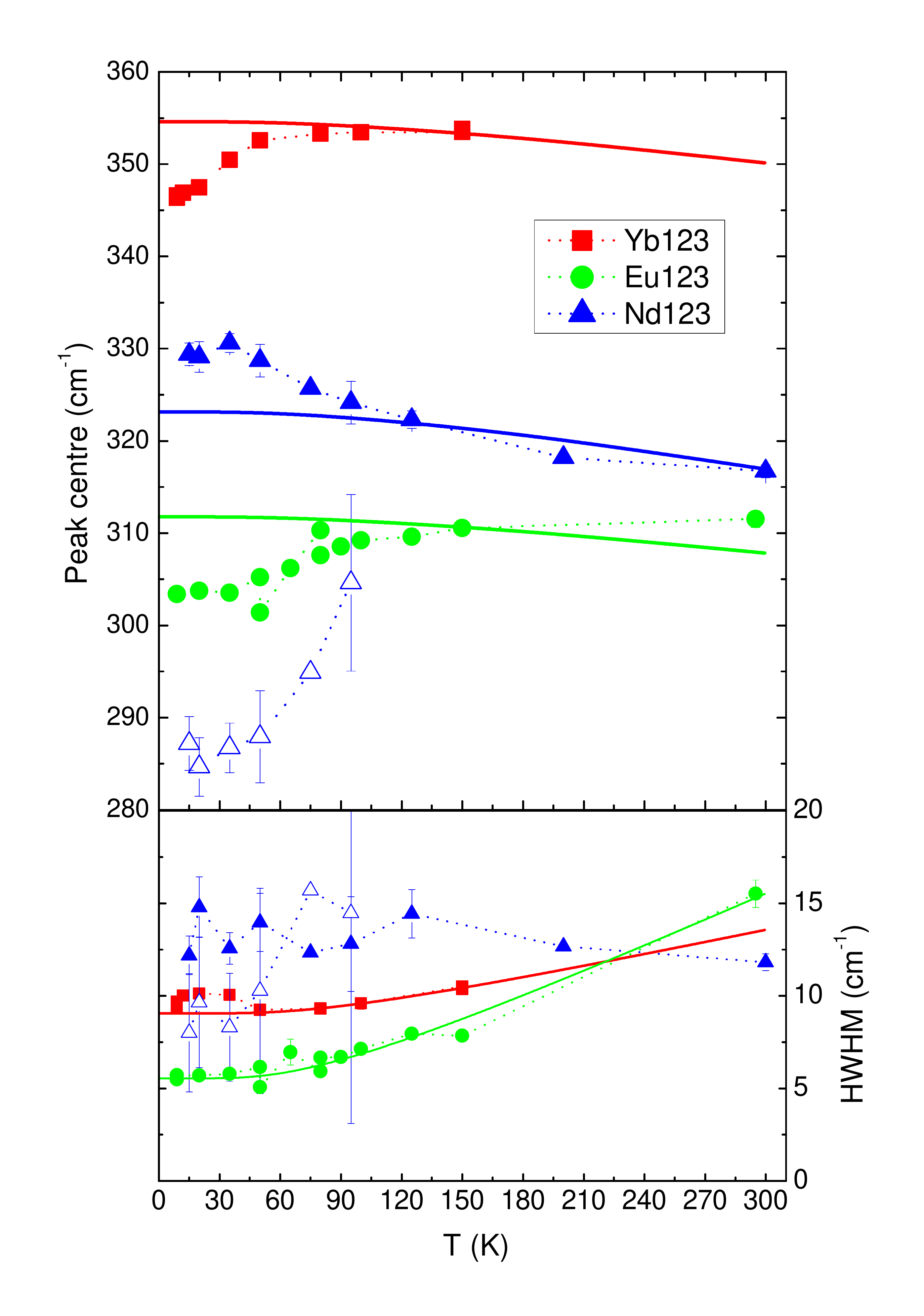}
		\includegraphics[width=0.45\textwidth]{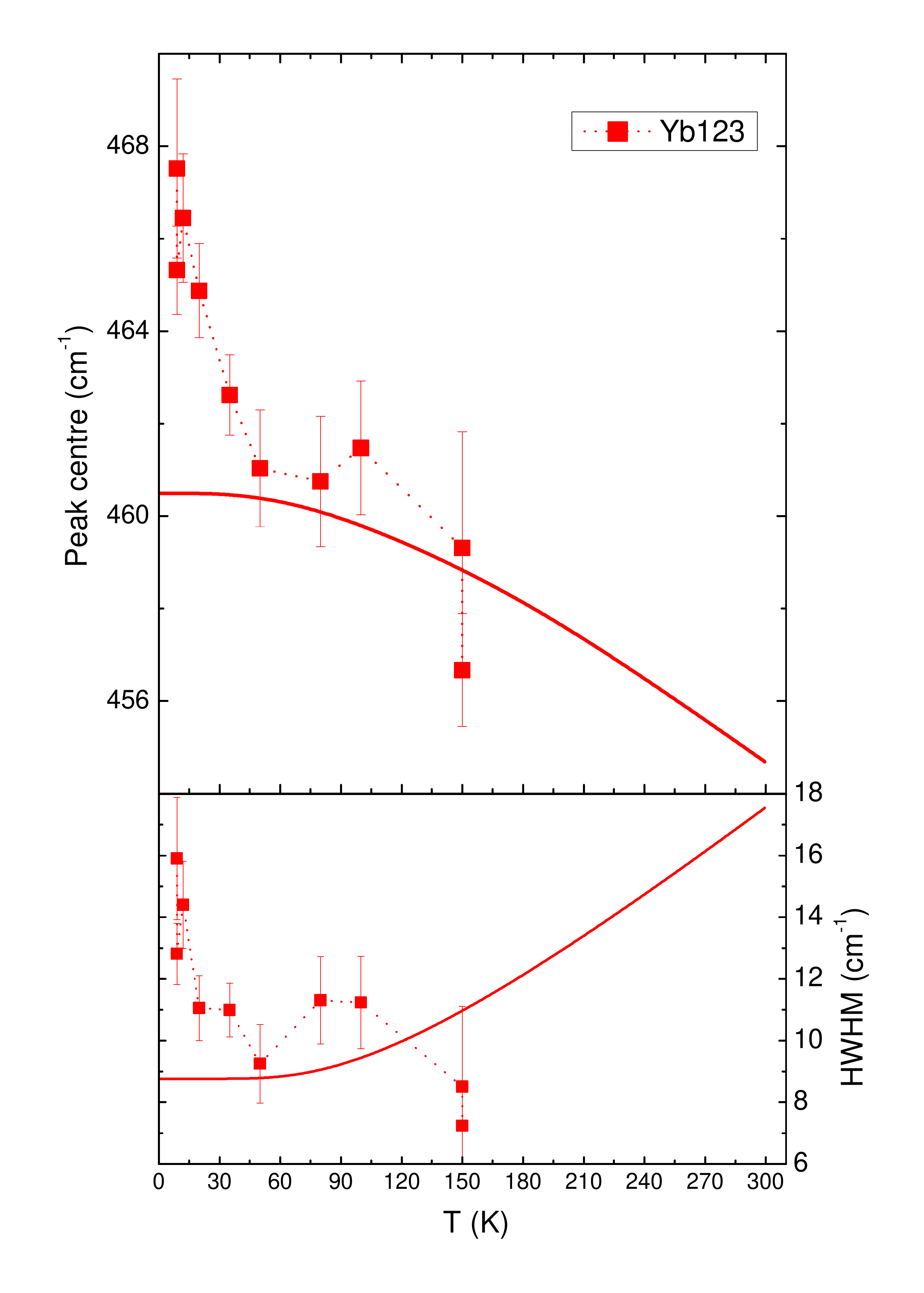}
		\includegraphics[width=0.45\textwidth]{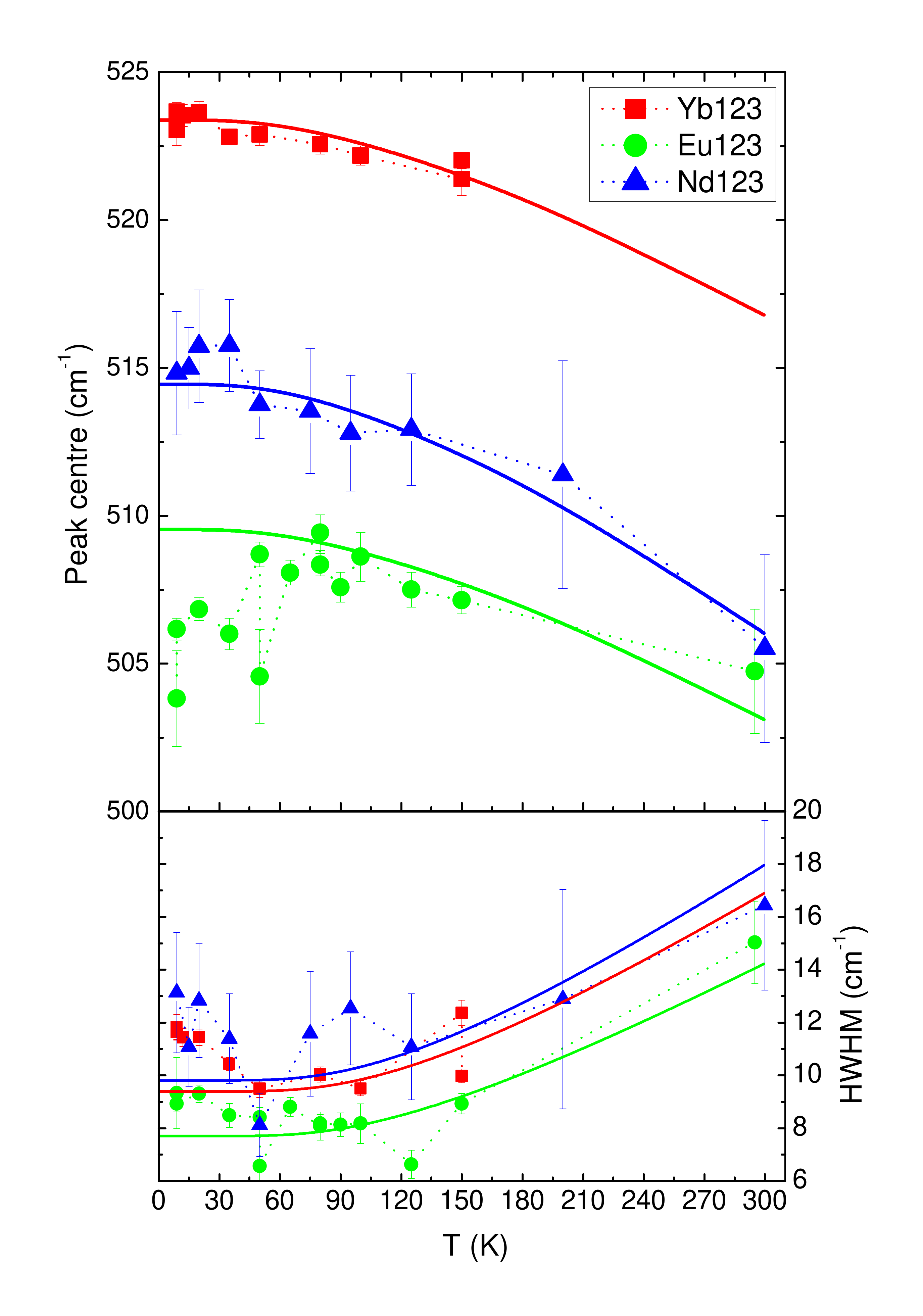}
		\includegraphics[width=0.45\textwidth]{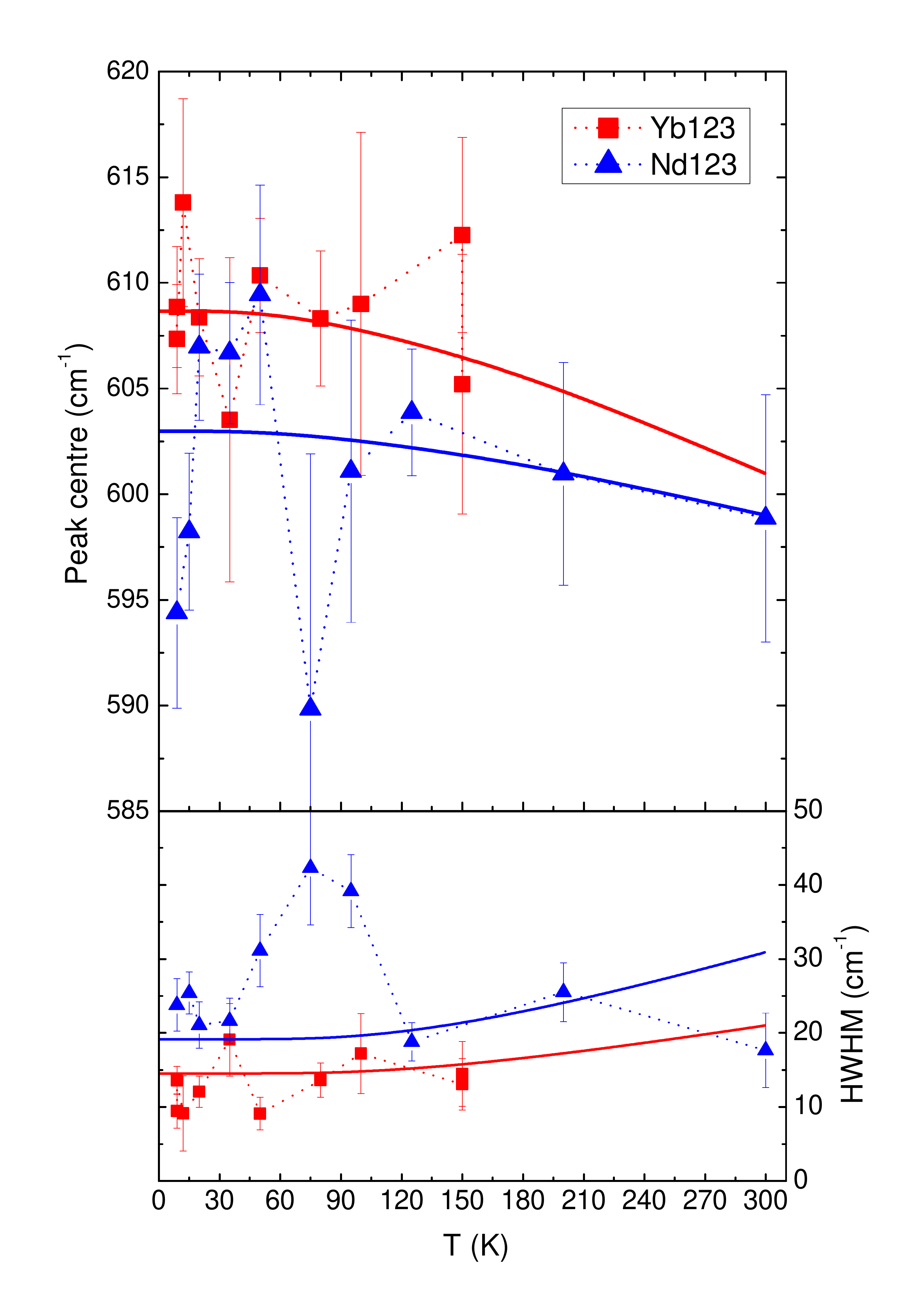}
	\caption[Temperature dependence of phonon mode energies and HWHM for optimally doped Yb123, Eu123 and Nd123.]{Phonon mode energies and HWHM determined from fitting to a Fano line shape.  Similar data for the 110, 150 and 675~\cm modes are presented in \fig~\ref{fig:otherpeakpos}.  Solid lines are fits of the data above \tc to the conventional temperature dependence for the mode energy, given by \eq~\ref{eq:modenormaltempdependence}, or HWHM, given by \eq~\ref{eq:anharmhwhmbroaden}.  }
	\label{fig:peakpos}
\end{figure}

Nicol \etal introduce a normalised chemical potential (or `filling factor'), $\bar{\mu}\equiv \mu/2t$ where $t$ is the hopping integral, that alters the `crossover energy' from renormalised softening to renormalised hardening.  In  \fig~\ref{fig:nicolfig1} we reproduce \fig~1 from their paper \cite{nicol1993} which illustrates this point (our annotations).  The solid line corresponds to $\bar{\mu}=0$ (half-filling or $p=0$), the long dashed curve is $\bar{\mu}=-1$ and the other curves are for intermediate values.  The upper panel plots the real part of the self-energy that relates to the energy of the mode, while the lower panel plots the imaginary part that relates to the HWHM.  For the isotropic $s$-wave case the crossing energy is $\omega = 2\Delta_0$ but for the \dxy case this energy now becomes $\omega= 4\Delta_0 - 2\Delta_0|\bar{\mu}|$.  At half-filling the crossing is in fact at $4\Delta_0$!  This reduces to $2\Delta_0$ predicted by Zeyher Zwicknagl \emph{if} $\bar{\mu}=-1$.  

Can we evaluate $\bar{\mu}$ for our compounds? Firstly, note $\bar{\mu} \sim t^{-1}$. The hopping integral in the Hubbard model, $t$, is related to the hopping integral between $p$ and $d$ orbitals in a two-band Hubbard model presumably as $t\sim t_{pd}^2$.  From Aronson \etal \cite{aronson1991}, $t_{pd}\sim r^{-n}$ where $r$ is the superexchange path length and $2.0<n<3.0$. $\bar{\mu}$ would appear to be very sensitive to our ion-size substitutions that result in lattice parameter variations (\refsec~\ref{sec:ln123systematics}): $\delta\bar{\mu}\sim -r^{-4}\delta r$. To the best of our knowledge, ion-size effect on $t$ remains to be tested and we discuss this issue in more detail later in \refsec~\ref{sec:incoherentpolar}.

With sufficient data all of these parameters could be extracted from a fitting procedure to the full theory.  Unfortunately, we are limited by the number of phonon-modes nature gives us (2-3 in the relevant region) and as such this approach would be inappropriate.  The solution to this problem for the authors of \cite{friedl1990} was to treat Y123 and various Ln123 as identical materials other than their shifted phonon peaks.  In this way they gained enough data points to fit to the theory!  This approach yielded the somewhat low estimate of $2\Delta_0= 39.1\pm0.7$ meV (c.f. the values in \tab~\ref{tab:ramangapsizesummary}).  These authors used the $s$-wave model of Zeyher and Zwicknagl - their work was published before the Nicol paper and at that time, 1990, it was far from clear whether the superconducting gap had $s$-wave symmetry or otherwise.  Nevertheless, they achieve good agreement between their measured phonon widths and the model using mostly experimental or calculated parameters.

\begin{figure}
	\centering
		\includegraphics[width=0.45\textwidth]{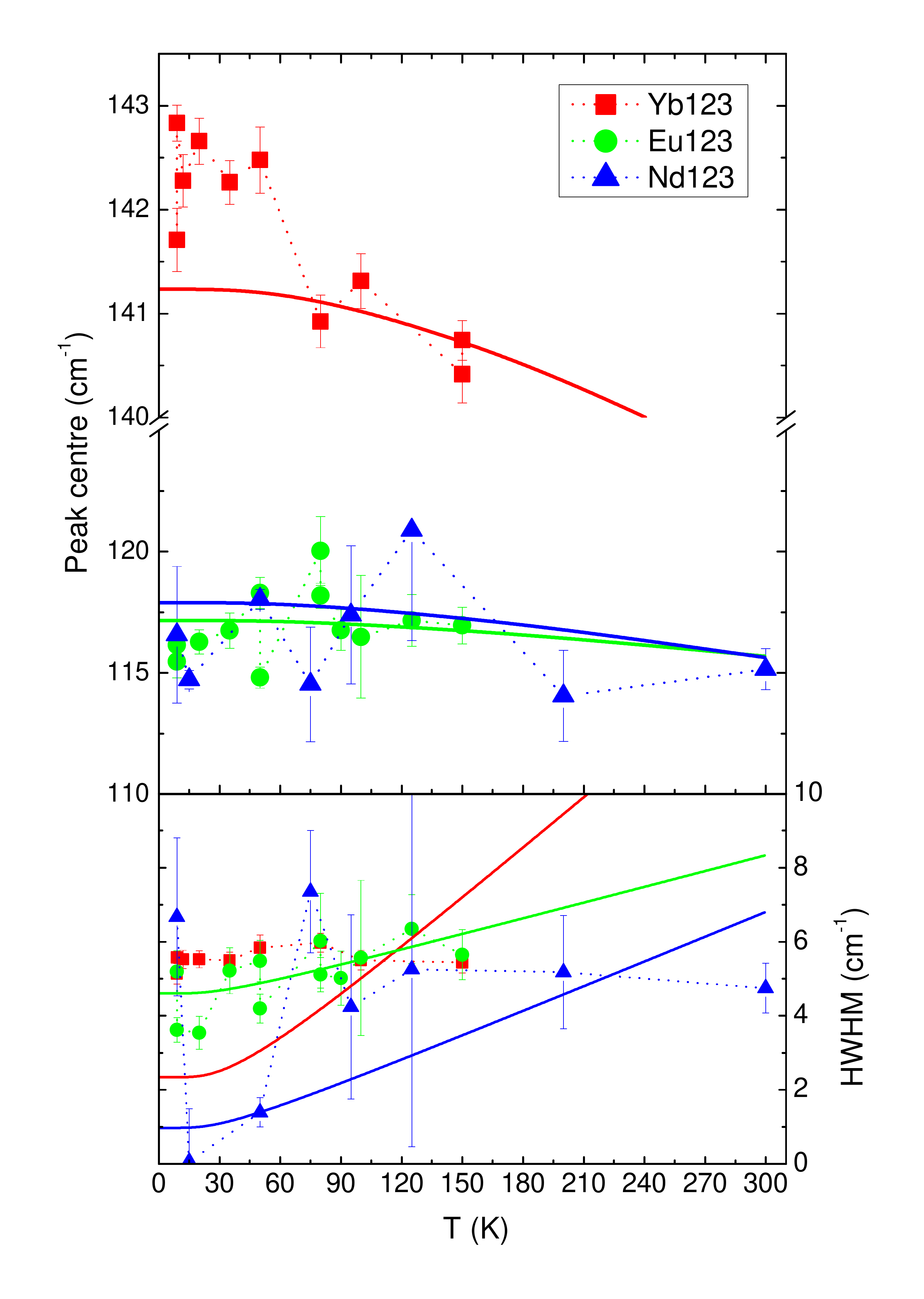}
		\includegraphics[width=0.45\textwidth]{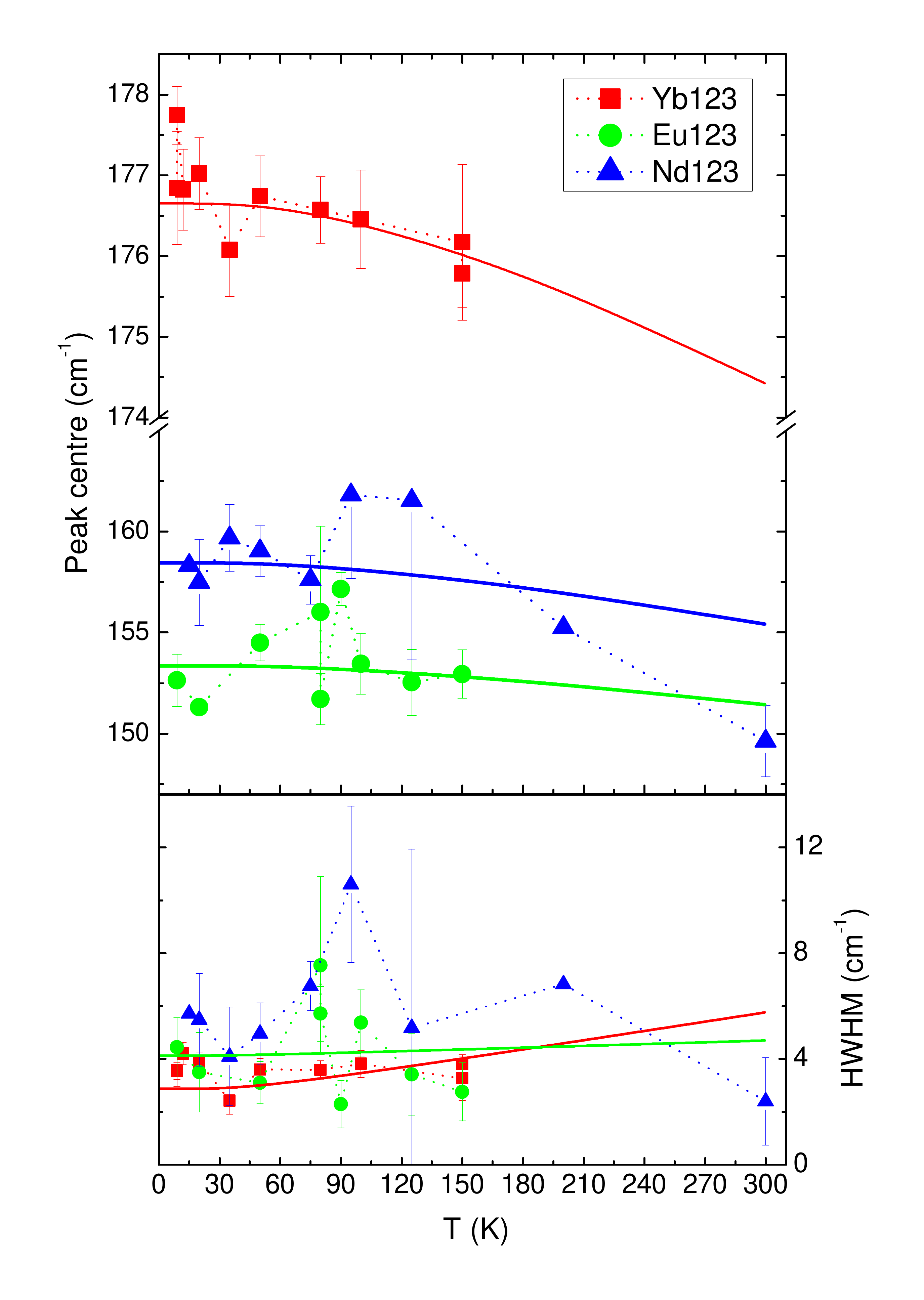}
		\includegraphics[width=0.45\textwidth]{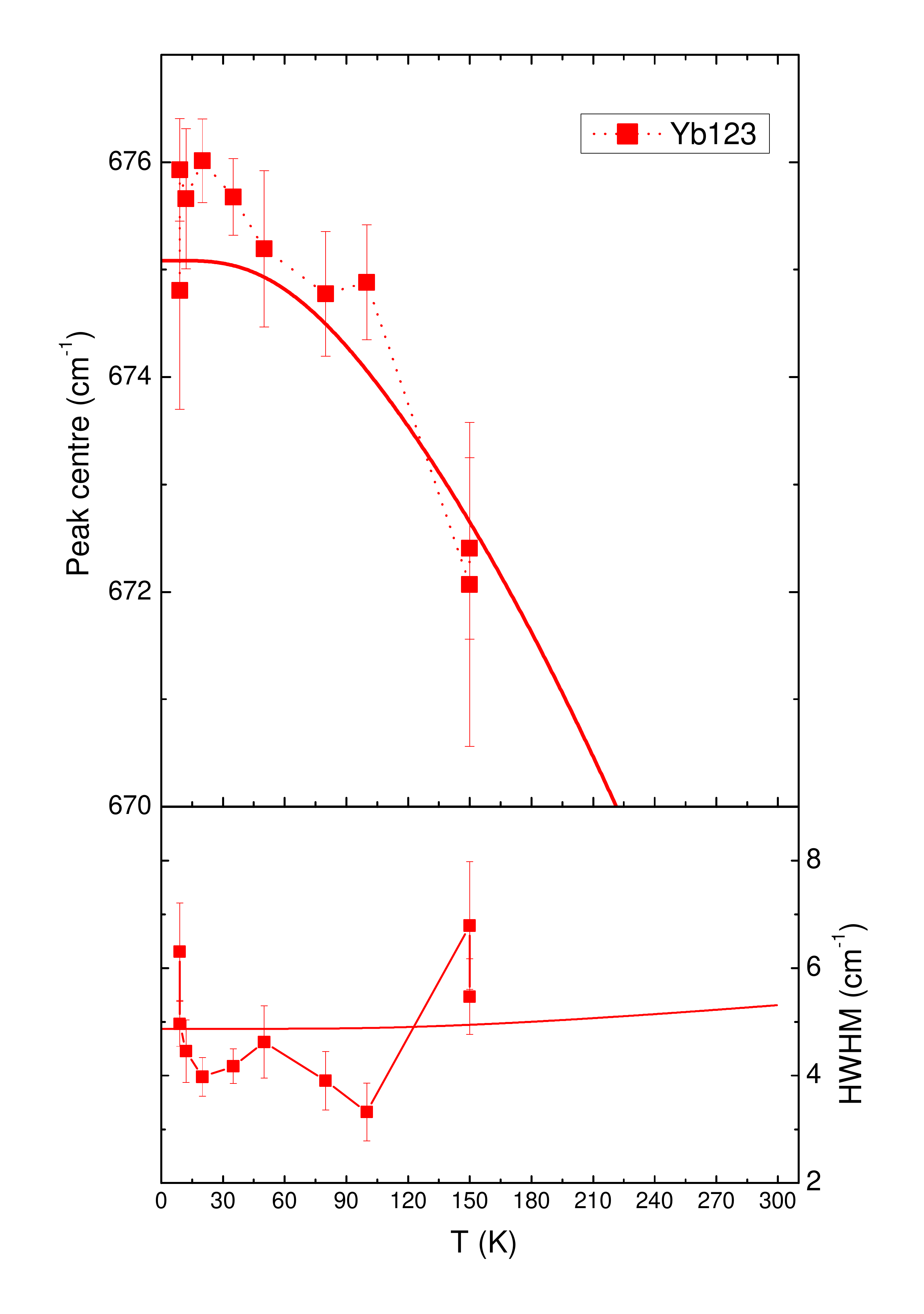}
	\caption[Temperature dependence of more phonon mode energies and HWHM for optimally doped Yb123, Eu123 and Nd123.]{Phonon mode energies and HWHM determined from fitting to a Fano line shape. Solid lines are fits of the data above \tc to conventional temperature dependences. }
	\label{fig:otherpeakpos}
\end{figure}

The aforementioned complications compel us to regard phonon renormalisation measurements as inappropriate for an accurate determination of any variation in \specgap between Nd123 and Yb123. However, we can place bounds on \specgap from these data. For example, renormalised softening only occurs for $\omega<4\hscgapm$ - from which we can place a strict lower bound on $\scgapm$ for Yb123 and Eu123.  Such a lower bound would be overly conservative because we know that $\bar{\mu}<0$ (our samples are doped).  It is likely that $\bar{\mu}\approx-1$ given the work of Friedl \etal \cite{friedl1990} described above and so we use the result that the transition from renormalised softening to hardening occurs at $\approx \specgapm$.  HWHM renormalised-broadening will be largest at $\approx \specgapm$, but present at lower and higher energies. 

We note the renormalisation of the 330~\cm \bog phonon shown in \fig~\ref{fig:peakpos} is largest only at $T\approx 60$~K, which is 30~K lower than \tc where the SC gap opens. Depending on the relative energy of the mode and the SC gap, this is not too surprising, see for example \cite{limonov2000}. 

\subsubsection{Yb123}


For Yb123 spectra we observe a conventional temperature dependence (solid line) of the `500~\cm mode' energy.  At low temperature there appears to be additional broadening of this mode, which signals a proximate energy gap.  The two higher energy phonon modes are completely conventional. On the other hand the 465~\cm (57.5 meV) mode \emph{hardens} at low temperatures, by $4\pm 1$~\cm more than one would expect lattice contraction alone (solid line).  In contrast, the `330~\cm (41~meV) mode' \emph{softens} by $-8.0\pm0.5$~\cm.  With the assumption the transition from softening to hardening occurs at $\approx \specgapm$ these observations lead to the follow plausible energy range for $\specgapm$: $39$ $\textnormal{meV}<\specgapm<58$~meV. 

\subsubsection{Eu123}
\begin{figure}
	\centering
		\includegraphics[width=0.66\textwidth]{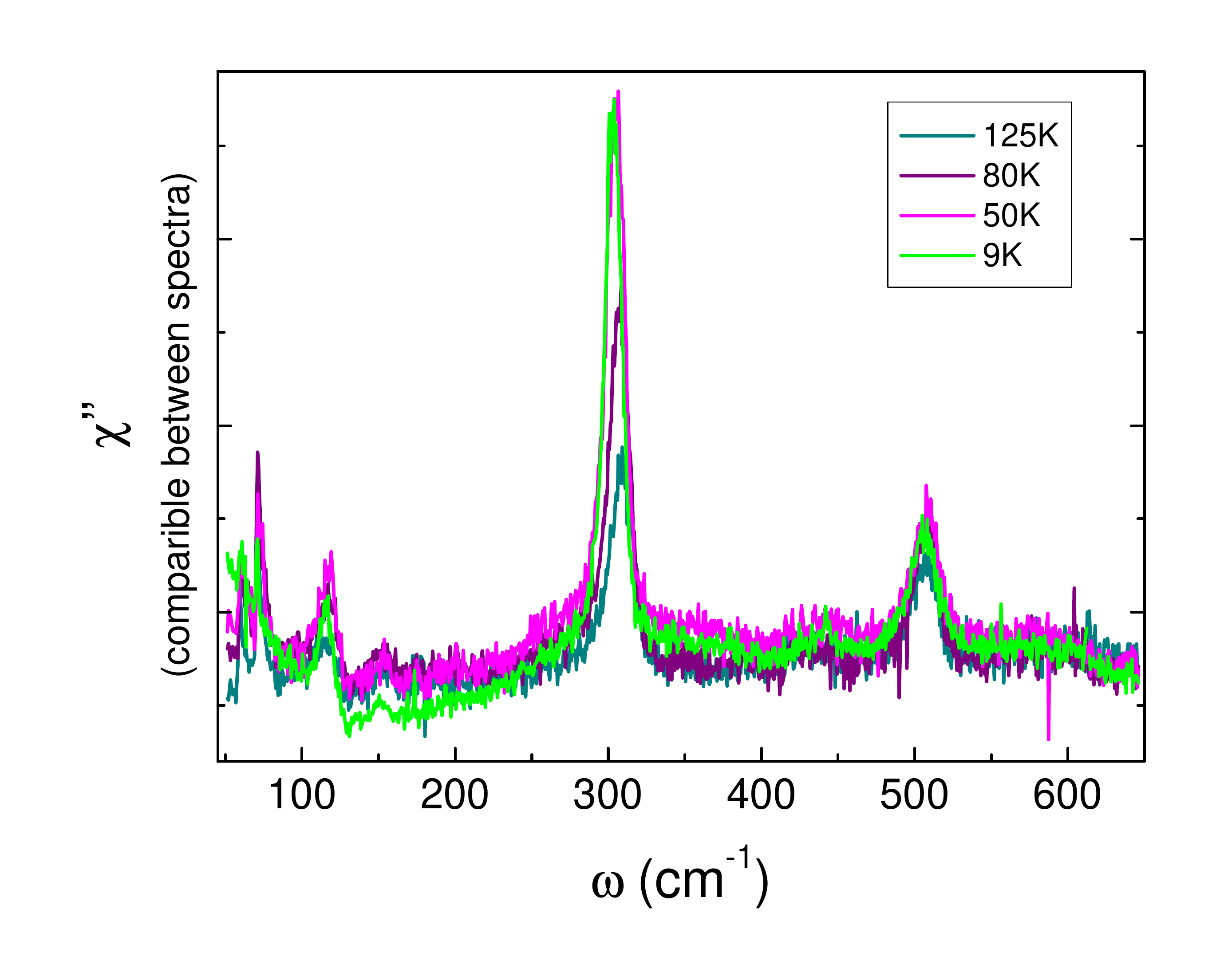}
	\caption[\aogbog spectra of optimal doped Eu123 at various temperatures.]{Representative \aogbog spectra of optimal doped Eu123.  Temperature is indicated in the legend.}
	\label{fig:eu123a1gb1g}
\end{figure}

The \aogbog spectra for Eu123 are shown in \fig~\ref{fig:eu123a1gb1g}.  There are fewer phonons of significant intensity in these data.  Of real interest is the $4\pm 1$~\cm renormalised softening of the 500~\cm (63 meV) mode, immediately suggesting $63$ $\textnormal{meV}<\specgapm$.  The 330~\cm mode also shows a $6\pm 2$~\cm softening (the large uncertainty given to this value reflects the poor fit from the unexpected softening of the phonon mode above \tc). In contrast to the 500~\cm mode, there is negligible additional broadening of the 330~\cm mode at low temperatures.  

Unfortunately, the higher energy phonon modes so clearly visible in the Yb123 spectra do not make an entrance in our Eu123 spectra which prevents us from putting an upper limit on \specgap based on these data.

\subsubsection{Nd123}
\label{sec:ramangapsnd123}

\begin{figure}
	\centering
		\includegraphics[width=0.66\textwidth]{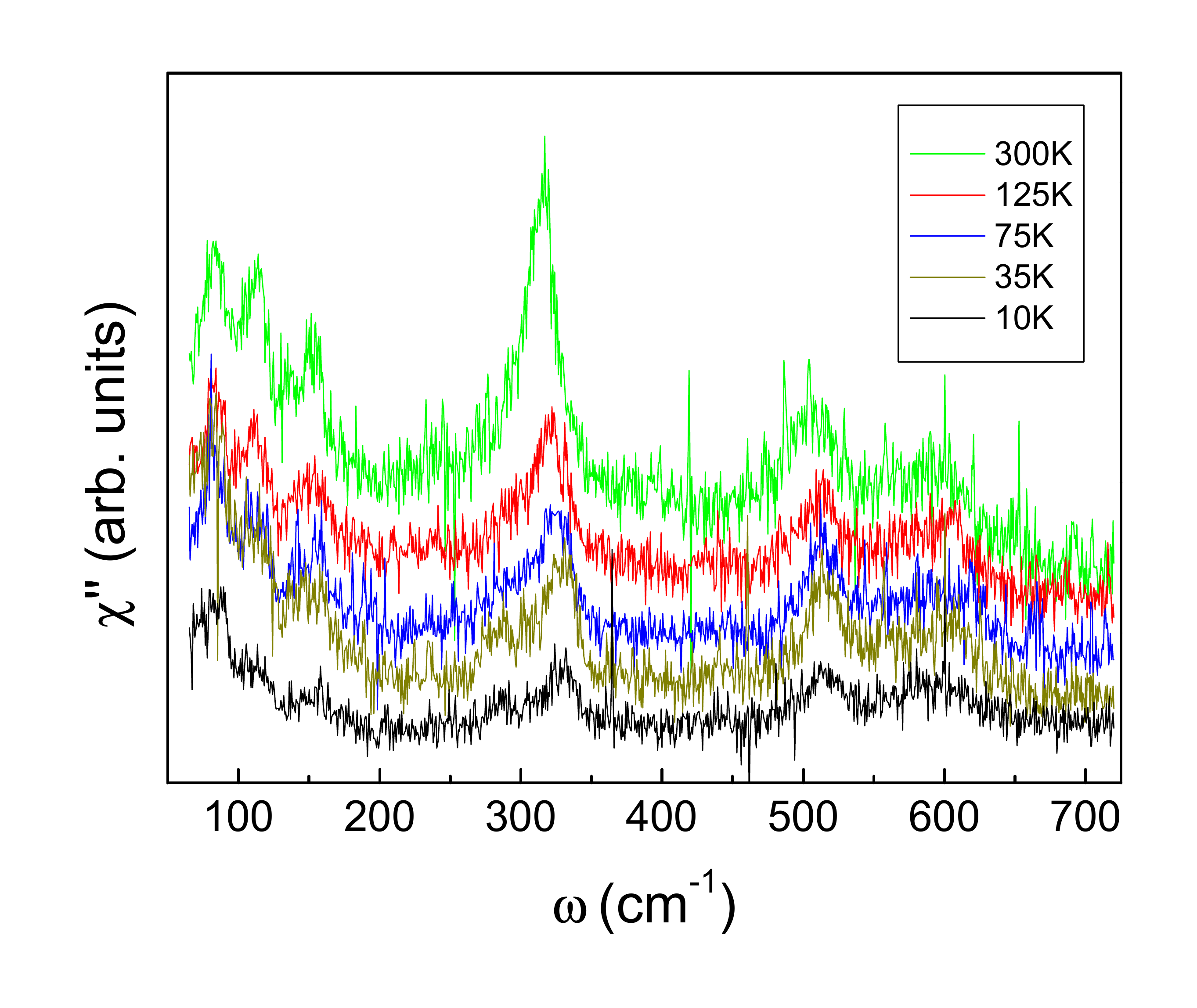}
	\caption[\aogbog spectra of optimal doped Nd123 at various temperatures.]{Representative \aogbog spectra of oxygen loaded Nd123.  Temperature is indicated in the legend \added{and the spectra have been vertically offset for clarity}.}
	\label{fig:nd123a1gb1g}
\end{figure}

The Nd123 spectra shown in \fig~\ref{fig:nd123a1gb1g} have several strange features which will be discussed shortly.  To begin with however we discuss our estimate of the gap energy. 

There is no anomalous temperature dependence of the 500~\cm mode energy (or for that matter the 610~\cm mode, although there is considerable scatter in fitting results here). It is possible however that there is some additional broadening at low temperatures.  That places an upper bound of $\specgapm \approx 63$~meV.

With a larger unit cell than both Eu123 and Yb123, one would not expect the 330~\cm mode of Nd123 (at 320~\cm) to be higher in energy than that of Eu123 (at 312~\cm) in a simple picture whereby the restoring force for the ion-displacement goes as Hooke's law, $F\propto -k(x-x_0)$. 
The higher energy cannot be explained by the Nd123 being underdoped (which is entirely possible and in fact likely) because the energy of this mode changes little with doping - for Nd123O$_6$ we measure this mode at 316~\cm at room temperature.

There is however an even bigger anomaly; the 330~\cm peak splits into two distinct peaks as the temperature is lowered! Below $T\approx 90$~K a second, lower energy peak becomes clearly distinguishable (at higher temperatures it is suggested by the asymmetric profile of the 330~\cm mode). As the temperature is further lowered the energy difference of these modes increases and the HWHM decreases from $15$~\cm to $10$~$ \cmm $.  At $T=10$~K we measure the two peaks to be centred on energies $330\pm3$~\cm and $285\pm4$~\cm respectively, a $45$~\cm difference. For comparison only, a conventional temperature dependence fitted to the high temperature data points is plotted as a solid line. By $T=9$~K the lower mode is softer by $40\pm 3$~\cm whilst the other is harder by $7\pm 1$~$ \cmm $.  

Both these observations are now understood as the result of coupling between and mixing of the out-of-phase O(2)-O(3) phonon mode and an Nd$^{3+}$ crystal-field excitation \cite{heyen1991, heyen1991prb}.

The crystal field (CF) of the almost tetragonal Nd123 lifts the degeneracy of the Nd$^{3+}$ 4$f$ electrons into five doublets. A particular transition between two of these has an energy of 304~\cm and symmetry \bog (or \ag in an orthorhombic crystal field) \cite{heyen1991prb}. In order for CF excitation and phonon to couple several requirements must be met. The CF electrons and ion involved in the phonon must be in close spatial proximity, in this case the Nd$^{3+}$ and O(2,3) ions are adjacent, separated by $2.45\pm0.04$~$\AA$ \cite{guillaume1994}.  Next both excitations must have similar symmetry, in this case both have quasi-\bog symmetry. Finally, the renormalisation effects are inversely proportional to the difference (or higher power of the difference) of their un-renormalised energies (perturbation theory). In this case that difference happens to be small; Heyen \etal observe these peaks at 334~\cm and 282~$\cmm $, a splitting of 52~$ \cmm $, at $T=10$~K. From these energies, which agree with our data, and relative peak intensity data they compute un-renormalised energies of the phonon and CF excitations as $308\pm3$~\cm and $304\pm4$~\cm respectively \cite{heyen1991prb}, a difference of $4\pm7$~$ \cmm $.  The un-renormalised phonon energy is slightly softer than the Eu123 mode.

The coupling between phonon and CF results in a mixed phonon-CF character to the 334~\cm and 282~\cm excitations. As the temperature is increased, it appears the effective coupling between the CF excitation and phonon decreases, as $(1-\kappa T^2)$, resulting in a weaker renormalisation of both energies and a greater phonon component to the higher energy mode \cite{heyen1991prb}. 



In fact, other than via ion-size variation, this represents the only observable effect of the variable Ln $f$-electron number in this thesis!


\subsection{\specgap estimates from ERS}

\subsubsection{\bog}

\begin{figure}
	\centering
		\includegraphics[width=0.78\textwidth]{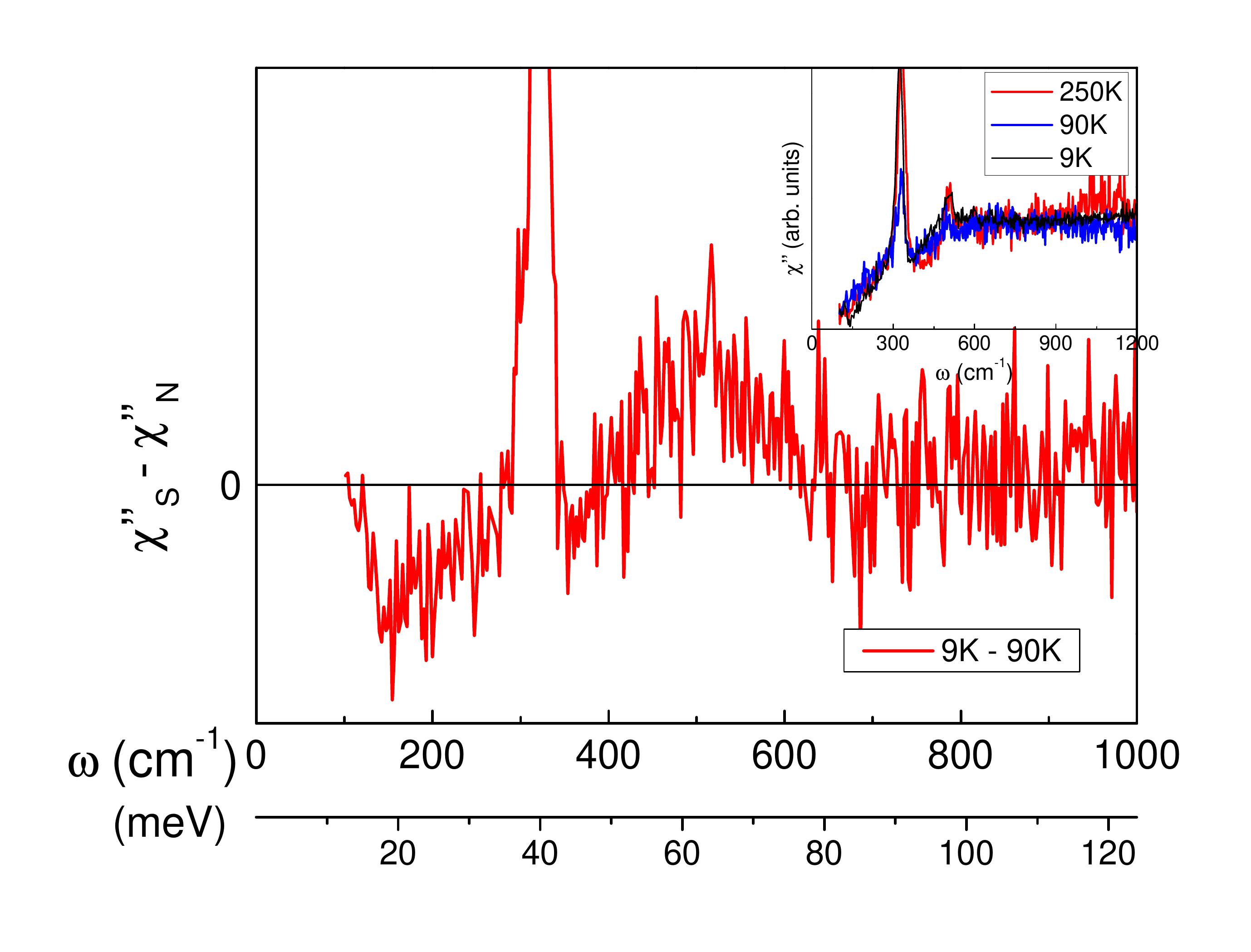}
		\includegraphics[width=0.78\textwidth]{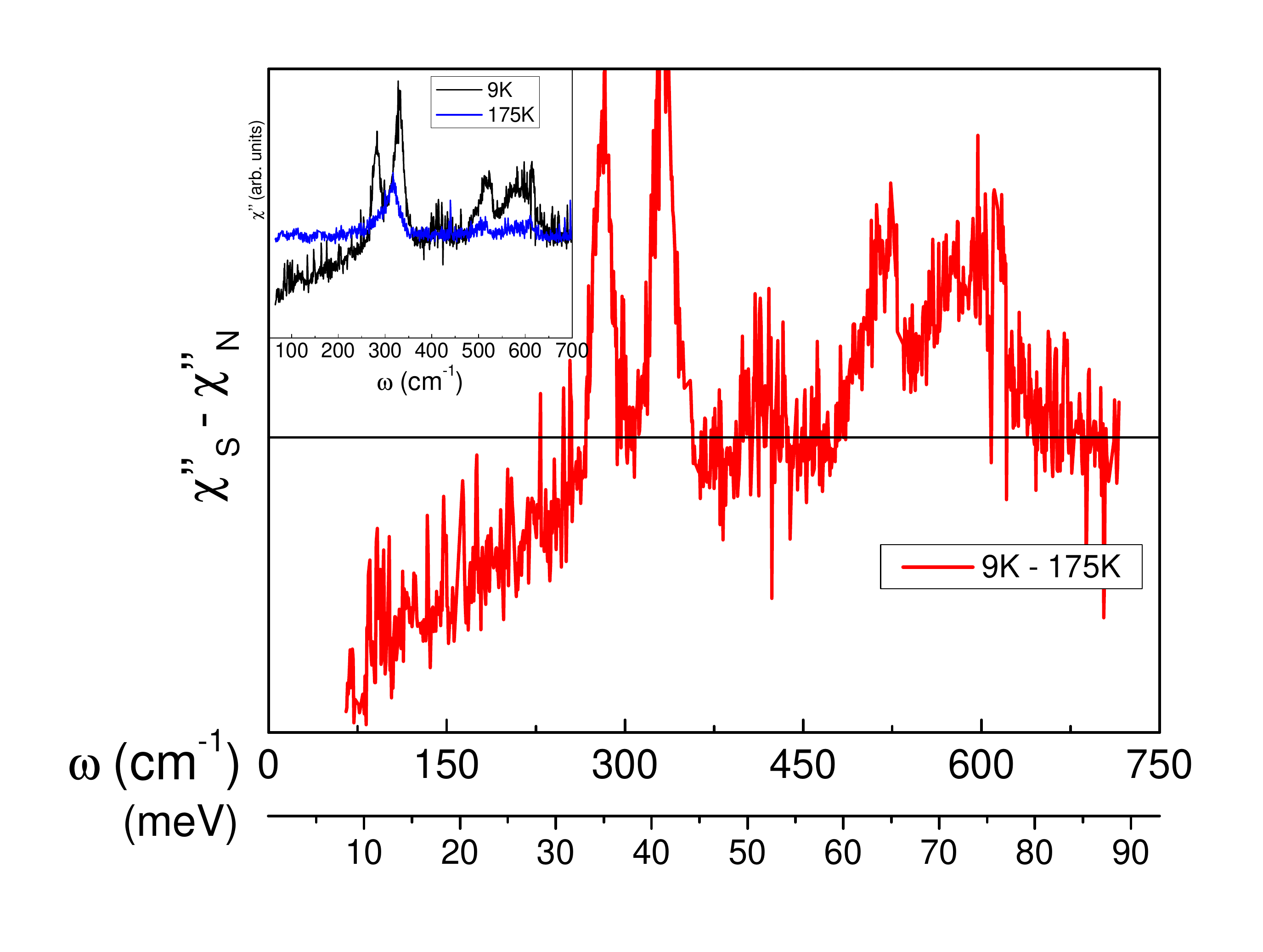}
	\caption[\bog electronic Raman scattering of Yb123 and Nd123.]{(Top panel) The difference between long count-time \bog spectra at 9 K and 90 K of optimally doped Yb123. Inset shows these spectra as well as data taken at $T=250$ K. (Bottom panel) The difference between long count-time \bog spectra, plotted individually in the inset, at 9~K and 175~K for optimally-doped Nd123.}
	\label{fig:yb123ersb1gdiffbctest}
\end{figure}

The inset to \fig~\ref{fig:yb123ersb1gdiffbctest} shows the results of very long count-time spectra on Yb123 at $T=9$~K$\ll T_c$, $T\approx T_c = 94.5$~K and $T=250$ K in \bog scattering geometry. The Raman shifts are shown in both \cm and meV units.  A constant dark count was subtracted prior to correction by the Bose-factor.  Next, each spectrum was scaled by a constant so that their intensities are equal at large $\omega$.

The top main panel of \fig~\ref{fig:yb123ersb1gdiffbctest} shows the difference between the 9 K and 90 K spectra. At low $\omega$ there is less spectral weight at 9 K with some of this weight shifting to the broad peak around 500~\cm (65 meV). We identify this broad peak of enhanced ERS as the `renormalisation peak' associated with an increase of states above the superconducting gap\footnote{We note however that this feature coincides with a phonon mode.}. These two features result from a redistribution of spectral weight from low to higher frequencies due to an opening of a gap in the electronic density of states.  Similar, but much weaker, features can be seen comparing 20 K data with the 250 K data.

Normally, the energy of \specgap is identified as position of this feature in the ERS spectra (see for example \cite{hewitt2002, sugai2003, letacon2006}). Thus, our estimate of the \bog \specgap for Yb123 from this ERS data is $65\pm4$ meV.

\begin{figure}
	\centering
		\includegraphics[width=0.78\textwidth]{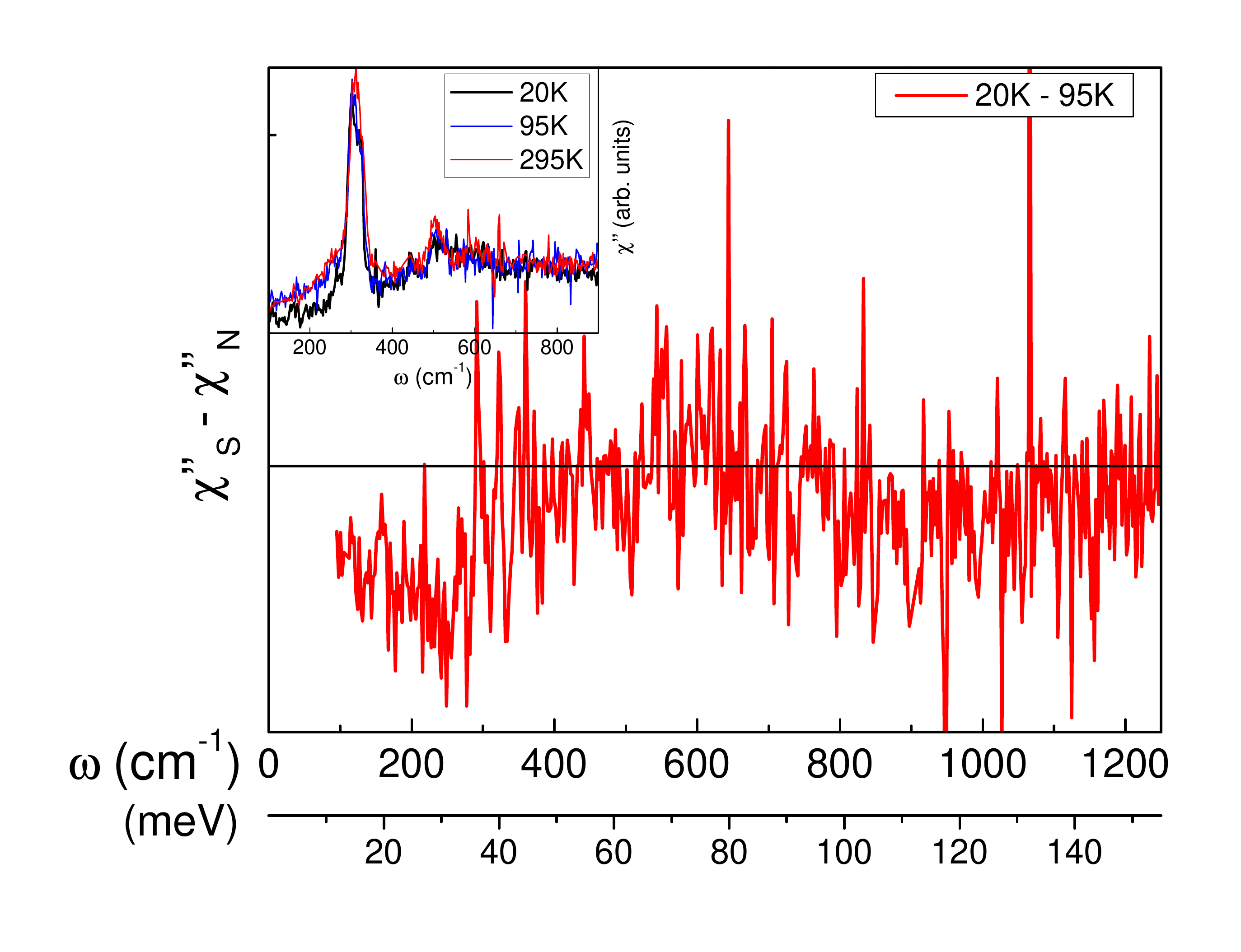}
		\includegraphics[width=0.78\textwidth]{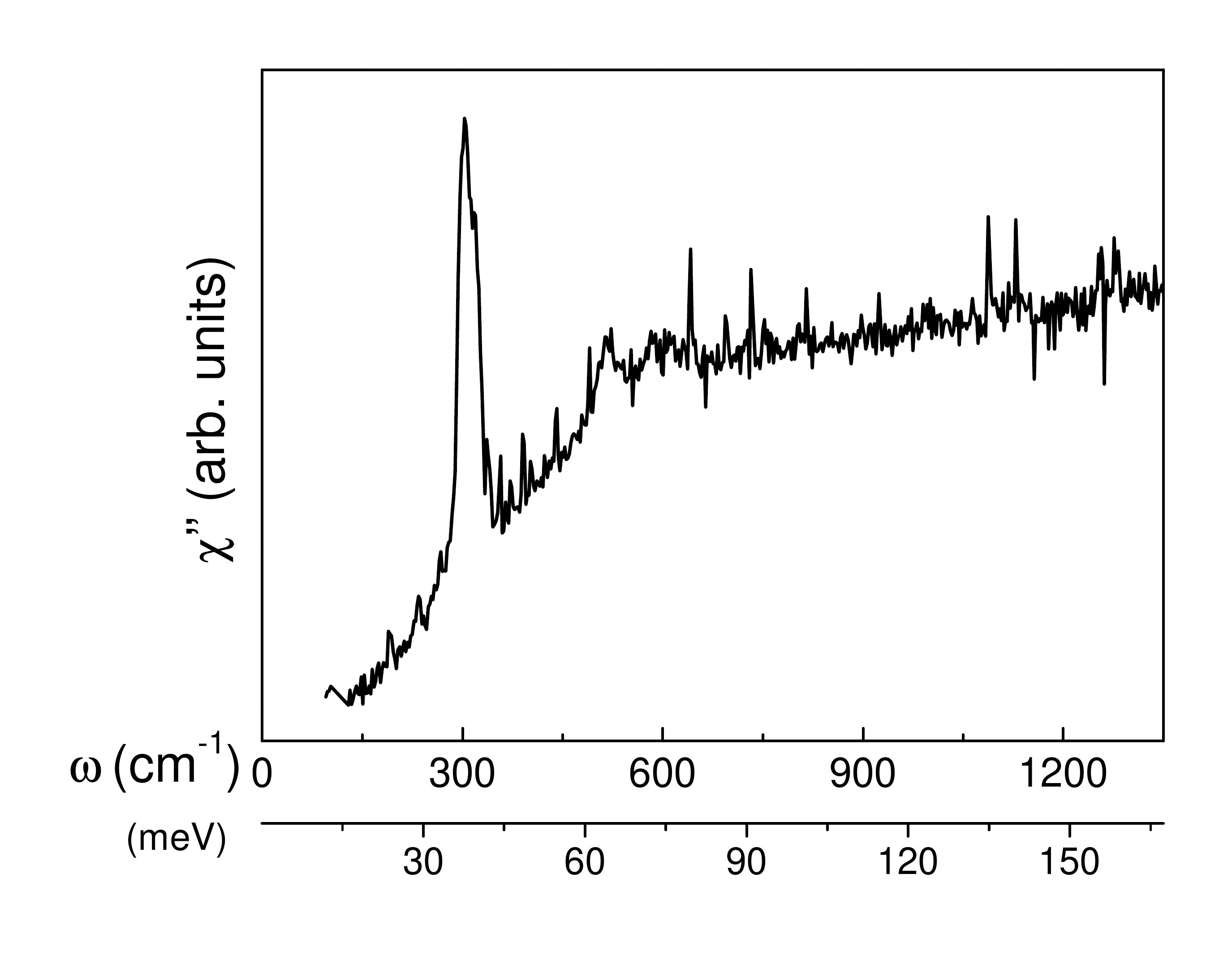}
	\caption[\bog electronic Raman scattering of Eu123.]{(Top) The difference between spectrum long count-time \bog spectrum at 20~K and 90~K of optimally-doped Eu123. The inset shows these spectra as well as data taken at $T=295$~K. (Bottom) \bog spectrum at 9~K on optimally-doped Eu123.}
	\label{fig:eu123ersb1g}
\end{figure}

\fig~\ref{fig:eu123ersb1g} shows similar data for Eu123.  Here a `renormalisation peak' candidate is not as clear, although spectral weight is lost in the 20~K data below $\approx 600$~\cm.  The figure also shows the data at 9~K on its own.  In this spectrum, spectral weight is clearly lost below $\approx 600$~\cm as indicated by a sudden change in gradient\footnote{With these data there is a significant background linear in $\omega$.  It is largest in the 9~K data and we suspect that it may be from vacuum grease on or around the sample.}. From these data we estimate $\specgapm \approx600$~$\cmm$, or $70\pm10$ meV.

Finally, we discuss the \bog ERS data for Nd123 which are shown in bottom panel of \fig~\ref{fig:yb123ersb1gdiffbctest}.  Other than \replaced{multiplication by the Bose-Einstein factor, $(n(\omega,T)+1)^{-1}$,}{division by the Bose statistical factor, $(n(\omega,T)+1)$,} these data have not been manipulated in any way. 
From the raw data shown in the inset, or the difference plot shown in the main panel, a `renormalisation peak' candidate is not as clear. The feature at $400$~\cm is possibly a weak phonon mode, slightly softened with respect to the equivalent mode in Yb123 due to Nd123's larger unit-cell, that has only been observable in this long count-time spectra. Note also the mixed phonon-CF excitations at 290~\cm and 330 $\cmm$. With similar reasoning to that used for the Eu123 data, we estimate $\specgapm \approx 400$~$\cmm$, or $50\pm 5$ meV for Nd123.  Alternatively, the extra spectral weight around 575~\cm (70 meV) may be the renormalisation peak, in which case these data would give us the estimate $\specgapm \approx 70$~meV.


\subsubsection{\aogbog} 
\label{sec:a1gb1gers}
\begin{figure}[tb]
	\centering
		\includegraphics[width=0.475\textwidth]{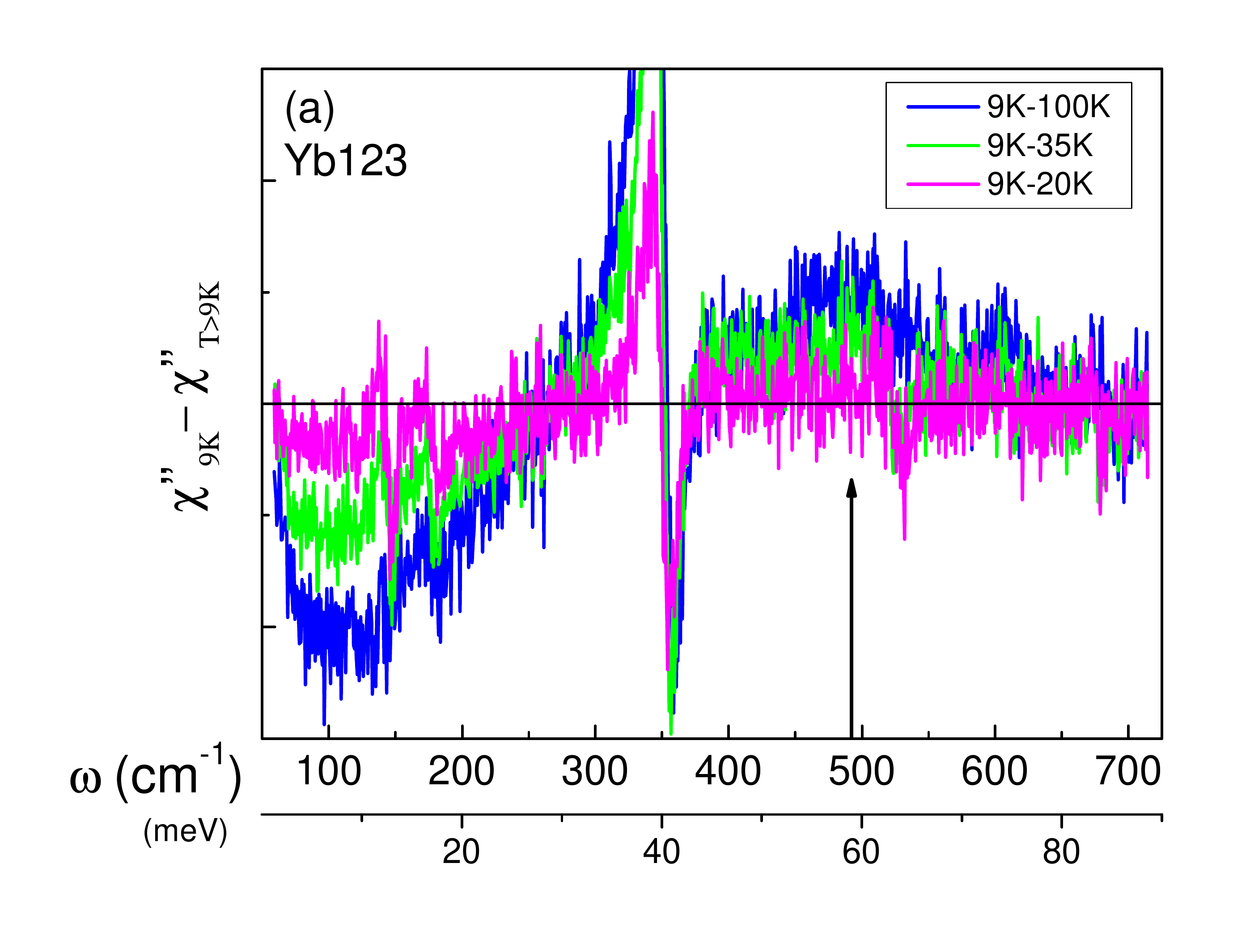}
		\includegraphics[width=0.475\textwidth]{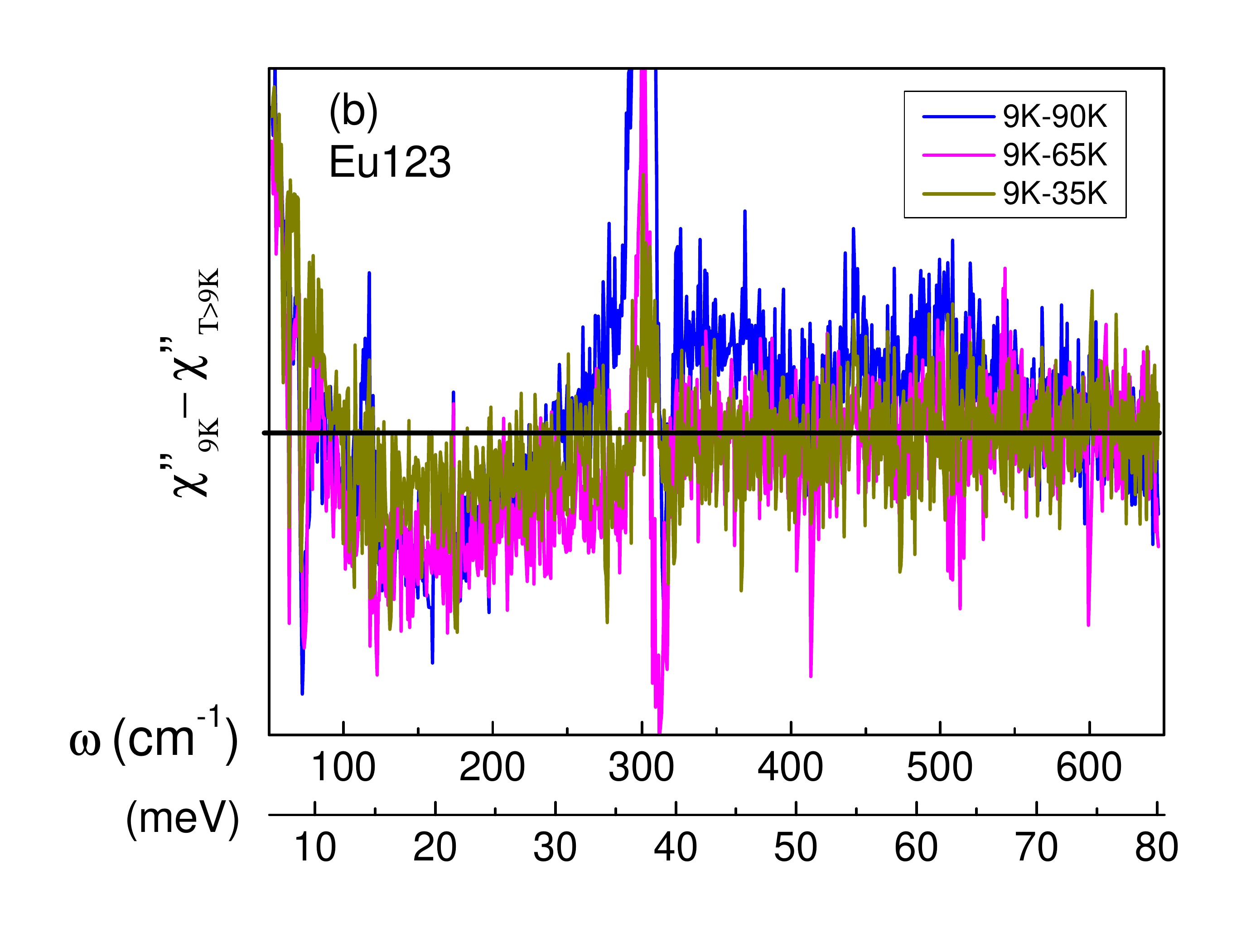}
	\caption[\aogbog electronic Raman scattering of Yb123 and Eu123.]{\aogbog difference spectra of (a) Yb123 and (b) Eu123 (with representative raw spectra shown in \fig~\ref{fig:yb123a1gb1graw} and \fig~\ref{fig:eu123a1gb1g} respectively).  These data clearly show the reduction in spectral weight for small $\omega$ at low temperatures due to the opening of an energy gap.  In (a) an arrow indicates a peak in the ERS. No such feature is clearly visible for the Eu123 data. }
	\label{fig:ersa1gb1gdiff}
\end{figure}

Similar to the long count-time \bog spectra discussed above, the \aogbog spectra in \fig~\ref{fig:yb123a1gb1graw} clearly show a suppression of the ERS at low $\omega$ for $T<T_c$. \fig~\ref{fig:ersa1gb1gdiff} plots the difference between the $T=9$ K data and several higher temperature data. In this geometry there are many more phonon modes complicating the analysis, nevertheless a peak feature can be identified in the Yb123 data presented in panel (a), as annotated with an arrow. This gives us the estimate for Yb123 of $\specgapm = 55\pm5$ meV.

Similarly, from the \aogbog ERS scattering of Eu123 in \fig~\ref{fig:ersa1gb1gdiff}b we estimate \specgap is between 40 meV and 50 meV - lower than expected from the previous two estimates for this material. The phonon at 303~\cm is significantly more intense at low temperatures, \fig~\ref{fig:eu123a1gb1g}, and its extra spectral weight may be the cause this feature we see in the difference data around 45 meV. 

\begin{figure}
	\centering
		\includegraphics[width=0.85\textwidth]{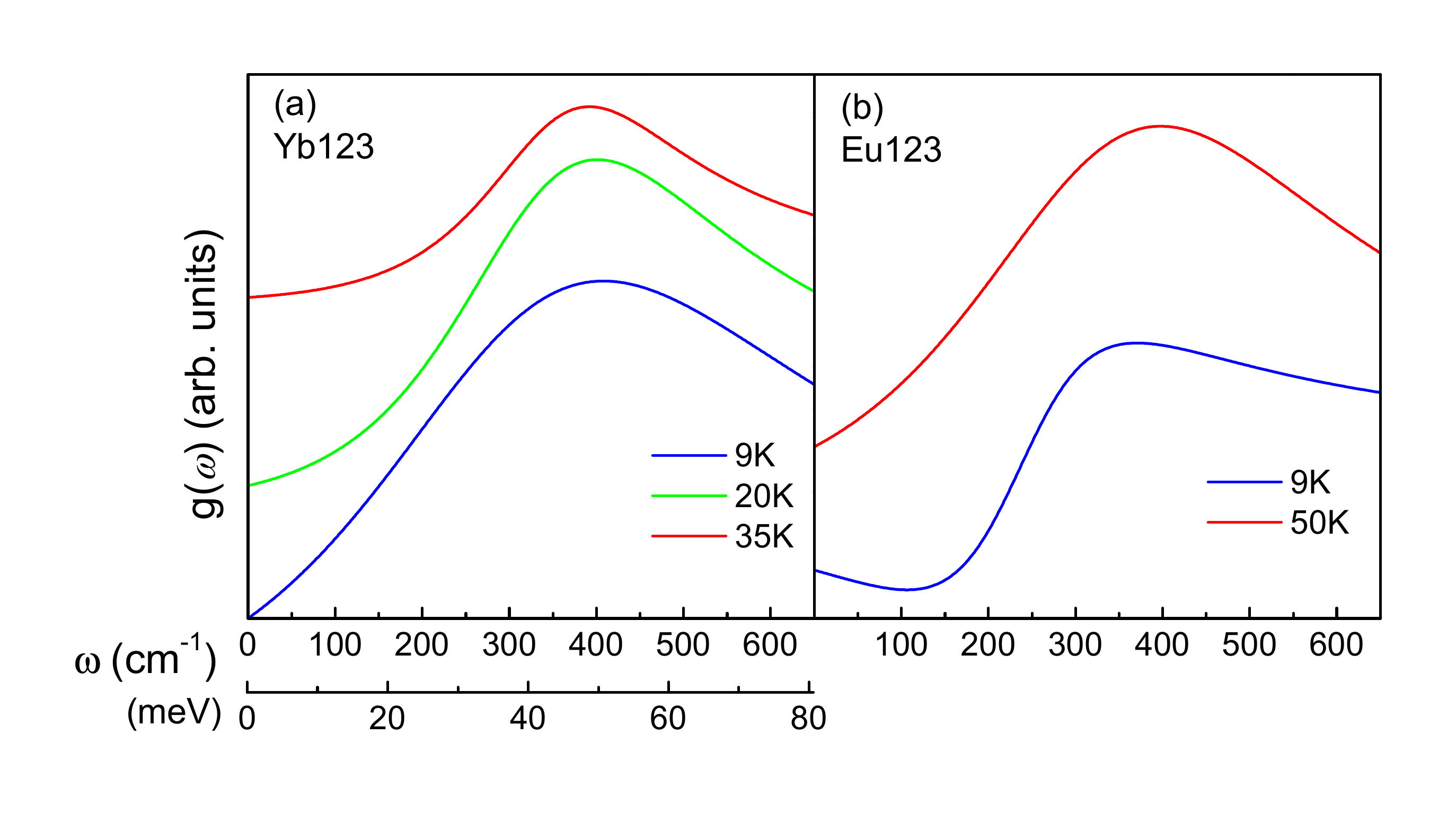}
	\caption[Fitting of Yb123 and Eu123 \aogbog electronic Raman scattering.]{The $g(\omega)$ parameter from the fitting procedure described in the text, \refsec~\ref{sec:greensfitting}, for data at various temperatures for (a) Yb123 and (b) Eu123. $g(\omega)$ is the imaginary part of the electronic response (\refsec~\ref{sec:ramaners}) and the energy where $g(\omega)$ is maximum can be interpreted as the energy of \specgap \cite{limonov2000}.}
	\label{fig:limonovfitting}
\end{figure}

Another approach is to fit the entire spectrum using a Green's function method described above (\refsec~\ref{sec:greensfitting}) that incorporates both phonon and electronic contributions to the Raman intensity as well as the interaction between them \cite{chen1993, panfilov1999, limonov2000}. \fig~\ref{fig:limonovfitting} shows fitted data and the $g(\omega)$ parameter derived from this procedure for Yb123 and Eu123 data at various temperatures.  The signal-to-noise ratio of Nd123 \aogbog data was not good enough for this analysis.  The energy where $g(\omega)$ is maximum can be interpreted as $\specgapm$.  As such, we obtain an estimate of the $\specgapm=50\pm 5$ meV for both Yb123 and Eu123 in agreement with the difference plot shown in \fig~\ref{fig:ersa1gb1gdiff}a. 


\subsubsection{\btg}
\begin{figure}
	\centering
		\includegraphics[width=0.70\textwidth]{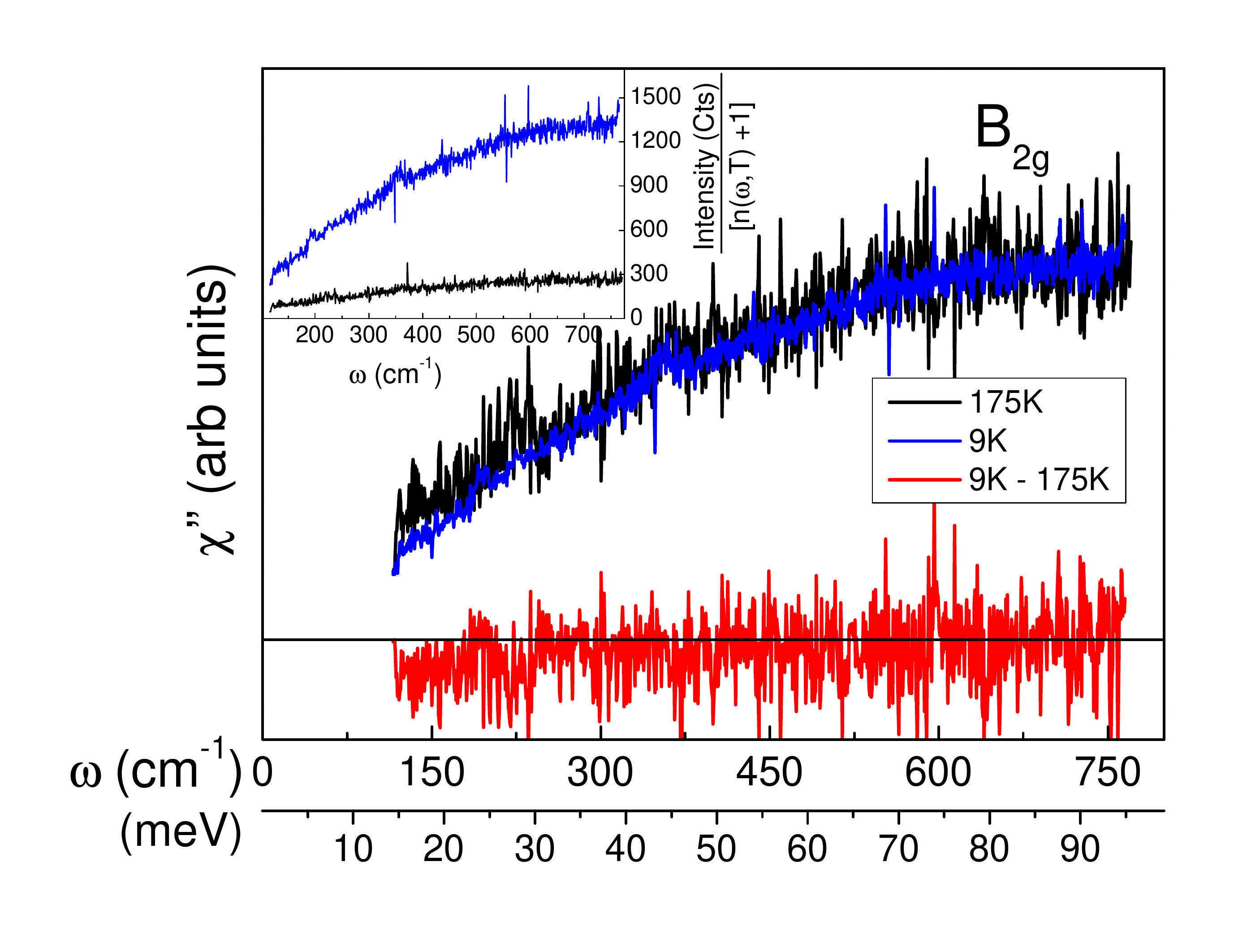}
	\caption[\btg ERS spectra for Eu123]{Scaled \btg spectra for optimally-doped Eu123 at $T=175$ K (black) and $T=9$ K (blue) and their difference (red).  The raw data is shown in the inset.  Any ERS from these data, and all other \btg data, are too weak to unambiguously identify.}
	\label{fig:eu123ersb2g}
\end{figure}

\fig~\ref{fig:eu123ersb2g} shows the results of long count-time scans in \btg scattering geometry on Eu123.   There is a weak suppression of spectral weight below 180~\cm which could be indicative of the gap, but it is not very compelling.  Unlike \bog scattering, in \btg we do not observe clearly identifiable evidence of an energy gap in any of the materials we have studied. These observations are similar to those of others on near-optimally-doped Y123\footnote{The \btg gap feature is much more prominent, for some reason, in optimally doped Bi2212 \cite{kendziora1995, opel2000, hewitt2002, sugai2003} and Hg1201 \cite{letacon2006, guyard2008}} \cite{opel2000, sugai2003}. This is a disappointing result as it means we are not able to disentangle the contributions to \specgap from \scgap and \epg using Raman spectroscopy alone.

\FloatBarrier 
\section{Summary and discussion of results}
\tab~\ref{tab:ramangapsizesummary} presents a summary of the various gap estimates on optimally-doped Yb123 and Eu123 and oxygen-loaded Nd123 single crystals. For comparison, literature results from optimally-doped Y123 are also included in the table. $xx$ and $yy$ for the two \aogbog Y123 values indicate the anisotropy in \specgap found by Limonov \etal \cite{limonov2000} and also indicates the potential precision of these measurements.

\begin{table}[ht]
	\centering
		\begin{tabular}{lcccr} \toprule
			Material & \tc (K) & $\specgapm$ (meV) & Symmetry & Method (Reference) \\ \midrule
			\multirow{3}{*}{Yb123} & \multirow{3}{*}{90.5} & $39<\specgapm<57$ & \aogbog & Phonon shifts \\ 
			  & & $60\pm4$ & \bog & ERS \\ 
			  & & $55\pm5$ & \aogbog & ERS \\ \midrule
			\multirow{3}{*}{Eu123} & \multirow{3}{*}{94.5} & $60<\specgapm$ & \aogbog & Phonon shifts \\ 
			  & & $70\pm10$ & \bog & ERS \\  
			  & & $50\pm5$ & \aogbog & ERS \\ \midrule 
			\multirow{3}{*}{Nd123} & \multirow{3}{*}{88.5} & $\specgapm<55$ & \aogbog & Phonon shifts \\ 
			  & & $57\pm4$ & \bog & ERS \\ \midrule  
			\multirow{3}{*}{Y123} & \multirow{3}{*}{93} & 52.7 ($xx$) 55.8 ($yy$) & \aogbog & \cite{limonov2000} \\ 
			  & & $67\pm1$ & \bog & \cite{sugai2003} \\ 
			  & & $67.5\pm0.5$ & \bog & \cite{gallais2002} \\ 
			  & & $40.0\pm0.5$ & \aog & \cite{gallais2002} \\ \bottomrule
		\end{tabular}
	\caption[Summary of measured spectral gaps for Yb123, Eu123 and Nd123]{\label{tab:ramangapsizesummary} A summary of estimates of the spectral gap, $\specgapm$, for Yb123, Eu123 and Nd123.  Recall \specgap is a pair-breaking peak, see \refsec~\ref{sec:gaps}.  The estimates are derived either from phonon energy renormalisation (labelled `Phonon shifts') or from electronic Raman scattering (labelled `ERS').  The symmetry condition by which the spectra used to estimate these values were taken is also listed.  For comparison, literature values of $\specgapm$ for optimally doped Y123 are included. For a given material, there is general agreement between various estimates $\specgapm$.}
\end{table}

We find that for a given material, our various estimates of \specgap are consistent with each other with the exception of the Eu123 \aogbog ERS estimate.  In our view the value from the \bog ERS data is more reliable. Further, our estimates are also similar to the equivalent values measured in Y123 \cite{limonov2000, gallais2002, sugai2003}. Nd123 is approximately $10$~meV lower than Y123, however it is our view that the precision of these estimates is generally inadequate, and in this case too poor to draw conclusions about the relevance of this discrepancy. Moreover, we are unable to comment on any systematic variation of \specgap with the Ln ion size.  For example; for Ln123 $\Delta T_c\approx10$\% at most and if we assume $T_c\propto\tcmfm\propto\specgapm$ the precision in our \specgap measurement must be better than $\approx5$\%.  Clearly, this is not the case. Nor does this estimate of the variation in $\specgapm$ consider the possible effect of the pseudogap.


\scgap is expected to scale with \tc (or more correctly, $\tcmfm$) whereas the pseudogap is thought to originate from AF fluctuations\footnote{It is interesting that a pseudogap like state has been seen in pnictide superconductors and is also correlated to anti-ferromagnetic fluctuations \cite{moon2012}.} and thus has energy scale $J$.  To see how these two quantities are affected by Ln substitution in Ln123 we express each as a function of the Cu(2)-O(2,3) bond length, $r$ (the superexchange path-length).

\begin{figure}
	\centering
		\includegraphics[width=0.45\textwidth]{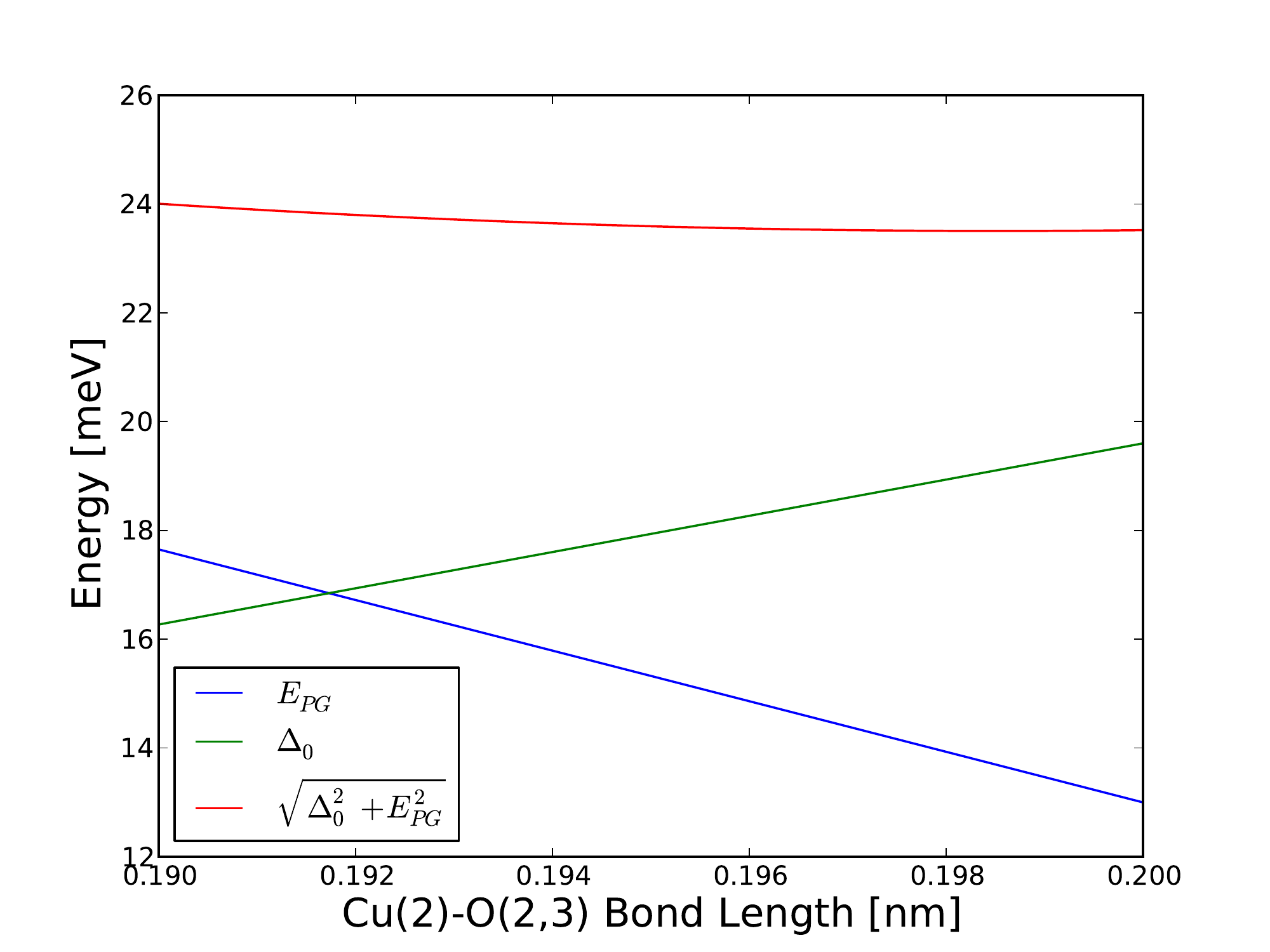}
		\includegraphics[width=0.45\textwidth]{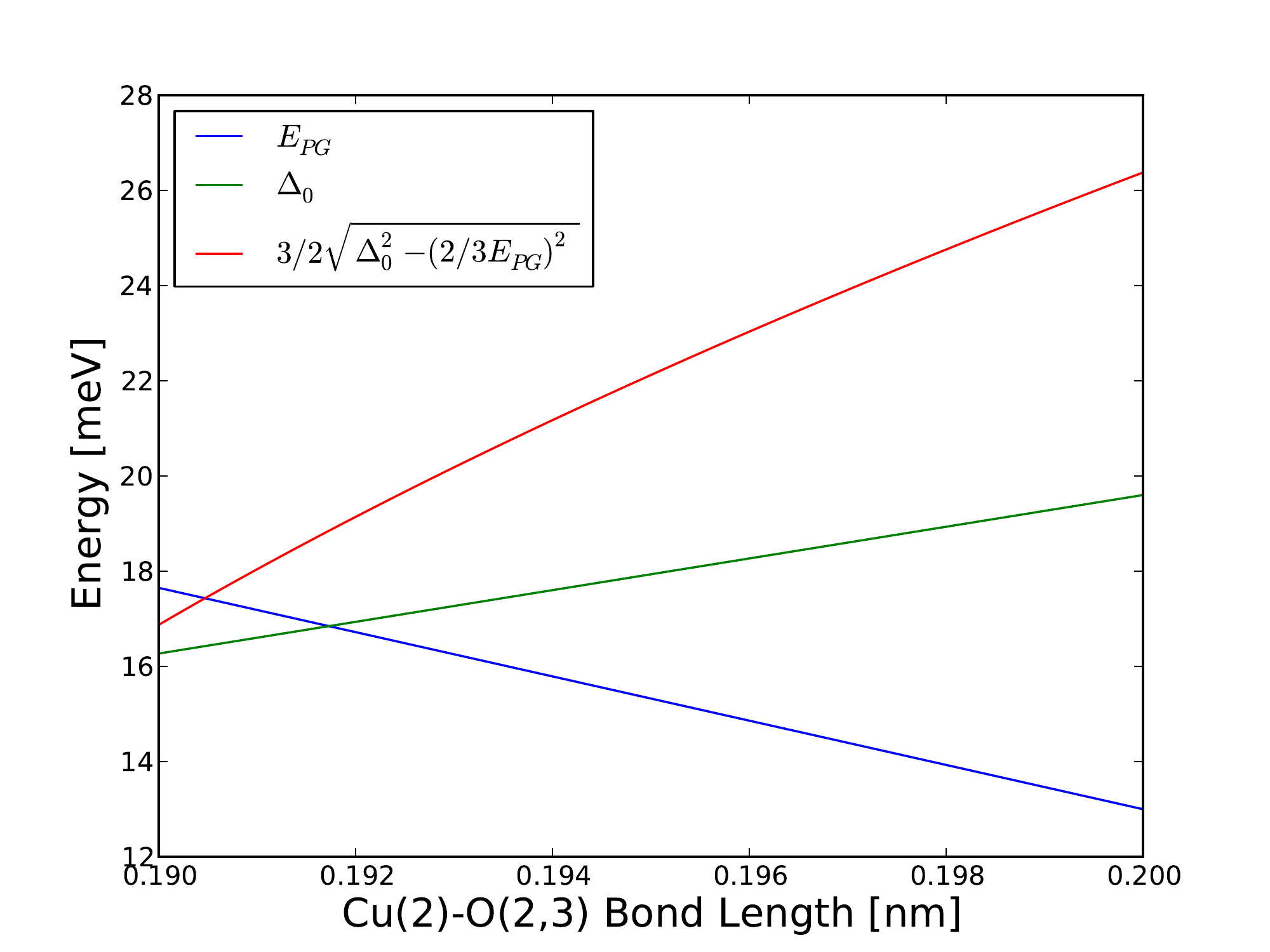}
	\caption[Possible ion-size dependence of the \bog and \btg spectral gap]{An estimate for the dependence of $\hscgapm$, \epg and $\hspecgapm[\bogm]$ (Left) and $\specgapm[\btgm]$ (Right) on the Cu(2)-O(2,3) bond length, $r$, (the super-exchange path length). These values are relevant to Ln123 at $p=0.16$. }
	\label{fig:gapsvsr}
\end{figure}

There is some evidence that $J$ sets the energy scale for \epg as $\epgm \approx J (1-p/0.19)$ \cite{loram2001}. We can combine this observation with the linear relation between $J$ and $r$ determined in Chapter~\ref{ch:twomag}, to arrive at the following estimate for the Ln123 series, $E_{PG}(r) \approx m_{PG}r + c_{PG}$ with $m_{PG}=-470\pm30$~meV/nm and $c_{PG}=105\pm5$~meV for $p=0.16$.  More recent data suggests the doping dependence is sub-linear with $\epgm(p)\approx (1-p/p_{\textnormal{crit}})^{0.8}$ \cite{tallonfluctuations}, in which case $m_{PG}=-710\pm40$~meV/nm, $c_{PG}=162\pm8$~meV for $p=0.16$.  We use the former values of $m_{PG}$ and $c_{PG}$ in \fig~\ref{fig:gapsvsr}.
We can estimate the dependence of \hscgap on $r$ by assuming that a similar relation to $\scgapm=4.2k_B\tcmfm$ holds across the Ln123 series \cite{tallonarXiv}. Note that at optimal doping $\tcm \approx \tcmfm$ \cite{dubroka2011}. Next we use the experimental dependence of \tc on $r$ (refer to \refsec~\ref{sec:ln123systematics}) to estimate $\hscgapm(r) = m_{\Delta}r+c_{\Delta}$ with $m_{\Delta}=330\pm30$~meV/nm and $c_{\Delta}=-47\pm5$~meV.

\fig~\ref{fig:gapsvsr} plots the results of these estimates for values of $r$ relevant to the Ln123 series. The energies of \epg and \hscgap have opposite dependences on $r$ such that, depending on the validity of the assumptions above, the spectral gap in \bog symmetry, $\specgapm=2\sqrt{\hscgapm^2+\epgm^2}$, is expected to be approximately independent of $r$ and therefore Ln ion substitution.

These ERS measurements have the potential to determine both the superconducting and pseudogap energies.  By a simple change in the polarisation selection and relative sample alignment, it is possible to measure \specgap either at the anti-nodes where $\specgapm = 2\sqrt{\hscgapm^2 + \epgm^2}$, or at the nodes where $\specgapm \approx \sfrac{3}{2}\sqrt{\hscgapm^2-(\sfrac{2}{3}\epgm)^2}$. 
With \btg data to accompany the \bog data, the \hscgap and \epg contributions to \specgap can in principle be isolated. Furthermore, \specgap in the \btg geometry is expected to be strongly dependent on ion-size as shown in \fig~\ref{fig:gapsvsr}. Unfortunately, we were not able to observe ERS in the \btg geometry.

  These polarisation selection rules were used in the same way to measure and unambiguously identify two-magnon scattering.  There is the possibility to combine ERS and two-magnon scattering measurements to study, in a very direct way, the relation between $J$ and the pseudogap, and indeed with the superconducting energy gap also. One can envisage then a combined two-magnon scattering and ERS study where doping and ion-size are systematically varied. Such a study would provide valuable evidence of the nature of the pairing interaction and the pseudogap.

What this chapter shows, in combination with the two-magnon scattering data for $p>0$ in \refsec~\ref{sec:jvsdoping}, is that such a study would be a major undertaking - perhaps sufficient work for another PhD?!
We have found these measurements are plagued by poor signal intensities which not only means having to collect data for a prohibitively long time, but also introduces errors and uncertainties into the data, such as contributions from stray light and sample movement.  For very long count-times subtracting all the cosmic ray events becomes problematic also. Scattering from two-magnons in undoped single crystals with good surface quality by micro-Raman is not problematic, however, the intensity of two-magnon scattering rapidly diminishes with increasing doping and moves to lower wavenumbers \cite{sugai2003} where it becomes increasingly difficult isolate. 

Such a study is possible.  Li \etal have very recently carried out such measurements on a single crystals of Hg1201 at three doping states \cite{li2012}.  Their study is exemplary and their results are fascinating. In \bog geometry they clearly observed temperature dependent two-magnon scattering from underdoped to a $p=0.19$ sample.  The two-magnon scattering intensity and energy appears to be correlated with the suppression in ERS at low $\omega$ that the authors associate with the pseudogap. Furthermore, both these features are evident above \tc in the underdoped samples.  The authors describe the correlation between these two features as a ``feedback effect'' such that magnon excitations are modified in the superconducting state.  Similar behaviour is observed for the prominent ERS renormalisation peak, the `pairing peak', in their data.


With the benefit of hindsight, we describe a simpler, yet still intensive, method for studying the evolution of the pseudogap and $J$ with ion-size in \refsec~\ref{sec:possibleexperiment} in the following chapter. The method we describe can be readily extended to study the doping dependence of $J$ and \epg as well.

\chapter{Conclusions and Discussion}
\label{ch:discussion}

\subsubsection{Summary}

We begin this chapter by summarising key results from previous chapters.  We then discuss three possibilities for the salient effect of ion size on $ \tcmaxm $.  The most attractive possibility is that the polarisability plays an important role in determining \tcmax because this can naturally resolve the opposite effect of internal pressure and external pressure.  

We then discuss how ion-size, polarisability and \tc might be related within the framework of the Hubbard model where superconductivity emerges from a strongly correlated electronic system. 
An alternative connection between polarisability and \tc is also proposed where Cooper pairing is based on the exchange of coherent, quantized waves of electronic polarisation (polarisation-waves).

\added{In \refsec~\ref{sec:furtherwork} we discuss and} suggest some studies that would utilize ion-size effects to further elucidate the nature of the pseudogap, parameters relevant to \tc and the pairing mechanism in the cuprates. 

\added{The chapter concludes with a very brief summary of some key findings from this work and how we see them relating to the wider field of HTS research.} 

	\section{Summary of results}
We have presented the argument that internal pressure, as induced by isovalent ion-substitution, decreases \tcmax in the cuprates.
In contra-distinction, external pressure increases $ \tcmaxm $.  This work seeks to understand the salient physical difference between these two pressures by studying ion-size effects on various important physical parameters.  

In Chapter~\ref{ch:bi2201} we used a simple materials variation approach to study the effect of ion-size and disorder on \tc Bi2201.  We found that ion substitution can increase $T_c$ despite increasing disorder and concluded that while disorder does play a role in decreasing \tc there is a comparable effect arising from changing ion-size.
		
In Chapter~\ref{ch:dft} we used DFT calculations for undoped ACuO$_2$ for A=\{Mg, Ca, Sr, Ba\} to investigate the effect of ion-size on the electronic properties in this model cuprate system.  We found that larger ions move the van Hove singularity (vHs) in the DOS closer to the Fermi-level.  This finding is consistent with an interpretation of the ion-size affecting \tc via the density of states.
		
In Chapter~\ref{ch:twomag} we measured the antiferromagnetic superexchange energy, $J$, by two-magnon Raman scattering in undoped cuprates while systematically altering the internal pressure by changing ion-size.  We then compared the internal-pressure dependence of $J$ with data in literature for the external pressure dependence of $J$. From these data we showed that $J$ is likely \textit{un}related to $\tcmaxm$: $J$ and \tcmax anti-correlate with internal pressure as the implicit variable and correlate with external pressure as the implicit variable. We concluded that some other physical property is dominant in setting the value of $\tcmaxm$. 
		
In Chapter~\ref{ch:musr} we tested the idea that ion-substitution-induced disorder is responsible for the variation in \tcmax with ion-size. We presented \musr measurements of the superfluid density in \ybasr showing the nice result that \ybasr is a boring, well-behaved material!  The superfluid densities are consistent with those previously reported for pure and Ca-doped Y123.  Consequently, the lower \tcmax seen in this compound, and those with lower Sr content, is a genuine `ion-size effect' (or `internal pressure effect'). 
				
In Chapter~\ref{ch:ramangaps} we sought to measure energy gaps in the single-electron dispersion of our ion-substituted cuprate materials.  Within the accuracy of our estimates, we did not find any systematic internal-pressure dependence of the energy gap.  We did find however that the gap values for our ion-substituted samples are similar to the well-characterised \ybco cuprate.  A better approach to addressing this important question needs to be found and our proposal for such an approach will be discussed later in this chapter.

As will be discussed in detail below, the cumulation of our studies leads us to believe that the polarisability plays a central role in superconductivity in the cuprates.

	\section{Discussion: The ion-size effect on \tc}

What is the salient effect of the ion size on $\tcm$? We identify several possibilities that we will discuss in turn below;
\begin{enumerate}
	\item Enhancement of a competing electronic state, e.g. the pseudogap.
	\item Modification of the DOS.
	\item Altering the dielectric properties of the material (polarisability). This in turn could play a role in altering the screening or modifying the spectrum of polarisation wave excitations. 
\end{enumerate}

	\subsection{Pseudogap}

As discussed in Chapter~\ref{ch:twomag}, the effect of ion-size on \tc may be indirect, such as via an electronic order that competes with superconductivity. The most likely candidate here is the pseudogap as there is evidence from inelastic neutron scattering \cite{storey2008pggroundstate} and specific heat data \cite{loram2001} that the pseudogap energy is set by $J$. As we have shown in Chapter~\ref{ch:twomag} and \refsec~\ref{sec:ln123systematics}, $J$ is in turn sensitive to the ion-size-modified superexchange path-length.  Alternately, it is possible at lower doping, that a charge ordered phase is the competing phase \cite{tranquada1997} especially modified by ion size.

However, when we consider the effect of external pressure on $J$ and \tc shown in \fig~\ref{fig:jvstc} and reproduced below in \fig~\ref{fig:tcmax}~(a), this interpretation becomes less likely. For example, if we took the view that the suppression of \tc with ion-size was due to an enhancement of the pseudogap phase (or, similarly a stripe phase) from an increasing $J$, we would need to explain why both \tc and $J$ increase under external pressure. 
To put it another way; if one expects \hscgap to scale with $J$ (magnetic pairing scenario \cite{letacon2011}) and \epg scales with $J$ as observed, then in the competing gap scenario after Bilbro and McMillan \cite{bilbromcmillan} we would expect $\tcm \sim \sqrt{\hscgapm^2 - \epgm^2}$ so that \tc will still scale with $J$ as both \hscgap and \epg themselves do.  We have shown that it does not for the Ln123 system: \tcmax anti-correlates with $J$ when ion-size is the implicit variable and only correlates with $J$ when external pressure is the implicit variable. 


\subsection{Density of States}
Another hypothesis is that these ion-size substitutions introduce structural distortion which in turn distorts the electronic dispersion in such a way to increase the DOS by shifting the vHs closer to $E_F$. In a BCS-like framework, a high DOS is an important requirement for high $ \tcm $, irrespective of the pairing mechanism \cite{bcspaper, abbpaper, surma1983}.  We tested this idea with DFT calculations of the band-structure for the infinite-layer cuprate ACuO$_2$ where A = \{Mg, Ca, Sr, Ba\} and the results are consistent with this hypothesis.

On the experimental side, Kim \etal showed that for optimally-doped Bi2201 \dos was suppressed with smaller ion-size \cite{kim2010}. However they also demonstrated that the decrease in \dos could be understood to result from an enhanced pseudogap as determined from $\epgm = k_B T^*$ with $T^*$ from resistivity measurements. We speculate that the changes in \epg in their Bi2201 system are caused by changes in $J$ which result from the smaller superexchange path-length with reduction in ion-size. From our own preliminary Zn-substitution studies on \ybasr presented in \fig~\ref{fig:tcsupp} it would also appear that smaller ion size depletes $ \dosm $.  But again this may be a pseudogap effect and further systematic studies, such as described below in \refsec~\ref{sec:possibleexperiment}, would clarify this. 

It is possible then that the increase of \tcmax with larger ion size\footnote{Or conversely, the decrease of \tcmax with internal pressure} is a result of an enhanced $ \dosm $. But does this offer a clear understanding of why internal and external pressure have opposing effects on $\tcmaxm$? We do not see how it could, but this is yet to be tested in our calculations.  Of course there remains the possibility that external pressure affects \tc in a manner distinct to internal pressure. Furthermore, the subtle relationship between crystallographic and electronic structure prevents us from commenting more broadly on correlations such as that between \tcmax and the bond valence sum $V_+$ shown in \fig~\ref{fig:bvs}.



\subsection{Polarisability: Resolution of a paradox}
\label{sec:polarisabilitydiscussion}

However, in all this we must bear in mind the simplicity of the systematic evolution of \tcmax with $ V_+ $, shown in \fig~\ref{fig:bvs}, that suggests we may have to look elsewhere to locate its underlying origins. One arena where ion-size plays a central role is in the dielectric properties, where the ionic polarisability varies roughly as the cube of the ion size \cite{shannon1993}. This suggests two possibilities (i) the material dependence of \tc results mainly from screening of longer-range repulsive interactions in a magnetic pairing scenario \cite{raghu2012, raghu2012optimaltc} (ii) an alternative and relatively unexplored idea that pairing in the multi-ion cuprates might be attributable to the exchange of bosons associated with polarisation waves \cite{atwal2004, ashcroft1987}. We consider these the key insights arising from this thesis and explore both in more detail below in \refsec~\ref{sec:incoherentpolar} and \ref{sec:coherentpolar} respectively. 

In an early treatment of the effect of polarisable ions on the dielectric response, Goldhammer \cite{goldhammer1913} (see also \cite{herzfeld1927}) showed for sufficiently symmetric systems that the dielectric constant is enhanced by a factor $\left[ 1-\frac{4\pi}{3}\sum_i{n_i \alpha _i}\right] ^{-1}$ where the sum is over all ions $i$, $n_i$ are the volume densities of these ions and $\alpha _i$ their polarisabilities.  The factor $\frac{4\pi}{3}$ is ultimately dependent on structure, the dielectric constant more generally being replaced by a dielectric matrix and, in a fuller treatment, the enhancement factor is replaced by frequency- and momentum-dependent terms \cite{atwal2004}. The dielectric constant for bound electrons can be written as;

\begin{equation}
\frac{\epsilon}{\epsilon_0} = 1 + \frac{{4\pi}\sum_i{n_i \alpha_i}}{1-\frac{4\pi}{3}\sum_i{n_i \alpha_i}}
\label{eq:dielectricfn}
\end{equation}

\begin{figure}
	\centering
		\includegraphics[width=0.75\textwidth]{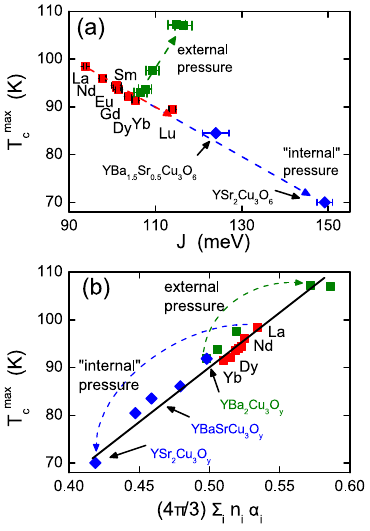}
		\caption[\tc vs $J$ and \tc vs the refractivity sum $(4\pi/3)\sum_i{n_i\alpha_i}$ for Ln123 under internal and external pressures.]{\label{fig:tcmax}(a) $T_c^{\textnormal{max}}$ plotted versus $J$ for single crystals of LnBa$_2$Cu$_3$O$_6$ (red squares) and for YBa$_{2-x}$Sr$_x$Cu$_3$O$_6$ (blue diamonds) with $x = 0.5$ and $2.0$. \tcmax anticorrelates with $J$ where ``internal pressure'' (blue trajectory) is the implicit variable. The green squares show that \tcmax versus $J$ under external pressure (green trajectory) is effectively orthogonal to the behaviour for internal pressure.  
(b) $T_c^{\textnormal{max}}$ plotted against the refractivity sum $(4\pi/3)\sum_i{n_i\alpha_i}$ for the ions in the non-cuprate layer for LnBa$_2$Cu$_3$O$_y$ (red squares, Ln = La, Nd, Sm, Eu, Gd, Dy, Yb) and YBa$_{2-x}$Sr$_x$Cu$_3$O$_y$ (blue diamonds, $x = 0, 0.5, 1.0, 1.25$ and 2) which summarize the effects of ``internal pressure''. The green squares show \tcmax versus $(4\pi/3)\sum_i{n_i\alpha_i}$ for YBa$_2$Cu$_3$O$_y$ under external pressure, revealing a correlation which remains consistent with that for internal pressure.}
\end{figure}

\noindent The product $ n_i \alpha_i $ is the \textit{ionic refractivity} and we call $\frac{4\pi}{3}\sum_i{n_i \alpha _i}$ the \textit{refractivity sum}. Here we focus just on the contributions to the refractivity sum from the non-cuprate layers and in \fig~\ref{fig:tcmax}~(b) we plot \tcmax versus $\frac{4\pi}{3}\sum_i{n_i \alpha _i}$ for Ln(Ba,Sr)$_2$Cu$_3$O$_7$ where the sum is over Ln, Ba, Sr, and the apical O(1) oxygen. Red squares summarise the effects of changing Ln and blue diamonds the effects of progressively replacing Ba by Sr. These are the internal pressure effects. Polarisabilities are from Shannon \cite{shannon1993}. 

Not only does \tcmax correlate with the $\frac{4\pi}{3}\sum_i{n_i \alpha _i}$	with internal pressure (or ion-size) as the implicit variable, but moreover, this also resolves the paradox of the opposing effects of internal and external pressure. On the one hand, increasing ion size (decreasing internal pressure) increases the polarisability whilst, on the other hand, increasing external pressure enhances the densities $n_i$, in both cases increasing the dielectric enhancement factor. To test this we plot in \fig~\ref{fig:tcmax}(b) $T_c^{\textnormal{max}}$ versus $\frac{4\pi}{3}\sum_i{n_i \alpha _i}$ for YBa$_2$Cu$_3$O$_7$ at 1 bar and 1.7, 4.5, 14.5 and 16.8 GPa (green squares) where we have assumed to first order that only the ion density $n_i$, and not the ionic polarisability, $\alpha _i$, alter under pressure. The correlation with polarisability is now preserved over a range of $T_c^{\textnormal{max}}$ from 70 K to 107 K, including both internal and external pressure. This is a key result and it also now links to the correlation with $V_+$ shown in Fig.~\ref{fig:bvs}.  The additional role of the apical oxygen bond length (which also contributes to the value of $V_+$) has yet to be clarified, but it possibly plays a supplementary role in controlling the large polarisability of the Zhang-Rice singlet \cite{zhang1988}.  Inclusion of the refractivities from the CuO$_2$ layers adds a further 0.4 to the refractivity sum bringing these systems close to the conditions for polarization catastrophe where $\frac{4\pi}{3} \sum_{i} n_i \alpha_i \rightarrow 1$. In practise this can implicate an insulator-to-metal transition or charge ordering, both of which are evident in the cuprates.

Based on this correlation and an inferred polarisability of $\alpha_{\textnormal{Ra}} = 8.3$ $\AA ^3$ for the Radium ion, Ra$^{2+}$, we deduce an implied $T_c^{\textnormal{max}}$ of $109 \pm 2$ K for YRa$_2$Cu$_3$O$_x$ and about $117 \pm  2$ K for LaRa$_2$Cu$_3$O$_x$. On similar grounds $T_c^{\textnormal{max}}$ for HgRa$_2$Ca$_2$Cu$_3$O$_8$ should be about 150 K. These inferences can be tested, though not without their challenges!

	\section{Implications for pairing mechanisms}
\label{sec:pairingmechanismdiscussion}	

We have presented the idea that the opposite effects of internal and external pressure on \tcmax can be understood by considering the polarisability. 
This implies that the polarisability of the material plays an important role in determining $ \tcmaxm $.  But, if there is a causal relationship between the two, how are we to understand it? 

In this section we present two broad ways in which to understand how the polarisability might effect $ \tcmaxm $. 

In the first approach, \refsec~\ref{sec:incoherentpolar}, we begin with the currently most successful microscopic model for superconductivity in the cuprates which is based on the Hubbard model of strongly correlated systems.  We then discuss how, via screening, the polarisability (modified by ion-size and external pressure) might impact relevant parameters in the model. A larger polarisability may enhance \tcmax by screening repulsive longer-range interactions in the \cuo layers.

The second approach invokes the alternative, and relatively unexplored, idea that pairing in the cuprates might be attributable to the exchange of so-called polarisation-waves. In this scenario, discussed in \refsec~\ref{sec:coherentpolar}, polarisation-waves are the boson that mediates Cooper pairing and the overall polarisability will set the energy scale of the pairing boson.  

In the following section, \refsec~\ref{sec:furtherwork}, we propose further studies to measure various parameters discussed here and, more generally, to further investigate ion-size and pressure effects in the cuprates.

		\subsection{Pairing from repulsive interactions}
		\label{sec:incoherentpolar}

It was first shown by Kohn and Luttinger \cite{kohnluttinger} that the superconducting state can be the ground state of a homogeneous electron liquid with purely \emph{repulsive} short-range interactions \cite{ashcroft2008}.  Substantial subsequent investigation has validated the idea of Kohn and Luttinger and extended it to more complex and realistic systems, see for example \cite{atwal2004, raghu2012} and references therein. 
The idea is that particle-hole fluctuations screen (or renormalise) the bare repulsive interaction and can lead to an effective attractive interaction.  The effective attractive interaction is weaker than the bare repulsive interaction, but if the attractive interaction is longer-range while the repulsive interaction is only short-range (or on-site), it can lead to pairing.

This can be called \textit{intrinsic} superconductivity as it does not involve any mediating bosons, e.g. phonons, or even the periodic potential of a lattice - simply the presence of `valence' electrons.

Intrinsic superconductivity has been found as a solution to the Hubbard model with only nearest-neighbour interactions - see \cite{raghu2012optimaltc, gull2012prb, gull2012} and references therein.  The Hubbard model is a model to treat correlations between electrons on a lattice. The Hamiltonian can be written as;

\begin{equation}
H = -t\sum_{\left\langle i,j \right\rangle,\sigma} c^{\dagger}_{i,\sigma} c_{j,\sigma} + U \sum_i n_{i,\uparrow} n_{i,\downarrow}
\label{eq:hubbardhamiltonian}
\end{equation}

\begin{figure}
	\centering
		{\includegraphics[width=0.75\textwidth]{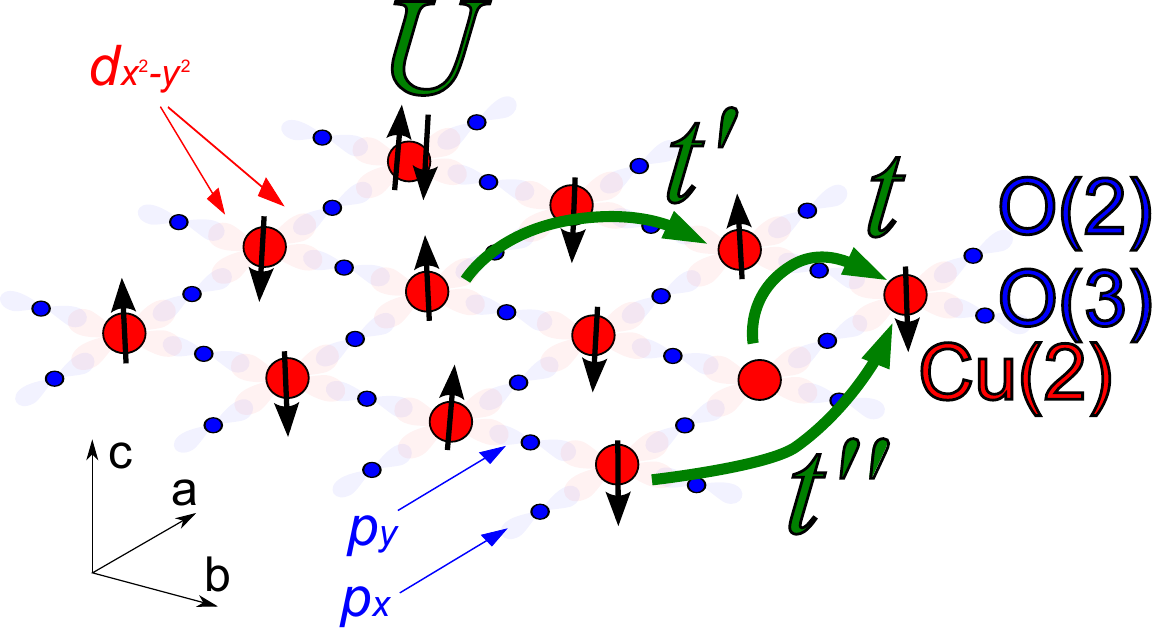}}
		\caption[Illustration of the higher-order hopping integrals relevant to Hubbard model.]{\label{fig:hubbardparams} Illustration of the higher-order hopping integrals relevant to Hubbard model. }
\end{figure}

$ i $ and $ j $ denote adjacent lattice sites and $ \sigma $ denotes the spin of the quasi-particle - spin up, $ \uparrow $, or spin down, $ \downarrow $. $c^{\dagger}$ ($c$) is a quasi-particle creation (annihilation) operator and $n$ a number operator. $ U $ is an on-site Coulomb repulsion term - the energy cost associated with doubly occupying a single lattice site. $ t $ is a hopping parameter and represents the energy associated with moving a quasi-particle, without a flipping its spin, to its nearest-neighbour site.  The Hamiltonian in \eq~\ref{eq:hubbardhamiltonian} can be extended to include hopping between next-nearest-neighbour sites and even longer-range interactions.  These hopping parameters are represented by $t'$, $t''$, etc. and are illustrated in \fig~\ref{fig:hubbardparams}. Note that here the nearest-neighbour superexchange energy is $J = 4t^2/U$ and an expression for the next-nearest-neighbour superexchange energy is $J'=4t'^2/U$.

Gull \etal \cite{gull2012} calculate exact numerical solutions to the Hubbard model on finite-size clusters of up to $ 16 \times 16 $ sites (which is about the practical size limit for exact numerical solutions).  They control the ratio $ U/t $ (the interaction strength) and the doped hole concentration, choosing parameter values relevant to the cuprates. The results of their calculations qualitatively capture most of the experimentally observed behaviour in the cuprates: with increasing doping there are transitions from an antiferromagnetic Mott insulator phase to a pseudogap phase, to a co-existing (and competing) superconducting and pseudogap phase, to a pure superconducting phase and finally to a Fermi-liquid-like phase\footnote{Refer to \fig~\ref{fig:phasediagram} for a schematic phase diagram of the cuprates.}.  Additionally, they find $ \tcmaxm \approx 100 $ K and an increasing $ \scgapm/\tcm $ ratio with underdoping.  

Thus the essential aspects of the phase behaviour of the cuprates are captured by the proper treatment of electron-electron interactions within the Hubbard model with nearest-neighbour interactions only ($ t'=0 $).  However, this leaves unanswered the question of why there is such large variability in superconducting properties between families of cuprates, and indeed within a single family of cuprates via ion-size and pressure effects. This is not too surprising given the many simplifying assumptions of a Hubbard model of the cuprates. If this significant materials variability is to be captured by the Hubbard model, it is clearly necessary to at least include longer-range interactions\footnote{Which are also called \textit{extended interactions}.} via $ t' $ and $ t'' $ or the related $ J'$ and $J''$.   

The idea that materials variability of \tcmax results from $ t' $ is not new.  For example, Pavarini \etal \cite{pavarini2001} correlated \tcmax with a composite parameter proportional to $ t'/t $ for a variety of cuprates.  In their work, the apical oxygen bond length was shown to significantly affect $ t'/t $.  $ t' $, $ t'' $, etc. alter the electronic dispersion (see e.g. \cite{damascelli2003, piazza2012}) and thus reshape the Fermi-surface.  The work presented in this thesis indirectly shows the importance of these extended interactions in providing an understanding of the systematic variation in $ \tcmaxm $ with ion-size. But what is the role of ion-size (and polarisability) and can it be understood within this Hubbard-model framework? 

Perhaps the role of $ t' $, $ t'' $, etc. is merely to modify the DOS by shifting the vHs - as discussed earlier and in Chapters~\ref{ch:bi2201} and \ref{ch:dft}. Alternatively, these parameters relating to longer-range interactions may be related to \tcmax in a more fundamental way and this will be subject of our discussion in the following section.

\subsection{Incoherent polarisation and longer-range interactions}		

A long-standing perceived difficulty with the Kohn and Luttinger picture has been the length scale of the repulsive interactions. For example, recently Raghu \etal \cite{raghu2012} show qualitatively that longer-(real-space)range interactions are always detrimental to superconducting pairing resulting from short-range (on-site) interactions. 

When we consider core electrons, or in the case of the cuprates, the additional ions outside of the \cuo layers, we introduce a complementary channel that adds additional screening and quite possibly a new pairing channel that involves excitations of this polarisable media as discussed below in \refsec~\ref{sec:coherentpolar}.  

Firstly we discuss implications of additional screening. The isolated CuO$_2$ planar array is electrostatically uncompensated and as such it cannot constitute a thermodynamic system. It is necessary to include the compensating charges lying outside of the CuO$_2$ plane in any thermodynamic treatment and these will also contribute to electron-electron interactions within the plane. These charges also reside on ions that are notably polarisable\footnote{In general, ions with larger atomic numbers, and hence a greater number `core' electrons which occupy a larger volume, are more polarisable \cite{shannon1993}.}, and they contribute to the high background dielectric constants observed in the cuprates \cite{reagor1989}. 


The highly-polarisable non-\cuo layers \textit{screen longer-range interactions} in the \cuo layer.  Recently, the effect of the charge reservoir layers on \tcmax via screening has been explored Raghu \etal \cite{raghu2012optimaltc}. Thus we can possibly understand the correlation of \tcmax with the refractivity sum shown in \fig~\ref{fig:tcmax} as resulting from more effective screening of longer-range interactions in the \cuo layer \cite{raghu2012optimaltc} that are detrimental to intrinsic superconductivity \cite{raghu2012}.  This is an attractive explanation for our data, especially as it connects with a large body of theoretical work on pairing interactions specific to the cuprates.


We do not yet have any measurements of how $U$, $t$, $t'$ and $t''$ vary with ion-size (or indeed with external pressure) and hence how the interaction strength $U/t$ varies. The role of future studies should focus on measuring and understanding the systematic variation of these parameters.  In \refsec~\ref{sec:furtherwork} below we propose studies to measure these important parameters. 
 
 	\subsubsection{Coherent polarisation excitations}
 	\label{sec:coherentpolar}

Building on the intrinsic pairing mechanism discussed above, it may be necessary to consider the more realistic \emph{in}homogeneous electron liquid where there is now a distinction between localised `core' electrons and the delocalised `valence' electrons.  As such, Atwal and Ashcroft include the effect of the lattice not by way of collective excitations of the ions (phonons) but instead by way of the coherent \emph{dynamics} of the `core' electrons associated with each ion \cite{atwal2004}.  This alternative, and relatively unexplored, idea is that pairing in the cuprates might be attributable to the exchange of bosons comprising coherent, quantised excitations of polarisation associated with the core electrons \cite{atwal2004, ashcroft1987} and/or the charge-reservoir-layer ions - so-called polarisation waves.  

This idea is analogous to the BCS-like phonon-mediated pairing of conventional superconductors, but with polarisation waves in place of phonons.  The energy scale of the polarisation waves 
is governed by the effective plasma frequency of the total ensemble of core electrons in and out of the \cuo layers and is of order $ 1.7 $ eV. This is two almost orders of magnitude larger than the energy scale for phonons, $ \approx 30 $ meV \cite{kresin2009}. 
What Atwal and Ashcroft find is that the exchange of polarisation waves, associated with the core electrons, by valence electrons can give rise to Cooper pairing and high transition temperatures, especially in the $d$-wave channel.  For example, they make a comparison of \tc values for homogeneous (valence only) electron system and inhomogeneous (valence and core) electron system calculated with BCS-Eliashberg theory and find \tc is up to 50 times larger in the latter case \cite{atwal2004}.  


While there is considerable scope to develop this idea, it is expected that the polarisability will set the energy scale of the pairing boson.  

Note the distinction between \tc enhancement from screening by incoherent polarisation from our first discussion, with the enhancement of \tc by coherent (quantised) polarisation waves in the second.  In both cases however the polarisability of the material is a key property.  The challenge remains to distinguish these.



%

\section{Outlook}

The cuprates are complicated but \emph{do} show some clear systematics. This thesis has shown there is good reason to believe that dielectric properties lie behind the systematic variations in $\tcmaxm$ associated with pressure and ion-size effects.  This is not to say that the polarisability is the only salient parameter, indeed the DOS is likely to set an additionally important energy scale, but it should prompt two responses;
\begin{itemize}
	\item To consolidate, or disprove, this interpretation with further experimental and theoretical studies. We briefly discuss some ideas in this regard below.
	\item To use this understanding in the search for quite novel superconducting materials. Perhaps superconductivity is to be found in a metallic medium with large \dos coupled with a highly polarisable medium.  The role of quasi-two-dimensionality may simply be to enhance the DOS through the introduction of a vHs. 
\end{itemize}

This thesis has also shown that by studying ion-size and pressure effects we can gain additional insights into the physics of the cuprates and unconventional superconductivity.

\section{Further studies}
\label{sec:furtherwork}

At the end of each chapter we discussed possible follow-on work.  We will not repeat that here, but instead discuss some additional studies we believe would be instructive and/or important.

\subsection{Quantities to measure}

Earlier in this chapter we discussed the recent work of Gull \etal \cite{gull2012} which demonstrates the ability of the Hubbard model in capturing essential aspects of the phase behaviour.  Here $ U/t $, the interaction strength, is the key variable, yet we do not know the systematic ion-size effect on $ U $ or $ t $.  From the work of Sawatzky \etal \cite{sawatzky2009} we expect enhanced screening from the more highly-polarisable larger-ions to lower $ U $, but this has not been measured. We see this as a priority and plan to undertake UV absorption measurements to determine the charge-transfer gap ($ \sim 2 $ eV) and $ U $ ($ \sim 6 $ eV) as a function of ion-size.  Since $ J = {4t^2}/{U} $ in the one-band Hubbard model, we can use our two-magnon scattering measurements of $J$ together with the UV absorption data to estimate $t$ as a function of ion-size as well. 

Furthermore, that discussion also made apparent the need to better understand the ion-size (and external pressure) effect on longer-range interactions as characterised by $ t', t'' $ or $ J', J'' $.  One possibility we are currently exploring is whether Raman \btg spectra\footnote{This idea is motivated by the separate region of the Brillouin Zone, the nodal region, probed by the \btg geometry. } contain information about these longer-range interactions.  See \refsec~\ref{sec:a1g} for a discussion of this idea, and in this regard we note two recent papers claiming that the two-magnon Raman scattering response is adequately described by a proper treatment of the Hubbard model, including longer-range interactions \cite{lin2012, piazza2012}.  However, it is likely a more sophisticated technique, such as RIXS (see \refsec~\ref{sec:rixs}) or laser-ARPES, will be needed to fully characterise these longer-range interactions.

Additional studies could focus on using more sensitive techniques and methods to measure \hscgap and \epg as a function of ion-size.  In this regard we discuss specific heat (\refsec~\ref{sec:specificheat}), ellipsometry (\refsec~\ref{sec:ellipsometry}) and impurity scattering (\refsec~\ref{sec:possibleexperiment}) studies below.  Finally the DOS could be investigated (in additional to a computational DFT approach) using specific heat (\refsec~\ref{sec:specificheat}), NMR and/or impurity scattering studies (\refsec~\ref{sec:possibleexperiment}).

	\subsection{Specific Heat}
	\label{sec:specificheat}
Collaborators at Cambridge University have a unique differential heat capacity rig.  The full technical description of their differential method can be found in reference \cite{loram2000}.  With this method the phonon contribution can be very accurately subtracted so that the electronic specific heat can be measured to within $0.1$ mJ.mol$^{-1}$.K$^{-2}$.  Their rig is capable of sample temperatures between $1.8$ K and 380 K in magnetic fields up to 13 T. 

Specific heat measurements would enable determination of the ion-size effect on: (i) the specific heat coefficient, $\gamma$, \added{and from this the \dos and any enhancement of $\gamma$ from the so called `bare band structure' value, $ \gamma_b $, that results from electron-boson coupling,} \deleted{(ii) the enhancement factor, $(1+ \Lambda)$, from electron-boson coupling} (ii) the condensation energy $U_0$, (iii) \epg \cite{loram2001} and (iv) the mean-field transition temperature \tcmf \cite{tallonfluctuations} \replaced{is believed to}{should} scale with $\scgapm$.  

	\subsection{RIXS}
	\label{sec:rixs}
Resonant, Inelastic, X-ray Scattering (RIXS) measurements \cite{guarise2010, letacon2011} allow one to explore the magnetic excitation spectrum across a wide area of the Brillouin zone.  The aim would be to probe with RIXS the full dispersion of paramagnon excitations (thus elucidating longer-range magnetic interactions $J'$, $ J'' $, etc.) to see if it is compatible with the observed ion-size effects on $T_c$.  Working from the full dispersion, \tc may be calculated under certain assumptions as shown by le Tacon \etal \cite{letacon2011}.  A key question is whether the variation in dispersion with ion size follows the observed \tc variation.  High quality (though not necessarily large, $1\times1$ mm$^2$ is sufficient) single crystals with clean surfaces, or high-quality epitaxial thin films are needed for these measurements.  In addition, there is the possibility of detecting polarisation waves with RIXS at the barium edge. Such a measurement of polarisation waves would be pioneering and would necessarily draw on a collaboration with RIXS experts.

	\subsection{Ellipsometry and Reflectometry}
	\label{sec:ellipsometry}

Infrared optical spectroscopy has previously provided crucial information about the charge and lattice excitations of the cuprate HTSC \cite{timusk1999, basov2005}. It has the advantage of a large probe depth of the light which ensures the bulk nature of the observed phenomena.  Significantly, the energy resolution and accuracy in the determination of the optical constants of infrared spectroscopy in combination with powerful sum rules enables one to identify the specific interactions of the charge carriers and their underlying energy scales. In particular, the ellipsometry technique, which is self-normalizing and does not require a Kramers-Kronig analysis procedure, allows one to determine with high accuracy even small changes of the dielectric function, $\epsilon$, for example versus temperature or doping \cite{azzam1978, bernhard2004thinfilms} or as we propose to do, ion-size.

	\subsubsection{$c$-axis conductivity measurements on Ln123}
The Bernhard group has previously established that measurements of the far-infrared (FIR) $c$-axis response allow one to identify the superconducting energy gap and pseudogap separately \cite{pimenov2005, yu2008, dubroka2010, dubroka2011}. They have shown that these two ordering phenomena have distinct spectral features. The superconducting energy gap yields a spectral weight shift toward a delta function at zero frequency that describes the response of the superconducting (SC) condensate. Its energy scale, $\scgapm$, can be identified from the onset of the suppression of the optical conductivity. In contrast the pseudogap (PG) gives rise to a spectral weight shift to higher energy into a very broad peak above the gap edge at $\epgm$. 

The $c$-axis conductivity is also a direct measure of the electronic coupling between the metallic and superconducting \cuo layers. In the normal state the anisotropy can be easily determined from the absolute value of the electronic part of the $c$-axis conductivity (which is usually only weakly frequency dependent).  In the superconducting state it can be deduced from the position of the so-called Josephson plasma edge (and for bi-layer materials such as our model system, of the transverse Josephson-plasma mode) which is proportional to the $c$-axis component of the superfluid density and the $c$-axis magnetic penetration depth. The importance of the apical oxygen, which couples the charge reservoir layer (CRL) and \cuo layer, has long been known \cite{pavarini2001} and may yet play an important role in controlling the huge polarisability of the Zhang-Rice singlet \cite{zhang1988}. Knowledge of systematics of the electronic coupling between the CRL and \cuo layers in our model system may prove an important piece in the puzzle.

Thus, it is would be desirable to undertake systematic measurements of the $c$-axis conductivity in our model LnA$_2$Cu$_3$O$_y$ system to reveal the effect of ion-size substitution on the important parameters of \hscgap and \epg and $c$-axis coupling. 

	\subsubsection{In-plane conductivity on crystals and thin films}
The in-plane conductivity represents the coherent dynamics of the charge carriers that are confined to the \cuo layers. From the normal-state data one can directly deduce the frequency-dependent scattering rate and mass enhancement due to the interaction of the charge carriers.  Also one can get the screened and, with some additional information, the unscreened plasma frequency of the charge carriers. These parameters are important measures of the interaction of the charge carriers, which may be strongly affected by ion-size substitution via the varying polarisability of the ions; e.g. screening due to large Ln-ions or reduced screening because of Sr substitution for Ba.  As discussed above, it has been suggested screening plays an important role in raising \tc and improving SC properties by suppressing longer-range, repulsive interactions \cite{raghu2012}. To our knowledge, no systematic measurements in this direction have been done. 

In the superconducting state, one can directly obtain the superconducting condensate density and its $a$-$b$ anisotropy (on de-twinned single crystals) from the in-plane data. These measurements would complement the further \musr studies discussed at the end of Chapter~\ref{ch:musr}.


\subsection{\ysco}

Sample quality was an issue in our \musr studies on YSr$_2$Cu$_3$O$_y$ (YSCO). There were magnetic impurities from K-Cl rich inclusion phases derived from the KClO$_3$ oxygen source used in the synthesis process. The magnetism from these phases masked the component arising from the vortex lattice in the superconducting YSCO phase in $ \mu $SR.  For our future \musr studies on YSCO it is clearly desirable to significantly improve the sample quality.  


In fact, it would be highly desirable to have well-oriented (preferably with the $c$-axis aligned perpendicular to the substrate) thin films of pure YSCO. To this end our MOD synthesis of YSCO on a (001) SrLaAlO$_4$ substrate at atmospheric pressures is an encouraging start, see \refsec~\ref{sec:yscothinfilmsynth}. YSCO is a member in the Ln(Ba,Sr)$_2$Cu$_3$O$_y$ system under a large `internal-pressure' (or rather, it has one of the smallest ion-sizes, LuSr$_2$Cu$_3$O$_y$ being the extreme).  Such extreme cases are important for determining systematics. 

Such samples would facilitate two important measurements of the YSCO material (i) Two-magnon Raman scattering, not only in the un-doped YSCO6 sample, but also doping dependent studies, and (ii) RIXS, a variant Raman-like technique involving synchrotron radiation that allows $\kk$-space selection and thus has the ability to map out the dispersion of magnetic excitations over almost the entire Brillouin zone \cite{letacon2011}.  Such thin films would also be amenable for precise resistivity measurements to investigate superconducting fluctuations above $\tcm$, from which the $c$-axis coherence length can be inferred.  High quality resistivity measurements can also provide an estimate of the pseudogap, $T^*=\epgm/k_B$, where $T^*$ is the temperature below which the pseudogap opens marked by a deviation from linearity in the resistivity. 

As such it is desirable and most likely feasible to grow well-oriented YSCO thin-films for on-going ion-size studies.  Clearly internal and external pressure effects on both fluctuations and the pseudogap should prove highly instructive.  Whilst the MOD synthesis processes might be tuned to grow high-quality, aligned films of YSCO, we believe it would be worth trying pulsed laser deposition (PLD) as well. PLD is a technique where one can have very fine control over the growth conditions.  However it requires a sophisticated set up and an experienced technician.  The attraction of PLD is the fine control of growth conditions and that, in principle, PLD can produce nearly \emph{atomically perfect} thin films \cite{malik2012}.  An alternative to PLD we may wish to try is Pulsed Electron Deposition for which we would use the expertise and equipment of our collaborator Dr. Gilioli at IMEM. 

\subsection{\epg and $J$ ion-size study}
\label{sec:possibleexperiment}
The previous section mentioned three important follow-up questions; (measurements of) the effect of ion-size on fluctuations, the pseudogap and on the doping dependence of $J$ (which we believe to be related). In principle, two-magnon scattering (and hence $J$) and the pseudogap can both be measured by Raman spectroscopy, along with \scgap and the density of states as discussed in Chapter \ref{ch:ramangaps}.  However, as we found, these are delicate measurements of weak effects. Below we describe a simpler study to determine \epg and $J$.   The study we describe requires only readily-available sample preparation and measurement apparatus.

The study involves both good quality polycrystalline (PC) and single crystals or well-oriented thin-films (SC/F) of the \emph{same material}, for a range of ion-size species, e.g. Nd123 or Sm123, Yb123, YBaSrCu$_3$O$_y$ and YSCO would represent ideal candidates that span a wide range of internal pressures (together with literature values from the Y123 material).  The polycrystalline samples, Ln(Ba,Sr)$_2$Cu$_{3-z}$Zn$_z$O$_y$, would be prepared for several concentrations of Zn impurities; $z$ = \{0.0, 0.04, 0.08, 0.12\} would be sufficient, although more values of $z$ will lead to statistically more accurate results.  The doping level of these samples will be controlled by the oxygen concentration, $y$, by way of \emph{co-annealing} all the PC samples with the SC/F samples.  Together with \tc from magnetisation or resistivity measurements, reliable measurements of \rttep can be readily made on the PC samples which will reflect the doping state of both the PC and SC/F samples.  In this way $p$-dependent studies can be effected.

A variety of measurements can now be made, but it would be of interest to measure both $J$ and the pseudogap energy \epg as a function of $p$ and ion-size.  With this set of samples $J$ can be determined from two-magnon Raman scattering using micro-Raman. Meanwhile $\epgm=k_bT^*$ can, in principle, be determined from resistivity measurements, see e.g. \cite{kim2010, naqib2005doping}. Normally if $T^* \leq T_c$ then it is not possible to determine \tstar from resistivity measurements.  However, Zn is known not to effect the doping state or pseudogap in the quantities we propose \cite{tallon1995, naqib2005, naqib2005doping, kim2010} but will suppress the \tc from its $z=0$ value, $T_{c0}$.  Thus, it would be possible to determine $T^*$ down to reasonably low temperatures (probably $\sim 25$ K) or conversely, \epg to doping levels quite close to $p=0.19$.

One possibility is that \epg scales with $J$ \cite{loram2001} and that $J(p)\sim J(p=0)\left[1-\sfrac{p}{0.19}\right]$, irrespective of ion-size, for the Ln123 system. But this remains to be tested!

Furthermore, such a study will provide quantitative estimates of the normal-state density of states (DOS) at the Fermi-energy, \dos, as a function of both ion-size and $p$ by use of the previously-validated \cite{tallon1997} Abrikosov-Gorkov model\footnote{STM studies showed scattering from Zn in the \cuo layers is close to the unitary limit \cite{pan2000}}. The linearized Abrikosov-Gorkov equation for small scattering rates, $\Gamma$, is \cite{abrikosovgorkov, sunmaki}; 
\begin{equation}
\frac{T_c}{T_{c0}}=1-0.69\frac{\Gamma}{\Gamma_c}
\label{eq:linearag}
\end{equation}
\noindent The scattering rate, $\Gamma$, is inversely proportional to the DOS as $\Gamma = {n}/{\left[\pi\dosm \right]}$ where $n=z_{ab}/V$ is the density of impurity scatterers in the \cuo layers, $V$ is the unit cell volume and $z_{ab}$ the concentration of Zn in a \cuo layer.  $\Gamma_c$ is the critical scattering rate required to fully suppress superconductivity and can be expressed as $\Gamma_c = 0.412\hscgapm$.  As such the suppression of \tc due to Zn substitution is inversely proportional to the DOS. 

The temperature coefficient of the normal-state heat capacity, $\gamma$, is related to \dos as $\gamma = \gamma_b(1+\Lambda) = (1+\Lambda) \sfrac{2}{3}\pi^2k_B^2\dosm$. Here $\gamma$ is enhanced by a factor $(1+\Lambda)$ from the so called `bare band structure' value, $\gamma_b$, by (screened) electron-electron interactions.  These interactions can be mediated by spin-fluctuations or phonons or more generally any other relevant bosonic interaction. For conventional superconductors, $\Lambda$ is a function of the same electron-phonon interaction that pairs electrons, see \cite{ashcroftmermin} or the Chapter 13 by Gladstone, Jensen and Schrieffer in \cite{parks} .  $(1+\Lambda) = {m^*}/{m} = 1 + \dosm V_{\textnormal{ph}} + \dosm V_{\textnormal{SF}}$ where $m^*$ is the effective mass of the electrons and $V_{\textnormal{ph}}$, $V_{\textnormal{SF}}$ are matrix elements of electron-electron interaction mediated by phonons and spin-fluctuations respectively.

\begin{figure}
	\centering
		\includegraphics[width=0.66\textwidth]{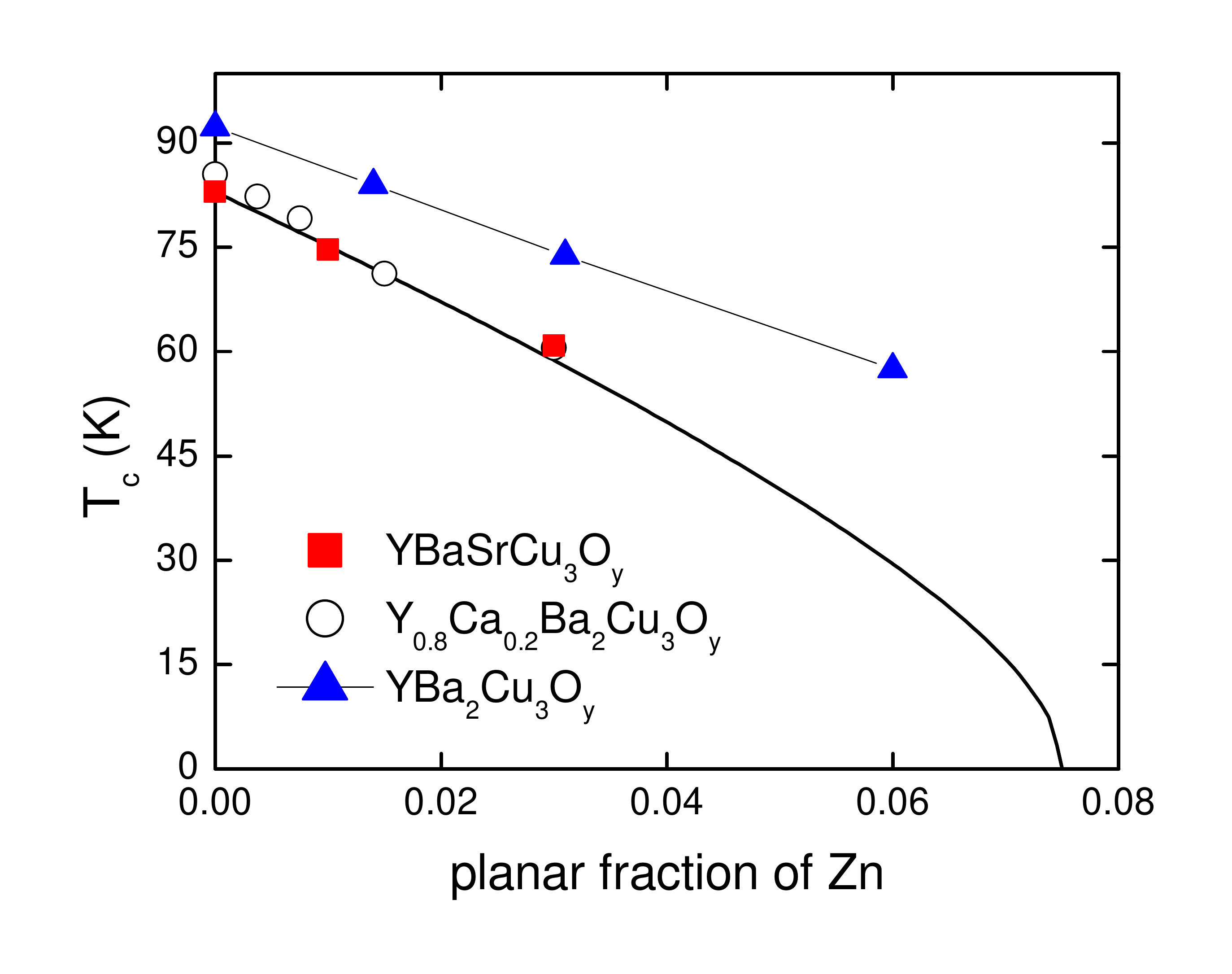}
	\caption[The suppression of \tc in Zn-doped \ybasr at optimal doping.]{The \tc suppression of optimally doped \ybasr (red squares), Ca-doped \ybco (open circles) and \ybco (blue triangles) due to Zn substitution in the \cuo layer.}
	\label{fig:tcsupp}
\end{figure}

The suppression of \tc with impurity content for optimally-doped YBaSrCu$_{3-z}$Zn$_z$O$_y$ is plotted in \fig~\ref{fig:tcsupp} along with similar data for Y$_{0.8}$Ca$_{0.2}$Ba$_2$Cu$_3$O$_y$ \cite{bernhard1996}.  Because these two data sets fit reasonably well to the same AG parameters\footnote{It appears possible that the \tc of the Ca-doped Y123 is suppressed more rapidly by Zn substitution than YBaSrCu$_{3-z}$Zn$_z$O$_y$.  A more thorough study is needed to explore this possibility.}, these initial measurements imply the $(1+\Lambda)$ enhancement in $\gamma$ is the same for these two systems with comparable $\tcm$. 

To this plot we add data for pure \ybco from \cite{semba1994}. $T_{c0}=93.5$ K $=T_{c0}^{\textnormal{max}}$ is slightly higher in the pure compound, however the slope, $\dd T_c/\dd z$, is reduced implying that the total specific heat $\gamma = \gamma_b (1+ \Lambda) $, is larger. This may be intrinsic to optimally doped \ybco and indicate an enhancement of $\Lambda$ because of its more highly polarisable matrix. This is indicative of how a comprehensive study such as this might proceed.  It would involve similar measurements for a number of different doping states (under- and over-doped) for different ion sizes in order to separate the ion-size effect from the doping effect on the DOS. But this is much easier than specific heat measurements and would be a powerful prelude to the latter if they were embarked upon.

Because the DOS changes dramatically around optimal doping whilst \tc does not, it is also possible the different slopes in \fig~\ref{fig:tcsupp} are a consequence of not comparing like doping states. If the authors had also reported thermopower values of their \ybco we could say more. In any case, with such small changes in the slope it is important to see if this trend continues for  the two end members Nd123 and YSCO, where the effects will be greatest.  From such studies we ought to be able to extract $(1+ \Lambda)$ and explore how $m^*$ varies with ion size and the refractivity sum.  In principle this will be measurable with well homogenised samples and with closely spaced $z$ values for good statistics.

\section{Closing remarks}
\label{closingremarks}

The study of ion-size effects on the cuprates presented in this thesis was motivated by the paradoxical opposite effect of external pressure and internal pressure, as induced by isovalent ion substitution, on $ \tcmaxm $. 

Motivated by this observation and wishing to further test its applicability, we synthesised a new variants of Bi2201 that were under positive and negative internal pressure.  We argued that in both Bi2201 and Ln(Ba,Sr)$ _{2} $Cu$ _{3} $O$ _{7-\delta} $ disorder introduced by the ion-substitutions studied here cannot solely explain the variation in \tc observed.  

We also explored several ideas relating to ion-size effects on;
\begin{itemize}
\item The density of states. Primarily this idea was explored via DFT calculations although in \refsec~\ref{sec:furtherwork} we suggest possible experimental routes to study ion-size effects on th DOS.
\item The anti-ferromagnetic superexchange energy, $ J $.  From these Raman-based studies we argued that $ J $ cannot explain the opposite effects of internal and external pressure on $ \tcmaxm $.
\item The superconducting gap and pseudogap.  Here, our Raman data were not precise enough to resolve any systematic ion-size effect on these two energies.
\end{itemize} 

In this final Chapter we described how by considering the polarisability one can naturally understand the seemingly paradoxical opposite effects of external and internal pressure on $ \tcm $.  Thus, this thesis highlights the relatively unexplored idea that the polarisability potentially plays a central role in superconductivity in the cuprates. 

These results do not necessarily conflict with the conventional understanding of Cooper pairing in the cuprates based on spin-fluctuations (although this picture should find difficulty with the Raman results presented in Chapter~\ref{ch:twomag}). For example as discussed above the primary role of the polarisability could be via the screening of longer-range Coulomb interactions which are detrimental to Cooper pairing.  

Alternatively, these results could be initial experimental evidence for a novel pairing mechanism based on the exchange of quantised, coherent waves of polarisation. Here the energy scale for such excitations is\footnote{Considering all valence electrons, this may in fact be more like $ \sim 10 $~eV!} $ \sim $~eV which is an order of magnitude larger than $ J $ and two larger than the Debye energy.  These two scenarios can be tested and this statement leads us an important aspect of this work: the ion-size variation approach we have utilised here, especially for Ln123, represents an under-utilised, systematic method for superconducting and pseudogap phenomenon to be explored theoretically and \textit{experimentally}. For example, we discussed the recent paper of Gull \etal that showed how a proper treatment of the Hubbard model could reproduce several key experimental observations in the cuprates. Now it is likely the Hubbard parameters $ U $, $ t $, $ t' $ etc. systematically vary in the Ln123 and this would provide for a presumably insightful comparison between theory and experiment. 

It appears that there remains many interesting, and most likely fruitful, avenues for further systematic ion-size studies such as those presented here in this thesis.

\addcontentsline{toc}{chapter}{Bibliography} 
\bibliographystyle{ieeetr} 
\bibliography{literature} 

\end{document}